\documentclass[12pt]{report}
\usepackage{amsmath,amssymb}
\usepackage{bm}
\usepackage{color}
\usepackage{graphicx}
\usepackage{cite}
\usepackage{mathrsfs}
\usepackage{here}
\usepackage{braket}

\usepackage{slashed}
\usepackage{comment}

\usepackage{amsthm}
\newtheoremstyle{break}
{\topsep}{\topsep}%
{\normalfont}{0pt}%
{\bfseries}{}%
{\newline}{}
\theoremstyle{break}

\newtheorem*{thmhom}{Fundamental homomorphism theorem}
\newtheorem*{thmcor}{Correspondence theorem}
\newtheorem{thmiso}{Isomorphism theorem}
\newtheorem*{thmAbel}{Fundamental structure theorem of finite Abelian group}
\newtheorem*{thmcirc}{The Chinese remainder theorem}
\newtheorem*{pf}{Proof}

\newtheorem{step}{Step}

\begin{document}

\title{
{\Large
Ph.D thesis\\*[50pt]
\bf
Flavor structure in magnetized orbifold and blow-up manifold compactification\\*[20pt]}
}

\author{
Hikaru Uchida\\*[20pt]
\centerline{
\begin{minipage}{\linewidth}
\begin{center}
{\it \normalsize
Department of Physics, Hokkaido University, Sapporo 060-0810, Japan} \\*[5pt]
\end{center}
\end{minipage}}
\\*[50pt]
}

\date{\centerline{\small Submitted to Department of Physics, Hokkaido University: January 2023}}

\begin{titlepage}
\maketitle
\thispagestyle{empty}
\end{titlepage}
\newpage


\section*{Acknowledgement}
I would like to express my greatest appreciation to my supervisor, Tatsuo Kobayashi, for a great research environment, 
valuable discussions, instructive suggestions, and helpful encouragement to my study.
I would also like to thank Hajime Otsuka, Morimitsu Tanimoto, and Osamu Seto for not only valuable discussions but also helpful supports and advices.
I am quite grateful to a lot of my collaborators, Takuya H. Tatsuishi, Shintaro Takada, Shio Tamba, Yuki Kariyazono, Shota Kikuchi, Kaito Nasu, Yuya Ogawa, Kouki Hoshiya, Shohei Takada, Shohei Uemura, Kenta Takagi, Yusuke Shimizu, Maki Takeuchi, Makoto Sakamoto, Tatsuta Yoshiyuki, Nobuchika Okada, and Kei Yamamoto, for useful discussions and comments.
I also thank to all members of theoretical particle and cosmological physics group at Hokkaido University and all members involved with me for supporting me in the Ph.D. course.
I have been financially supported by Grant-in-Aid for JSPS Research Fellows No. JP20J20388.

\thispagestyle{empty}
\newpage


\section*{Abstract}

As the effective field theory of the superstring theory, ten-dimensional ${\cal N}=1$ supersymmetric Yang-Mills theory is induced.
We consider the ten-dimensional space-time ${\cal M}_{10}$ as direct products of our four-dimensional space-time ${\cal M}_{4}$ and three two-dimensional compact spaces $X_i\ (i=1,2,3)$,~i.e. ${\cal M}_{10} = {\cal M}_4 \times X_2 \times X_2 \times X_3$.
In particular, we consider $X_i$ as a torus, the torus orbifold, and also the blow-up manifold of the torus orbifold with background magnetic fluxes.
We discuss the modular symmetry in the magnetized torus and its orbifold compactifications.
We find that the modular flavor groups and the representations as well as modular weights of the chiral fields such as quarks and leptons obtained from the magnetized torus as well as the orbifold compactifications can be uniquely determined with the magnetic fluxes.
We also discuss the magnetized blow-up manifold compactification in which orbifold singularities are replaced by parts of sphere as a smooth manifold.
We find that the chiral zero mode numbers on the blow-up manifold as well as the orbifold can be determined only by the magnetic fluxes including localized fluxes due to the Atiyah-Singer index theorem, and then the degree of freedom of localized fluxes gives new additional chiral zero modes.
We also find that the new chiral zero modes correspond to localized modes around the orbifold singularities.

\thispagestyle{empty}
\newpage


\tableofcontents

\newpage


\chapter{Introduction}
\label{chap:intro}

In the particle physics, the standard model (SM) is the successful theory to explain a lot of observations related to the strong and electroweak (EW) interactions by $G_{SM} = SU(3)_C \times SU(2)_L \times U(1)_Y$ gauge symmetry for their gauge bosons and the local symmetry for matter fermions as well as Higgs fields in four-dimensional (4D) space-time.
The SM particle contents are listed in Table~\ref{Tab:SMcontents}.
Among them, there are three types of renormalizable couplings: gauge couplings among gauge bosons and matter fermions as well as Higgs fields, Yukawa couplings among matter fermions and Higgs fields, and Higgs self-couplings.
Note that weak boson masses and matter fermion masses are obtained by the EW spontaneously symmetry breaking (SSB) of Higgs fields through the gauge couplings and Yukawa couplings with the Higgs fields, respectively.
On the other hand, before the EW symmetry breaking, gauge bosons and chiral matter fermions cannot have their masses because of the gauge symmetries and the chiral symmetries, respectively.
Although neutrinos have much tinier masses than the other quarks and charged leptons, they can be explained by considering the Weinberg operators among lepton doublets and Higgs doublets with mass dimension $5$ if neutrinos are Majorana neutrinos.
Then, by appropriately fitting their SM parameters: gauge coupling constants, Yukawa coupling constants (as well as Weinberg operator coefficients), and Higgs self-coupling constants, the SM predictions are consistent with a lot of observations.

However, there remain a lot of issues to be solved: quantum gravitational phenomena cannot be explained in the SM, there are still some discrepancies between the SM predictions and the experimental results, there are some mysteries such as the origins of the values of the SM parameters to explain observations, and so on.
In particular, to clear the origins of the SM parameters may give us hints to discover more  fundamental theories.
For example, the three gauge coupling constants may be unified around $O(10^{16})$GeV, (see Ref.~\cite{Buttazzo:2013uya} for the current renormalization group (RG) flows of the SM parameters,) which implies that there is a more fundamental theory to unify the three gauge symmetries into one larger gauge symmetry including $G_{SM}$ such as $SU(5)$ and $SO(10)$ with one gauge coupling constant above the scale and the large gauge symmetry is spontaneously broken at the scale like the EW symmetry breaking by the SM Higgs fields.
Such a fundamental theory is called a grand unified theory (GUT)~\cite{Georgi:1974sy,Fritzsch:1974nn,Georgi:1974my}.
In such a GUT, quarks and leptons can be also unified into several multiplets.
For example, $i$ th generational quarks and leptons including right-handed neutrinos are minimally unified into one multiplet ${\mathbf{16}_i}$ in $SO(10)$ GUT.
In other words, a GUT can answer the origin of three gauge couplings as well as quarks and leptons.
However, three-generational quarks and leptons have quite different generational structure (flavor structure); both three-generational up-type and down-type quarks have quite hierarchical masses among the generations and their flavor mixing through the weak interaction, written by Cabbibo-Kobayashi-Masukawa (CKM) matrix, is small, while charged leptons also have hierarchical masses but neutrinos have much tinier masses than quarks, and charged leptons and their flavor mixing, written by Pontecorvo-Maki-Nakagawa-Skata (PMNS) matrix, is large.
Note that their flavor structure can be determined by the structure of Yukawa coupling constants as well as Weinberg operator coefficients and there are about 20 parameters related to quark and lepton masses, flavor mixing angles, and $CP$ phases.
Then, it is significant to clarify the origin of the flavor structure, and it is useful to consider a flavor symmetry among the generations by which some Yukawa coupling coefficients as well as the Weinberg operator coefficients are related to each other.
In particular, non-Abelian discrete flavor symmetries such as $S_n$, $A_n$, $\Delta(3n^2)$, and $\Delta(6n^2)$ have been well studied~\cite{Altarelli:2010gt,Ishimori:2010au,Ishimori:2012zz,Hernandez:2012ra,King:2013eh,King:2014nza,Tanimoto:2015nfa,King:2017guk,Petcov:2017ggy}.
We note that the flavor symmetry must exist at a certain scale and the flavor structure can be determined.
To break the flavor symmetry, we usually introduce additional SM gauge singlet scalar fields, so-called flavons.
Of course, it is also important how the Weinberg operators are generated at the cut-off scale.
One of the famous scenarios is the type-I see-saw mechanism~\cite{Minkowski:1977sc,Gell-Mann:1979vob,Yanagida:1979as,Glashow:1979nm,Mohapatra:1979ia} by heavy right-handed Majorana neutrinos.
The typical scale of right-handed neutrino Majorana masses around $O(10^{14})$GeV if the Yukawa couplings are $O(1)$.
Furthermore, we have not known yet why the EW symmetry breaking occurs at $v \simeq 246$GeV, which is the vacuum expectation value (VEV) of the Higgs fields.
To understand it, it may be useful to consider the supersymmetry between chiral fermions and bosons such as the minimal supersymmetric standard model (MSSM)~\cite{Fayet:1976et,Fayet:1977yc,Farrar:1978xj}\footnote{See, for a review, e.g. Ref.~\cite{Martin:1997ns}.} or/and consider gauge-Higgs unification (GHU)~\cite{Manton:1979kb,Fairlie:1979at,Hosotani:1983vn,Hosotani:1983xw}, in which Higgs fields are regarded as extra-dimensional vector fields (4D scalar fields) of higher dimensional vector fields with the gauge symmetries.
In these ways, to solve mysteries of the SM, it is useful to consider some symmetries.
However, the above symmetries are just assumptions and we have not known the reason why the symmetries are chosen.

The superstring theory, which is the string theory with the supersymmetry, is a promising candidate for explaining all interactions including gravitational interaction above the Plank scale $O(10^{18})$GeV.\footnote{See Refs.~\cite{Blumenhagen:2006ci,Ibanez:2012zz} for phenomenological aspects of string theory.}
Note that there are 5 types of superstring theories: type-I, type-IIA, type-IIB, $SO(32)$ heterotic, and $E_8 \times E_8$ heterotic superstring theories.
In particular, we consider type-IIB superstring theory.
In addition, there are two types of strings: closed string and open string, and particles are regarded as excitation modes of them in the string theory;
from the closed string, graviton and gravitino which is the superpartner of the graviton appear as massless particles, while from the open string whose edge moves on D-branes, gauge bosons and gauginos which are the superpartners of the gauge bosons appear as massless particles.
In more detail, from open strings both of whose edges move on the same kind of $N$-stack D-branes, $U(N)$ gauge bosons and gauginos appear, that is, $U(N)$ supersymmetrc Yang-Mills (SYM) theory is induced as the effective field theory (EFT), while from open strings whose edges move on different $N_a$-stack D-branes and $N_b$-stack D-branes, respectively, chiral super fields with $(N_a, \bar{N}_b)$ representation ($(\bar{N}_a, N_b))$ representation) under $U(N_a) \times U(N_b)$ gauge symmetry, corresponding to chiral matter fields as well as Higgs fields, appear.
Thus, the superstring theory is a promising theory to explain not only the strong and EW interactions but also the gravitational interaction.

The superstring theory requires 10D space-time ${\cal M}_{10}$, which means that there is an extra 6D compact space ${\cal M}_{6}$ in addition to our observable 4D space-time ${\cal M}_{4}$.
Hence, the superstring theory leads to the 10D EFT and a 4D EFT can be led below the compactification scale, which is typically considered as $O(10^{16}){\rm GeV} \sim O(10^{18}){\rm GeV}$.
Namely, the 4D EFT depends on the compactifications.
Thus, it is important to reveal the property of the 4D EFT by making use of the properties of a 6D compactification and whether the 4D EFT is consistent with the SM.
In particular, a Calabi-Yau (CY) manifold is promising compact space to derive 4D ${\cal N}=1$ EFT such as the MSSM.
However, it is difficult to calculate physical quantities such as Yukawa couplings analytically.
On the other hand, toroidal orbifolds are supposed to be singular limits of certain CY manifolds, where only orbifold fixed points have curvature, and the analytical calculation can be possible.
Furthermore, certain toroidal orbifolds have a kind of the geometrical symmetry called the modular symmetry.
In particular, a 2D torus orbifold has $\bar{\Gamma} \equiv PSL(2,\mathbb{Z})$ modular symmetry.
Interestingly, the modular group contains certain non-Abelian discrete groups such as $S_3$, $A_4$, $S_4$, and $A_5$~\cite{deAdelhartToorop:2011re}, and new type flavor models, called modular flavor models, in which such non-Abelian discrete groups derived from the modular group are assumed as flavor groups have been investigated recently~\cite{Feruglio:2017spp,Kobayashi:2018vbk,Penedo:2018nmg,Criado:2018thu,Kobayashi:2018scp,Novichkov:2018ovf,Novichkov:2018nkm,deAnda:2018ecu,Okada:2018yrn,Kobayashi:2018wkl,Novichkov:2018yse}.
Then, we mainly consider toroidal orbifold compactifications.
We also assume that a 6D compact space ${\cal M}_6$ can be decomposed as three 2D compact space $X_i\ (i=1,2,3)$ for simplicity,~i.e. ${\cal M}_6 \simeq X_1 \times X_2 \times X_3$, and one 2D compact space as 2D torus orbifold,~i.e. $X = T^2/\mathbb{Z}_N$.

Therefore, in this paper,we consider 10D ${N}=1$ $U(N)$ SYM theory on ${\cal M}_{4} \times X_1 \times X_2 \times X_3$ (mainly $X=T^2/\mathbb{Z}_N$) with homogeneous background magnetic fluxes on each $X_i\ (i=1,2,3)$ as the EFT of magnetized D-brane models in the type-IIB superstring theory, and the 4D EFT can be obtained after overlap integration of wave functions on $X_1 \times X_2 \times X_3$.
Due to the magnetic fluxes, the $U(N)$ gauge symmetry is broken into $\prod_{A}U(N_A)$ including $G_{SM}$.
In particular, the broken component fields, which become the fields with bi-fundamental representations under $\prod_{A}U(N_A)$, have multi-generational chiral fermions as well as their superpartners~\cite{Cremades:2004wa,Abe:2008fi,Abe:2013bca,Abe:2014noa,Kobayashi:2017dyu,Sakamoto:2020pev}~\footnote{Originally, it has been discussed in Refs.~\cite{Bachas:1995ik,Blumenhagen:2000wh,Angelantonj:2000hi,Blumenhagen:2000ea}.} such as matter fermions as well as Higgs fields.
In other words, the 10D SYM theory can be regarded as a kind of GUT as well as GHU model, and considering appropriate 6D compactifications including magnetic fluxes can lead MSSM-like models.
Especially, the three-generational modes have been classified in Refs.~\cite{Abe:2008sx,Abe:2015yva,Hoshiya:2020hki}.
Their Yukawa couplings~\cite{Cremades:2004wa,Fujimoto:2016zjs} as well as higher-order couplings~\cite{Abe:2009dr} have been calculated.
In addition, Majorana neutrino mass terms derived from D-brane instanton effects have been calculated in Refs.~\cite{Kobayashi:2015siy,Hoshiya:2021nux}.
Actually, realistic quark and lepton masses and mixing angle as well as the $CP$ phase can be obtained in Refs.~\cite{Abe:2012fj,Abe:2014vza,Fujimoto:2016zjs,Kobayashi:2016qag,Buchmuller:2017vho,Buchmuller:2017vut,Kikuchi:2021yog,Kikuchi:2022geu,Hoshiya:2022qvr}.

Since certain torus orbifolds have the modular symmetry, it can induce flavor symmetries among the multi-generational chiral fermions as well as the superpartners.
Originally, the behavior of modular transformation for wave functions on the magnetized (torus and) torus orbifold has been studied in Refs.~\cite{Kobayashi:2017dyu,Kobayashi:2018rad}.
Furthermore, in Refs.~\cite{Kikuchi:2020frp,Kikuchi:2021ogn}, details of the modular flavor structure was revealed.
Then, the modular flavor structure of the 4D fields can also be found in Ref.~\cite{Kikuchi:2020frp}.
In particular, it was found that the three-generational 4D chiral fields obtained from magnetized $T^2/\mathbb{Z}_N$ orbifold compactification transform under the modular transformation as three-dimensional representation of the modular $\widetilde{\Delta}(6M^2)$ group, which is the quadruple covering group of $\Delta(6M^2)$, or $PSL(2,\mathbb{Z}_M) \times \mathbb{Z}_8$ group with modular weight $1/2$~\cite{Kikuchi:2021ogn}.
In addition, not only 4D fields but also coupling coefficients in 4D EFT such as Yukawa couplings~\cite{Hoshiya:2020hki}~\footnote{See also Ref.~\cite{Kobayashi:2018bff}.} as well as Majorana mass terms~\cite{Hoshiya:2021nux,Kikuchi:2022bkn} transform under the modular transformation.
To see these modular transformation behaviors is one of the main parts of this paper.

Moreover, not only magnetized (2D) torus orbifold compactification but also its magnetized blow-up manifold compactification were studied in Refs.~\cite{Kobayashi:2019fma,Kobayashi:2019gyl,Kobayashi:2022xsk,Kobayashi:2022tti}, in which orbifold singularities are replaced by the parts of the magnetized 2D sphere~\cite{Conlon:2008qi,Dolan:2020sjq}.
Through these studies, one can understand the magnetized torus orbifold compactification deeply; certain magnetic fluxes are localized at orbifold singularities and the total magnetic fluxes including the localized fluxes decide the chiral zero mode numbers due to the Atiyah-Singer (AS) index theorem.
In addition, the degree of freedom of localized fluxes gives new chiral zero modes corresponding to localized modes at the orbifold singularities.
These studies will be useful to explore phenomena on the CY compactification, for the future.
To see behavior on the magnetized blow-up manifold is the other of main parts of this paper.

This paper is organized as follows.
In chapter~\ref{chap:Mtorusmodel}, we discuss magnetized torus models.
In particular, we review magnetized torus compactification in section~\ref{sec:toruscomp} and then we discuss the modular symmetry in magnetized torus compactification in section~\ref{sec:MSmagnetizedT2}.
In chapter~\ref{chap:Mtorusorbifold}, we discuss magnetized torus orbifold models.
In particular, we review magnetized $T^2/\mathbb{Z}_N$ compactification in section~\ref{sec:T2ZN} and then we discuss the modular symmetry in magnetized $T^2/\mathbb{Z}_N$ compactification (and also magnetized $(T^2_1 \times T^2_2)/(\mathbb{Z}^{(t)}_2 \times \mathbb{Z}^{(p)}_2)$ compactification) in section~\ref{sec:MSmagnetizedT2ZN}.
In chapter~\ref{chap:blowup}, we discuss magnetized blow-up manifold of magnetized $T^2/\mathbb{Z}_N$ orbifold.
In particular, we review magnetized $S^2$ compactification in section~\ref{sec:S2} and then we discuss the magnetized blow-up manifold compactification in section~\ref{sec:Mblowupman}.
In chapter~\ref{chap:sumdis}, we summarize this paper.
In Appendix~\ref{app:MTT2T2Z2}, we show the detailed calculation of modular symmetry on magnetized $T^2$ and $T^2/\mathbb{Z}_2$ orbifold.
In Appendix~\ref{ap:NWMBU}, we show the detailed calculation of the normalization of wave functions on magnetized blow-up manifold.
In Appendix~\ref{ap:Anomaly}, we discuss the detailed anomaly structure of discrete symmetries.

\begin{table}[H]
\centering
{\tiny
    \begin{tabular}{|c|c|c|c|c|} \hline 
    $\begin{array}{c}
    {\rm Matter\ fermions} \\ {\rm (Spinor\ fields)}
    \end{array}$ & left-handed quarks & left-handed leptons & right-handed quarks & right-handed leptons \\ \hline
    Representation & $q^i_L = (\mathbf{3}, \mathbf{2}, +\frac{1}{6})$ & $l^i_L = (\mathbf{1}, \mathbf{2}, -\frac{1}{2})$
    &$\begin{array}{c}
    u^i_R = (\mathbf{3}, \mathbf{1}, +\frac{2}{3}) \\
    d^i_R = (\mathbf{3}, \mathbf{1}, -\frac{1}{3})
    \end{array}$& $e^i_R = (\mathbf{1}, \mathbf{1}, -1)$ \\ \hline
    $\begin{array}{c}
	{\rm The\ first} \\ {\rm generation}
	\end{array}$ &
    $q^1_L =
    \begin{pmatrix}
    u^r_L & u^g_L & u^b_L \\
    d^r_L & d^g_L & d^b_L
    \end{pmatrix}$ &
    $l^1_L =
    \begin{pmatrix}
    \nu_{eL} \\
    e
    \end{pmatrix}$ &
    $\begin{array}{c}
    u^1_R =
    \begin{pmatrix}
    u^r_R & u^g_R & u^b_R
    \end{pmatrix} \\
    d^1_R =
    \begin{pmatrix}
    d^r_R & d^g_R & d^b_R
    \end{pmatrix}
    \end{array}$ &
    $e^1_R = e_R$ \\ \hline
    $\begin{array}{c}
	{\rm The\ second} \\ {\rm generation} 
	\end{array}$ &
    $q^2_L =
    \begin{pmatrix}
    c^r_L & c^g_L & c^b_L \\
    s^r_L & s^g_L & s^b_L
    \end{pmatrix}$ &
    $l^2_L =
    \begin{pmatrix}
    \nu_{\mu L} \\
    \mu
    \end{pmatrix}$ &
    $\begin{array}{c}
    u^2_R =
    \begin{pmatrix}
    c^r_R & c^g_R & c^b_R
    \end{pmatrix} \\
    d^2_R =
    \begin{pmatrix}
    s^r_R & s^g_R & s^b_R
    \end{pmatrix}
    \end{array}$ &
    $e^2_R = \mu_R$ \\ \hline
    $\begin{array}{c}
	{\rm The\ third} \\ {\rm generation} 
	\end{array}$ &
    $q^3_L =
    \begin{pmatrix}
    t^r_L & t^g_L & t^b_L \\
    b^r_L & b^g_L & b^b_L
    \end{pmatrix}$ &
    $l^3_L =
    \begin{pmatrix}
    \nu_{\tau L} \\
    \tau
    \end{pmatrix}$ &
    $\begin{array}{c}
    u^3_R =
    \begin{pmatrix}
    t^r_R & t^g_R & t^b_R
    \end{pmatrix} \\
    d^3_R =
    \begin{pmatrix}
    b^r_R & b^g_R & b^b_R
    \end{pmatrix}
    \end{array}$ &
    $e^3_R = \tau_R$ \\ \hline\hline
    $\begin{array}{c}
    {\rm Gauge\ bosons} \\ {\rm (Vector\ fields)}
    \end{array}$ & $SU(3)_C$ & $SU(2)_L$ & $U(1)_Y$ &
    $\begin{array}{c}
    {\rm Higgs\ fields} \\ {\rm (Scalar\ fields)}
    \end{array}$ \\ \hline
    Representation & $G=(\mathbf{8}, \mathbf{1}, 0)$ & $W_L=(\mathbf{1}, \mathbf{3}, 0)$ & $B=(\mathbf{1}, \mathbf{1}, 0)$ & $\phi=(\mathbf{1}, \mathbf{2}, +\frac{1}{2})$ \\
    \par & \par & $\Rightarrow \ W_L^{\pm},Z_L$ & $\Rightarrow \ \gamma \ (A)$ & $\Rightarrow \ h$ \\ \hline
    Interaction & 
	$\begin{array}{c}
	{\rm Strong} \\ {\rm interaction}
	\end{array}$ & 
	$\begin{array}{c}
	{\rm Weak} \\ {\rm interaction}
	\end{array}$ & 
	$\begin{array}{c}
	{\rm Electromagnetic\ (EM)} \\ {\rm interaction}
	\end{array}$ & 
	$\left( \begin{array}{c}
	{\rm Yukawa} \\ {\rm interaction}
	\end{array} \right)$ \\ \hline
    \end{tabular}}
    \caption{The standard model particle contents with $G_{SM}=SU(3)_C \times SU(2)_L \times U(1)_Y$ symmetry}
    \label{Tab:SMcontents}
\end{table}


\chapter{Magnetized torus models}
\label{chap:Mtorusmodel}

First, in this chapter, we discuss magnetized torus models.
We consider magnetized torus compactifications in which homogeneous background magnetic fluxes are inserted in a torus.
The review is in section~\ref{sec:toruscomp}.
Moreover, since a torus has a geometrical symmetry, called the modular symmetry, we discuss the modular symmetry in magnetized torus compactification in section~\ref{sec:MSmagnetizedT2}.
In particular, we can find that multi-generational chiral bi-fundamental fields such as quarks and leptons, which are obtained in the compactification, transform non-trivially under the modular transformation.


\section{Magnetized torus compactification}
\label{sec:toruscomp}

In this section, we review magnetized torus compactifications.


\subsection{10D ${\cal N}=1$ Super Yang-Mills theory}
\label{subsec:10DSYM}

First of all, as the EFT of magnetized D-brane models in the superstring theory, let us start from 10D ${\cal N}=1$ $U(N)$ SYM theory.
We denote a 10D space-time as ${\cal M}_{10}$.
The coordinate of ${\cal M}_{10}$ is $X^M\ (M=0,...,9)$.
The metric on ${\cal M}_{10}$ is $G_{MN}$.
It gives the gamma matrices for $2^{\frac{10}{2}} = 32$D spinor space such that $\{\Gamma^M, \Gamma^N\} = 2G^{MN} \mathbb{I}_{32}$.
We consider $U(N)$ gauge bosons $A_M^{IJ}(X)$, which are $U(N)$ adjoint vector fields, and $U(N)$ gauginos $\lambda^{IJ}(X)$, which are $U(N)$ adjoint spinor fields and the superpartners of the gauge bosons, where $I,J=1,...,N$.
The 10D ${\cal N}=1$ SYM action $S_{{\rm SYM}}$ is given by
\begin{align}
S_{{\rm SYM}} =& \int_{{\cal M}_{10}} d^{10}X \sqrt{|{\rm det}(G_{MN})|} \left[- \frac{1}{4g_{10}^2} F^{MN}_{IJ}(X)F_{MN}^{JI}(X) + \frac{1}{2g_{10}^2} \bar{\lambda}_{IJ}(X) i\Gamma^{M} D_{M}\lambda^{JI}(X)\} \right], \label{eq:actionSYM}
\end{align}
where 
$g_{10}$ denotes the $U(N)$ gauge coupling constant in ${\cal M}_{10}$ with the mass dimension $-3$,
$F_{MN}^{IJ}$ denotes the field strength of the gauge bosons,
\begin{align}
F_{MN}^{IJ}(X) = \partial_{M} A_{N}^{IJ}(X) - \partial_{N} A_{M}^{IJ} -i [A_{M}(X), A_{N}(X)]^{IJ}, \label{eq:FinSYM}
\end{align}
and $D_M$ denotes the covariant derivative for the gauginos,
\begin{align}
D_{M} \lambda^{IJ}(X) = \partial_{M} \lambda^{IJ}(X) -i [A_{M}(X), \lambda(X)]^{IJ}. \label{eq:DinSYM}
\end{align}
Note that the chirality of $\lambda$ on ${\cal M}_{10}$ is fixed by $\Gamma_{11} \lambda = - \lambda$.

In this paper, let us assume the 10D space-time ${\cal M}_{10}$ as product of 4D Minkowski space-time ${\cal M}_4$ and three 2D compact spaces $X_i\ (i=1,2,3)$,~i.e. ${\cal M}_{10} = {\cal M}_{4} \times X_1 \times X_2 \times X_3$.
In particular, in this chapter, let us consider $X_i=T^2_i$ simply.
In the following subsection, we discuss wave functions on one $T^2$ with background magnetic fluxes, called magnetized $T^2$.


\subsection{Magnetized $T^2$ compactification}
\label{subsec:MT2}

In this subsection, we review magnetized $T^2$ compactification\footnote{See also Ref.~\cite{Sakamoto:2020vdy} for $T^2$ compactification without any magnetic fluxes.}~\cite{Cremades:2004wa,Hamada:2012wj}.


\subsubsection{Geometry of $T^2$}
\label{subsubsec:geoT2}

First, let us review the geometry of a 2D torus, $T^2$.
A 2D torus $T^2$ can be constructed by dividing the complex plane $\mathbb{C}$ into 2D lattice $\Lambda=\{ \sum_{i=1,2} n_i e_i | n_i \in \mathbb{Z} \}$,~i.e. $T^2 \simeq \mathbb{C}/\Lambda$.

When we denote the complex coordinate of $\mathbb{C}$ as $u$, we define the complex coordinate of $T^2$ as $z \equiv u/e_1$.
Then, the coordinate of the point of one lattice vector $e_1$ is $z=1$ while that of the point of the other lattice vector $e_2$ is $z=e_2/e_1 \equiv \tau$ (${\rm Im}\tau > 0$). Note that $\tau$ is called the complex structure (CS) modulus of $T^2$.
Thus, $T^2$ is defined by the identifications: $z+1 \sim z$ and $z+\tau \sim z$.
In addition, $dz$ is defined as $dz = dy^1 + \tau dy^2$, where $dy^i\ (i=1,2)$ denote the dual one-forms satisfying $dy^i(e_j)=\delta^i_j$.
Similarly, $\partial_z$ is defined as $\partial_z = (2{\rm Im}\tau)^{-1} ( i\bar{\tau} \partial_{y^1} - i \partial_{y^2} )$.

The metric of $T^2$ is given by
\begin{align}
ds^2 = g_{ij} dy^i dy^j =& 2 h_{\mu\nu} dz^{\mu} d\bar{z}^{\nu}, \notag \\
g = |e_1|^2
\begin{pmatrix}
1 & {\rm Re}\tau \\
{\rm Re}\tau & |\tau|^2
\end{pmatrix},
&\quad
h = |e_1|^2
\begin{pmatrix}
0 & \frac{1}{2} \\
\frac{1}{2} & 0
\end{pmatrix}.
\label{eq:metricT2}
\end{align}
Then, the area of $T^2$, ${\cal A}$, is calculated as
\begin{align}
{\cal A} = \int_{T^2} dy^1dy^2 \sqrt{|{\rm det}g|} = \int_{T^2} dz d\bar{z} \sqrt{|{\rm det}(2h)|} = |e_1|^2 {\rm Im}\tau.
\end{align}
The gamma matrices $\gamma^z$ and $\gamma^{\bar{z}}$ satisfying $\{ \gamma^z, \gamma^{\bar{z}} \} = 2h^{z\bar{z}}$ are defined as follows, where $h^{z\bar{z}}$ is the inverse matrix of $h_{z\bar{z}}$.
First, the gamma matrices $\gamma^1$ and $\gamma^2$ satisfying 2D Clifford algebra $\{ \gamma^a, \gamma^b \} = 2\delta^{ab}$ ($a,b=1,2$) are given by
\begin{align}
\gamma^1 = \sigma^1 =
\begin{pmatrix}
0 & 1 \\
1 & 0
\end{pmatrix},
\quad
\gamma^2 = \sigma^2 =
\begin{pmatrix}
0 & -i \\
i & 0
\end{pmatrix}.
\label{eq:2Dclifford}
\end{align}
Next, we introduce two vielbeins $e$ and $f$ such that $g_{ij} = e_i^a e_j^b \delta_{ab}$ and $h_{\mu\nu} = f_{\mu}^i f_{\nu}^j g_{ij}$, respectively.
They are given by
\begin{align}
&e = |e_1|
\begin{pmatrix}
1 & {\rm Re}\tau \\
0 & {\rm Im}\tau
\end{pmatrix},
\quad
f^{-1} = |e_1|^{-1}
\begin{pmatrix}
e_1 & e_2 \\
\bar{e}_1 & \bar{e}_2
\end{pmatrix}, \label{eq:ef} \\
\Rightarrow \ &
(\tilde{e}^{-1}) ^{\mu}_a \equiv  ( f^{-1} )^{\mu}_i ( e^{-1} )^i_a =
|e_1|^{-2} ({\rm Im}\tau)^{-1}
\begin{pmatrix}
e_1 {\rm Im}\tau &-  e_1 {\rm Re}\tau + e_2 \\
\bar{e}_1 {\rm Im}\tau & - \bar{e}_1 {\rm Re}\tau + \bar{e}_2
\end{pmatrix}.
\label{eq:etilde}
\end{align}
Then, the gamma matrices $\gamma^z$ and $\gamma^{\bar{z}}$ satisfying $\{ \gamma^z, \gamma^{\bar{z}} \} = 2h^{z\bar{z}}$ are given as
\begin{align}
\gamma^{z} = (\tilde{e}^{-1}) ^{z}_a \gamma^a = e_1^{-1}
\begin{pmatrix}
0 & 2 \\
0 & 0
\end{pmatrix},
\quad
\gamma^{\bar{z}} = (\tilde{e}^{-1}) ^{\bar{z}}_b \gamma^b = \bar{e}_1^{-1}
\begin{pmatrix}
0 & 0 \\
2 & 0
\end{pmatrix}.
\label{eq:gammaz}
\end{align}

The non-trivial Levi-Civita connection is given by
\begin{align}
\begin{array}{l}
\Gamma^{z}_{zz} = h^{z\bar{z}} \partial_{z} h_{z\bar{z}} = 0, \\ \Gamma^{\bar{z}}_{\bar{z}\bar{z}} = h^{\bar{z}z} \partial_{\bar{z}} h_{\bar{z}z} = 0.
\end{array}
\label{eq:LeviCivita}
\end{align}
Moreover, the spin connection is given by
\begin{align}
\begin{array}{l}
\omega^a_{z b} = (\tilde{e}^{-1})^{z}_b \Gamma^{z}_{zz} \tilde{e}^a_{z} - (\tilde{e}^{-1})^{z}_b \partial_{z} \tilde{e}^a_{z} - (\tilde{e}^{-1})^{\bar{z}}_b \partial_{z} \tilde{e}^a_{\bar{z}} = 0, \\
\omega^a_{\bar{z} b} = (\tilde{e}^{-1})^{\bar{z}}_b \Gamma^{\bar{z}}_{\bar{z}\bar{z}} \tilde{e}^a_{\bar{z}} - (\tilde{e}^{-1})^{\bar{z}}_b \partial_{z} \tilde{e}^a_{\bar{z}} - (\tilde{e}^{-1})^{z}_b \partial_{\bar{z}} \tilde{e}^a_{z} = 0.
\end{array}
\label{eq:spincon}
\end{align}
Note that the Lorentz generator is $\Sigma^{ab} = \frac{1}{4} [\gamma^a, \gamma^b]$.

The curvature of $T^2$ is given by
\begin{align}
\frac{1}{2\pi i} \int_{T^2} R^{z}_{zz\bar{z}} dz \wedge d\bar{z} = \chi(T^2) = 0, \label{eq:curvT2}
\end{align}
where $R^{z}_{zz\bar{z}}$ is obtained by
\begin{align}
R^{z}_{zz\bar{z}} = \partial_{\bar{z}} \Gamma^{z}_{zz} = 0, \label{eq:RiemannT2}
\end{align}
while $\chi(T^2)$ denotes the Euler number of $T^2$.


\subsubsection{Wave functions on magnetized $T^2$}
\label{subsubsec:waveMT2}

Here, let us consider that the following Abelian homogeneous background magnetic flux,
\begin{align}
\frac{1}{2\pi} \int_{T^2} \langle F_{z\bar{z}} \rangle dz \wedge d\bar{z} =
\begin{pmatrix}
M^a \mathbb{I}_{N_a} & \ \\
\ & M^b \mathbb{I}_{N_b}
\end{pmatrix}, \ 
\left( \frac{1}{2\pi} \int_{T^2} \langle F^{aa,bb}_{z\bar{z}} \rangle dz \wedge d\bar{z} = M^{a,b} \right),
\label{eq:flux}
\end{align}
is inserted on $T^2$, where $N_a+N_b=N$ and $M^{a,b}$ must be integers\footnote{It comes from the fact that the fist homotopy group of $U(1)$, $\pi_1(U(1))$, is $\pi_1(U(1))=\mathbb{Z}$. On the other hand, $\pi_1(SU(N)) =\{e\}$ means that off-diagonal components of the magnetic flux are zeros.} (Dirac's quantization).
Due to this magnetic flux, $U(N)$ symmetry is broken to $U(N_a) \times U(N_b)$ symmetry.\footnote{The magnetic flux is inserted along directions of the diagonal $U(1)$ of $U(N) \simeq U(1) \times SU(N)$ and the base of Cartan subalgebra of $SU(N)$. Then, $U(N) \simeq U(1) \times SU(N)$ symmetry is broken to $U(1)_a \times SU(N_a) \times U(1)_b \times SU(N_b) \simeq U(N_a) \times U(N_b)$ symmetry.}
The magnetic flux is given by the 2-form field strength, 
\begin{align}
\frac{1}{2\pi} \langle F^{aa,bb} \rangle = \frac{1}{2\pi} \langle F^{aa,bb}_{z\bar{z}} \rangle dz \wedge d\bar{z} = \frac{M^{a,b}}{{\rm Im}\tau} \frac{i}{2} dz \wedge d\bar{z},
\label{eq:FVEV}
\end{align}
which satisfies the Yang-Mills equations
\begin{align}
\partial_z \langle F^{aa,bb}_{z\bar{z}} \rangle = \partial_{\bar{z}} \langle F^{aa,bb}_{z\bar{z}} \rangle = 0. \label{eq:YMEFVEV}
\end{align}
Moreover, the field strength is obtained from the 1-form background gauge field,
\begin{align}
\langle A^{aa,bb}(z) \rangle
&= \langle A^{aa,bb}_{z}(z) \rangle dz + \langle A^{aa,bb}_{\bar{z}}(z) \rangle d\bar{z} \notag \\
&= - \frac{i}{4} \frac{2\pi M^{a,b}}{{\rm Im}\tau} (\bar{z}+\bar{\zeta}^{a,b}) dz + \frac{i}{4} \frac{2\pi M^{a,b}}{{\rm Im}\tau} (z+\zeta^{a,b}) d\bar{z} \label{eq:AVEV} \\
&= \frac{2\pi M^{a,b}}{2{\rm Im}\tau} {\rm Im}((\bar{z}+\bar{\zeta}^{a,b})dz), \notag
\end{align}
by $\langle F^{aa,bb} \rangle =d \langle A^{aa,bb} \rangle$, where $\zeta^{a,b}$ denote Wilson line (WL) phases.
The boundary conditions (BCs) of the gauge field are
\begin{align}
\langle A^{aa,bb}(z+1) \rangle &= \langle A^{aa,bb}(z) \rangle + d\left( \frac{2\pi M^{a,b}}{2{\rm Im}\tau} {\rm Im}z \right),
\label{eq:BCAVEVz1} \\
\langle A^{aa,bb}(z+\tau) \rangle &= \langle A^{aa,bb}(z) \rangle + d\left(  \frac{2\pi M^{a,b}}{2{\rm Im}\tau} {\rm Im}\bar{\tau}z \right), \label{eq:BCAVEVztau}
\end{align}
which correspond to the Abelian gauge transformation.
The covariant derivative is defined as
\begin{align}
\hat{D} &= d -i\langle A(z) \rangle \notag \\
\hat{D}_zdz + \hat{D}_{\bar{z}}d{\bar{z}} &= (\partial_z -i \langle A_z(z) \rangle )dz + (\partial_{\bar{z}} - i \langle A_{\bar{z}}(z) \rangle )d{\bar{z}},
\label{eq:D}
\end{align}
%
Along with BCs of the gauge field in Eqs.~(\ref{eq:BCAVEVz1}) and (\ref{eq:BCAVEVztau}), $U(N)$ adjoint spinor and scalar fields on the magnetized $T^2$,
\begin{align}
\Phi_{T^2}(z) =
\begin{pmatrix}
\Phi_{T^2}^{aa}(z) & \Phi_{T^2}^{ab}(z) \\
\Phi_{T^2}^{ba}(z) & \Phi_{T^2}^{bb}(z)
\end{pmatrix},
\label{eq:Phi}
\end{align}
should satisfy the following BCs,
\begin{align}
&\Phi_{T^2}(z+1) = U_1(z) \Phi_{T^2}(z) U_1^{-1}(z), \ U_1(z) =
\begin{pmatrix}
e^{\pi iM^a\frac{{\rm Im}(z+\zeta^a)}{{\rm Im}\tau}+2\pi i\alpha^a_1} \mathbb{I}_{N_a} & 0 \\
0 & e^{\pi iM^b\frac{{\rm Im}(z+\zeta^b)}{{\rm Im}\tau}+2\pi i\alpha^b_1} \mathbb{I}_{N_b}
\end{pmatrix},
\label{eq:BCPhiz1} \\
&\Leftrightarrow \left\{
\begin{array}{l}
\Phi_{T^2}^{aa}(z+1) = \Phi_{T^2}^{aa}(z) \\
\Phi_{T^2}^{bb}(z+1) = \Phi_{T^2}^{bb}(z) \\
\Phi_{T^2}^{ab}(z+1) = e^{\pi iM\frac{{\rm Im}(z+\zeta)}{{\rm Im}\tau}+2\pi i\alpha_1} \Phi_{T^2}^{ab}(z) \\
\Phi_{T^2}^{ba}(z+1) = e^{-\pi iM\frac{{\rm Im}(z+\zeta)}{{\rm Im}\tau}-2\pi i\alpha_1} \Phi_{T^2}^{ba}(z) \\
\end{array}
\right.,
\label{eq:BCPhicompz1} \\
&\Phi_{T^2}(z+\tau) = U_{\tau}(z) \Phi_{T^2}(z) U_{\tau}^{-1}(z), \ U_{\tau}(z) =
\begin{pmatrix}
e^{\pi iM^a\frac{{\rm Im}(\bar{\tau}(z+\zeta^a))}{{\rm Im}\tau}+2\pi i\alpha^a_{\tau}} \mathbb{I}_{N_a} & 0 \\
0 & e^{\pi iM^b\frac{{\rm Im}(\bar{\tau}(z+\zeta^b))}{{\rm Im}\tau}+2\pi i\alpha^b_{\tau}} \mathbb{I}_{N_b}
\end{pmatrix},
\label{eq:BCPhiztau} \\
&\Leftrightarrow \left\{
\begin{array}{l}
\Phi_{T^2}^{aa}(z+\tau) = \Phi_{T^2}^{aa}(z) \\
\Phi_{T^2}^{bb}(z+\tau) = \Phi_{T^2}^{aa}(z) \\
\Phi_{T^2}^{ab}(z+\tau) = e^{\pi iM\frac{{\rm Im}(\bar{\tau}(z+\zeta))}{{\rm Im}\tau}+2\pi i\alpha_{\tau}} \Phi_{T^2}^{ab}(z) \\
\Phi_{T^2}^{ba}(z+\tau) = e^{-\pi iM\frac{{\rm Im}(\bar{\tau}(z+\zeta))}{{\rm Im}\tau}-2\pi i\alpha_{\tau}} \Phi_{T^2}^{ba}(z) \\
\end{array}
\right.,
\label{eq:BCPhicompztau}
\end{align}
where $\alpha^{a,b}_1$ and $\alpha^{a,b}_{\tau}$ denote the Scherk-Schwarz (SS) phases.
Here, $\Phi^{aa}$ and $\Phi^{bb}$ denote fields with $U(N_a)$ and $U(N_b)$ adjoint representations, respectively, while $\Phi^{ab}$ and $\Phi^{ba}$ denote fields with bi-fundamental representations, $(N_a, \bar{N}_b)$ and $(\bar{N}_a, N_b)$, under $U(N_a) \times U(N_b)$, respectively.
In particular, $\Phi^{ab}$ feel the magnetic flux $M \equiv M^a - M^b$, the WL phase $M\zeta \equiv M^a\zeta^a - M^b\zeta^b$, and the SS phases $\alpha_{1,\tau} \equiv \alpha^a_{1,\tau} - \alpha^b_{1,\tau}$, while $\Phi^{ba}$ feel the opposite sign of them.
On the other hand, the BCs of $U(N)$ adjoint 1-form vector field on the magnetized $T^2$,
\begin{align}
A_{T^2}(z) =
\begin{pmatrix}
A_{T^2}^{aa}(z) & A_{T^2}^{ab}(z) \\
A_{T^2}^{ba}(z) & A_{T^2}^{bb}(z)
\end{pmatrix},
\label{eq:A}
\end{align}
can be written as
\begin{align}
&A_{T^2}(z+1) = U_1(z) A_{T^2}(z) U_1^{-1} + i U_1(z) d U_1^{-1}(z),
\label{eq:BCAz1} \\
\Leftrightarrow& \left\{
\begin{array}{l}
A_{T^2}^{aa}(z+1) = A_{T^2}^{aa}(z) + d\left( \frac{2\pi M^a}{2{\rm Im}\tau}{\rm Im}z \right)  \\
A_{T^2}^{bb}(z+1) = A_{T^2}^{bb}(z) + d\left( \frac{2\pi M^b}{2{\rm Im}\tau}{\rm Im}z \right)  \\
A_{T^2}^{ab}(z+1) = e^{\pi iM\frac{{\rm Im}(z+\zeta)}{{\rm Im}\tau}+2\pi i\alpha_1} A_{T^2}^{ab}(z) \\
A_{T^2}^{ba}(z+1) = e^{-\pi iM\frac{{\rm Im}(z+\zeta)}{{\rm Im}\tau}-2\pi i\alpha_1} A_{T^2}^{ba}(z) \\
\end{array}
\right.,
\label{eq:BCAcompz1} \\
&A_{T^2}(z+\tau) = U_{\tau}(z) A_{T^2}(z) U_{\tau}^{-1} + i U_{\tau}(z) d U_{\tau}^{-1}(z),
\label{eq:BCAztau} \\
\Leftrightarrow& \left\{
\begin{array}{l}
A_{T^2}^{aa}(z+\tau) = A_{T^2}^{aa}(z) + d\left( \frac{2\pi M^a}{2{\rm Im}\tau}{\rm Im}\bar{\tau}z \right)  \\
A_{T^2}^{bb}(z+\tau) = A_{T^2}^{bb}(z) + d\left( \frac{2\pi M^b}{2{\rm Im}\tau}{\rm Im}\bar{\tau}z \right)  \\
A_{T^2}^{ab}(z+\tau) = e^{\pi iM\frac{{\rm Im}(\bar{\tau}(z+\zeta))}{{\rm Im}\tau}+2\pi i\alpha^a_{\tau}} A_{T^2}^{ab}(z) \\
A_{T^2}^{ba}(z+\tau) = e^{-\pi iM\frac{{\rm Im}(\bar{\tau}(z+\zeta))}{{\rm Im}\tau}-2\pi i\alpha_{\tau}} A_{T^2}^{ba}(z) \\
\end{array}
\right..
\label{eq:BCAcompztau}
\end{align}
When we write $A_{T^2}(z) = \langle A(z) \rangle + A_{T^2}(z)$, BCs of the first term corresponds to Eqs.~(\ref{eq:BCAVEVz1}) and (\ref{eq:BCAVEVztau}) while BCs of the second term corresponds to (\ref{eq:BCPhicompz1}) and (\ref{eq:BCPhicompztau}).
We also note that WL phases $\zeta^{a,b}=\zeta^{a,b}_1+\tau \zeta^{a,b}_2$ can be converted into SS phases, $\alpha^{a,b}_1 \rightarrow {\alpha'}^{a,b}_1 = \alpha^{a,b}_1+M\zeta^{a,b}_2$ and $\alpha^{a,b}_{\tau} \rightarrow {\alpha'}^{a,b}_{\tau} = \alpha^{a,b}_{\tau} - M\zeta^{a,b}_1$, by the following gauge transformation
\begin{align}
\begin{array}{l}
\langle A(z) \rangle \rightarrow \langle \widetilde{A}(z) \rangle = \langle A(z) \rangle + i U^{-1}_{\zeta}(z) d U_{\zeta}(z), \\
\Phi_{T^2}(z) \rightarrow \widetilde{\Phi}_{T^2}(z) = U^{-1}_{\zeta}(z) \Phi_{T^2}(z) U_{\zeta}(z),
\end{array}
\label{eq:WLSS}
\end{align}
with
\begin{align}
U^{-1}_{\zeta}(z) =
\begin{pmatrix}
e^{-\pi iM^a\left( \frac{{\rm Im}(\bar{\zeta}^a z)}{{\rm Im}\tau}+\zeta^a_1\zeta^a_2\right)} \mathbb{I}_{N_a} & 0 \\
0 & e^{-\pi iM^b\left(\frac{{\rm Im}(\bar{\zeta}^b z)}{{\rm Im}\tau}+\zeta^b_1\zeta^b_2\right)} \mathbb{I}_{N_b}
\end{pmatrix}.
\label{eq:V}
\end{align}
Hence, hereafter, we consider vanishing WL phases $\zeta^{a,b} = 0$ and then consider only SS phases $(\alpha^{a,b}_1, \alpha^{a,b}_{\tau})$.
In the following, let us see wave functions of $U(N)$ adjoint spinor, scalar, and vector fields on $T^2$ with the magnetic flux in Eq.~(\ref{eq:flux}), which satisfy individual equation of motions 
under the BCs in Eqs.~(\ref{eq:BCPhiz1}) ((\ref{eq:BCAz1})) and (\ref{eq:BCPhiztau}) ((\ref{eq:BCAz1})).

\begin{itemize}
\item {\large Spinor fields}
\end{itemize}

First, let us see wave functions of $U(N)$ adjoint 2D Majorana-Weyl (MW) spinor fields on the magnetized $T^2$,
\begin{align}
\begin{array}{c}
\psi_{T^2}^{(2)}(z) =
\begin{pmatrix}
\psi_{T^2,+}(z) \\ \psi_{T^2,-}(z)
\end{pmatrix}, \ 
\psi_{\pm}(z) =
\begin{pmatrix}
\psi^{aa}_{T^2,\pm}(z) & \psi^{ab}_{T^2,\pm}(z) \\
\psi^{ba}_{T^2,\pm}(z) & \psi^{bb}_{T^2,\pm}(z)
\end{pmatrix}, \\
\left( \psi_{T^2,\mp}^{JI}(z) = \overline{\psi_{T^2,\pm,n,j}^{IJ}(z)} \ (I,J=a,b) \right), 
\end{array}
\label{eq:spinor} 
\end{align}
which satisfy the Dirac equation,
\begin{align}
\left( i (\gamma^{z} \hat{D}_{z} + \gamma^{\bar{z}} \hat{D}_{\bar{z}}) - m_{n} \right) \psi_{T^2,n}^{(2)}(z) = 0,
\label{eq:DiraceqX}
\end{align}
under the BCs in Eqs.~(\ref{eq:BCPhiz1}) and (\ref{eq:BCPhiztau}).
Here, we define the Dirac operator,
\begin{align}
i\hat{\slashed{D}} \equiv i (\gamma^z \hat{D}_z + \gamma^{\bar{z}} \hat{D}_{\bar{z}}) =
\begin{pmatrix}
0 & 2ie^{-1}_1(\partial_{z} -i \langle A_{z}(z) \rangle ) \\
2i\bar{e}^{-1}_1 (\partial_{\bar{z}} -i \langle A_{\bar{z}}(z) \rangle ) & 0
\end{pmatrix}
\equiv
\begin{pmatrix}
0 & -i{\cal D}^{\dagger} \\
i{\cal D} & 0
\end{pmatrix},
\label{eq:DiracOP}
\end{align}
and then 
the Dirac equation for component fields
can be written as
\begin{align}
&i{\cal D} \psi_{T^2,+,n}(z) = 
2i\bar{e}^{-1}_1 (\partial_{\bar{z}} \psi_{T^2,+,n}(z) -i [\langle A_{\bar{z}}(z) \rangle, \psi_{T^2,+,n}(z)]) = m_n \psi_{T^2,-,n}(z), \label{eq:Dirac+}  \\
\Leftrightarrow&
\left\{
\begin{array}{l}
i{\cal D}_{aa} \psi^{aa}_{T^2,+,n}(z) = 2i\bar{e}^{-1}_1\partial_{\bar{z}} \psi^{aa}_{T^2,+,n}(z) = m_n \psi^{aa}_{T^2,-,n}(z) \\
i{\cal D}_{bb} \psi^{bb}_{T^2,+,n}(z) = 2i\bar{e}^{-1}_1\partial_{\bar{z}} \psi^{bb}_{T^2,+,n}(z) = m_n \psi^{bb}_{T^2,-,n}(z) \\
i{\cal D}_{ab} \psi^{ab}_{T^2,+,n}(z) = 2i\bar{e}^{-1}_1\left( \partial_{\bar{z}} + \frac{\pi M}{2{\rm Im}\tau} z \right) \psi^{ab}_{T^2,+,n}(z) = m_n \psi^{ab}_{T^2,-,n}(z) \\
-i{\cal D}_{ba} \psi^{ba}_{T^2,+,n}(z) = 2i\bar{e}^{-1}_1\left( \partial_{\bar{z}} - \frac{\pi M}{2{\rm Im}\tau} z \right) \psi^{ba}_{T^2,+,n}(z) = m_n \psi^{ba}_{T^2,-,n}(z)
\end{array}
\right., \notag \\
\Leftrightarrow&
\left\{
\begin{array}{l}
2i\bar{e}^{-1}_1\partial_{\bar{z}} \psi^{aa}_{T^2,+,n}(z) = m_n \psi^{aa}_{T^2,-,n}(z) \\
2i\bar{e}^{-1}_1\partial_{\bar{z}} \psi^{bb}_{T^2,+,n}(z) = m_n \psi^{bb}_{T^2,-,n}(z) \\
2i\bar{e}^{-1}_1 e^{-\frac{\pi M}{2{\rm Im}\tau}|z|^2} \partial_{\bar{z}} \left[ e^{\frac{\pi M}{2{\rm Im}\tau}|z|^2} \psi^{ab}_{T^2,+,n}(z) \right] = m_n \psi^{ab}_{T^2,-,n}(z) \\
2i\bar{e}^{-1}_1 e^{\frac{\pi M}{2{\rm Im}\tau}|z|^2} \partial_{\bar{z}} \left[ e^{-\frac{\pi M}{2{\rm Im}\tau}|z|^2} \psi^{ba}_{T^2,+,n}(z) \right] = m_n \psi^{ba}_{T^2,-,n}(z)
\end{array}
\right., \notag \\
&-i{\cal D}^{\dagger} \psi_{T^2,-,n}(z) = 
2ie_1 (\partial_{z} \psi_{T^2,-,n}(z) -i [\langle A_{z}(z) \rangle, \psi_{T^2,-,n}(z)]) = m_n \psi_{T^2,+,n}(z), \label{eq:Dirac-} \\
\Leftrightarrow&
\left\{
\begin{array}{l}
-i{\cal D}^{\dagger}_{aa} \psi^{aa}_{T^2,-,n}(z) = 2i e^{-1}_1 \partial_{z} \psi^{aa}_{T^2,-,n}(z) = m_n \psi^{aa}_{T^2,+,n}(z) \\
-i{\cal D}^{\dagger}_{bb} \psi^{bb}_{T^2,-,n}(z) = 2i e^{-1}_1 \partial_{z} \psi^{bb}_{T^2,-,n}(z) = m_n \psi^{bb}_{T^2,+,n}(z) \\
-i{\cal D}^{\dagger}_{ab} \psi^{ab}_{T^2,-,n}(z) = 2i e^{-1}_1 \left( \partial_{z} - \frac{\pi M}{2{\rm Im}\tau} \bar{z} \right) \psi^{ab}_{T^2,-,n}(z) = m_n \psi^{ab}_{T^2,+,n}(z) \\
i{\cal D}^{\dagger}_{ba} \psi^{ba}_{T^2,-,n}(z) = 2i e^{-1}_1 \left( \partial_{z} + \frac{\pi M}{2{\rm Im}\tau} \bar{z} \right) \psi^{ba}_{T^2,-,n}(z) = m_n \psi^{ba}_{T^2,+,n}(z)
\end{array}
\right., \notag \\
\Leftrightarrow&
\left\{
\begin{array}{l}
2i e^{-1}_1 \partial_{z} \psi^{aa}_{T^2,-,n}(z) = m_n \psi^{aa}_{T^2,+,n}(z) \\
2i e^{-1}_1 \partial_{z} \psi^{bb}_{T^2,-,n}(z) = m_n \psi^{bb}_{T^2,+,n}(z) \\
2i e^{-1}_1 e^{\frac{\pi M}{2{\rm Im}\tau}|z|^2} \partial_{z} \left[ e^{-\frac{\pi M}{2{\rm Im}\tau}|z|^2} \psi^{ab}_{T^2,-,n}(z) \right] = m_n \psi^{ab}_{T^2,+,n}(z) \\
2i e^{-1}_1 e^{-\frac{\pi M}{2{\rm Im}\tau}|z|^2} \partial_{z} \left[ e^{\frac{\pi M}{2{\rm Im}\tau}|z|^2} \psi^{ab}_{T^2,-,n}(z) \right] = m_n \psi^{ab}_{T^2,+,n}(z)
\end{array}
\right., \notag
\end{align}
%
where $n$ denotes the Landau level.

In particular, the lowest modes ($n=0$) satisfying the above Dirac equation with $m_0=0$ are expressed as
\begin{align}
\begin{array}{ll}
\psi^{aa}_{T^2,+,0}(z) = h^{aa}_0(z), & \psi^{aa}_{T^2,-,0}(z) = \overline{\psi^{aa}_{T^2,+,0}(z)} = \bar{h}^{aa}_0(\bar{z}), \\
\psi^{bb}_{T^2,+,0}(z) = h^{bb}_0(z), & \psi^{bb}_{T^2,-,0}(z) = \overline{\psi^{bb}_{T^2,+,0}(z)} = \bar{h}^{bb}_0(\bar{z}), \\
\psi^{ab}_{T^2,+,0}(z) = e^{-\frac{\pi M}{2{\rm Im}\tau}|z|^2} h^{ab}_0(z), & \psi^{ab}_{T^2,-,0}(z) = \overline{\psi^{ba}_{T^2,+,0}(z)} = e^{\frac{\pi M}{2{\rm Im}\tau}|z|^2} \bar{h}^{ba}_0(\bar{z}), \\
\psi^{ba}_{T^2,+,0}(z) = e^{\frac{\pi M}{2{\rm Im}\tau}|z|^2} h^{ba}_0(z), & \psi^{ab}_{T^2,-,0}(z) = \overline{\psi^{ab}_{T^2,+,0}(z)} = e^{-\frac{\pi M}{2{\rm Im}\tau}|z|^2} \bar{h}^{ab}_0(\bar{z}),
\end{array}
\label{eq:Diraczeromode}
\end{align}
where $h(z)$ denotes a holomorphic function.
Since they should also satisfy the BCs in Eqs.~(\ref{eq:BCPhicompz1}) and (\ref{eq:BCPhicompztau}), the holomorphic functions can be determined as
\begin{align}
\begin{array}{l}
h^{aa}_0(z) = {\cal N}^{aa}_0, \\
h^{bb}_0(z) = {\cal N}^{bb}_0, \\
h^{ab}_0(z) = \sum_{j=0}^{|M|} {\cal N}^{ab}_{0,j} e^{\frac{\pi M}{2{\rm Im}\tau}z^2} \vartheta
\begin{bmatrix}
\frac{j+\alpha_1}{M} \\ -\alpha_{\tau}
\end{bmatrix}
\left( Mz, M\tau \right)
\equiv \sum_{j=0}^{|M|} h^{(j+\alpha_1,\alpha_{\tau}),M}_{T^2,0}(z), \\
h^{ba}_0(z) = \sum_{j=0}^{|M|-1} {\cal N}^{ba}_{0,j} e^{-\frac{\pi M}{2{\rm Im}\tau}z^2} \vartheta
\begin{bmatrix}
\frac{-(j+\alpha_1)}{-M} \\ -(-\alpha_{\tau})
\end{bmatrix}
\left( -Mz, -M\tau \right)
\equiv \sum_{j=0}^{|M|-1} h^{-(j+\alpha_1,\alpha_{\tau}),-M}_{T^2,0}(z),
\end{array}
\label{eq:holT2}
\end{align}
where ${\cal N}$ denotes a constant and $\vartheta$ denotes the Jacobi-theta function,
\begin{align}
\vartheta
\begin{bmatrix}
a \\ b
\end{bmatrix}
\left( \nu, \tau \right)
=
\sum_{l \in \mathbb{Z}} e^{\pi i(a+l)^2\tau} e^{2\pi i (a+l)(\nu+b)}.
\end{align}
When we also define
\begin{align}
\psi^{(j+\alpha_1,\alpha_{\tau}),M}_{T^2,0}(z,\tau) \equiv e^{-\frac{\pi M}{2{\rm Im}\tau}|z|^2} h^{(j+\alpha_1,\alpha_{\tau}),M}_{T^2,0}(z),
\label{eq:psizero}
\end{align}
$\psi^{ab}_{+,0}$ and $\psi^{ba}_{+,0}$ can be written as
\begin{align}
\psi^{ab}_{T^2,+,0}(z) = \sum_{j=0}^{|M|-1} \psi^{(j+\alpha_1,\alpha_{\tau}),M}_{T^2,0}(z,\tau), \quad
\psi^{ba}_{T^2,+,0}(z) = \sum_{j=0}^{|M|-1} \psi^{-(j+\alpha_1,\alpha_{\tau}),-M}_{T^2,0}(z,\tau).
\label{eq:lambdazero}
\end{align}
Note that each $j$ th wave function written by Eq.~(\ref{eq:psizero}) satisfies the lowest mode (zero mode) Dirac equation and the BCs, which means that there are $|M|$ number of independent zero mode solutions.
Finally, these solutions must be normalizable.
Thus, when $M>0$ ($M<0$), $\psi^{ab}_{+}$ ($\psi^{ba}_{+}$) as well as their anti-fields $\psi^{ba}_{-} = \overline{\psi^{ab}_{+}}$ ($\psi^{ab}_{-} = \overline{\psi^{ba}_{+}}$) have well-defined $|M|$ number of degenerate zero modes while $\psi^{ba}_{+}$ ($\psi^{ab}_{+}$) as well as their anti-fields $\psi^{ab}_{-} = \overline{\psi^{ba}_{+}}$ ($\psi^{ba}_{-} = \overline{\psi^{ab}_{+}}$) have no physical relevant zero modes.
Indeed, this result is consistent with the AS index theorem,
\begin{align}
n^{ab}_{+} - n^{ab}_{-} &= \frac{1}{2\pi} \int_{T^2} F_{ab} = M, \label{eq:indexFab} \\
n^{ba}_{+} - n^{ba}_{-} &= \frac{1}{2\pi} \int_{T^2} F_{ba} = -M, \label{eq:indexFba}
\end{align}
where $n^{ab}_{+}$, $n^{ab}_{-}$, $n^{ba}_{+}$, $n^{ba}_{-}$ denote zero mode numbers of $\psi^{ab}_{+}$, $\psi^{ab}_{-}$, $\psi^{ba}_{+}$, $\psi^{ba}_{-}$, respectively, and $F_{ab}$, $F_{ba}$ denote the magnetic fluxes which $\psi^{ab}_{\pm}$, $\psi^{ba}_{\pm}$ feel, respectively.
Therefore, we can obtain $|M|$ generational bi-fundamental chiral fermions from the magnetized $T^2$ compactification.
Hereafter, let us consider $M>0$.
In addition, we set the SS phases as $0 \leq \alpha_{1,\tau} < 1$ because of the periodicity.
Now, let us discuss the normalization of wave functions in Eq.~(\ref{eq:psizero}) from the following inner product,
\begin{align}
&\int_{T^2} dz d\bar{z} \sqrt{|{\rm det}(2h)|} \overline{\psi^{(j+\alpha_1,\alpha_{\tau}),M}_{T^2,0}(z,\tau)} \psi^{(k+\alpha_1,\alpha_{\tau}),M}_{T^2,0}(z,\tau) \notag \\
=& \delta_{j,k} |{\cal N}^{ab}_{0,j}|^2 {\cal A} \sum_{l \in \mathbb{Z}}\int_{0}^{1} d\left( \frac{{\rm Im}z}{{\rm Im}\tau} \right) e^{-2\pi M{\rm Im}\tau \left( \frac{{\rm Im}z}{{\rm Im}\tau} + \frac{j+\alpha_1}{M} + l \right)^2} \label{eq:innerproduct} \\
=& |{\cal N}^{ab}_{0,j}|^2 {\cal A} (2{\rm Im}\tau M)^{-1/2} \delta_{j,k}. \notag
\end{align}
Here, we assume that the normalization factor ${\cal N}^{ab}_{0,j}$ does not depend on the geometrical parameters such as ${\cal A}$ and $\tau$, that is, we set the normalization condition as
\begin{align}
\int_{T^2} dz d\bar{z} \sqrt{|{\rm det}(2h)|} \overline{\psi^{(j+\alpha_1,\alpha_{\tau}),M}_{T^2,0}(z,\tau)} \psi^{(k+\alpha_1,\alpha_{\tau}),M}_{T^2,0}(z,\tau) = {\cal A} (2{\rm Im}\tau)^{-1/2} \delta_{j,k},
\label{eq:normalizationT2}
\end{align}
and then the normalization factor is determined by $|{\cal N}^{ab}_{0,j}| = (M)^{1/4}$.
Although the phase of ${\cal N}^{ab}_{0,j}$ is not determined, it seems to be natural\footnote{When we set the phase at $z=0$ as $1$, from the BC (\ref{eq:BCPhicompztau}), the phase at $z=\tau$ becomes $e^{2\pi i\alpha_{\tau}}$. It implies that the phase at $z=y^2\tau$ is $e^{2\pi iy^2\tau}$. Moreover, from Eq.~(\ref{eq:innerproduct}), the wave function's density $|\psi^{(j+\alpha_1,\alpha_{\tau}),M}_{T^2,0}(z,\tau)|$ behaves as a gaussian whose peak at $y^2={\rm Im}z/{\rm Im}\tau = (j+\alpha_1)/M$. Hence, it seems to be natural that the wave function $\psi^{(j+\alpha_1,\alpha_{\tau}),M}_{T^2,0}(z,\tau)$ has the phase $e^{2\pi i(j+\alpha_1)\alpha_{\tau}/M}$.} to set ${\cal N}^{ab}_{0,j} = |{\cal N}^{ab}_{0,j}| e^{2\pi i \frac{j+\alpha_1}{M} \alpha_{\tau}}$.
Therefore, we consider the following $M$ number of zero mode wave functions on the magnetized $T^2$,
\begin{align}
\begin{array}{l}
\psi^{(j+\alpha_1,\alpha_{\tau}),M}_{T^2,0}(z,\tau) = e^{-\frac{\pi M}{2{\rm Im}\tau}|z|^2} h^{(j+\alpha_1,\alpha_{\tau}),M}_{T^2}(z), \\
h^{(j+\alpha_1,\alpha_{\tau}),M}_{T^2,0}(z) = ( M )^{1/4} e^{2\pi i \frac{j+\alpha_1}{M} \alpha_{\tau}} e^{\frac{\pi M}{2{\rm Im}\tau}z^2} \vartheta
\begin{bmatrix}
\frac{j+\alpha_1}{M} \\ -\alpha_{\tau}
\end{bmatrix}
\left( Mz, M\tau \right).
\end{array}
\label{eq:zeroT2}
\end{align}

Next, let us see wave functions of the $n$ th excited modes.
By further acting the Dirac operator on the Dirac equation, we can obtain the characteristic equations,
\begin{align}
\begin{pmatrix}
{\cal D}^{\dagger} {\cal D} & 0 \\
0 & {\cal D} {\cal D}^{\dagger}
\end{pmatrix}
\begin{pmatrix}
\psi_{T^2,+,n}(z) \\ \psi_{T^2,-,n}(z)
\end{pmatrix}
= m_n^2
\begin{pmatrix}
\psi_{T^2,+,n}(z) \\ \psi_{T^2,-,n}(z)
\end{pmatrix},
\label{eq:charactereqspinor}
\end{align}
where ${\cal D}$ and ${\cal D}^{\dagger}$ satisfy the commutation relations:
\begin{align}
[ {\cal D}_{aa}, {\cal D}^{\dagger}_{aa} ] = 0, &\quad [ {\cal D}_{bb}, {\cal D}^{\dagger}_{bb} ] = 0, \label{eq:DDdaggeraabb} \\
[ {\cal D}_{ab}, {\cal D}^{\dagger}_{ab} ] = \frac{4\pi M}{{\cal A}}, &\quad [ {\cal D}_{ba}, {\cal D}^{\dagger}_{ba} ] = \frac{-4\pi M}{{\cal A}}. \label{eq:DDdaggerabba}
\end{align}
For $\psi^{aa}_{T^2,\pm}$ and $\psi^{bb}_{T^2,\pm}$,
the solutions satisfying the BCs in Eqs.~(\ref{eq:BCPhiz1}) and (\ref{eq:BCPhiztau}) are
\begin{align}
\begin{array}{l}
\psi^{aa}_{T^2,+,\mathbf{n}}(z) = {\cal N}^{aa}_{\mathbf{n}} e^{2\pi i\frac{{\rm Im}(\bar{\mathbf{n}}z)}{{\rm Im}\tau}}, \quad \psi^{aa}_{T^2,-,\mathbf{n}} (z) = \overline{\psi^{aa}_{T^2,+,\mathbf{n}}(z)}, \\
\psi^{bb}_{T^2,+,\mathbf{n}}(z) = {\cal N}^{bb}_{\mathbf{n}} e^{2\pi i\frac{{\rm Im}(\bar{\mathbf{n}}z)}{{\rm Im}\tau}}, \quad \psi^{bb}_{T^2,-,\mathbf{n}} (z) = \overline{\psi^{bb}_{T^2,+,\mathbf{n}}(z)},
\end{array}
\label{eq:Diracmassiveaabb}
\end{align}
with
\begin{align}
m_{\mathbf{n}}^2 = \frac{(2\pi)^2}{{\cal A}}\frac{|\mathbf{n}|^2}{{\rm Im}\tau} = \left| \frac{2\pi}{{\cal A}} \mathbf{n}_u \right|^2, \label{eq:massaabb}
\end{align}
where $\mathbf{n} = n_1 + \tau n_2 = (n_1e_1+n_2e_2)/e_1 \equiv \mathbf{n}_u/e_1$, which corresponds to the coordinate of the lattice point.
In order for the normalization factors ${\cal N}^{aa,bb}_{\mathbf{n}}$ not to depend on the geometrical parameters, we set the normalization conditions as
\begin{align}
\int_{T^2} dz d\bar{z} \sqrt{|{\rm det}(2h)|} \overline{\psi^{aa,,bb}_{T^2,+,\mathbf{n}'}(z)} \psi^{aa,bb}_{T^2,+,\mathbf{n}}(z) = {\cal A} \delta_{\mathbf{n}',\mathbf{n}}.
\label{eq:normalizationaabbT2}
\end{align}
On the other hand, let us consider
$\psi^{ab}_{\pm}$ and $\psi^{ba}_{\pm}$.
Note that when $M>0$, $j$ th zero mode of $\psi^{ab}_{T^2,+,0}$ in Eq.~(\ref{eq:zeroT2}) satisfies
\begin{align}
{\cal D}_{ab} \psi^{(j+\alpha_1,\alpha_{\tau}),M}_{T^2,0}(z,\tau) = 0, \label{eq:lowestDab}
\end{align}
while it also becomes of the $j$ th zero mode of $\psi^{ba}_{T^2,-,0} = \overline{\psi^{ab}_{T^2,+,0}}$, that is, it also satisfies
\begin{align}
{\cal D}^{\dagger}_{ba} \overline{\psi^{(j+\alpha_1,\alpha_{\tau}),M}_{T^2,0}(z,\tau)} = 0. \label{eq:lowestDdaggerba}
\end{align}
Here, when we define
$a^{(\dagger)} \equiv \sqrt{\frac{{\cal A}}{4\pi M}} {\cal D}^{(\dagger)}$,
they satisfy
\begin{align}
\begin{array}{lll}
\ [a_{ab}, a^{\dagger}_{ab}] = 1, & [a^{\dagger}_{ab}a_{ab}, a_{ab}] = -a_{ab}, & [a^{\dagger}_{ab}a_{ab}, a^{\dagger}_{ab}] = a^{\dagger}_{ab}, \\
\ [a^{\dagger}_{ba}, a_{ba}] = 1, & [a_{ba}a^{\dagger}_{ba}, a^{\dagger}_{ba}] = -a^{\dagger}_{ba}, &  [a_{ba}a^{\dagger}_{ba}, a_{ba}] = a_{ba}.
\end{array}
\label{eq:aadaggerabba}
\end{align}
This means that $a_{ab}$ and $a^{\dagger}_{ba}$ can be regarded as lowering operators while $a^{\dagger}_{ab}$ and $a_{ba}$ can be regarded as raising operators, and then $a^{\dagger}_{ab}a_{ab}$ and $a_{ba}a^{\dagger}_{ba}$ can be regarded as number operators.
Thus, we can obtain
\begin{align}
&a^{\dagger}_{ab} a_{ab} \psi^{(j+\alpha_1,\alpha_{\tau}),M}_{T^2,n}(z,\tau) = n \psi^{(j+\alpha_1,\alpha_{\tau}),M}_{T^2,n}(z,\tau), \quad a_{ab} \psi^{(j+\alpha_1,\alpha_{\tau}),M}_{T^2,0}(z,\tau) = 0, \label{eq:aadaggerab} \\
\Rightarrow \ &
\psi^{(j+\alpha_1,\alpha_{\tau}),M}_{T^2,n}(z,\tau) = \frac{1}{\sqrt{n!}} (a^{\dagger}_{ab})^n \psi^{(j+\alpha_1,\alpha_{\tau}),M}_{T^2,0}(z,\tau), \label{eq:nthT2} \\
&a_{ba} a^{\dagger}_{ba} \overline{\psi^{(j+\alpha_1,\alpha_{\tau}),M}_{T^2,n}(z,\tau)} = n \overline{\psi^{(j+\alpha_1,\alpha_{\tau}),M}_{T^2,n}(z,\tau)}, \quad a^{\dagger}_{ba} \overline{\psi^{(j+\alpha_1,\alpha_{\tau}),M}_{T^2,0}(z,\tau)} = 0,
\label{eq:adaggeraba} \\
\Rightarrow \ &
\overline{\psi^{(j+\alpha_1,\alpha_{\tau}),M}_{T^2,n}(z,\tau)} = \frac{1}{\sqrt{n!}} (a_{ba})^n \overline{\psi^{(j+\alpha_1,\alpha_{\tau}),M}_{T^2,0}(z,\tau)}, \label{eq:massiveba-}
\end{align}
which correspond to the $j$ th mode of $n$ th excited modes of $\psi^{ab}_{T^2,+}$ and $\psi^{ba}_{T^2,-}$,~i.e.,
\begin{align}
\psi^{ab}_{T^2,+,n}(z) = \sum_{j=0}^{M-1} \psi^{(j+\alpha_1,\alpha_{\tau}),M}_{T^2,n}(z,\tau), \quad \psi^{ba}_{T^2,-,n}(z) = \overline{\psi^{ab}_{+,n}(z)} = \sum_{j=0}^{M-1} \overline{\psi^{(j+\alpha_1,\alpha_{\tau}),M}_{T^2,n}(z,\tau)},
\label{eq:Diracmassiveab+ba-}
\end{align}
with
\begin{align}
m_n^2 = \frac{4\pi M}{{\cal A}}n. \label{eq:massabba}
\end{align}
Note that when we rewrite zero mode wave functions as
\begin{align}
\begin{array}{l}
\psi^{(j+\alpha_1,\alpha_{\tau}),M}_{T^2,0}(z,\tau) = {\cal N}^{ab}_{0,j} \sum_{l \in \mathbb{Z}} \Theta^{(j+\alpha_1,\alpha_{\tau}),M}_{l}, \\
\Theta^{(j+\alpha_1,\alpha_{\tau}),M}_{l} = e^{2\pi i\frac{j+\alpha_1}{M}\alpha_{\tau}} e^{\pi iM{\rm Re}\tau \left(\frac{j+\alpha_{1}}{M}\right)^2} e^{\pi iM{\rm Re}z\frac{{\rm Im}z}{{\rm Im}\tau}} e^{2\pi i(M{\rm Re}z-\alpha_{\tau})\left(\frac{j+\alpha_1}{M}+l\right)} e^{-\pi iM{\rm Im}\tau \left( \frac{{\rm Im}z}{{\rm Im}\tau} + \frac{j+\alpha_1}{M} + l\right)^2},
\end{array}
\notag
\end{align}
the $n$ th excited mode wave functions in Eq.~(\ref{eq:nthT2}) can be explicitly written by using the Hermite function,
\begin{align}
H_n(x) = (-1)^n e^{x^2} \frac{d^n}{dx^n} [e^{-x^2} ], \label{eq:Hermitefunc}
\end{align}
as
\begin{align}
\psi^{(j+\alpha_1,\alpha_{\tau}),M}_{T^2,n}(z,\tau) =  {\cal N}^{ab}_{n,j} \sum_{l \in \mathbb{Z}} \Theta^{(j+\alpha_1,\alpha_{\tau}),M}_{l} H_n\left( \sqrt{2\pi M{\rm Im}\tau} \left( \frac{{\rm Im}z}{{\rm Im}\tau} + \frac{j+\alpha_1}{M} + l \right) \right), \label{eq:nthT2explicit}
\end{align}
where normalization factors ${\cal N}^{ab}_{n,j}$ are determined by the normalization condition,
\begin{align}
\int_{T^2} dz d\bar{z} \sqrt{|{\rm det}(2h)|} \overline{\psi^{(j+\alpha_1,\alpha_{\tau}),M}_{T^2,n'}(z,\tau)} \psi^{(k+\alpha_1,\alpha_{\tau}),M}_{T^2,n}(z,\tau) = {\cal A} (2{\rm Im}\tau)^{-1/2} \delta_{j,k} \delta_{n',n},
\label{eq:normalizationmassiveT2}
\end{align}
and then they are related to those of zero modes: ${\cal N}^{ab}_{n,j} = (2^n n!)^{-1/2} {\cal N}^{ab}_{0,j}$.
On the other hand,
from Eq.~(\ref{eq:Dirac+}),
\begin{align}
{\cal D}_{ab} \psi^{ab}_{T^2,+,n}(z) = m_n \psi^{ab}_{T^2,-,n}(z) \ \Leftrightarrow \ a_{ab} \psi^{ab}_{T^2,+,n}(z) = \sqrt{n} \psi^{ab}_{T^2,-,n}(z), \label{eq:ab+-}
\end{align}
and the fact that
\begin{align}
a_{ab} \psi^{(j+\alpha_1,\alpha_{\tau}),M}_{T^2,n}(z,\tau)
= a_{ab} \left( \frac{1}{\sqrt{n}} a^{\dagger}_{ab} \psi^{(j+\alpha_1,\alpha_{\tau}),M}_{T^2,n-1}(z,\tau) \right)
= \sqrt{n} \psi^{(j+\alpha_1,\alpha_{\tau}),M}_{T^2,n-1}(z,\tau),
\label{eq:ann1}
\end{align}
we can also find the solutions for $\psi^{ab}_{T^2,-,n}$ and $\psi^{ba}_{T^2,+,n}$ as
\begin{align}
\psi^{ab}_{T^2,-,n}(z) = \sum_{j=0}^{M-1} \psi^{(j+\alpha_1,\alpha_{\tau}),M}_{T^2,n-1}(z,\tau), \quad \psi^{ba}_{T^2,+,n}(z) = \overline{\psi^{ab}_{-,n}(z)} = \sum_{j=0}^{M-1} \overline{\psi^{(j+\alpha_1,\alpha_{\tau}),M}_{T^2,n-1}(z,\tau)},
\label{eq:Diracmassiveab-ba+}
\end{align}
with the eigenvalue $m_n^2$ in Eq.~(\ref{eq:massabba}).
We comment that Eq.~(\ref{eq:ann1}) is also related to the property of the Hermite function:
\begin{align}
\frac{d}{dx} H_n(x) = 2n H_{n-1}(x). \label{eq:propHermitefunc}
\end{align}

\begin{itemize}
\item {\large Scalar fields}
\end{itemize}

Second, let us see wave functions of $U(N)$ adjoint scalar fields on the magnetized $T^2$,
\begin{align}
\phi_{T^2}(z) =
\begin{pmatrix}
\phi_{T^2}^{aa}(z) & \phi_{T^2}^{ab}(z) \\
\phi_{T^2}^{ba}(z) & \phi_{T^2}^{ba}(z)
\end{pmatrix},
\label{eq:scalar}
\end{align}
which satisfy the Klein-Gordon equation,
\begin{align}
\left( h^{z\bar{z}} (\hat{D}_{z} \hat{D}_{\bar{z}} + \hat{D}_{\bar{z}} \hat{D}_{z}) + m_{n}^2 \right) \phi_{T^2,n}(z) = 0,
\label{eq:KGeqX}
\end{align}
under the BCs in Eqs~(\ref{eq:BCPhiz1}) and (\ref{eq:BCPhiztau}).
Here, we define the Laplace operator,
\begin{align}
\hat{\Delta} \equiv -h^{z\bar{z}} (\hat{D}_z \hat{D}_{\bar{z}} + \hat{D}_{\bar{z}} \hat{D}_z ) = \frac{1}{2} \left\{ {\cal D}, {\cal D}^{\dagger} \right\} =  {\cal D} {\cal D}^{\dagger} - \frac{1}{2} \left[ {\cal D}, {\cal D}^{\dagger} \right] =  {\cal D}^{\dagger} {\cal D} + \frac{1}{2} \left[  {\cal D}, {\cal D}^{\dagger} \right],
\label{eq:LaplaceOP}
\end{align}
and then the Klein-Gordon equation (\ref{eq:KGeqX}) can be rewritten as
\begin{align}
\hat{\Delta} \phi_{T^2,n}(z) = m_n^2 \phi_{T^2,n}(z).
\label{eq:KG}
\end{align}
From Eq.~(\ref{eq:LaplaceOP}) and Eqs.~(\ref{eq:charactereqspinor})-(\ref{eq:Diracmassiveab-ba+}),
the well-defined solutions are
\begin{align}
\phi^{aa}_{T^2,n}(z) = \psi^{aa}_{T^2,+,\mathbf{n}}(z), &\quad m_{\mathbf{n}}^2 = \frac{(2\pi)^2}{{\cal A}}\frac{|\mathbf{n}|^2}{{\rm Im}\tau} = \left| \frac{2\pi}{{\cal A}} \mathbf{n}_u \right|^2 \ (|\mathbf{n}| \geq 0), \label{eq:KGmassiveaa} \\
\phi^{bb}_{T^2,n}(z) = \psi^{bb}_{T^2,+,\mathbf{n}}(z), &\quad m_{\mathbf{n}}^2 = \frac{(2\pi)^2}{{\cal A}}\frac{|\mathbf{n}|^2}{{\rm Im}\tau} = \left| \frac{2\pi}{{\cal A}} \mathbf{n}_u \right|^2 \ (|\mathbf{n}| \geq 0), \label{eq:KGmassivebb} \\
\phi^{ab}_{T^2,n}(z) = \psi^{ab}_{T^2,+,n}(z), &\quad m_n^2 = \frac{4\pi M}{{\cal A}}\left(n+\frac{1}{2}\right) \ (n \geq 0), \label{eq:KGmassiveab} \\
\phi^{ba}_{T^2,n}(z) = \psi^{ba}_{T^2,+,n+1}(z), &\quad m_n^2 = \frac{4\pi M}{{\cal A}}\left(n+\frac{1}{2}\right) \ (n \geq 0). \label{eq:KGmassiveba}
\end{align}
We note that even the lowest modes of both $\phi^{ab}_{T^2}$ and $\phi^{ba}_{T^2}$ are massive while those of $\phi^{aa}_{T^2}$ and $\phi^{bb}_{T^2}$ are massless and constants.

\begin{itemize}
\item {\large Vector fields}
\end{itemize}

Third, let us see wave functions of $U(N)$ adjoint vector fields on magnetized $T^2$,
\begin{align}
&A_{T^2}(z) \notag \\
=& A_{T^2,z}(z) dz + A_{T^2,\bar{z}} d\bar{z} \notag \\
=&
\begin{pmatrix}
\langle A^{aa}_{z}(z) \rangle + A^{aa}_{T^2,z}(z) & A^{ab}_{T^2,z}(z) \\
A^{ba}_{T^2,z}(z) & \langle A^{bb}_{z}(z) \rangle + A^{bb}_{T^2,z}(z)
\end{pmatrix}
dz \notag \\ 
&+
\begin{pmatrix}
\langle A^{aa}_{\bar{z}}(z) \rangle + A^{aa}_{T^2,\bar{z}}(z) & A^{ab}_{T^2,\bar{z}}(z) \\
A^{ba}_{T^2,\bar{z}}(z) & \langle A^{bb}_{\bar{z}}(z) \rangle + A^{bb}_{T^2,\bar{z}}(z)
\end{pmatrix}
d\bar{z}, 
\notag \\
&\left( A_{T^2\bar{z}} = (A_{T^2,z}(z))^{\dagger} \right),
\label{eq:vector}
\end{align}
which satisfy the Yang-Mills-Proca equation,
\begin{align}
h^{z\bar{z}} \hat{D}_{z} F_{T^2,\bar{z}z} + m_{n}^2 A_{T^2,z,n}(z) = 0, \quad \left( h^{\bar{z}z} \hat{D}_{\bar{z}} F_{T^2,z\bar{z}} + m_{n}^2 A_{T^2,\bar{z},n}(z) = 0 \right),
\label{eq:YMPeqX}
\end{align}
with the gauge-fixing condition,
\begin{align}
h^{z\bar{z}} \left(\hat{D}_{z}A_{T^2,\bar{z},n}(z) + \hat{D}_{\bar{z}} A_{T^2,z,n}(z) \right) = 0,
\label{eq:GFCX}
\end{align}
under the BCs in Eqs.~(\ref{eq:BCAz1}) and (\ref{eq:BCAztau}).
The field strength in Eq.~(\ref{eq:YMPeqX}) can be rewritten as
\begin{align}
F_{T^2,z\bar{z}}(z)
&= \langle F_{z\bar{z}} \rangle + \hat{D}_z A_{T^2,\bar{z}}(z) - \hat{D}_{\bar{z}} A_{T^2,z}(z) \notag\\
&= \langle F_{z\bar{z}} \rangle + 2 \hat{D}_z A_{T^2,\bar{z}}(z) \label{eq:Fre} \\
&= \langle F_{z\bar{z}} \rangle - 2 \hat{D}_{\bar{z}} A_{T^2,z}(z), \notag
\end{align}
where we use the gauge fixing condition (\ref{eq:GFCX}) in the second and third lines.
Then, the Yang-Mills-Proca equation in Eq.~(\ref{eq:YMPeqX}) can be rewritten by
\begin{align}
&{\cal D}^{\dagger} {\cal D} A_{T^2,z,n}(z) + ih^{z\bar{z}} [\langle F_{z\bar{z}} \rangle, A_{T^2,z,n}(z)] = m_n^2 A_{T^2,z,n}(z), \notag \\
\Leftrightarrow \ &
\left( \hat{\Delta} - [{\cal D}, {\cal D}^{\dagger}] \right) A_{T^2,z,n}(z) = m_n^2 A_{T^2,z,n}(z), \label{eq:YMPeqT2} \\
&{\cal D} {\cal D}^{\dagger} A_{T^2,\bar{z},n} -i h^{z\bar{z}} [\langle F_{z\bar{z}} \rangle, A_{T^2,\bar{z},n}(z)] = m_n^2 A_{T^2,\bar{z},n}(z), \notag \\
\Leftrightarrow \ &
\left( \hat{\Delta} + [{\cal D}, {\cal D}^{\dagger}] \right) A_{T^2,\bar{z},n}(z) = m_n^2 A_{T^2,\bar{z},n}(z). \label{eq:YMPeqT2}
\end{align}
Therefore, by using the results in Eqs.~(\ref{eq:KGmassiveaa})-(\ref{eq:KGmassiveba}), the well-defined solutions are
\begin{align}
A^{aa}_{T^2,z,n}(z) = \psi^{aa}_{T^2,+,\mathbf{n}}(z), &\quad A^{aa}_{T^2,\bar{z},n}(z) = \overline{A^{aa}_{T^2,z,n}(z)}, \quad m_{\mathbf{n}}^2 = \frac{(2\pi)^2}{{\cal A}}\frac{|\mathbf{n}|^2}{{\rm Im}\tau} = \left| \frac{2\pi}{{\cal A}} \mathbf{n}_u \right|^2 \ (|\mathbf{n}| \geq 0), \label{eq:YMPmassiveaa} \\
A^{bb}_{T^2,z,n}(z) = \psi^{bb}_{T^2,+,\mathbf{n}}(z), &\quad A^{bb}_{T^2,\bar{z},n}(z) = \overline{A^{bb}_{T^2,z,n}(z)}, \quad m_{\mathbf{n}}^2 = \frac{(2\pi)^2}{{\cal A}}\frac{|\mathbf{n}|^2}{{\rm Im}\tau} = \left| \frac{2\pi}{{\cal A}} \mathbf{n}_u \right|^2 \ (|\mathbf{n}| \geq 0), \label{eq:YMPmassivebb} \\
A^{ab}_{T^2,z,n}(z) = \psi^{ab}_{T^2,+,n}(z), &\quad A^{ba}_{T^2,\bar{z},n}(z) = \overline{A^{ab}_{T^2,z,n}(z)}, \quad m_n^2 = \frac{4\pi M}{{\cal A}}\left(n-\frac{1}{2}\right) \ (n \geq 0), \label{eq:YMPmassiveab} \\
A^{ba}_{T^2,z,n}(z) = \psi^{ba}_{T^2,+,n-1}(z), &\quad A^{ab}_{T^2,\bar{z},n}(z) = \overline{A^{ba}_{T^2,z,n}(z)}, \quad m_n^2 = \frac{4\pi M}{{\cal A}}\left(n-\frac{1}{2}\right) \ (n \geq 2). \label{eq:YMPmassiveba}
\end{align}
We note that the lowest mode of $A^{ab}_{T^2,z}$ ($A^{ba}_{T^2,\bar{z}}$) becomes tachyonic while that of $A^{ba}_{T^2,z}$ ($A^{ab}_{\bar{z}}$) becomes massive.
On the other hand, the lowest modes of $A^{aa}_{T^2,z}$ ($A^{aa}_{T^2,\bar{z}}$) and $A^{bb}_{T^2,z}$ ($A^{bb}_{T^2,\bar{z}}$) are massless but they are constants, which correspond to the WL phases.


\subsection{4D EFT}
\label{subsec:4DEFT}

Now, let us see 4D EFT of 10D ${\cal N}=1$ SYM theory, in which the action is given by Eq.~(\ref{eq:actionSYM}), on ${\cal M}_{10} = {\cal M}_4 \times T^2_1 \times T^2_2 \times T^2_3$, where the magnetic fluxes are inserted on each $T^2_i\ (i=1,2,3)$.
First, we denote the real coordinate of ${\cal M}_4$ as $x^{\mu}\ (\mu=0,...,3)$ and the complex coordinate of $T^2_i$ as $(z_i, \bar{z}_i)$.
The metric of ${\cal M}_{10}$ is defined as
\begin{align}
ds^2 &= G_{MN} dX^M dX^N \notag \\
&= g_{\mu\nu} dx^{\mu} dx^{\nu} + \sum_{i=1,2,3} 2h_{z_i \bar{z}_i} dz_id\bar{z}_i. \label{eq:10Dmetrix}
\end{align}
$\Gamma^M$ matrices
are decomposed as
\begin{align}
\begin{array}{ll}
\Gamma^{\mu} = \gamma^{\mu} \otimes \mathbb{I}_2 \otimes \mathbb{I}_2 \otimes \mathbb{I}_2, \\
\Gamma^{z_1} = \gamma_5 \otimes \gamma^{z_1} \otimes \mathbb{I}_2 \otimes \mathbb{I}_2, & \Gamma^{\bar{z}_1} = \gamma_5 \otimes \gamma^{\bar{z}_1} \otimes \mathbb{I}_2 \otimes \mathbb{I}_2, \\
\Gamma^{z_2} = \gamma_5 \otimes \sigma_3 \otimes \gamma^{z_2} \otimes \mathbb{I}_2, & \Gamma^{\bar{z}_2} = \gamma_5 \otimes \sigma_3 \otimes \gamma^{\bar{z}_2} \otimes \mathbb{I}_2, \\
\Gamma^{z_3} = \gamma_5 \otimes \sigma_3 \otimes \sigma_3 \otimes \gamma^{z_3}, & \Gamma^{\bar{z}_3} = \gamma_5 \otimes \sigma_3 \otimes \sigma_3 \otimes \gamma^{\bar{z}_3},
\end{array}
\label{eq:decgamma}
\end{align}
where $\{ \gamma^{\mu}, \gamma^{\nu} \} = 2 \eta^{\mu\nu} \mathbb{I}_{4}$ and $\{ \gamma^{z_i}, \gamma^{\bar{z}_i} \} = 2 h^{z_i\bar{z}_i} \mathbb{I}_2$ are satisfied.
The spinor fields $\lambda^{IJ}$ are decomposed as
\begin{align}
\lambda^{IJ}(X)
&= \sum_{n_1, n_2, n_3} \psi^{IJ(4)}_{n_1n_2n_3}(x) \otimes \psi^{IJ(2)}_{n_1}(z_1) \otimes \psi^{IJ(2)}_{n_2}(z_2) \otimes \psi^{IJ(2)}_{n_3}(z_3) \notag \\
&= \sum_{n_1, n_2, n_3}
\begin{pmatrix}
\psi_{L,n_1n_2n_3}^{IJ}(x) \\ \psi_{R,n_1n_2n_3}^{IJ}(x)
\end{pmatrix}
\otimes
\begin{pmatrix}
\psi_{+,n_1}^{IJ}(z_1) \\ \psi_{-,n_1}^{IJ}(z_1)
\end{pmatrix}
\otimes
\begin{pmatrix}
\psi_{+,n_2}^{IJ}(z_2) \\ \psi_{-,n_2}^{IJ}(z_2)
\end{pmatrix}
\otimes
\begin{pmatrix}
\psi_{+,n_3}^{IJ}(z_3) \\ \psi_{-,n_3}^{IJ}(z_3)
\end{pmatrix},
\label{eq:decspin}
\end{align}
where $\gamma_5 \psi_L^{IJ}(x) = - \psi_L^{IJ}(x)$, $\gamma_5 \psi_R^{IJ}(x) = + \psi_R^{IJ}(x)$, $\sigma_3 \psi_{\pm}^{IJ}(z_i) = \pm \psi_{\pm}^{IJ}(z_I)$, and $\psi^{IJ(2)} = \sigma_1 \overline{\psi^{JI(2)}}$ are satisfied.
On the other hand, the vector fields can be decomposed as
\begin{align}
\begin{array}{rcl}
A_{\mu}^{IJ}(X) &=& \sum_{n_1,n_2,n_3} A_{\mu,n_1n_2n_3}^{IJ}(x) \phi_{n_1}^{IJ}(z_1) \phi_{n_2}^{IJ}(z_2) \phi_{n_3}^{IJ}(z_3), \\
A_{z_1}^{IJ}(X) &=& \sum_{n_1,n_2,n_3} \phi^{(z_1)IJ}_{n_1n_2n_3}(x) A_{z_1,n_1}^{IJ}(z_1) \phi_{n_2}^{IJ}(z_2) \phi_{n_3}^{IJ}(z_3), \\
A_{\bar{z}_1}^{IJ}(X) &=& \sum_{n_1,n_2,n_3} \phi^{(\bar{z}_1)IJ}_{n_1n_2n_3}(x) A_{\bar{z}_1,n_1}^{IJ}(z_1) \phi_{n_2}^{IJ}(z_2) \phi_{n_3}^{IJ}(z_3) \\
&=& \sum_{n_1,n_2,n_3} \overline{ \phi^{(z_1)JI}_{n_1n_2n_3}(x) A_{z_1,n_1}^{JI}(z_1) \phi_{n_2}^{JI}(z_2) \phi_{n_3}^{JI}(z_3)  } = \overline{ A_{z_1}^{JI}(X) }, \\
A_{z_2}^{IJ}(X) &=& \sum_{n_1,n_2,n_3} \phi^{(z_2)IJ}_{n_1n_2n_3}(x) \phi_{n_1}^{IJ}(z_1)  A_{z_2,n_2}^{IJ}(z_2) \phi_{n_3}^{IJ}(z_3), \\
A_{\bar{z}_2}^{IJ}(X) &=& \sum_{n_1,n_2,n_3} \phi^{(\bar{z}_2)IJ}_{n_1n_2n_3}(x) \phi_{n_1}^{IJ}(z_1) A_{\bar{z}_2,n_2}^{IJ}(z_2) \phi_{n_3}^{IJ}(z_3) \\
&=& \sum_{n_1,n_2,n_3} \overline{ \phi^{(z_2)JI}_{n_1n_2n_3}(x) \phi_{n_1}^{JI}(z_1) A_{z_2,n_2}^{JI}(z_2) \phi_{n_3}^{JI}(z_3) } = \overline{ A_{z_2}^{JI}(X) }, \\
A_{z_3}^{IJ}(X) &=& \sum_{n_1,n_2,n_3} \phi^{(z_3)IJ}_{n_1n_2n_3}(x) \phi_{n_1}^{IJ}(z_1) \phi_{n_2}^{IJ}(z_2) A_{z_3,n_3}^{IJ}(z_3), \\
A_{\bar{z}_3}^{IJ}(X) &=& \sum_{n_1,n_2,n_3} \phi^{(z_3)IJ}_{n_1n_2n_3}(x) \phi_{n_1}^{IJ}(z_1) \phi_{n_2}^{IJ}(z_2) A_{z_3,n_3}^{IJ}(z_3) \\
&=& \sum_{n_1,n_2,n_3} \overline{ \phi^{(\bar{z}_3)JI}_{n_1n_2n_3}(x) \phi_{n_1}^{JI}(z_1) \phi_{n_2}^{JI}(z_2) A_{\bar{z}_3,n_3}^{JI}(z_3) } =  \overline{ A_{z_3}^{JI}(X) }.
\end{array}
\label{eq:decA}
\end{align}

Here, we consider the following Abelian homogeneous background magnetic fluxes
\begin{align}
\frac{1}{2\pi} \int_{T^2_i} \langle F_{z_i\bar{z}_i} \rangle dz_i \wedge d\bar{z}_i
=
\begin{pmatrix}
M^a_i \mathbb{I}_{N_a} &  &  \\
 & M^b_i \mathbb{I}_{N_b} & \\
 & & \ddots
\end{pmatrix}
\quad (N_a + N_b + \cdots = N),
\label{eq:MFX}
\end{align}
are inserted in each $T^2_i$.
Due to the magnetic fluxes $U(N)$ gauge symmetry is broken to $\prod_{A=a,b,...}U(N_{A})$.

Then, by applying wave functions on the magnetized $T^2_i\ (i=1,2,3)$ obtained in the previous subsection and calculating overlap integration of them, we can obtain the following 4D EFT:
\begin{align}
S_{4D} = &\int_{{\cal M}_4} d^4x \sqrt{|{\rm det}(g_{\mu\nu})|} \notag \\
&\times \Biggl[ \sum_{n_1,n_2,n_3} \left( -\frac{C^{IJ}}{4g_{4}^2} (F_{IJ}^{\mu\nu,n_1n_2n_3}(x))^2 + \frac{C^{IJ}}{2g_{4}^2}(\sum_{i=1,2,3}m_{n_i}^2) (A_{IJ}^{\mu,n_1n_2n_3}(x))^2 \right) \notag \\
&+ \sum_{i=1,2,3} \sum_{n_1,n_2,n_3} \Biggl( -\frac{C^{IJ}}{g_{4}^2} |D_{\mu}\phi^{(z_i)IJ}_{n_1n_2n_3}(x)|^2 + \frac{C^{IJ}}{g_4^2} (\sum_{i=1,2,3}m_{n_i}^2) |\phi^{(z_i)IJ}_{n_1n_2n_3}(x)|^2 \notag \\
&\qquad \qquad \quad \ 
+ \left( \frac{\hat{Y}^{(3)}_{n_1n_2n_3,-1_i}m_{n_i}}{2g_{4}^2} \sqrt{\left(\frac{h^{z_i\bar{z}_i}}{2}\right)^3} \left\{ \left[ \phi^{(z_i)}(x), \phi^{(\bar{z}_i)}(x) \right] \phi^{(z_i)}(x) \right\}_{n_1n_2n_3,-1_i} +{\rm h.c.} \right) \notag \\
&\qquad \qquad \quad \ 
+ \frac{\hat{Y}^{(4)}_{n_1n_2n_3}}{g_{4}^2}\left| \frac{h^{z_i\bar{z_i}}}{2} \left[ \phi^{(z_i)}(x), \phi^{(\bar{z}_i)}(x) \right]^{IJ}_{n_1n_2n_3} \right|^2 \Biggl) \notag \\
&+\sum_{n_1,n_2,n_3} \Biggl( \frac{C^{IJ}}{2g_4^2} \overline{\psi^{(4)IJ}_{n_1n_2n_3}(x)} (i \gamma^{\mu}D_{\mu} + \sum_{i=1,2,3}m_{n_i}) \psi^{(4)IJ}_{n_1n_2n_3}(x) \notag \\
&\quad \quad \quad \ 
+ \frac{\hat{Y}^{(3)}_{n_1n_2n_3}}{2g_4^2} \overline{\psi^{(4)IJ}_{n_1n_2n_3}(x)} \left[ \phi^{(\slashed{z_i})}(x), \psi^{(4)}(x) \right]^{JI}_{n_1n_2n_3} \Biggl) 
+ \sum_{i=1,2,3} \frac{1}{4g_4^2} \left( \sum_{A} \frac{2\pi M^A}{{\cal A}_i} \right)^2 \Biggl],
\label{eq:4DEFT}
\end{align}
where, each coefficient can be obtained from overlap integration of wave functions on the magnetized $T^2_i$~\footnote{In the 4D scalar three-point couplings, $2i\sqrt{\frac{h^{z_i\bar{z_i}}}{2}}\hat{D}_{\bar{z}_i} A_{z_i,n_i} = m_{n_i} A_{z_i,n_i-1}$ is used.}.
In more detail, we need to calculate two, three, and four point couplings of them.

The two-point couplings are given by normalization conditions in Eqs.~(\ref{eq:normalizationaabbT2}) and (\ref{eq:normalizationmassiveT2}).
Both of them include the are of $T^2_i$, ${\cal A}_i$, which has mass dimension $-2$.
Then, we can obtain the mass-dimensionless gauge coupling on ${\cal M}_4$, $g_4^{-2} \equiv {\cal A}_1{\cal A}_2{\cal A}_3g_{10}^{-2}$.
The two-point couplings between bi-fundamental fields under the remaining gauge group are also proportional to $C^{IJ} =\prod_{i=1,2,3}  (2{\rm Im}\tau_i)^{-1/2} = \prod_{i=1,2,3} (i\bar{\tau}_i-i\tau_i)^{-1/2}$.

Next, let us see three-point couplings.
The non-trivial three-point couplings are such as couplings among bi-fundamental fields with $(N_a, \bar{N}_b)$, $(N_b, \bar{N}_c)$, and $(N_c, \bar{N}_a)$, which feel the magnetic fluxes $M^{ab}_i \equiv M^a_i - M^b_i$, $M^{bc}_i \equiv M^b_i - M^c_i$, and $M^{ca}_i \equiv M^c_i - M^a_i$, respectively.
Obviously, $M^{ab}_i + M^{bc}_i + M^{ca}_i = 0$ and then one of them whose absolute value is largest has different sign from the others.
For example, let us consider that on one $T^2$, $M^{ab}, M^{bc}>0$, $M^{ca}<0$, and $|M^{ca}| = M^{ac} = M^{ab}+M^{bc}$ are satisfied.
In this case, the non-trivial three-point couplings are written by
\begin{align}
\hat{y}^{IJK}_{T^2,n^I n^J n^K}
=
\int_{T^2} dz d\bar{z} \sqrt{|{\rm det}(2h_{z\bar{z}})|} \overline{\psi^{(K+\alpha^{ac}_{1},\alpha^{ac}_{\tau}),M^{ac}}_{T^2,n^K}(z,\tau)} \psi^{(I+\alpha^{ab}_{1},\alpha^{ab}_{\tau}),M^{ab}}_{T^2,n^I}(z,\tau) \psi^{(J+\alpha^{bc}_{1},\alpha^{bc}_{\tau}),M^{bc}}_{T^2,n^J}(z,\tau), \label{eq:3pointintegral} 
\end{align}
where $\alpha^{AB}_{1,\tau} \equiv \alpha^{A}_{1,\tau} - \alpha^{B}_{1,\tau}\ (A,B=a,b,c)$ and they satisfy $\alpha^{ab}_{1,\tau}+\alpha^{bc}_{1,\tau}=\alpha^{ac}_{1,\tau}$.
By applying the multiple relation~\cite{Cremades:2004wa,Hamada:2012wj,Fujimoto:2016zjs}\footnote{See also Ref.~\cite{Almumin:2021fbk}.},
\begin{align}
&\psi^{(I+\alpha^{ab}_{1},\alpha^{ab}_{\tau}),M^{ab}}_{T^2,n^I}(z,\tau) \psi^{(J+\alpha^{bc}_{1},\alpha^{bc}_{\tau}),M^{bc}}_{T^2,n^J}(z,\tau) \notag \\
&= \sum_{p=0}^{M^{ac}-1} \sum_{s^I=0}^{n^I} \sum_{s^J=0}^{n^J} y^{IJp,M^y}_{T^2,n^In^J(s^I,s^J)}(\tau) \psi^{(I+J+pM^{ab}+\alpha^{ac}_{1},\alpha^{ac}_{\tau}),M^{ac}}_{T^2,s^I+s^J}(z,\tau), \label{eq:multiplerel} \\
&y^{IJp,M^y}_{T^2,n^In^J(s^I,s^J)}(\tau) = {}_{n^I}C_{s^I} {}_{n^J}C_{s^J} (-1)^{n^J-s^J} \sqrt{\frac{(s^I+s^J)!(n^I-s^I+n^J-s^J)!}{n^I!n^J!}} \notag \\
&\times \sqrt{\frac{(M^{ab})^{s^I+n^J-s^J}(M^{bc})^{n^I-s^I+s^J}}{(M^{ac})^{n^I+n^J+1}}} \psi^{(L^{IJp}+\alpha^y_1,\alpha^y_{\tau}),M^y}_{T^2,n^y}(0,\tau),
\label{eq:ypM} \\
&\Biggl(
\begin{array}{l}
L^{IJp}+\alpha^y_1=M^{bc}(I+pM^{ab}+\alpha^{ab}_1)-M^{ab}(J+\alpha^{bc}_1), \ 
\alpha^y_{\tau} = M^{bc}\alpha^{ab}_{\tau} - M^{ab}\alpha^{bc}_{\tau} \\
M^y=M^{ab}M^{bc}M^{ac}, \quad n^y=n^I-s^I+n^J-s^J,
\end{array}\Biggl),
\label{eq:Lalpha1tauMn}
\end{align}
we can calculate Eq.~(\ref{eq:3pointintegral}) as
\begin{align}
\hat{y}^{IJK}_{T^2,n^I n^J n^K} =& \sum_{p=0}^{M^{ac}-1} \sum_{s^I=0}^{n^I} \sum_{s^J=0}^{n^J} y^{IJp,M^y}_{T^2,n^In^J(s^I,s^J)}(\tau) \notag \\
&\times \int_{T^2} dz d\bar{z} \sqrt{|{\rm det}(2h_{z\bar{z}})|} \overline{\psi^{(K+\alpha^{ac}_{1},\alpha^{ac}_{\tau}),M^{ac}}_{T^2,n^K}(z,\tau)} \psi^{(I+J+pM^{ab}+\alpha^{ac}_{1},\alpha^{ac}_{\tau}),M^{ac}}_{T^2,s^I+s^J}(z,\tau) \notag \\
=& {\cal A}(2{\rm Im}\tau)^{-1/2} \sum_{p=0}^{M^{ac}-1} \sum_{s^I=0}^{n^I} \sum_{s^J=0}^{n^J} y^{IJp,M^y}_{T^2,n^In^J(s^I,s^J)}(\tau) \delta_{s^I+s^J,n^K} \delta_{I+J+pM^{ab},K+qM^{ac}} \notag \\
=& {\cal A}(2{\rm Im}\tau)^{-1/2} y^{IJK}_{T^2,n^In^Jn^K}(\tau), \label{eq:yhat}
\end{align}
with
\begin{align}
y^{IJK}_{T^2,n^In^Jn^K}(\tau) 
=& \sum_{p=0}^{M^{ac}-1} \sum_{s^I=0}^{n^I} \sum_{s^J=0}^{n^J} y^{IJp,M^y}_{T^2,n^In^J(\ell^I,\ell^J)}(\tau) \delta_{s^I+s^J,n^K} \delta_{I+J+pM^{ab},K+qM^{ac}} \notag \\
=& \sum_{p'=0}^{g-1} \sum_{s^J={\max}(0,n^K-n^I)}^{{\rm min}(n^J,n^K)} y^{IJ(p_0+(M^{ac}/g)p'),M^y}_{T^2,n^In^J(n^K-\ell,\ell^J)}(\tau) \notag \\
=& \sum_{s^J={\max}(0,n^K-n^I)}^{{\rm min}(n^J,n^K)} y^{IJp_0,\ell^y}_{T^2,n^In^J(n^K-\ell,\ell^J)}(\tau),
\label{eq:yK}
\end{align}
where $g^y \equiv {\rm gcd}(M^{ab},M^{bc},M^{ac})$, $\ell^y \equiv {\rm lcm}(M^{ab},M^{bc},M^{ac})$, $M^y=\ell^y (g^y)^2$, and $(p_0,q_0)$ is the solution of
\begin{align}
&\frac{M^{ac}}{g} q_0 - \frac{M^{ab}}{g} p_0 = \frac{I+J-K}{g} \in \mathbb{Z}, \quad \left( {\rm gcd}\left( \frac{M^{ac}}{g}, \frac{M^{ab}}{g}\right) =1 \right). \label{eq:pq}
\end{align}
In particular, when $n^I=n^J=0$, $n^K$ must be also $n^K=0$.
The 4D three-point coupling coefficient $\hat{Y}^{IJK}$ is given by $\hat{Y}^{IJK} = \prod_{i=1,2,3} (\hat{y}^{IJK}_{T^2_i,n_i^In_i^Jn_i^K}/{\cal A}_i)$, where $\prod_{i=1,2,3} {\cal A}_i$ is absorbed in the gauge coupling constant.
Now, let us see in detail the relation of $\hat{y}^{IJK}_{T^2,n^In^Jn^K}/{\cal A}$ and $y^{IJK}_{T^2,n^In^Jn^K}(\tau)$;
$y^{IJK}_{T^2,n^In^Jn^K}(\tau)$ becomes a holomorphic function of $\tau$, which can be regarded as the holomorphic three-point coupling in the supergravity theory in the following, while $\hat{y}^{IJK}_{T^2,n^I n^J n^K}/{\cal A}$ is the three-pint coupling in the global supersymmetric theory.
Since a superpotential in the global supersymmetric theory, $\hat{W}$, and one in the supergravity theory, $W(\tau)$, are related as
\begin{align}
|\hat{W}|^2 = e^K |W(\tau)|^2, \label{eq:WhatW}
\end{align}
with the K\"{a}hler potential, $K$ ,of the modulus $\tau$,
\begin{align}
K = - \ln{[i(\bar{\tau}-\tau)]} = - \ln{(2{\rm Im}\tau)}, \label{eq:Kahler}
\end{align}
we can write
\begin{align}
\frac{\hat{y}^{IJK}_{T^2,n^I n^J n^K}}{\cal A} = e^{K/2} y^{IJK}_{T^2,n^In^Jn^K}(\tau) = (2{\rm Im}\tau)^{-1/2} y^{IJK}_{T^2,n^In^Jn^K}(\tau).
\label{eq:yhaty}
\end{align}
Thus, $\hat{y}^{IJK}_{T^2,n^I n^J n^K}/{\cal A}$ and $y^{IJK}_{T^2,n^In^Jn^K}(\tau)$ are regarded as three-point couplings in the global supersymmetric theory and the supergravity theory, respectively.
We note that Eq.~(\ref{eq:Kahler}) is derived from the metric on the CS modulus space:
\begin{align}
ds^2_{CS} = 2\partial_{\tau}\partial_{\bar{\tau}}K d\tau d\bar{\tau} = \frac{1}{2{\cal A}} \int dzd\bar{z} \sqrt{|{\rm det}(2h_{z\bar{z}})|} (h^{z\bar{z}})^2 \delta h_{zz} \delta h_{\bar{z}\bar{z}} = 2\frac{-1}{(\tau-\bar{\tau})^2} d\tau d\bar{\tau},
\label{eq:modulusmetric}
\end{align}
where $\delta h_{zz}$ and $\delta h_{\bar{z}\bar{z}}$ can be obtained from the modulus shift $\tau \rightarrow \tau + d\tau$ for $dz = dy^1 + \tau dy^2$ and $d\bar{z} = dy^1+\bar{\tau}dy^2$ in the metric of $T^2$, $ds^2=2h_{z\bar{z}}dzd\bar{z}$,
\begin{align}
&dz \rightarrow \left( 1+ \frac{d\tau}{\tau-\bar{\tau}} \right) dz - \frac{d\tau}{\tau - \bar{\tau}} d\bar{z}, \quad
d\bar{z} \rightarrow \left( 1- \frac{d\bar{\tau}}{\tau-\bar{\tau}} \right) d\bar{z} + \frac{d\bar{\tau}}{\tau - \bar{\tau}} dz, \notag \\
\Rightarrow \ &
\delta h_{zz} = \frac{2h_{z\bar{z}}d\bar{\tau}}{\tau - \bar{\tau}}, \quad \delta h_{\bar{z}\bar{z}} = - \frac{2h_{z\bar{z}}d\tau}{\tau - \bar{\tau}}.
\label{eq:deltah}
\end{align}

Similarly, the four-point coupling (with $M^{ab}, M^{bc}, M^{cd}>0$, $M^{da}<0$, and $|M^{da}| = M^{ad} = M^{ab}+M^{bc}+M^{cd}$) can be also calculated as
\begin{align}
&\hat{y}^{IJKL}_{T^2,n^I n^J n^Kn^L} \notag \\
=&
\int_{T^2} dzd\bar{z} \sqrt{|{\rm det}(2h_{z\bar{z}})|} \overline{\psi^{(L+\alpha^{ad}_{1},\alpha^{ad}_{\tau}),M^{ad}}_{T^2,n^L}(z)} \psi^{(I+\alpha^{ab}_{1},\alpha^{ab}_{\tau}),M^{ab}}_{T^2,n^I}(z) \psi^{(J+\alpha^{bc}_{1},\alpha^{bc}_{\tau}),M^{bc}}_{T^2,n^J}(z) \psi^{(K+\alpha^{cd}_{1},\alpha^{cd}_{\tau}),M^{cd}}_{T^2,n^K}(z), 
\notag \\
=& \int_{T^2} d^2z d^2\hat{z} \sqrt{|{\rm det}(2h_{z\bar{z}})|} \overline{\psi^{(L+\alpha^{ad}_{1},\alpha^{ad}_{\tau}),M^{ad}}_{T^2,n^L}(\hat{z})} \psi^{(I+\alpha^{ab}_{1},\alpha^{ab}_{\tau}),M^{ab}}_{T^2,n^I}(\hat{z}) \notag \\
&\times \delta^2(\hat{z}-z) \psi^{(J+\alpha^{bc}_{1},\alpha^{bc}_{\tau}),M^{bc}}_{T^2,n^J}(z) \psi^{(K+\alpha^{cd}_{1},\alpha^{cd}_{\tau}),M^{cd}}_{T^2,n^K}(z), \notag \\
=& {\cal A}^{-1} (2{\rm Im}\tau)^{1/2} \times \notag \\
&\sum_{H,n^H} \left( \int_{T^2} d^2\hat{z} \sqrt{|{\rm det}(2h_{\hat{z}\bar{\hat{z}}})|} \overline{\psi^{(L+\alpha^{ad}_{1},\alpha^{ad}_{\tau}),M^{ad}}_{T^2,n^L}(\hat{z})} \psi^{(I+\alpha^{ab}_{1},\alpha^{ab}_{\tau}),M^{ab}}_{T^2,n^I}(\hat{z}) \psi^{(H+\alpha^{bd}_{1},\alpha^{bd}_{\tau}),M^{bd}}_{T^2,n^H}(\hat{z}) \right) \notag \\
&\times \left( \int_{T^2} d^2z \sqrt{|{\rm det}(2h_{z\bar{z}})|} \overline{\psi^{(H+\alpha^{bd}_{1},\alpha^{bd}_{\tau}),M^{bd}}_{T^2,n^H}(z,\tau)} \psi^{(J+\alpha^{bc}_{1},\alpha^{bc}_{\tau}),M^{bc}}_{T^2,n^J}(\hat{z},\tau) \psi^{(K+\alpha^{cd}_{1},\alpha^{cd}_{\tau}),M^{cd}}_{T^2,n^K}(z,\tau) \right) \notag \\
=& {\cal A} (2{\rm Im}\tau)^{-1/2} \sum_{H,n^H} y^{LIH}_{T^2,n^Ln^In^H}(\tau) y^{HJK}_{T^2,n^Hn^Jn^K}(\tau).
\label{eq:4pointintegral}
\end{align}
Thus, once we obtain three-point coupling, we can also calculate four-point couplings as well as higher order couplings~\cite{Abe:2009dr,Hamada:2012wj}.

Now, let us go back to the 4D effective action in Eq.~(\ref{eq:4DEFT}).
Here, since the mass scale ${\cal A}^{-2}$ is around $(O(10^{16}){\rm GeV})^2 \sim (O(10^{18}){\rm GeV})^2$, we focus on 4D massless modes.
For unbroken $U(N_A)\ (A=a,b,c,...)$ adjoint fields, $\Phi^{AA}$, the lowest constant modes are massless.
By reminding the chirality on ${\cal M}_{10}$ is fixed, there are four number of massless gauginos and ${}_4C_2 = 6$ number of massless scalar fields for one massless gauge boson.
That is, there remains 4D ${\cal N}=4$ supersymmetry for the unbroken $U(N_A)$ adjoint sector.
Next, let us see bi-fundamental fields under $U(N_A) \times U(N_B)\ (A,B=a,b,c,...,\ {\rm and}\ A \neq B)$, $\Phi^{AB}$.
First, 4D vector gauge bosons become massive, that is, the gauge symmetry between $U(N_A)$ and $U(N_B)$ is broken.
On the other hand, there are $\prod_{i=1,2,3} |M^{AB}_i|$ number of degenerate bi-fundamental massless chiral fermions like quarks and leptons as well as Higgsinos.
The chirality is determined by the signs of magnetic fluxes on each $T^2_i\ (i=1,2,3)$.
In addition, if the following condition,
\begin{align}
M^{AB}_{1,2,3} > 0, \quad &\frac{M^{AB}_1}{{\cal A}_1} + \frac{M^{AB}_2}{{\cal A}_2} = \frac{M^{AB}_3}{{\cal A}_3}, \notag \\
\Rightarrow \ &
\frac{M^{AB}_1}{{\cal A}_1} + \frac{M^{AB}_2}{{\cal A}_2} - \frac{M^{AB}_3}{{\cal A}_3} = 0, \label{eq:SUSYcond}
\end{align}
is satisfied, only $\phi^{(Z_3)AB}(x)$ has $\prod_{i=1,2,3} |M^{AB}_i|$ number of massless modes, and they become superpartners of the above chiral fermions like squarks and sleptons as well as Higgs bosons.
Thus, in such cases, there remains 4D ${\cal N}=1$ supersymmetry for the bi-fundamental sector.

Therefore, 4D effective action for the lowest fields is given as
\begin{align}
S_{4D}^0 =&\int_{{\cal M}_4} d^4x \sqrt{|{\rm det}(g_{\mu\nu})|} g_4^{-2} \notag \\
&\times \Biggl[ \sum_{A} \left( - \frac{1}{4} (F_{\mu\nu}^{AA}) - \sum_{i=1,2,3} |D_{\mu} \phi^{(z_i)AA}(x)|^2+ \sum_{i=1}^{4} \overline{\psi_i^{(4)AA}(x)} i\gamma^{\mu}D_{\mu} \psi_i^{(4)A}(x) \right) \label{eq:4Dgaugeadj} \\
&+\sum_{j=0}^{\prod_{i} |M^{AB}_i|} \left(- \frac{\left|D_{\mu} \phi_{j}^{(z_i)AB}\right|^2}{\prod_{i} (i\bar{\tau}_i-i\tau_i)^{1/2}} + \frac{\overline{\psi^{(4)AB}_j(x)}i\gamma^{\mu}D_{\mu}\psi^{(4)AB}_j(x)}{\prod_{i} (i\bar{\tau}_i-i\tau_i)^{1/2}} \right) \label{eq:4Dbifundkinetic} \\
&+ \sum_{j,k,l} \left( \hat{Y}^{ijl} \overline{\psi^{(4)IJ}_{l}(x)} \left[ \phi_j^{(\slashed{z_i})}(x), \psi_k^{(4)}(x) \right]^{IJ} + \left| \hat{Y}^{jkl} \frac{h^{z_i\bar{z_i}}}{2} \left[ \phi_j^{(z_i)}(x), \phi_k^{(\bar{z}_i)}(x) \right]^{IJ} \right|^2 \right)
\Biggl], \label{eq:4Dcoupling}
\end{align}
where Eqs.~(\ref{eq:4Dgaugeadj}), (\ref{eq:4Dbifundkinetic}), and (\ref{eq:4Dcoupling}) show that the 4D effective actions for the kinetic terms of unbroken gauge adjoint sector, the kinetic terms of bi-fundamental fields, and their coupling terms: Yukawa couplings and four-point scalar couplings, respectively.
In particular, to identify the $\prod_{i=1,2,3} |M^{AB}_i|$ degenerate chiral bi-fundamental fermions $\psi^{AB}$ as three-generational chiral fermions such as quarks and leptons, it is needed that $M^{AB}_1=3$ and $M^{AB}_2=M^{AB}_3$, which means that the three-generational structure comes from only $X_1=T^2_1$ and the contributions on $X_2=T^2_2$ and $X_3=T^2_3$ are just constants.
Then, we mainly focus on one 2D compact space to search flavor structure.

\section{Modular symmetry in magnetized $T^2$ models}
\label{sec:MSmagnetizedT2}

In this section, we discuss the modular symmetry in magnetized $T^2$ models.


\subsection{Modular symmetry on $T^2$}
\label{subsec:MS}

In this subsection, let us review the modular symmetry on $T^2$ and modular forms.~\footnote{See e.g. Refs.~\cite{Gunning:1962,Schoeneberg:1974,Koblitz:1984,Bruinier:2008} for the modular symmetry on $T^2$ and modular forms of even weight. See Ref.~ \cite{Liu:2019khw} for modular forms of odd modular weight. See Refs.~\cite{Schoeneberg:1974,Koblitz:1984,shimura,Duncan:2018wbw,Liu:2020msy} for modular forms of half-integral weight.}


\subsubsection{Modular symmetry}
\label{subsubsec:MS}

As mentioned before, $T^2$ is constructed by identifying the lattice $\Lambda=\{ \sum_{i=1,2} n_i e_i | n_i \in \mathbb{Z} \}$,~i.e. $T^2 \simeq \mathbb{C}/\Lambda$.
The same lattice can be spanned by lattice vectors $\gamma(e_i)\ (i=1,2)$ transformed by
\begin{align}
\Gamma=SL(2,\mathbb{Z})=\left\{ \gamma =
\begin{pmatrix}
a & b \\
c & d
\end{pmatrix}
\biggl| a,b,c,d \in \mathbb{Z}, {\rm det}\gamma=ad-bc=1 \right\},
\label{eq:Gamma}
\end{align}
such that
\begin{align}
\gamma:
\begin{pmatrix}
e_2 \\ e_1
\end{pmatrix}
\rightarrow
\begin{pmatrix}
\gamma(e_2) \\ \gamma(e_1)
\end{pmatrix}
=
\begin{pmatrix}
a & b \\
c & d
\end{pmatrix}
\begin{pmatrix}
e_2 \\ e_1
\end{pmatrix}
=
\begin{pmatrix}
a e_2 + b e_1 \\ c e_2 +d e_1
\end{pmatrix}.
\label{eq:SL2Zlatticevector}
\end{align}
Here, the generators of $SL(2,\mathbb{Z})$ are
\begin{align}
S=
\begin{pmatrix}
0 & 1 \\
-1 & 0
\end{pmatrix}, \quad
T=
\begin{pmatrix}
1 & 1 \\
0 & 1
\end{pmatrix},
\label{eq:STofSL2Z}
\end{align}
and they satisfy
\begin{align}
Z \equiv S^2=-\mathbb{I}_2, \quad Z^2=S^4=(ST)^3=\mathbb{I}_2, \quad ZT=TZ.
\label{eq:algSL2Z}
\end{align}
Under the lattice transformation, the CS modulus $\tau$ transforms as
\begin{align}
\gamma : \tau=\frac{e_2}{e_1} &\rightarrow \gamma(\tau)=\frac{\gamma(e_2)}{\gamma(e_1)}=\frac{a\tau+b}{c\tau+d}, \label{eq:modulartrans} \\
S : \tau &\rightarrow S(\tau) = -\frac{1}{\tau}, \label{eq:Stau} \\
T : \tau &\rightarrow T(\tau) = \tau +1, \label{eq:Ttau}
\end{align}
where $S$ and $T$ transformations for $\tau$ satisfy
\begin{align}
S^2 = (ST)^3 = \mathbb{I}. \label{eq:algPSL2Z}
\end{align}
This transformation is called the modular transformation, and the (inhomogeneous) modular group is $\bar{\Gamma} \equiv SL(2,\mathbb{Z})/\{\pm\mathbb{I}\} = PSL(2,\mathbb{Z})$.
Thus, there is a symmetry between a torus with a modulus $\tau$ and another torus with the modulus $\gamma(\tau)$.
This $\bar{\Gamma}$ symmetry for the modulus is called the modular symmetry.
Then, the fundamental region of the modulus $\tau$ becomes $|{\rm Re}\tau| \leq 1/2$ and ${\rm Im}\tau \geq \sqrt{1-({\rm Re}\tau)^2}$.

Although, in a 4D EFT with the modular symmetry, the modular transformed parameter is only the modulus $\tau$, not only the modulus $\tau$ but also the coordinate $z$ transform under the modular transformation as follows:
\begin{align}
\gamma : z=\frac{u}{e_1} &\rightarrow \gamma(z) = \frac{u}{\gamma(e_1)} = \frac{z}{c\tau+d}, \label{eq:modulartransz} \\
S : z &\rightarrow S(z) = -\frac{z}{\tau}, \label{eq:Sz} \\
T : z &\rightarrow T(z) = z. \label{eq:Sz}
\end{align}
Hereafter, we call $\gamma:(z,\tau) \rightarrow (\gamma(z,\tau))$ as the ``modular transformation''.
In particular, $(S^2(z,\tau)) = (-z,\tau)$.
Thus, $\Gamma = SL(2,\mathbb{Z})$ is important and it is called the full modular group (or homogeneous modular group).


\subsubsection{Modular forms}
\label{susec:MF}

Now, let us see modular transformation for functions of modular parameters.
When holomorphic functions of $\tau$, $f^j(\tau)$, satisfy
\begin{align}
&\gamma : f^j(\tau) \rightarrow
f^j(\gamma(\tau)) = J_k(\gamma, \tau) f^i(\tau), \quad J_k(\gamma, \tau) = (c\tau+d)^k, \quad \gamma =
\begin{pmatrix}
a & b \\
c & d
\end{pmatrix}
\in \Gamma,
\label{eq:MFGamma} \\
&S : f^j(\tau) \rightarrow
f^i(S(\tau)) = J_k(S, \tau) f^j(\tau), \quad J_k(S, \tau) = (-\tau)^k, \quad S =
\begin{pmatrix}
0 & 1 \\
-1 & 0
\end{pmatrix}
\in \Gamma,
\label{eq:MFGammaS} \\
&T : f^j(\tau) \rightarrow
f^i(T(\tau)) = J_k(T, \tau) f^j(\tau), \quad J_k(T, \tau) = 1, \qquad \quad T =
\begin{pmatrix}
1 & 1 \\
0 & 1
\end{pmatrix}
\in \Gamma,
\label{eq:MFGammaT}
\end{align}
$f^j(\tau)$ are called modular forms of weight $k$ for $\Gamma$, where $J_k(\gamma,\tau)$ denotes the automorphy factor with weight $k$ and satisfies
\begin{align}
J_k(\gamma_2\gamma_1,\tau) = J_k(\gamma_2,\gamma_1(\tau)) J_k(\gamma_1,\tau), \quad \gamma_1, \gamma_2 \in \Gamma. \label{eq:automorphyfactor}
\end{align}
Note that $k$ must be even since they transform under $S^2=-\mathbb{I}$ transformation as
\begin{align}
f^j(S^2(\tau)) = f^j(\tau) = (-1)^k f^j(\tau).
\label{eq:MFGammaS2}
\end{align}

Here, let us consider the principal congruence subgroup of level $N$, defined by
\begin{align}
\Gamma(N) \equiv \left\{ h=
\begin{pmatrix}
a' & b' \\
c' & d'
\end{pmatrix}
\in \Gamma \biggl|
\begin{pmatrix}
a' & b' \\
c' & d'
\end{pmatrix}
\equiv
\begin{pmatrix}
1 & 0 \\
0 & 1
\end{pmatrix}
\ ({\rm mod}\ N) \right\},
\label{eq:Gamma(N)}
\end{align}
which is a normal subgroup of $\Gamma$ ($\Gamma(N) \triangleleft \Gamma$).
Obviously, $\Gamma(1) \simeq \Gamma$.
$S^2=-\mathbb{I}$ is included in $\Gamma(N)$ only with $N=1,2$.
Then we also define $\bar{\Gamma} \equiv \Gamma/\{\pm\mathbb{I}\}$.
Similarly, holomorphic functions $f^j(\tau)$ satisfying
\begin{align}
h : f^j(\tau) \rightarrow
f^j(h(\tau)) = J_k(h, \tau) f^j(\tau), \quad J_k(h, \tau) = (c'\tau+d')^k, \quad h =
\begin{pmatrix}
a' & b' \\
c' & d'
\end{pmatrix}
\in \Gamma(N),
\label{eq:MFGamma(N)}
\end{align}
are called modular forms of weight $k$ for $\Gamma(N)$.
When $N=2$, $k$ must be even because of the same reason for $\Gamma(1) \simeq \Gamma$.
When $N>2$, on the other hand, $k$ must be integer because of $(1)^k=1$.
They transform under $\gamma \in \Gamma$ transformation as
\begin{align}
&\gamma : f^j(\tau) \rightarrow
f^j(\gamma(\tau)) = J_k(\gamma, \tau) \rho(\gamma)_{jj'} f^{j'}(\tau), \quad J_k(\gamma, \tau) = (c\tau+d)^k, \quad \gamma =
\begin{pmatrix}
a & b \\
c & d
\end{pmatrix}
\in \Gamma,
\label{eq:MFGamma(N)gamma} \\
&S : f^j(\tau) \rightarrow
f^j(S(\tau)) = J_k(S, \tau) \rho(S)_{jj'} f^{j'}(\tau), \quad J_k(S, \tau) = (-\tau)^k, \quad S =
\begin{pmatrix}
0 & 1 \\
-1 & 0
\end{pmatrix}
\in \Gamma,
\label{eq:MFGamma(N)S} \\
&T : f^i(\tau) \rightarrow
f^j(T(\tau)) = J_k(T, \tau) \rho(T)_{jj'} f^{j'}(\tau), \quad J_k(T, \tau) = 1, \qquad \quad T =
\begin{pmatrix}
1 & 1 \\
0 & 1
\end{pmatrix}
\in \Gamma,
\label{eq:MFGammaT}
\end{align}
where $\rho(\gamma)$ denotes unitary representation satisfying
\begin{align}
\rho(\gamma_2\gamma_1) = \rho(\gamma_2) \rho(\gamma_1), \quad \gamma_1, \gamma_2 \in \Gamma, \label{eq:rho}
\end{align}
and $\rho(h)=\mathbb{I}$ for $h \in \Gamma(N)$,~i.e.
\begin{align}
\rho(Z)^2 = \rho(S)^4 = [ \rho(S) \rho(T) ]^3 = \rho(T)^N = \mathbb{I}, \quad \left( \rho(S)^2=\mathbb{I} \ (N=2) \right).
\label{eq:algGamma(N)}
\end{align}
In addition, for $\gamma=S^2=-\mathbb{I}$, the following should be satisfied:
\begin{align}
f^j(S^2(\tau)) = f^i(\tau) = (-1)^k \rho(S)_{jj'}^2 f^{j'}(\tau) \ \Leftrightarrow \ \rho(Z) = \rho(S)^2 = (-1)^k \mathbb{I}, \label{eq:MFGamma(N)S2}
\end{align}
which also satisfies
\begin{align}
\rho(Z) \rho(T) = \rho(T) \rho(Z). \label{eq:ZT}
\end{align}
Thus, when $k$ is even, because of $\rho(S)^2=\mathbb{I}$, $\rho$ becomes the representation of the quotient group $\Gamma_N \equiv \bar{\Gamma}/\bar{\Gamma}(N)$, where $\bar{\Gamma}(N)$ is defined as $\bar{\Gamma}(N) \equiv \Gamma(N)/\{\pm\mathbb{I}\}$ for $N=1,2$ and $\bar{\Gamma}(N) \equiv \Gamma$ for $N>2$.
Here, $\Gamma_N$ is called the finite modular group since its order becomes finite.
Interestingly, $\Gamma_N$ with $N=2$, $3$, $4$, and $5$ are isomorphic to well-known non-Abelian discrete groups, $S_3$, $PSL(2,\mathbb{Z}_3) \simeq A_4$, $S_4$, and $PSL(2,\mathbb{Z}_5) \simeq A_5$, respectively.
Moreover, for $N>5$, (although other algebraic relations in addition to those in Eqs.~(\ref{eq:algGamma(N)}) and (\ref{eq:MFGamma(N)S2}) are imposed,) $\Gamma_7$ is isomorphic to $PSL(2,\mathbb{Z}_7)$ while $\Gamma_8$ and $\Gamma_{16}$ contain $\Delta(96)$ and $\Delta(384)$ as their subgroups, respectively~\cite{deAdelhartToorop:2011re}.
On the other hand, when $k$ is odd, $\rho$ becomes the representation of the quotient group $\Gamma'_N \equiv \Gamma/\Gamma(N)$, called the homogeneous finite modular group~\cite{Liu:2019khw}.
It is the double covering group of $\Gamma_N$.
Therefore, modular forms of odd (even) weight for $\Gamma(N)$ transform under $\Gamma'_N = \Gamma/\Gamma(N)$ ($\Gamma_N = \bar{\Gamma}/\bar{\Gamma}(N)$) non-trivially.

In order to treat modular forms of half-integral weight in the next section, we introduce the metapletic double covering group of $\Gamma = SL(2,\mathbb{Z})$, defined as
\begin{align}
\widetilde{\Gamma} = \widetilde{SL}(2,\mathbb{Z}) = \left\{ \widetilde{\gamma} = [\gamma, \epsilon] | \gamma \in \Gamma, \epsilon \in \{\pm1\} \right\}. 
\end{align}
The multiplication is given by
\begin{align}
\widetilde{\gamma}_1 \widetilde{\gamma}_2 = [\gamma_1, \epsilon_1] [\gamma_2, \epsilon_2] = [\gamma_1\gamma_2, A(\gamma_1,\gamma_2) \epsilon_1 \epsilon_2], \quad \widetilde{\gamma}_{1,2} = [\gamma_{1,2}, \epsilon_{1,2}] \in \widetilde{\Gamma}, \label{eq:multipl}
\end{align}
where $A(\gamma_1,\gamma_2)$ denotes the Kubota's twisted 2-cocycle~\cite{Kubota} for $\Gamma$, satisfying the following the relation,
\begin{align}
(\widetilde{\gamma}_1\widetilde{\gamma}_2)\widetilde{\gamma}_3 &= \widetilde{\gamma}_1 (\widetilde{\gamma}_2\widetilde{\gamma}_3), \notag \\
\Leftrightarrow \ 
A(\gamma_1,\gamma_2) A(\gamma_1\gamma_2, \gamma_3) &= A(\gamma_1, \gamma_2\gamma_3) A(\gamma_2,\gamma_3). \label{eq:cocyclerelation}
\end{align}
It is defined as follows.
First, let us introduce Kubota's function $\chi : \Gamma \rightarrow \mathbb{Z}$, defined by
\begin{align}
\chi(\gamma) =
\left\{
\begin{array}{ll}
c & (c \neq 0) \\
d & (c=0)
\end{array}
\right., \quad
\gamma = 
\begin{pmatrix}
a & b \\
c & d
\end{pmatrix}
\in \Gamma.
\label{eq:Kubotafunc}
\end{align}
We also introduce the Hilbert symbol, defined by
\begin{align}
(a,b)_H =
\left\{
\begin{array}{ll}
-1 & (a<0\ {\rm and}\ b<0) \\
1 & ({\rm otherwise})
\end{array}
\right..
\label{eq:Hilbertsymb}
\end{align}
Then, the 2-cocycle $A : \Gamma \times \Gamma \rightarrow \{\pm 1\}$ is defined as
\begin{align}
A(\gamma_1,\gamma_2) = \left({\rm det}\gamma_1, {\rm det}\gamma_2 \right)_H \left( \frac{\chi(\gamma_1\gamma_2)}{\chi(\gamma_1)}, \frac{\chi(\gamma_1\gamma_2)}{\chi(\gamma_2){\rm det}\gamma_1} \right)_H.
\label{eq:Kubotacocycle}
\end{align}
Here, let us set generators of $\widetilde{\Gamma}$ as
\begin{align}
\widetilde{S} \equiv [S,-1] \quad (\widetilde{S}^{-1} = [S^{-1}, -1]), \quad \widetilde{T} \equiv [T,1] \quad (\widetilde{T}^{-1} = [T^{-1}, 1]). \label{eq:StildeTtilde}
\end{align}
They satisfy the following algebraic relations;
\begin{align}
\widetilde{S}^2 = [-\mathbb{I},1] \equiv \widetilde{Z}, \quad \widetilde{Z}^2 = \widetilde{S}^4 = [\mathbb{I},-1], \quad \widetilde{Z}^4 = \widetilde{S}^8 = (\widetilde{S}\widetilde{T})^3 = [\mathbb{I},1] \equiv \widetilde{\mathbb{I}}, \quad \widetilde{Z}\widetilde{T} = \widetilde{T}\widetilde{Z}.  \label{eq;algtildeSL2Z}
\end{align}
The action of $\widetilde{\gamma} \in \widetilde{\Gamma}$ on $\tau$ as well as $z$ is the same as one of $\gamma \in \Gamma$,~i.e. $\widetilde{\gamma}(z,\tau) = \gamma(z,\tau)$.
Furthermore, we introduce the metapletic congruence subgroup of level $N \in 4\mathbb{Z}$, defined by
\begin{align}
\widetilde{\Gamma}(N) \equiv \{ [h, \epsilon] \in \widetilde{\Gamma} | h \in \Gamma(N), \epsilon=1 \}, \label{eq:Gammatilde(N)}
\end{align}
which is a normal subgroup of $\widetilde{\Gamma}$ ($\widetilde{\Gamma}(N) \triangleleft \widetilde{\Gamma}$) and isomorphic to $\Gamma(N)$.
Now, modular forms of half-integral weight $k/2$ for $\widetilde{\Gamma}(N)$ ($N \in 4\mathbb{Z}$)\footnote{It is required to define modular forms of half-integral weight mathematically.}, $f^j(\tau)$, are holomorphic functions which transform under $\widetilde{\gamma} \in \widetilde{\Gamma}$ as
\begin{align}
&\begin{array}{r}
\widetilde{\gamma} : f^j(\tau) \rightarrow f^j(\widetilde{\gamma}(\tau)) = \widetilde{J}_{k/2}(\widetilde{\gamma},\tau) \rho(\widetilde{\gamma})_{jj'} f^{j'}(\tau) \\
\widetilde{J}_{k/2}(\widetilde{\gamma},\tau) = \epsilon^k (c\tau+d)^{k/2} 
\end{array},
\quad \widetilde{\gamma} = \left[ 
\begin{pmatrix}
a & b \\
c & d
\end{pmatrix}, \epsilon \right] \in \widetilde{\Gamma},
\label{eq:MFGamma(N)tilde} \\
&\begin{array}{r}
\widetilde{S} : f^j(\tau) \rightarrow f^j(\widetilde{S}(\tau)) = \widetilde{J}_{k/2}(\widetilde{S},\tau) \rho(\widetilde{S})_{jj'} f^{j'}(\tau) \\
\widetilde{J}_{k/2}(\widetilde{S},\tau) = (-1)^k (-\tau)^{k/2}
\end{array},
\quad \widetilde{S} = \left[ 
\begin{pmatrix}
0 & 1 \\
-1 & 0
\end{pmatrix}, -1 \right] \in \widetilde{\Gamma},
\label{eq:MFGamma(N)tildeS} \\
&\begin{array}{r}
\widetilde{T} : f^j(\tau) \rightarrow f^j(\widetilde{T}(\tau)) = \widetilde{J}_{k/2}(\widetilde{T},\tau) \rho(\widetilde{T})_{jj'} f^{j'}(\tau) \\
\widetilde{J}_{k/2}(\widetilde{T},\tau) =  1
\end{array},
\quad \widetilde{T} = \left[ 
\begin{pmatrix}
1 & 1 \\
0 & 1
\end{pmatrix}, 1 \right] \in \widetilde{\Gamma},
\label{eq:MFGamma(N)tildeT}
\end{align}
and for $\widetilde{h} \in \widetilde{\Gamma}(N) \subset \widetilde{\Gamma}$, in particular, satisfy
\begin{align}
\widetilde{h} : f^j(\tau) \rightarrow f^j(\widetilde{h}(\tau)) = \widetilde{J}_{k/2}(\widetilde{h},\tau) f^j(\tau), \quad \widetilde{J}_{k/2}(\widetilde{h},\tau) = J_{k/2}(h,\tau), \quad \widetilde{h} = [ h,1] \in \widetilde{\Gamma}(N).
\label{eq:MFGamma(N)tilde}
\end{align}
The automorphy factor $J_{k/2}(\widetilde{\gamma},\tau) = \epsilon^k J_{k/2}(\gamma,\tau)$ satisfies
\begin{align}
\widetilde{J}_{k/2}(\widetilde{\gamma}_2\widetilde{\gamma}_1,\tau) = \left( A(\gamma_2,\gamma_1) \right)^k \widetilde{J}_{k/2}(\widetilde{\gamma}_2,\widetilde{\gamma}_1(\tau)) \widetilde{J}_{k/2}(\widetilde{\gamma}_1,\tau), \quad \widetilde{\gamma}_1, \widetilde{\gamma}_2 \in \widetilde{\Gamma}, \label{eq:automorphyfactortilde}
\end{align}
while the unitary representation should satisfy $\rho(\widetilde{h}) = \mathbb{I}$ for $\widetilde{h} \in \widetilde{\Gamma}(N)$~i.e.
\begin{align}
\rho(\widetilde{Z})^4 = \rho(\widetilde{S})^8 = [\rho(\widetilde{S})\rho(\widetilde{T})]^3 = \rho(\widetilde{T})^N = \mathbb{I}. \label{eq:Gamma(N)tilde}
\end{align}
Here, we take $(-1)^k=e^{-\pi ik/2}$.
In addition, for $\widetilde{Z}^2 = \widetilde{S}^4 = [\mathbb{I},-1]$ and $\widetilde{Z} = \widetilde{S}^2 = [-\mathbb{I},1]$, the followings should be satisfied:
\begin{align}
&f^j(\widetilde{S}^4(\tau)) = f^j(\tau) = - \rho(\widetilde{S})^4_{jj'} f^{j'}(\tau) \ \Leftrightarrow \ \rho(\widetilde{Z})^2 = \rho(\widetilde{S})^4 = - \mathbb{I}, \label{eq:MFGamma(N)tildeS4} \\
&f^j(\widetilde{S}^2(\tau)) = f^j(\tau) = e^{-\pi ik/2} \rho(\widetilde{S})^2_{jj'} f^{j'}(\tau) \ \Leftrightarrow \ \rho(\widetilde{Z}) = \rho(\widetilde{S})^2 = e^{\pi ik/2} \mathbb{I}, \label{eq:MFGamma(N)tildeS2}
\end{align}
which also satisfies
\begin{align}
\rho(\widetilde{Z}) \rho(\widetilde{T}) = \rho(\widetilde{T}) \rho(\widetilde{Z}). \label{eq:ZTtilde}
\end{align}
Thus, $\rho$ becomes the representation of the quotient group $\widetilde{\Gamma}_N \equiv \widetilde{\Gamma}/\widetilde{\Gamma}(N)$, called the metapletic finite modular group.
It is the further double covering group of $\Gamma'_N$, that is, the quadruple covering group of $\Gamma_N$.
In other words, modular forms of half-integral weight for $\widetilde{\Gamma}(N)$ transform under $\widetilde{\Gamma}_N = \widetilde{\Gamma}/\widetilde{\Gamma}(N)$ non-trivially.
Note that these results are consistent with an integral weight case.

Furthermore, we treat modular transformation of not only the CS modulus $\tau$ but also the complex coordinate $z$ in the next section, we extend the definition of modular forms; in stead of modular forms $f^i(\tau)$, when wave functions $\psi^j(z,\tau)$ transform under $\widetilde{\gamma} \in \widetilde{\Gamma}$ as
\begin{align}
&\widetilde{\gamma} : \psi^j(z,\tau) \rightarrow \psi^j(\widetilde{\gamma}(z,\tau)) = \widetilde{J}_{k/2}(\widetilde{\gamma},\tau) \rho(\widetilde{\gamma})_{jj'} \psi^{j'}(z,\tau), \quad \widetilde{J}_{k/2}(\widetilde{\gamma},\tau) = \epsilon^k (c\tau+d)^{k/2}, \quad \widetilde{\gamma} \in \widetilde{\Gamma},
\label{eq:wavMFGamma(N)tilde} \\
&\widetilde{S} : \psi^j(z,\tau) \rightarrow \psi^j(\widetilde{S}(z,\tau)) = \widetilde{J}_{k/2}(\widetilde{S},\tau) \rho(\widetilde{S})_{jj'} \psi^{j'}(z,\tau), \quad \widetilde{J}_{k/2}(\widetilde{S},\tau) = (-1)^k (-\tau)^{k/2}, \quad \widetilde{S} \in \widetilde{\Gamma},
\label{eq:wavMFGamma(N)tildeS} \\
&\widetilde{T} : \psi^j(z,\tau) \rightarrow \psi^j(\widetilde{T}(z,\tau)) = \widetilde{J}_{k/2}(\widetilde{T},\tau) \rho(\widetilde{T})_{jj'} \psi^{j'}(z,\tau), \quad \widetilde{J}_{k/2}(\widetilde{T},\tau) =  1, \qquad \qquad \quad  \widetilde{T} \in \widetilde{\Gamma},
\label{eq:wavMFGamma(N)tildeT}
\end{align}
and for $\widetilde{h} \in \widetilde{\Gamma}(N)$, in particular, they satisfy
\begin{align}
\widetilde{h} : \psi^j(z,\tau) \rightarrow \psi^j(\widetilde{h}(z,\tau)) = \widetilde{J}_{k/2}(\widetilde{h},\tau) \psi^j (z,\tau), \quad \widetilde{h} \in \widetilde{\Gamma}(N), \label{eq:wavMF}
\end{align}
we call $\psi^j(z,\tau)$ as ``modular forms'' of weight $k/2$ for $\widetilde{\Gamma}(N)$ and they transform under $\widetilde{\Gamma}_N = \widetilde{\Gamma}/\widetilde{\Gamma}(N)$ non-trivially.
Note that Eq.~(\ref{eq:MFGamma(N)tildeS2}) does not have to be satisfied since $(\widetilde{S}^2(z,\tau)) = (-z,\tau)$.


\subsection{Modular symmetry in magnetized $T^2$ compactification}
\label{subsec:MSmagnetizedT2}

In this section, let us discuss the modular symmetry on the magnetized $T^2$.
The following analysis is based on Refs.~\cite{Kikuchi:2020frp,Kikuchi:2021ogn,Hoshiya:2020hki,Kikuchi:2022bkn}.

In addition to the ``modular transformation'' for $(z,\tau)$ in Eqs.~(\ref{eq:modulartrans}) and (\ref{eq:modulartransz}), by considering the transformation,
\begin{align}
\begin{array}{lll}
\gamma : {\rm Im}\tau \rightarrow \frac{{\rm Im}\tau}{|c\tau+d|^2}, & \gamma : e_1 \rightarrow (c\tau+d) e_1,
\end{array}
\label{eq:useful}
\end{align}
we obtain the following transformation:
\begin{align}
\begin{array}{ll}
\gamma : h_{\mu\nu} \rightarrow |c\tau+d|^2 h_{\mu\nu}, & \\
\gamma : \gamma^z \rightarrow (c\tau+d)^{-1} \gamma^z, & \gamma : \gamma^{\bar{z}} \rightarrow (c\bar{\tau}+d)^{-1} \gamma^{\bar{z}}, \\
\gamma : \langle F_{z\bar{z}} \rangle \rightarrow |c\tau+d|^2 \langle F_{z\bar{z}} \rangle, & \\
\gamma : \langle A_z \rangle \rightarrow (c\tau+d) \langle A_z \rangle, & \gamma : \langle A_{\bar{z}} \rangle \rightarrow (c\bar{\tau}+d) \langle A_{\bar{z}} \rangle, \\
\gamma : D_z \rightarrow (c\tau+d) D_z, & \gamma : D_{\bar{z}} \rightarrow (c\bar{\tau}+d) D_{\bar{z}}. 
\end{array}
\label{eq:MThgammaFAD}
\end{align}
Then, we find that the square of the line element $ds^2$, the area of $T^2$ ${\cal A}$, the constant 2-form field strength $\langle F \rangle$, 1-form vector potential $\langle A \rangle$, the covariant derivative $D$ and the derivative operators such as ${\cal D}^{(\dagger)}$, ($a^{(\dagger)}$,) and mass eigenvalues $m_n^2$, ($m_{\mathbf{n}}^2$,) are modular invariant.
Hence, both wave functions of $(z,\tau)$ and $(\gamma(z,\tau))$ satisfy the same equation of motions.
Indeed, non-holomorphic parts of wave functions are modular invariant.
Next, let us see the ``modular transformation'' for the BCs by $U_1$ and $U_{\tau}$.
We note that the shifts on the coordinate $S(z)$, $S(z) \rightarrow S(z) + 1$ and $S(z) \rightarrow S(z) + S(\tau)$, correspond to the shifts on the coordinate $z$, $z \rightarrow z - \tau$ and $z \rightarrow z + 1$, respectively, while the shifts on the coordinate $T(z)$, $T(z) \rightarrow T(z) + 1$ and $T(z) \rightarrow T(z) + T(\tau)$, correspond to the shifts on the coordinate $z$, $z \rightarrow z+1$, $z \rightarrow z+\tau+1$, respectively.
Indeed, we can find that $S(U_1) = U_{-\tau}$ and $S(U_{\tau}) = U_{1}$ by mapping $S(\alpha^{a,b}_1, \alpha^{a,b}_{\tau}) = (\{1- \alpha^{a,b}_{\tau}\}, \alpha^{a,b}_1)$, while $T(U_1) = U_1$ and $T(U_{\tau}) = U_{\tau+1}$ by mapping $T(\alpha^{a,b}_{1},\alpha^{a,b}_{\tau}) = (\alpha^{a,b}_1,\{\alpha^{a,b}_{\tau} + \alpha^{a,b}_1 + M^{a,b}/2\})$.
Here, we define $\{x\}$ for a number $x$ such that $0 \leq \{x\} \equiv x - [x] <1$, where $[x]$ denotes the floor function. 
In particular, when $(\alpha^{a,b}_1,\alpha^{a,b}_{\tau})=(0,0)\ [(1/2,1/2)]$ for $M^{a,b}=$ even [odd], the SS phases are modular invariant.
In these cases, wave functions of $(z,\tau)$ and $(\gamma(z,\tau))$ satisfy same BCs as well as equation of motions.

Now, let us see the following ``modular transformation'' for wave functions in Eq.~(\ref{eq:zeroT2}):
\begin{align}
&S:\psi^{(j+\alpha_1,\alpha_{\tau}),M}_{T^2,0}(z,\tau) \rightarrow \psi^{(j+S(\alpha_1,\alpha_{\tau})),M}_{T^2,0}(S(z,\tau)) = \psi^{(j+\{1-\alpha_{\tau}\},\alpha_{1}),M}_{T^2,0}\left( -\frac{z}{\tau},-\frac{1}{\tau} \right), \label{eq:waveS} \\
&T:\psi^{(j+\alpha_1,\alpha_{\tau}),M}_{T^2,0}(z,\tau) \rightarrow \psi^{(j+T(\alpha_1,\alpha_{\tau})),M}_{T^2,0}(T(z,\tau)) = \psi^{(j+\alpha_{\tau},\{\alpha_{\tau}+\alpha_{1}+M/2\}),M}_{T^2,0}(z,\tau+1). \label{eq:waveT}
\end{align}
They can be rewritten by $\psi^{(j+\alpha_1,\alpha_{\tau}),M}_{T^2,0}(z,\tau)$ as follows:
\begin{align}
&\psi^{(j+\{1-\alpha_{\tau}\},\alpha_{1}),M}_{T^2,0}\left( -\frac{z}{\tau},-\frac{1}{\tau} \right)
= -(-\tau)^{1/2} \sum_{j'=0}^{M-1} \frac{-e^{\pi i/4}}{\sqrt{M}} e^{2\pi i\frac{(j+\{1-\alpha_{\tau}\})(j'+\alpha_1)}{M}} \psi^{(j'+\alpha_1,\alpha_{\tau}),M}_{T^2,0}(z,\tau), \label{eq:waveS2} \\
&\psi^{(j+\alpha_{1},\{\alpha_{\tau}+\alpha_{1}+M/2\}),M}_{T^2,0}(z,\tau+1)
= \sum_{j'=0}^{M-1} e^{\pi i\frac{(j+\alpha_1)^2}{M}} \delta_{j,j'} \psi^{(j+\alpha_1,\alpha_{\tau}),M}_{T^2,0}(z,\tau). \label{eq:waveT2}
\end{align}
The detailed calculations are in Appendix~\ref{app:MTW}.
In particular, for $M=$ even and $(\alpha_1,\alpha_{\tau})=(0,0)$, they can be expressed as
\begin{align}
&\begin{array}{c}
\widetilde{S}:\psi^{(j+0,0),M}_{T^2,0}(z,\tau) \rightarrow \psi^{(j+0,0),M}_{T^2,0}(\widetilde{S}(z,\tau)) = \widetilde{J}_{1/2}(\widetilde{S},\tau) \sum_{j'=0}^{M-1} \rho_{T^2}^{(0,0)}(\widetilde{S})_{jj'} \psi^{(j'+0,0),M}_{T^2,0}(z,\tau), \\
\widetilde{J}_{1/2}(\widetilde{S},\tau) = - (-\tau)^{1/2}, \quad \rho_{T^2}^{(0,0)}(\widetilde{S})_{jj'} = \frac{-e^{\pi i/4}}{\sqrt{M}} e^{2\pi i\frac{jj'}{M}},
\end{array} \label{eq:MFwave00S} \\
&\begin{array}{c}
\widetilde{T}:\psi^{(j+0,0),M}_{T^2,0}(z,\tau) \rightarrow \psi^{(j+0,0),M}_{T^2,0}(\widetilde{T}(z,\tau)) = \widetilde{J}_{1/2}(\widetilde{T},\tau) \sum_{j'=0}^{M-1} \rho(\widetilde{T})_{jj'} \psi^{(j'+0,0),M}_{T^2,0}(z,\tau), \\
\widetilde{J}_{1/2}(\widetilde{T},\tau) = 1, \quad \rho_{T^2}^{(0,0)}(\widetilde{T})_{jj'} = e^{\pi i\frac{j^2}{M}} \delta_{j,j'},
\end{array} \label{eq:MFwave00T}
\end{align}
where $\widetilde{J}_{1/2}$ and $\rho_{T^2}^{(0,0)}$ satisfy the following relations,
\begin{align}
\widetilde{J}_{1/2}(\widetilde{Z},\tau) = \widetilde{J}_{1/2}(\widetilde{S}^2,\tau) = (-1)^{1/2} = e^{-\pi i/2}, \quad& \rho_{T^2}^{(0,0)}(\widetilde{Z})_{jj'} = \rho_{T^2}^{(0,0)}(\widetilde{S})^2_{jj'} = e^{\pi i/2} \delta_{M-j,j'}, \label{eq:JrhoT200S2} \\
\widetilde{J}_{1/2}(\widetilde{Z}^2,\tau) = \widetilde{J}_{1/2}(\widetilde{S}^4,\tau) = -1, \quad& \rho_{T^2}^{(0,0)}(\widetilde{Z})^2_{jj'} = \rho_{T^2}^{(0,0)}(\widetilde{S})^4_{jj'} =- \delta_{j,j'}, \label{eq:JrhoT200S4} \\
\widetilde{J}_{1/2}(\widetilde{Z}^4,\tau) = \widetilde{J}_{1/2}(\widetilde{S}^8,\tau) = 1, \quad& \rho_{T^2}^{(0,0)}(\widetilde{Z})^4_{jj'} = \rho_{T^2}^{(0,0)}(\widetilde{S})^8_{jj'} = \delta_{j,j'}, \label{eq:JrhoT200S8} \\
\widetilde{J}_{1/2}((\widetilde{S}\widetilde{T})^3,\tau) = 1, \quad& [ \rho_{T^2}^{(0,0)}(\widetilde{S})\rho_{T^2}^{(0,0)}(\widetilde{T}) ]^3_{jj'} = \delta_{j,j'}, \label{eq:JrhoT200ST3} \\
\widetilde{J}_{1/2}(\widetilde{T}^{2M},\tau) = 1, \quad& \rho_{T^2}^{(0,0)}(\widetilde{T})^{2M}_{jj'} = \delta_{j,j'}, \label{eq:JrhoT200T2M} \\
& \rho_{T^2}^{(0,0)}(\widetilde{Z})\rho_{T^2}^{(0,0)}(\widetilde{T}) = \rho_{T^2}^{(0,0)}(\widetilde{T}) \rho_{T^2}^{(0,0)}(\widetilde{Z}). \label{eq:rhoT200ZT}
\end{align}
Therefore, by comparing the definition of ``modular forms'' in the previous subsection, $M (\in 2\mathbb{Z})$ number of degenerate wave functions of bi-fundamental fields on the magnetized $T^2$ behave as ``modular forms'' of weight $1/2$ for $\widetilde{\Gamma}(2M)$ and they transform non-trivially under $\widetilde{\Gamma}_{2M} = \widetilde{\Gamma}/\widetilde{\Gamma}(2M)$.
The detailed calculations are in Appendix~\ref{app:MTW}.
On the other hand, for $M=$ odd and $(\alpha_1,\alpha_{\tau})=(1/2,1/2)$, they can be expressed as
\begin{align}
&\begin{array}{c}
\widetilde{S}:\psi^{(j+\frac{1}{2},\frac{1}{2}),M}_{T^2,0}(z,\tau) \rightarrow \psi^{(j+\frac{1}{2},\frac{1}{2}),M}_{T^2,0}(\widetilde{S}(z,\tau)) = \widetilde{J}_{1/2}(\widetilde{S},\tau) \sum_{j'=0}^{M-1} \rho_{T^2}^{(\frac{1}{2},\frac{1}{2})}(\widetilde{S})_{jj'} \psi^{(j'+\frac{1}{2},\frac{1}{2}),M}_{T^2,0}(z,\tau), \\
\widetilde{J}_{1/2}(\widetilde{S},\tau) = - (-\tau)^{1/2}, \quad \rho_{T^2}^{(\frac{1}{2},\frac{1}{2})}(\widetilde{S})_{jj'} = \frac{-e^{\pi i/4}}{\sqrt{M}} e^{2\pi i\frac{(j+\frac{1}{2})(j'+\frac{1}{2})}{M}},
\end{array} \label{eq:MFwave11S} \\
&\begin{array}{c}
\widetilde{T}:\psi^{(j+\frac{1}{2},\frac{1}{2}),M}_{T^2,0}(z,\tau) \rightarrow \psi^{(j+\frac{1}{2},\frac{1}{2}),M}_{T^2,0}(\widetilde{T}(z,\tau)) = \widetilde{J}_{1/2}(\widetilde{T},\tau) \sum_{j'=0}^{M-1} \rho_{T^2}^{(\frac{1}{2},\frac{1}{2})}(\widetilde{T})_{jj'} \psi^{(j'+\frac{1}{2},\frac{1}{2}),M}_{T^2,0}(z,\tau), \\
\widetilde{J}_{1/2}(\widetilde{T},\tau) = 1, \quad \rho_{T^2}^{(\frac{1}{2},\frac{1}{2})}(\widetilde{T})_{jj'} = e^{\pi i\frac{(j+\frac{1}{2})^2}{M}} \delta_{j,j'},
\end{array} \label{eq:MFwave11T}
\end{align}
where $\widetilde{J}_{1/2}$ and $\rho_{T^2}^{(\frac{1}{2},\frac{1}{2})}$ satisfy the following relations,
\begin{align}
\widetilde{J}_{1/2}(\widetilde{Z},\tau) = \widetilde{J}_{1/2}(\widetilde{S}^2,\tau) = (-1)^{1/2} = e^{-\pi i/2}, \ & \rho_{T^2}^{(\frac{1}{2},\frac{1}{2})}(\widetilde{Z})_{jj'} = \rho_{T^2}^{(\frac{1}{2},\frac{1}{2})}(\widetilde{S})^2_{jj'} = - e^{\pi i/2} \delta_{M-(j+\frac{1}{2}),(j'+\frac{1}{2})}, \label{eq:JrhoT211S2} \\
\widetilde{J}_{1/2}(\widetilde{Z}^2,\tau) = \widetilde{J}_{1/2}(\widetilde{S}^4,\tau) = -1, \ & \rho_{T^2}^{(\frac{1}{2},\frac{1}{2})}(\widetilde{Z})^2_{jj'} = \rho_{T^2}^{(\frac{1}{2},\frac{1}{2})}(\widetilde{S})^4_{jj'} =- \delta_{j,j'}, \label{eq:JrhoT211S4} \\
\widetilde{J}_{1/2}(\widetilde{Z}^4,\tau) = \widetilde{J}_{1/2}(\widetilde{S}^8,\tau) = 1, \ & \rho_{T^2}^{(\frac{1}{2},\frac{1}{2})}(\widetilde{Z})^4_{jj'} = \rho_{T^2}^{(\frac{1}{2},\frac{1}{2})}(\widetilde{S})^8_{jj'} = \delta_{j,j'}, \label{eq:JrhoT211S8} \\
\widetilde{J}_{1/2}((\widetilde{S}\widetilde{T})^3,\tau) = 1, \ & [\rho_{T^2}^{(\frac{1}{2},\frac{1}{2})}(\widetilde{S})\rho_{T^2}^{(\frac{1}{2},\frac{1}{2})}(\widetilde{T}) ]^3_{jj'} = \delta_{j,j'}, \label{eq:JrhoT211ST3} \\
\widetilde{J}_{1/2}(\widetilde{T}^{M},\tau) = 1, \ & \rho_{T^2}^{(\frac{1}{2},\frac{1}{2})}(\widetilde{T})^{M}_{jj'} = e^{\pi i/4} \delta_{j,j'}, \label{eq:JrhoT211TM} \\
\widetilde{J}_{1/2}(\widetilde{T}^{8M},\tau) = 1, \ & \rho_{T^2}^{(\frac{1}{2},\frac{1}{2})}(\widetilde{T})^{8M}_{jj'} = \delta_{j,j'}, \label{eq:JrhoT211T8M} \\
& \rho_{T^2}^{(\frac{1}{2},\frac{1}{2})}(\widetilde{Z})\rho_{T^2}^{(\frac{1}{2},\frac{1}{2})}(\widetilde{T}) = \rho_{T^2}^{(\frac{1}{2},\frac{1}{2})}(\widetilde{T}) \rho_{T^2}^{(\frac{1}{2},\frac{1}{2})}(\widetilde{Z}). \label{eq:rhoT211ZT}
\end{align}
Therefore, similarly, $M (\in 2\mathbb{Z}+1)$ number of degenerate wave functions of bi-fundamental fields on the magnetized $T^2$ behave as ``modular forms'' of weight $1/2$ for $\widetilde{\Gamma}(8M)$ and they transform non-trivially under $\widetilde{\Gamma}_{8M} = \widetilde{\Gamma}/\widetilde{\Gamma}(8M)$.
Note that since $\rho_{T^2}^{(\frac{1}{2},\frac{1}{2})}(\widetilde{T})^{M} = e^{\pi i/4} \mathbb{I}$ commutes with $\forall \rho_{T^2}^{(\frac{1}{2},\frac{1}{2})}(\widetilde{\gamma})$, it becomes the generator of the center group $\mathbb{Z}_8$ of $\widetilde{\Gamma}_{8M}$.
The detailed calculations are in Appendix~\ref{app:MTW}.
For example, when $M=2$ and $(\alpha_1,\alpha_{\tau}) = (0,0)$, $\rho_{T^2}^{(0,0)}$ can be expressed as
\begin{align}
\rho_{T^2}^{(0,0)}(\widetilde{S}) = \frac{e^{\pi i/4}}{\sqrt{2}}
\begin{pmatrix}
-1 & -1 \\
-1 & 1
\end{pmatrix}, \quad
\rho_{T^2}^{(0,0)}(\widetilde{T}) =
\begin{pmatrix}
1 & 0 \\
0 & i
\end{pmatrix}.
\label{eq:M2rho}
\end{align}
Actually, according to Ref.~\cite{Liu:2020msy}, it corresponds to the 2D irreducible unitary representation\footnote{When we denote $\hat{\mathbf{2}}$ in Ref.~\cite{Liu:2020msy} as $\hat{\mathbf{2}}_{\rm ref}$, Eq.~(\ref{eq:M2rho}) can be obtained by the unitary transformation, $\sigma_3 \hat{\mathbf{2}}_{\rm ref} \sigma_3^{-1}$.} $\hat{\mathbf{2}}$ of $\widetilde{\Gamma}_4 \simeq \widetilde{S}_4$, which is further double covering group of $\Gamma'_4 \simeq S'_4$ (quadruple covering group of $\Gamma_4 \simeq S_4$).
Here, we have a few comments.
First, not only the above zero mode wave functions but also any $n$ th excited mode wave functions in Eq.~(\ref{eq:nthT2}) have same characters since $a^{(\dagger)}$ is modular invariant and it commutes with $\rho$ as well as $\widetilde{J}_{1/2}$.
Second, as for wave functions of the remaining $U(N_{a,b})$ adjoint fields on the magnetized $T^2$ in Eqs.~(\ref{eq:Diracmassiveaabb}), they are modular invariant by considering the modular transformation of $\mathbf{n} = n_1 + \tau n_2 = (n_1e_1+n_2e_2)/e_1 \equiv \mathbf{n}_u/e_1$ which corresponds to the coordinate of the lattice point,~i.e. $\gamma : \mathbf{n} \rightarrow \mathbf{n}/(c\tau+d)$.

Then, let us consider the modular symmetry in the 4D EFT.
Notice that the modular transformation comes from $SL(2,\mathbb{Z})$ transformation of lattice vectors and the lattice itself does not change, that is, the modular transformation induces just basis transformation:
\begin{align}
\Phi_n(X) = \sum_{j} \Phi_{n,j}(x) \otimes \Phi^{j}_{T^2,n}(z,\tau) = \sum_{j'} \Phi_{n,j'}(x) \otimes \Phi^{j'}_{T^2,n}(z,\tau).
\label{eq:basetransform}
\end{align}
From the modular transformation for wave functions on the magnetized $T^2$ in Eqs.~(\ref{eq:MFwave00S})-(\ref{eq:MFwave00T}) and Eqs.~(\ref{eq:MFwave11S})-(\ref{eq:MFwave11T}), the modular transformation induces the following transformation for $j$ th 4D bi-fundamental fields $\Phi_{n,j}(x)$;
\begin{align}
\widetilde{\gamma} : \Phi_{n,j}(x) \rightarrow \widetilde{J}_{-1/2}(\widetilde{\gamma},\tau) \sum_{j'=0}^{M-1} \bar{\rho}^{(\alpha,\alpha)}_{T^2}(\widetilde{\gamma})_{jj'} \Phi_{n,j'}(x),
\label{eq:4DMT}
\end{align}
with $\alpha=0$ for $M \in 2\mathbb{Z}$ and $\alpha=1/2$ for $M \in 2\mathbb{Z}+1$, where we used the unitarity $(\rho^{-1})^{T}=\bar{\rho}$.
Therefore, $M$-generational 4D bi-fundamental fields transform under the modular transformation as $M$-dimensional representation of $\widetilde{\Gamma}_{2M}\ (M \in 2\mathbb{Z})$ or $\widetilde{\Gamma}_{8M}\ (M \in 2\mathbb{Z}+1)$ with modular weight $-k=-1/2$.
Actually, by combining the modular transformation for the coefficient $(i\bar{\tau} - i\tau)^{-1/2}$, $(i\bar{\tau} - i\tau)^{-1/2} \rightarrow |c\tau+d| (i\bar{\tau} - i\tau)^{-1/2}$, we can find that the kinetic terms in the 4D action in Eq.~(\ref{eq:4Dbifundkinetic}):
\begin{align}
K_{\Phi} = \sum_{n,j} \frac{\Phi_{n,j}(x)^{\dagger}D\Phi_{n,j}(x)}{(i\bar{\tau}-i\tau)}, \label{eq:matterKahler}
\end{align}
are modular invariant, where $D$ denotes the covariant derivative operator.
On the other hand, remaining $U(N_{a,b})$ adjoint 4D fields are modular invariant since their wave functions on the magnetized $T^2$ are modular invariant.
It is also consistent with the 4D action.
Next, let us see the modular symmetry for three-point couplings.
In particular, let us see the modular symmetry for holomorphic three-point couplings in Eq.~(\ref{eq:multiplerel}) as well as Eq.~(\ref{eq:yK}).
Here, we note that $\ell^y = {\rm lcm}(M^{ab}, M^{bc}, M^{ac}) \in 2\mathbb{Z}$ and $(\alpha^y_1,\alpha^y_{\tau}) \equiv (0,0)\ ({\rm mod}\ 1)$ whether each wave function has $(M;\alpha_1,\alpha_{\tau})=({\rm even};0,0)$ or $({\rm odd}; 1/2,1/2)$.
We also note that the results of the modular transformation for wave functions in Eqs.~(\ref{eq:MFwave00S}) and (\ref{eq:MFwave00T}) do not depend on the coordinate $z$.
Thus, holomorphic three-point couplings become modular forms of weight $1/2$ for $\widetilde{\Gamma}(2\ell^y)$ and they transform non-trivially under $\widetilde{\Gamma}_{2\ell^y}$.
Actually, the representation of the holomorphic three-point couplings, $\rho^{y(0,0)}_{T^2}(\widetilde{\gamma})_{(ijk)(i'j'k')}$, is consistent with tensor products of the representations of matter fields, $\rho^{i(\alpha^{i},\alpha^{i})}_{T^2}(\widetilde{\gamma})_{ii'}$:
\begin{align}
\rho^{y(0,0)}_{T^2}(\widetilde{\gamma})_{(ijk)(i'j'k')} = \rho^{i(\alpha^{i},\alpha^{i})}_{T^2}(\widetilde{\gamma})_{ii'} \otimes \rho^{j(\alpha^{j},\alpha^{j})}_{T^2}(\widetilde{\gamma})_{jj'} \otimes \bar{\rho}^{k(\alpha^{k},\alpha^{k})}_{T^2}(\widetilde{\gamma})_{kk'}. \label{eq:Yukawatensor}
\end{align}
It can be also understood from Eq.~(\ref{eq:3pointintegral}).
Then, combining the modular transformation for 4D fields, the holomorphic superpotential of the three-point couplings
\begin{align}
W_y(\tau) = y^{ijk}_{T^2,n^in^jn^k}(\tau) \Phi_{n^i,i}(x) \Phi_{n^j,j}(x) \Phi_{n^k,k}(x), \label{eq:holoYukawa}
\end{align}
is transformed under the modular transformation as
\begin{align}
\gamma: W_y(\tau) \rightarrow J_{-1}(\gamma,\tau) W_y(\tau) = (c\tau+d)^{-1} W_y(\tau), \label{eq:holosuperYukawa}
\end{align}
which means that the holomorphic superpotential has modular weight $-1$.
This is consistent within the supergravity theory.
In the supergravity theory, the following combination of the K\"{a}hler potential $K$ and a holomorphic superpotential $W(\tau)$,
\begin{align}
G \equiv K + \ln |W(\tau)|^2, \label{eq:GKW}
\end{align}
called the supergravity K\"{a}hler function,
must be invariant under the modular transformation.
In $T^2$ case, the K\"{a}hler potential is given by Eq.~(\ref{eq:Kahler}) and it transforms under the modular transformation as
\begin{align}
\gamma: K \rightarrow K + \ln|c\tau+d|^2. \label{eq:MTK}
\end{align}
Thus, a holomorphic superpotential $W(\tau)$ should transform as
\begin{align}
\gamma: W(\tau) \rightarrow (c\tau+d)^{-1}W(\tau). \label{eq:holosuperpot}
\end{align}
(This is a K\"{a}hler transformation.)
Hence, the transformation in Eq.~(\ref{eq:holosuperYukawa}) is consistent with Eq.~(\ref{eq:holosuperpot}).
However, we consider global supersymmetric theory.
Remind that the relations of superpotentials as well as three-point couplings in the supergravity theory and the global supersymmetric theory are discussed in Eqs.~(\ref{eq:WhatW}) and (\ref{eq:yhaty}).
Thus, the super potential in the global supersymmetric theory,
\begin{align}
\hat{W}_y(\tau) = \hat{y}^{ijk}_{T^2,n^in^jn^k}(\tau) \Phi_{n^i,i}(x) \Phi_{n^j,j}(x) \Phi_{n^k,k}(x), \label{eq:globalYukawa}
\end{align}
is modular invariant.
Therefore, the 4D effective actions in Eqs.~(\ref{eq:4Dgaugeadj})-(\ref{eq:4Dcoupling}) as well as Eq.~(\ref{eq:4DEFT}) are modular invariant.
In other words, the 4D EFT obtained from the magnetized torus compactification has $\widetilde{\Gamma}_{N}$ modular flavor symmetry, where $N$ is determined by magnetic fluxes.

Here, we comment
on the ``modular transformation'' for
wave functions on magnetized $T^2_1 \times T^2_2$,
\begin{align}
\Psi^{(j_1+\alpha^{(1)},j_2+\alpha^{(2)}),M_1M_2}_{T^2_1 \times T^2_2,0}((z_1,\tau_1),(z_2,\tau_2)) \equiv \psi^{(j_1+\alpha^{(1)},\alpha^{(1)}),M_1}_{T^2_1,0}(z_1,\tau_1) \psi^{(j_2+\alpha^{(2)},\alpha^{(2)}),M_2}_{T^2_2,0}(z_2,\tau_2), \label{eq:waveT2T2} 
\end{align}
where $\alpha^{(1,2)}_1 = \alpha^{(1,2)}_{\tau} = \alpha^{(1,2)} = 0$, $1/2$ for $M_{1,2}=$even, odd, respectively.
In general, each torus $T^2_i\ (i=1,2)$ has the ``modular symmetry'' $SL(2,\mathbb{Z}_i)$ independently and then wave functions on magnetized $T^2_i$ transform under $SL(2,\mathbb{Z}_i)$ independently.
However, when the moduli $\tau_1$ and $\tau_2$ are identified,~i.e. $\tau_1=\tau_2 \equiv \tau$, the ``modular symmetry on $T^2_1 \times T^2_2$ is reduced from $SL(2,\mathbb{Z}_1) \times SL(2,\mathbb{Z}_2)$ to $SL(2,\mathbb{Z})$.
Then, in this case, the ``modular symmetry'' for wave functions in Eq.~(\ref{eq:waveT2T2}) is non-trivial;
we can find that the automorphy factor becomes $(\widetilde{J}_{1/2}(\widetilde{\gamma},\tau))^2 = J_1(\gamma,\tau)$ and the unitary representation,
\begin{align}
\rho^{(\alpha^{(1)},\alpha^{(2)})}_{T^2_1 \times T^2_2}(\gamma)_{(j_1j_2)(j'_1j'_2)} \equiv \rho^{(\alpha^{(1)},\alpha^{(1)})}_{T^2_1}(\widetilde{\gamma})_{j_1j'_1} \rho^{(\alpha^{(2)},\alpha^{(2)})}_{T^2_1}(\widetilde{\gamma})_{j_2j'_2}, \label{eq:rhogammaT2T2}
\end{align}
satisfies
\begin{align}
&\rho^{(\alpha^{(1)},\alpha^{(2)})}_{T^2_1 \times T^2_2}(Z) = \rho^{(\alpha^{(1)},\alpha^{(2)})}_{T^2_1 \times T^2_2}(S)^2 = - e^{2\pi i(\alpha^{(1)} + \alpha^{(2)} )} \delta_{M-(j_1+\alpha^{(1)}),(j'_1+\alpha^{(1)})} \delta_{M-(j_2+\alpha^{(2)}),(j'_2+\alpha^{(2)})}, \label{eq:T2T2ZS2} \\
&\rho^{(\alpha^{(1)},\alpha^{(2)})}_{T^2_1 \times T^2_2}(Z)^2 = \rho^{(\alpha^{(1)},\alpha^{(2)})}_{T^2_1 \times T^2_2}(S)^4 = [\rho^{(\alpha^{(1)},\alpha^{(2)})}_{T^2_1 \times T^2_2}(S) \rho^{(\alpha^{(1)},\alpha^{(2)})}_{T^2_1 \times T^2_2}(T)]^3 = \ \delta_{(j_1,j_2),(j'_1,j'_2)}, \label{eq:T2T2Z2S4ST3} \\
&[\rho^{(\alpha^{(1)},\alpha^{(2)})}_{T^2_1 \times T^2_2}(Z) \rho^{(\alpha^{(1)},\alpha^{(2)})}_{T^2_1 \times T^2_2}(T)] = [\rho^{(\alpha^{(1)},\alpha^{(2)})}_{T^2_1 \times T^2_2}(T) \rho^{(\alpha^{(1)},\alpha^{(2)})}_{T^2_1 \times T^2_2}(Z)], \label{eq:T2T2ZTTZ}
\end{align}
and
\begin{align}
\begin{array}{ll}
\rho^{(\alpha^{(1)},\alpha^{(2)})}_{T^2_1 \times T^2_2}({T})^{N=2\ell_{12}} = \delta_{(j_1,j_2),(j'_1,j'_2)}, & (M_1=2s_1, M_2=2s_2), \\
\rho^{(\alpha^{(1)},\alpha^{(2)})}_{T^2_1 \times T^2_2}({T})^{N=2\ell_{12}} = \delta_{(j_1,j_2),(j'_1,j'_2)}, & (M_1=4s_1, M_2=2s_2-1), \\
\rho^{(\alpha^{(1)},\alpha^{(2)})}_{T^2_1 \times T^2_2}({T})^{2\ell_{12}} = -\delta_{(j_1,j_2),(j'_1,j'_2)}, & (M_1=2(2s_1-1), M_2=2s_2-1), \\
\ \Rightarrow \rho^{(\alpha^{(1)},\alpha^{(2)})}_{T^2_1 \times T^2_2}({T})^{N=4\ell_{12}} = \delta_{(j_1,j_2),(j'_1,j'_2)},  \\
\rho^{(\alpha^{(1)},\alpha^{(2)})}_{T^2_1 \times T^2_2}({T})^{\ell_{12}} = e^{\pi i\frac{M_1+M_2}{4g_{1,2}}} \delta_{(j_1,j_2),(j'_1,j'_2)}, & (M_1=2s_1-1, M_2=2s_2-1), \\
\ \Rightarrow \rho^{(\alpha^{(1)},\alpha^{(2)})}_{T^2_1 \times T^2_2}({T})^N = \delta_{(j_1,j_2),(j'_1,j'_2)}, & N= \left\{
  \begin{array}{l}
    \ell_{12} \quad (M_1+M_2 \in 8\mathbb{Z}) \\
    2\ell_{12} \quad (M_1+M_2 \in 4\mathbb{Z}) \\
    4\ell_{12} \quad (M_1+M_2 \in 2\mathbb{Z}) \\
  \end{array}
  \right., 
\end{array}
\label{eq:T2T2TN}
\end{align}
where $\ell_{12} \equiv {\rm lcm}(M_1,M_2)$, $g_{12} \equiv {\rm gcd}(M_1,M_2)$, and $s_{1,2} \in \mathbb{Z}$.
Hence, the $M_1M_2$ number of wave functions (\ref{eq:waveT2T2}) behave as ``modular forms'' of weight $1$ for $\Gamma(N)$ and they transform non-trivially under $\Gamma'_N = \Gamma/\Gamma(N)$~\cite{Kikuchi:2020nxn}.


\subsection{$CP$ symmetry in magnetized $T^2$ compactification}
\label{subsec:CPmagnetizedT2}

In this subsection, let us discuss $CP$ symmetry in magnetized $T^2$ compactification~\cite{Kikuchi:2020frp}.

First of all, we consider the 6D space-time, ${\cal M}_6 = R^{1,3} \times T^2$, and a spinor field on ${\cal M}_6$,
\begin{align}
\begin{array}{l}
\psi^{(4)IJ}(X) = \sum_{n,j} 
\begin{pmatrix}
\psi^{(2)IJ}_{L,n,j}(x) \otimes \psi^{IJ}_{+,n,j}(z,\tau) \\
\psi^{(2)IJ}_{R,n,j}(x) \otimes \psi^{IJ}_{-,n,j}(z,\tau) 
\end{pmatrix}, \\
\psi^{(\bar{4})IJ}(X) = \sum_{n,j} 
\begin{pmatrix}
\psi^{(2)IJ}_{L,n,j}(x) \otimes \psi^{IJ}_{-,n,j}(z,\tau) \\
\psi^{(2)IJ}_{R,n,j}(x) \otimes \psi^{IJ}_{+,n,j}(z,\tau)
\end{pmatrix},
\end{array}
\label{eq:6Dspinor}
\end{align}
where $IJ$ corresponds to the charge in an internal space symmetry such as gauge symmetry.
Let us consider the simultaneous CP transformation: 4D CP transformation and the following 2D transformation~\cite{Novichkov:2019sqv,Kikuchi:2020frp} (see also Refs.~\cite{Baur:2019kwi,Baur:2019iai}),
\begin{align}
&CP:
\begin{pmatrix}
e_2 \\ e_1
\end{pmatrix}
\rightarrow
\begin{pmatrix}
CP(e_2) \\ CP(e_1)
\end{pmatrix}
=
\begin{pmatrix}
1 & 0 \\
0 & -1
\end{pmatrix}
\begin{pmatrix}
\bar{e}_2 \\ \bar{e}_1
\end{pmatrix}
=
\begin{pmatrix}
\bar{e}_2 \\ -\bar{e}_1
\end{pmatrix}, \quad
CP =
\begin{pmatrix}
1 & 0 \\
0 & -1 
\end{pmatrix},
\label{eq:CPlatticetrans} \\
\Rightarrow \ &
CP: \left( z=\frac{u}{e_1}, \tau=\frac{e_2}{e_1} \right) \rightarrow \left( CP(z) = \frac{CP(u)}{CP(e_1)}, CP(\tau) = \frac{CP(e_2)}{CP(e_1)} \right) \notag \\
&\qquad \qquad \qquad \qquad \qquad \qquad = \left( -\bar{z} = \frac{\bar{u}}{-\bar{e}_1}, -\bar{\tau} = \frac{\bar{e}_2}{-\bar{e}_1} \right),
\label{eq:CPztau} \\
\bigl(&CP: \left( z=y^1+\tau y^2, \tau={\rm Re}\tau+i{\rm Im}\tau \right) \rightarrow \left( -\bar{z} = -y^1 -\bar{\tau} y^2, -\bar{\tau} =-{\rm Re}\tau - i (-{\rm Im}\tau) \right) \bigl),
\label{eq:CPztaure}
\end{align}
where it satisfies that ${\rm Im}(-\bar{\tau}) = {\rm Im}\tau>0$, $(CP)^2=\mathbb{I}$, and ${\rm det}(CP)=-1$.
That is the complex conjugate transformation and the parity transformation for $e_1$.
Hereafter, we call this 2D transformation 2D ``$CP$ transformation''.
Then, this simultaneous transformation can be embedded in a 6D proper Lorentz transformation~\cite{Green:1987mn,Strominger:1985it,Dine:1992ya,Choi:1992xp,Lim:1990bp,Kobayashi:1994ks}, that is, this transformation is the same as 6D $C$ transformation.
Under 6D $C$ transformation,
\begin{align}
C:\psi^{(4)IJ}(X) \rightarrow \psi^{(\bar{4})JI}(X). \label{eq:6DCtrans}
\end{align}
In addition, under 4D $CP$ transformation,
\begin{align}
CP: \psi^{(2)IJ}_{L,n,j}(x) \rightarrow \psi^{(2)JI}_{L,n,j}(x). \label{eq:4DCPtrans}
\end{align}
Hence, under 2D ``$CP$ transformation'' in Eq.~(\ref{eq:CPztau}), 
\begin{align}
CP: \psi^{IJ}_{+,n,j}(z,\tau) \rightarrow \psi^{IJ}_{+,n,j}(-\bar{z},-\bar{\tau}) = \psi^{JI}_{-,n,j}(z,\tau) = \overline{\psi^{IJ}_{+,n,j}(z,\tau)}, \label{eq:2DCPtrans}
\end{align}
will be obtained, where we also consider 2D MW condition: $\psi^{JI}_{\mp,n,j}(z,\tau)=\overline{\psi^{IJ}_{\pm,n,j}(z,\tau)}$.

Now, let us see the ``$CP$ symmetry'' in the magnetized $T^2$ compactification.
Along with the ``$CP$ transformation'', the magnetic fluxes $M^{a,b}$ in 2-form field strength as well as 1-form gauge field are also transformed as
\begin{align}
CP: M^{A} \rightarrow -M^{A}\ (A=a,b,...). \label{eq:CPM}
\end{align}
Then, we can obtain the following ``$CP$ transformation''
\begin{align}
\begin{array}{ll}
CP: \gamma^{z} \rightarrow - \gamma^{\bar{z}}, & CP: \gamma^{\bar{z}} \rightarrow -\gamma^{z}, \\
CP: \langle F_{z\bar{z}} \rangle \rightarrow \langle F_{\bar{z}z} \rangle = - \langle F_{z\bar{z}} \rangle, \\
CP: \langle A_{z} \rangle \rightarrow - \langle A_{\bar{z}} \rangle, & CP: \langle A_{\bar{z}} \rangle \rightarrow - \langle A_{z} \rangle, \\
CP: D_{z} \rightarrow - D_{\bar{z}}, & CP: D_{\bar{z}} \rightarrow - D_{z}.
\end{array}
\label{eq:CPgammaFAD}
\end{align}
From those ``$CP$ transformation'', in particular, the Dirac operator $i{\cal D}_{ab}$ transforms into $-i{\cal D}^{\dagger}_{ba}$.
Indeed, wave functions of $\psi^{ab}_{+,n,j}(z,\tau) = \psi^{(j+\alpha_1,\alpha_{\tau}),M}_{T^2,n}$ transform as
\begin{align}
CP: \psi^{(j+\alpha_1,\alpha_{\tau}),M}_{T^2,n}(z,\tau) \rightarrow \overline{\psi^{(-(j+\alpha_1),\alpha_{\tau}),-M}_{T^2,n}}(-\bar{z},-\bar{\tau}) = \overline{\psi^{(j+\alpha_1,\alpha_{\tau}),M}_{T^2,n}(z,\tau)}. \label{eq:CPT2wave}
\end{align}
That is, we can obtain Eq.~(\ref{eq:2DCPtrans}): wave functions of $\psi^{ab}_{+,n,j}(z,\tau) = \psi^{(j+\alpha_1,\alpha_{\tau}),M}_{T^2,n}$ transform into $\psi^{ba}_{-,n,j}(z,\tau) = \overline{\psi^{(j+\alpha_1,\alpha_{\tau}),M}_{T^2,n}(z,\tau)}$ under the ``$CP$ transformation''.

Next, let us extend to the generalized ``$CP$ symmetry'' consistent with the modular symmetry~\cite{Novichkov:2019sqv}.
First, among the ``$CP$ transformation'' in Eq.~(\ref{eq:CPlatticetrans}) and the $\Gamma = SL(2,\mathbb{Z})$ transformation in Eq.~(\ref{eq:STofSL2Z}), they have the following relations;
\begin{align}
(CP)^{-1} S (CP)^{-1} = S^{-1}, \quad (CP) T (CP)^{-1} = T^{-1}. \label{eq:CPSTCP}
\end{align}
Then, in this case, the symmetry can be extended as $\Gamma^{\ast} \equiv SL(2,\mathbb{Z}) \rtimes \mathbb{Z}_2^{CP} \simeq GL(2,\mathbb{Z})$, called the extended modular group.
Under the extended modular transformation by $\gamma^{\ast} \in \Gamma^{\ast}$, $(z,\tau)$ transforms as
\begin{align}
\gamma^{\ast}: (z,\tau) \rightarrow (\gamma^{\ast}(z,\tau)) =
\left\{
\begin{array}{cc}
\left( \frac{z}{c\tau+d}, \frac{a\tau+b}{c\tau+d} \right), & ({\rm det} \gamma^{\ast} = 1) \\
\left( \frac{\bar{z}}{c\bar{\tau}+d}, \frac{a\bar{\tau}+b}{c\bar{\tau}+d} \right), & ({\rm det} \gamma^{\ast} = -1)
\end{array}
\right. , \quad
\gamma^{\ast}=
\begin{pmatrix}
a & b \\
c & d
\end{pmatrix} \in \Gamma^{\ast},
\label{eq:gammaast}
\end{align}
where the above in Eq.~(\ref{eq:gammaast}) contains even number of $CP$ transformation, while the below in Eq.~(\ref{eq:gammaast}) contains odd number of $CP$ transformation.
Then, the automorphy factor with weight $k \in \mathbb{Z}$ is also extended as
\begin{align}
J^{\ast}_k(\gamma^{\ast}, \tau) \equiv
\left\{
\begin{array}{cc}
(c\tau+d)^k, & ({\rm det} \gamma^{\ast} = 1) \\
(c\bar{\tau}+d)^k, & ({\rm det} \gamma^{\ast} = -1)
\end{array}
\right. , \quad
\gamma^{\ast} =
\begin{pmatrix}
a & b \\
c & d 
\end{pmatrix}
\in \Gamma^{\ast}. \label{eq:reautomorphyfact}
\end{align}
Furthermore, we can consider the metapletic double covering group of the extended modular group $\Gamma^{\ast} = GL(2,\mathbb{Z})$, $\widetilde{\Gamma^{\ast}} \equiv \widetilde{GL}(2,\mathbb{Z})$~\cite{Duncan:2018wbw,Kikuchi:2020frp}, by simply replacing $\gamma \in \Gamma$ with $\gamma^{\ast} \in \Gamma^{\ast}$ and setting the generator $\widetilde{CP}$ as
\begin{align}
\widetilde{CP} \equiv [CP, 1], \quad ( \widetilde{CP}^{-1} = [CP, -1] ). \label{eq:CPtilde}
\end{align}
In this case, the generators $\widetilde{S}$, $\widetilde{T}$, and $\widetilde{CP}$ satisfy the following algebraic relations:
\begin{align}
\begin{array}{c}
\widetilde{CP}^2 = [ \mathbb{I},-1] =\widetilde{Z}^2, \quad \widetilde{CP}^4 = [\mathbb{I},1] = \widetilde{Z}^4 = \widetilde{I}, \\
(\widetilde{CP}) \widetilde{S} (\widetilde{CP})^{-1} = [S^{-1}, -1] = \widetilde{S}^{-1}, \quad (\widetilde{CP}) \widetilde{T} (\widetilde{CP})^{-1} = [T^{-1}, 1] = \widetilde{T}^{-1},
\end{array}
\end{align}
and also Eq.~(\ref{eq;algtildeSL2Z}).
The automorphy factor with weight $k/2$ is similarly written as Eq.~(\ref{eq:MFGamma(N)tilde}) by replacing $J_{k/2}(\gamma,\tau)$ with $J^{\ast}_{k/2}(\gamma^{\ast}, \tau)$.

Now, let us see the generalized $CP$ symmetry consistent with the modular symmetry in magnetized $T^2$ compactification.
We can rewrite the ``$CP$ transformation'' for wave functions on the magnetized $T^2$ in Eq.~(\ref{eq:CPT2wave}) as
\begin{align}
\widetilde{CP}: \psi^{(-(j+\alpha),\alpha),M}_{T^2,n}(z,\tau) \rightarrow \overline{\psi^{(j+\alpha,\alpha),-M}_{T^2,n}}(-\bar{z},-\bar{\tau}) &= \widetilde{J}^{\ast}_{1/2}(\widetilde{CP},\tau) \sum_{j'=0}^{M-1} \rho^{(\alpha,\alpha)}_{T^2}(\widetilde{CP})_{jj'} \overline{\psi^{(j'+\alpha,\alpha),M}_{T^2,n}(z,\tau)}, \notag \\
\widetilde{J}^{\ast}_{1/2}(\widetilde{CP},\tau) = (-1)^{1/2} = e^{-\pi i/2},& \quad \rho^{(\alpha,\alpha)}_{T^2}(\widetilde{CP})_{jj'} = e^{\pi i/2} \delta_{j,j'},
\label{eq:CPwave}
\end{align}
where $\alpha = 0$ for $M \in 2\mathbb{Z}$ or $\alpha=1/2$ for $M \in 2\mathbb{Z}+1$.
Note that we can check that
\begin{align}
\widetilde{J}^{\ast}_{1/2}(\widetilde{CP},\tau) = - (-1)^{1/2} = -e^{-\pi i/2} = e^{\pi i/2}, \quad \rho^{(\alpha,\alpha)}_{T^2}(\widetilde{CP}^{-1})_{jj'} = \rho^{(\alpha,\alpha)}_{T^2}(\widetilde{CP})^{-1}_{jj'} = e^{-\pi i/2} \delta_{j,j'}. \label{eq:waveCPinv}
\end{align}
By combining Eqs.~(\ref{eq:MFwave00S})-(\ref{eq:MFwave00T}) or Eqs.~(\ref{eq:MFwave11S})-(\ref{eq:MFwave11T})
, we can obtain the following relations
\begin{align}
\widetilde{J}^{\ast}_{1/2}((\widetilde{CP})^2,\tau)=\widetilde{J}^{\ast}_{1/2}(\widetilde{Z}^2,\tau)=-1, &\quad \rho^{(\alpha,\alpha)}_{T^2}(\widetilde{CP})^2_{jj'}=\rho^{\alpha,\alpha}_{T^2}(\widetilde{Z})^2_{jj'}=-\delta_{j,j'}, \label{eq:CP2} \\
\widetilde{J}^{\ast}_{1/2}((\widetilde{CP})^4,\tau)=\widetilde{J}^{\ast}_{1/2}(\widetilde{Z}^4,\tau)=1, &\quad \rho^{(\alpha,\alpha)}_{T^2}(\widetilde{CP})^4_{jj'}=\rho^{(\alpha,\alpha)}_{T^2}(\widetilde{Z})^4_{jj'}=\delta_{j,j'}, \label{eq:CP4} \\
\widetilde{J}^{\ast}_{1/2}((\widetilde{CP})\widetilde{S}(\widetilde{CP})^{-1},\tau)=\widetilde{J}^{\ast}_{1/2}(\widetilde{S}^{-1},\tau), &\quad [\rho^{(\alpha,\alpha)}_{T^2}(\widetilde{CP})\overline{\rho^{(\alpha,\alpha)}_{T^2}(\widetilde{S})}\rho^{(\alpha,\alpha)}_{T^2}(\widetilde{CP})^{-1}]_{jj'}=\rho^{(\alpha,\alpha)}_{T^2}(\widetilde{S})^{-1}_{jk}, \label{eq:CPSCP-1} \\
\widetilde{J}^{\ast}_{1/2}((\widetilde{CP})\widetilde{T}(\widetilde{CP})^{-1},\tau)=\widetilde{J}^{\ast}_{1/2}(\widetilde{T}^{-1},\tau), &\quad [\rho^{(\alpha,\alpha)}_{T^2}(\widetilde{CP})\overline{\rho^{(\alpha,\alpha)}_{T^2}(\widetilde{T})}\rho^{(\alpha,\alpha)}_{T^2}(\widetilde{CP})^{-1}]_{jj'}=\rho^{(\alpha,\alpha)}_{T^2}(\widetilde{T})^{-1}_{jk}, \label{eq:CPTCP-1}
\end{align}
in addition to Eqs.~(\ref{eq:JrhoT200S2})-(\ref{eq:rhoT200ZT}) or Eqs.~(\ref{eq:JrhoT211S2})-(\ref{eq:rhoT211ZT}).
Therefore, we can find that the wave functions, which behave as ``modular forms'' of weight $1/2$ for $\widetilde{\Gamma}(N)$ with $N=2M\ (M \in 2\mathbb{Z})$ or $N=8M\ (M \in 2\mathbb{Z}+1)$, transform non-trivially under $\widetilde{\Gamma^{\ast}}_{N} \equiv \widetilde{\Gamma^{\ast}}/\widetilde{\Gamma}(N)$.


\chapter{Magnetized torus orbifold models}
\label{chap:Mtorusorbifold}

Next, in this chapter, we discuss magnetized torus orbifold models.
In the previous section, we have seen that due to background magnetic fluxes, the original gauge symmetry is broken to the smaller gauge subsymmetries,~e.g. $U(N) \rightarrow \prod_{A} U(N_{A})$, and particularly we can obtain multi-generational chiral bi-fundamental fields, which can correspond to three-generational quarks and leptons.
On the other hand, for unbroken $U(N_A)$ adjoint sector, there remains 4D ${\cal N}=4$ supersymmetry. (Even if we start from 6D space-time,~i.e. ${\cal M}_{6} = R^{1,3} \times T^2$, there remains 4D ${\cal N}=2$ supersymmetry.)
It means there are massless adjoint scalar fields.
However, at least in the low energy scale,~e.g. below a TeV scale, we have not observed their existence.
Thus, in the following, we consider magnetized torus orbifold compactification to project out those 4D gauge adjoint scalar fields by orbifold projection.
In addition, although three-generational chiral bi-fundamental fields can be obtained by only one case that the magnetic flux which they feel is just equal to $3$, there are various cases to obtain them in the magnetized torus orbifold models. 
In section~\ref{sec:T2ZN}, we review magnetized $T^2/\mathbb{Z}_N$ compactification.
Then, in section~\ref{sec:MSmagnetizedT2ZN}, we discuss the modular symmetry in magnetized $T^2/\mathbb{Z}_N$ compactification.
We also discuss the modular symmetry in magnetized $(T^2_1 \times T^2_2)/(\mathbb{Z}^{(t)}_2 \times \mathbb{Z}^{(p)}_2)$ compactification.
In particular, we can find that the three-generation modes, in particular, transform as three-dimensional representations of certain non-Abelian modular flavor symmetries.


\section{Magnetized $T^2/\mathbb{Z}_N$ compactification}
\label{sec:T2ZN}

In this section, we review magnetized $T^2/\mathbb{Z}_N$ twisted orbifold compactification\footnote{See also Ref.~\cite{Sakamoto:2020vdy} for $T^2/\mathbb{Z}_N$ twisted orbifold compactification without any magnetic fluxes.}~\cite{Abe:2008fi,Abe:2013bca,Kobayashi:2017dyu,Sakamoto:2020pev}.


\subsection{Construction of $T^2/\mathbb{Z}_N$ orbifolds}
\label{subsec:constT2ZN}

First, let us construct $T^2/\mathbb{Z}_N$ twisted orbifolds.
A $T^2/\mathbb{Z}_N$ twisted orbifold can be constructed by further identifying a point of $T^2$, $z$, with the $\mathbb{Z}_N$ twisted point, $\rho z\ (\rho = e^{2\pi i/N})$,~i.e. $\rho z \sim z$, in addition to $z+1 \sim z$ and $z+\tau \sim z$.
In particular, a lattice point $z=m_1+m_2\tau\ (\forall m_1,m_2 \in \mathbb{Z})$ must transform into another lattice point $n_1+n_2\tau\ (\exists n_1,n_2 \in \mathbb{Z})$ under the $\mathbb{Z}_N$ transformation,~i.e. $\rho (m_1+m_2\tau) = n_1+ n_2\tau$.
Here, since $(m_2,m_1)$ and $(n_2,n_1)$ are 2D integer vectors, $(n_2,n_1)$ can be obtained by a $SL(2,\mathbb{Z})$ transformation of $(m_2,m_1)$ as
\begin{align}
\begin{pmatrix}
n_2 & n_1
\end{pmatrix}
=
\begin{pmatrix}
m_2 & m_1
\end{pmatrix}
\begin{pmatrix}
a & b \\
c & d
\end{pmatrix}
=
\begin{pmatrix}
a m_2 + c m_1 & b m_2 + d m_1
\end{pmatrix}, \quad a,b,c,d \in \mathbb{Z}, \ ad-bc=1.
\label{eq:mn}
\end{align}
Thus, for $\forall m_1, m_2 \in \mathbb{Z}$, the following relation should be satisfied;
\begin{align}
&\rho (m_1 + m_2 \tau) = (b m_2 + d m_1) + (a m_2 + c m_1)\tau, \label{eq:twist} \\
\Leftrightarrow \ &
(\rho  - c\tau - d) m_1 + (\rho \tau  -a\tau - b) m_2 = 0, \notag \\
\Leftrightarrow \ &
\left\{
\begin{array}{l}
\rho = c \tau + d, \\
\rho \tau = a\tau + b,
\end{array}
\right. \notag \\
\Leftrightarrow \ &
\left\{
\begin{array}{l}
c\tau^2 - (a-d)\tau - b = 0, \ (\leftarrow \tau = \frac{a\tau+b}{c\tau+d}), \\
\rho = c \tau + d.
\end{array}
\right. \label{eq:taurho}
\end{align}
Here, we consider the fundamental region of $\tau$: $|{\rm Re}\tau| \leq 1/2,\ {\rm Im}\tau \geq \sqrt{1-({\rm Re}\tau)^2}$.
When $c=0$, the solution is
\begin{align}
\begin{pmatrix}
a & b \\
c & d
\end{pmatrix}
=
\begin{pmatrix}
-1 & 0 \\
0 & -1
\end{pmatrix}
= S^{-2}
\ \Rightarrow \ 
\rho = -1\ (N=2), \ \forall \tau.
\label{eq:N2}
\end{align}
When $c \neq 0$, the solutions are
\begin{align}
&\tau = \frac{a-d}{2c} + i \frac{\sqrt{4-(a+d)^2}}{2c}, \ \rho = \frac{a+d}{2} + i \frac{\sqrt{4-(a+d)^2}}{2}, \label{eq:solcneq0}
\end{align}
with
\begin{align}
a+d = 0, \pm1,\ |a-d| \leq c \leq -b, \ ad-bc=1, \label{eq:condsol}
\end{align}
that is,
\begin{align}
\begin{pmatrix}
a & b \\
c & d
\end{pmatrix}
=
\begin{pmatrix}
0 & -1 \\
1 & 0
\end{pmatrix}
= S^{-1}
\ &\Rightarrow \ 
\tau = i, \ \rho =  i \ (N=4), \label{eq:N4} \\
\begin{pmatrix}
a & b \\
c & d
\end{pmatrix}
=
\begin{pmatrix}
-1 & -1 \\
1 & 0
\end{pmatrix}
= (ST)^{-1}
\ &\Rightarrow \ 
\tau = -\frac{1}{2} + i \frac{\sqrt{3}}{2}, \ \rho = -\frac{1}{2} + i \frac{\sqrt{3}}{2} \ (N=3), \label{eq:N3} \\
\Biggl(
\begin{pmatrix}
a & b \\
c & d
\end{pmatrix}
=
\begin{pmatrix}
0 & -1 \\
1 & -1
\end{pmatrix}
= (TS)^{-1}
\ &\Rightarrow \ 
\tau = \frac{1}{2} + i \frac{\sqrt{3}}{2}, \ \rho = -\frac{1}{2} + i \frac{\sqrt{3}}{2} \ (N=3)
\Biggl), \notag \\
\begin{pmatrix}
a & b \\
c & d
\end{pmatrix}
=
\begin{pmatrix}
1 & -1 \\
1 & 0
\end{pmatrix}
= (ST^{-1})^{-1}
\ &\Rightarrow \ 
\tau = \frac{1}{2} + i \frac{\sqrt{3}}{2}, \ \rho =  \frac{1}{2} + i \frac{\sqrt{3}}{2} \ (N=6), \label{eq:N6} \\
\Biggl(
\begin{pmatrix}
a & b \\
c & d
\end{pmatrix}
=
\begin{pmatrix}
0 & -1 \\
1 & 1
\end{pmatrix}
= (T^{-1}S)^{-1}
\ &\Rightarrow \ 
\tau = -\frac{1}{2} + i \frac{\sqrt{3}}{2}, \ \rho = \frac{1}{2} + i \frac{\sqrt{3}}{2} \ (N=6)
\Biggl). \notag
\end{align}
Therefore, there are only four types of $T^2/\mathbb{Z}_N$ twisted orbifolds with $N=2,3,4,6$.
In detail, $T^2/\mathbb{Z}_2$ twisted orbifold is allowed for any modulus $\tau$, while the other $T^2/\mathbb{Z}_N$ twisted orbifolds are allowed for $\tau = \rho$.
Here, Eq.~(\ref{eq:twist}) can be reinterpreted as
\begin{align}
\rho(m_1+m_2\tau)
=& (n_1+n_2\tau) \notag \\
=&
\begin{pmatrix}
n_2 & n_1
\end{pmatrix}
\begin{pmatrix}
\tau \\ 1
\end{pmatrix} \notag \\
=&
\begin{pmatrix}
m_2 & m_1
\end{pmatrix}
\begin{pmatrix}
a & b \\
c & d
\end{pmatrix}
\begin{pmatrix}
\tau \\ 1
\end{pmatrix} \notag \\
=&
\begin{pmatrix}
m_2 & m_1
\end{pmatrix}
\begin{pmatrix}
a\tau+b \\ c\tau+d
\end{pmatrix} \notag \\
=& (c\tau+d)
\begin{pmatrix}
m_2 & m_1
\end{pmatrix}
\begin{pmatrix}
\frac{a\tau+b}{c\tau+d} \\ 1
\end{pmatrix} \notag \\
=& (c\tau+d) (m_1+m_2\tau),
\label{eq:reinterrhotau} \\
\Leftrightarrow \ 
\frac{(m_1+m_2\tau)}{c\tau+d}
=& \rho^{-1} (m_1+m_2\tau).
\end{align}
Thus, moduli $\tau$ of $T^2/\mathbb{Z}_N$ orbifolds with $N=2,3,4,6$ are invariant under $[S^2]^{(-1)}$, $[ST]^{(-1)}$, $S^{(-1)}$, $[ST^{-1}]^{(-1)}$ transformations, respectively, and these transformations with the moduli induce the $\mathbb{Z}_N$ twists $z \rightarrow \rho^{(-1)}z$.

In the $T^2/\mathbb{Z}_N$ orbifolds, furthermore, there are some fixed points $z^{{\rm fp}}_I = y^{{\rm fp}}_{1I} + y^{{\rm fp}}_{2I}\tau$, satisfying
\begin{align}
\rho z^{{\rm fp}}_I + u + v\tau = z^{{\rm fp}}_I \quad (\exists u,v \in \mathbb{Z}). \label{eq:condfp}
\end{align}
Note that from the above analysis, the following relation is satisfied:
\begin{align}
\rho (m_1+m_2\tau) + \left( (1-d)m_1-bm_2 \right) + \left(-cm_1+(1-a)m_2 \right) \tau = m_1 + m_2\tau.
\label{eq:latticefixedpt}
\end{align}
Then, if there exists $\exists m_1,m_2, k, u, v \in \mathbb{Z}$ satisfying
\begin{align}
\left\{
\begin{array}{l}
(1-d)m_1-bm_2 = k u \\
-cm_1+(1-a)m_2 = kv \\
0 \leq \frac{m_{1,2}}{k} < 1
\end{array}
\right., \label{eq:condfp2}
\end{align}
$z^{{\rm fp}}_I = \frac{m_1+m_2\tau}{k}$ satisfies Eq.~(\ref{eq:condfp}).
Therefore, fixed points $z^{{\rm fp}}_I $ and $(u,v)$ are given as
\begin{align}
z^{{\rm fp}}_I = \left\{
\begin{array}{ll}
0,\frac{1}{2},\frac{\tau}{2},\frac{1+\tau}{2} & (N=2) \\
0,\frac{2+\tau}{3},\frac{1+2\tau}{3} & (N=3) \\
0,\frac{1+\tau}{2} & (N=4) \\
0 & (N=6)
\end{array}
\right., \quad
(u,v) = \left\{
\begin{array}{ll}
(0,0),(1,0),(0,1),(1,1) & (N=2) \\
(0,0),(1,0),(1,1) & (N=3) \\
(0,0),(1,0) & (N=4) \\
(0,0) & (N=6)
\end{array}
\right..
\label{eq:fp}
\end{align}
In addition, $T^2/\mathbb{Z}_4$ has one $\mathbb{Z}_2$ fixed point, and $T^2/\mathbb{Z}_6$ has one $\mathbb{Z}_2$ fixed point and one $\mathbb{Z}_3$ fixed point.
Here, since the deficient angle around a $\mathbb{Z}_N$ fixed point becomes $2\pi - \frac{2\pi}{N} = 2\pi \frac{N-1}{N}$, there is the localized curvature,
\begin{align}
\frac{\xi^R_I}{N} = \frac{N-1}{N}, \label{eq:localcurvature}
\end{align}
at the $\mathbb{Z}_N$ fixed point $z^{{\rm fp}}_I$.
The total curvatures of $T^2/\mathbb{Z}_N$ orbifolds become
\begin{align}
&T^2/\mathbb{Z}_2 : \sum_{I} \frac{\xi^R_I}{N} = \frac{1}{2} \times 4 = 2, \label{eq:curvN2} \\
&T^2/\mathbb{Z}_3 : \sum_{I} \frac{\xi^R_I}{N} = \frac{2}{3} \times 3 = 2, \label{eq:curvN3} \\
&T^2/\mathbb{Z}_4 : \sum_{I} \frac{\xi^R_I}{N} = \frac{3}{4} \times 2 + \frac{1}{2} = 2, \label{eq:curvN4} \\
&T^2/\mathbb{Z}_6 : \sum_{I} \frac{\xi^R_I}{N} = \frac{5}{6} + \frac{1}{2} + \frac{2}{3} = 2, \label{eq:curvN6}
\end{align}
and they are the same as that of $S^2$ written by the Euler number, $\chi(S^2) = 2$.


\subsection{Wave functions on magnetized $T^2/\mathbb{Z}_N$}
\label{subsec:waveMT2ZN}

Now, let us see wave functions on the $T^2/\mathbb{Z}_N$ twisted orbifolds.
Because of identifications under $\mathbb{Z}_N$ twists, a field on the $T^2/\mathbb{Z}_N$, $\Phi(z)$, should also satisfies the following BC(s),
\begin{align}
\Phi_{T^2/\mathbb{Z}_N}(\rho z) &= {\cal S}_{\Phi} U_{0,\rho} \Phi_{T^2/\mathbb{Z}_N}(z) U_{0,\rho}^{-1}, \label{eq:BCPhirhoz} \\
\Phi_{T^2/\mathbb{Z}_N}(\rho^Nz) &= {\cal S}_{\Phi}^N U_{0,\rho}^N \Phi_{T^2/\mathbb{Z}_N}(z) U_{0,\rho}^{-N} = \Phi_{T^2/\mathbb{Z}_N}(z), \label{eq:BCPhirhoNz}
\end{align}
in addition to BCs in Eqs.~(\ref{eq:BCPhiz1}) and (\ref{eq:BCPhiztau}),
where $U_{\rho}$ acts on the gauge space of $\Phi$ while ${\cal S}_{\Phi}$ acts on a scalar $\Phi=\phi$, a vector $\Phi=A$, and a spinor $\Phi=\psi$, as follows, respectively.
First, ${\cal S}_{\phi}$ acts on a scalar $\phi$ as
\begin{align}
{\cal S}_{\phi} : \phi^{IJ}_{T^2/\mathbb{Z}_N}(z) \rightarrow 1 \cdot \phi^{IJ}_{T^2/\mathbb{Z}_N}(z). \label{eq:Sphi}
\end{align}
Accordingly, $U_{0,\rho}$ should satisfy
\begin{align}
U_{0,\rho}^N = \mathbb{I}, \label{eq:UN1}
\end{align}
and then ${\cal S}_{\Phi}$ should also satisfy
\begin{align}
{\cal S}_{\Phi}^N = \mathbb{I}. \label{eq:SPhiN1}
\end{align}
Second, from the result in Eq.~(\ref{eq:Sphi}) and the fact that $dz$ and $d\bar{z}$ are transformed under $\mathbb{Z}_N$ twist as $dz \rightarrow \rho dz$ and $d\bar{z} \rightarrow \bar{\rho} d\bar{z}$, respectively, ${\cal S}_{A}$ acts on an 1-form vector $(A_{T^2/\mathbb{Z}_N,z}, A_{T^2/\mathbb{Z}_N,\bar{z}})^T$ as
\begin{align}
{\cal S}_{A} :
\begin{pmatrix}
A^{IJ}_{T^2/\mathbb{Z}_N,z}(z) \\ A^{IJ}_{T^2/\mathbb{Z}_N,\bar{z}}(z)
\end{pmatrix}
\rightarrow
\begin{pmatrix}
\bar{\rho} & \\
 & \rho
\end{pmatrix}
\begin{pmatrix}
A^{IJ}_{T^2/\mathbb{Z}_N,z}(z) \\ A^{IJ}_{T^2/\mathbb{Z}_N,\bar{z}}(z)
\end{pmatrix},
\label{eq:SA}
\end{align}
where it satisfies Eq.~(\ref{eq:SPhiN1}).
Third, ${\cal S}_{\psi}$ should satisfy
\begin{align}
\rho \gamma^z = {\cal S}_{\psi} \gamma^z {\cal S}_{\psi}^{-1}, \quad
\bar{\rho} \gamma^{\bar{z}} =  {\cal S}_{\psi} \gamma^{\bar{z}} {\cal S}_{\psi}^{-1}.
\label{eq:SASpsi}
\end{align}
In addition, it should also satisfy Eq.~(\ref{eq:SPhiN1}).
Hence, ${\cal S}_{\psi}$ acts on a spinor $(\psi_{T^2/\mathbb{Z}_N,z}, \psi_{T^2/\mathbb{Z}_N,-})^T$ as
\begin{align}
{\cal S}_{\psi} :
\begin{pmatrix}
\psi^{IJ}_{T^2/\mathbb{Z}_N,+}(z) \\ \psi^{IJ}_{T^2/\mathbb{Z}_N,-}(z)
\end{pmatrix}
\rightarrow \rho^{\varphi_N^{IJ}+1/2}
\begin{pmatrix}
\rho^{-1/2} & \\
 & \rho^{1/2}
\end{pmatrix}
\begin{pmatrix}
\psi^{IJ}_{T^2/\mathbb{Z}_N,+}(z) \\ \psi^{IJ}_{T^2/\mathbb{Z}_N,-}(z)
\end{pmatrix}, \quad \varphi^{IJ}_N \in \mathbb{Z}/N\mathbb{Z}.
\label{eq:Spsi}
\end{align}
Note that there remains $\mathbb{Z}_N \subset U(1)_s$ uncertainty\footnote{It may depend on information on another $X_2 \times X_3$ compact space.} and Eq.~(\ref{eq:Spsi}) implies $\psi^{IJ}$ has the $U(1)_s$ charge $\varphi^{IJ}_N+1/2$.
Finally, we assume that $U_{0,\rho}$ does not further break $U(N_a) \times U(N_b)$ symmetry.
Thus, $U_{0,\rho}$, satisfying Eq.~(\ref{eq:UN1}), forms as
\begin{align}
U_{0,\rho} =
\begin{pmatrix}
\rho^{m^a} \mathbb{I}_{N_a} & \\
 & \rho^{m^b} \mathbb{I}_{N_b}
\end{pmatrix}, \quad m^a,m^b \in \mathbb{Z}/N\mathbb{Z}.
\label{eq:Urho}
\end{align}
For the component fields, the $\mathbb{Z}_N$ twisted BCs are as follows:
\begin{align}
&\left\{
\begin{array}{l}
\phi_{T^2/\mathbb{Z}_N}^{aa}(\rho z) = \phi_{T^2/\mathbb{Z}_N}^{aa}(z), \\
\phi_{T^2/\mathbb{Z}_N}^{bb}(\rho z) = \phi_{T^2/\mathbb{Z}_N}^{bb}(z), \\
\phi_{T^2/\mathbb{Z}_N}^{ab}(\rho z) = \rho^{m} \phi_{T^2/\mathbb{Z}_N}^{ab}(z), \\
\phi_{T^2/\mathbb{Z}_N}^{ba}(\rho z) = \rho^{-m} \phi_{T^2/\mathbb{Z}_N}^{aa}(z),
\end{array}
\right. \label{eq:BCphicomprhoz} \\
&\left\{
\begin{array}{ll}
A_{T^2/\mathbb{Z}_N,z}^{aa}(\rho z) = \rho^{-1} A_{T^2/\mathbb{Z}_N,z}^{aa}(z), & A_{T^2/\mathbb{Z}_N,\bar{z}}^{aa}(\rho z) = \rho A_{T^2/\mathbb{Z}_N,\bar{z}}^{aa}(z), \\
A_{T^2/\mathbb{Z}_N,z}^{bb}(\rho z) = \rho^{-1} A_{T^2/\mathbb{Z}_N,z}^{bb}(z), & A_{T^2/\mathbb{Z}_N,\bar{z}}^{bb}(\rho z) = \rho A_{T^2/\mathbb{Z}_N,\bar{z}}^{bb}(z), \\
A_{T^2/\mathbb{Z}_N,z}^{ab}(\rho z) = \rho^{m-1} A_{T^2/\mathbb{Z}_N,z}^{ab}(z), & A_{T^2/\mathbb{Z}_N,\bar{z}}^{ab}(\rho z) = \rho^{m+1} A_{T^2/\mathbb{Z}_N,\bar{z}}^{ab}(z), \\
A_{T^2/\mathbb{Z}_N,z}^{ba}(\rho z) = \rho^{-m-1} A_{T^2/\mathbb{Z}_N,z}^{ba}(z), & A_{T^2/\mathbb{Z}_N,\bar{z}}^{ba}(\rho z) = \rho^{-m+1} A_{T^2/\mathbb{Z}_N,\bar{z}}^{aa}(z),
\end{array}
\right. \label{eq:BCAcomprhoz} \\
&\left\{
\begin{array}{ll}
\psi_{T^2/\mathbb{Z}_N,+}^{aa}(\rho z) = \psi_{T^2/\mathbb{Z}_N,+}^{aa}(z), & \psi_{T^2/\mathbb{Z}_N,-}^{aa}(\rho z) = \rho \psi_{T^2/\mathbb{Z}_N,-}^{aa}(z), \\
\psi_{T^2/\mathbb{Z}_N,+}^{bb}(\rho z) = \psi_{T^2/\mathbb{Z}_N,+}^{bb}(z), & \psi_{T^2/\mathbb{Z}_N,-}^{bb}(\rho z) = \rho \psi_{T^2/\mathbb{Z}_N,-}^{bb}(z), \\
\psi_{T^2/\mathbb{Z}_N,+}^{ab}(\rho z) = \rho^m \psi_{T^2/\mathbb{Z}_N,+}^{ab}(z), & \psi_{T^2/\mathbb{Z}_N,-}^{ab}(\rho z) = \rho^{m+1} \psi_{T^2/\mathbb{Z}_N,-}^{ab}(z), \\
\psi_{T^2/\mathbb{Z}_N,+}^{ba}(\rho z) = \rho^{-m-1} \psi_{T^2/\mathbb{Z}_N,+}^{ba}(z), & \psi_{T^2/\mathbb{Z}_N,-}^{ba}(\rho z) = \rho^{-m} \psi_{T^2/\mathbb{Z}_N,-}^{ba}(z),
\end{array}
\right. \label{eq:BCpsicomprhoz}
\end{align}
where $m \equiv m^a - m^b \in \mathbb{Z}/N\mathbb{Z}$ and we adopted $\varphi^{aa}_N = \varphi^{bb}_N = \varphi^{ab}_N=0, \ \varphi^{ba}_N=-1$.

Furthermore, $U_1(z)$ and $U_{\tau}(z)$ are constrained by the $\mathbb{Z}_N$ identification as follows:
\begin{align}
&\left\{
\begin{array}{lll}
U_{-1}(-z) = U_{1}(z), & U_{-\tau}(-z) = U_{\tau}(z), & (N=2) \\
U_{\tau}(\tau z) = U_{1}(z), & U_{\tau^2}(\tau z) = U_{\tau}(z), & (N=3,4,6)
\end{array}
\right., \label{eq:U1Utaurhoz} \\
\Rightarrow \ &
\left\{
\begin{array}{lll}
2\alpha^{a,b}_1 \equiv 0 \ ({\rm mod}\ 1), & \quad 2\alpha^{a,b}_{\tau} \equiv 0 \ ({\rm mod}\ 1), & (N=2) \\
\alpha^{a,b}_1= \alpha^{a,b}_{\tau} \equiv \alpha^{a,b}, & 3\alpha^{a,b} +M^{a,b}/2 \equiv 0 \ ({\rm mod}\ 1), & (N=3) \\
\alpha^{a,b}_1= \alpha^{a,b}_{\tau} \equiv \alpha^{a,b}, & 2\alpha^{a,b} \equiv 0 \ ({\rm mod}\ 1), & (N=4) \\ 
\alpha^{a,b}_1= \alpha^{a,b}_{\tau} \equiv \alpha^{a,b}, & \alpha^{a,b}+M^{a,b}/2 \equiv 0 \ ({\rm mod}\ 1), & (N=6)
\end{array}
\right., \notag \\
\Leftrightarrow \ &
(\alpha^{a,b}_1,\alpha^{a,b}_{\tau}) = 
\left\{
\begin{array}{ll}
(0,0), (0,1/2), (1/2,0), (1/2,1/2) & (N=2) \\
\left\{
\begin{array}{ll}
(0,0), (1/3,1/3), (2/3,2/3) & (M^{a,b} \in 2\mathbb{Z}) \\
(1/6,1/6), (3/6,3/6), (5/6,5/6) & (M^{a,b} \in 2\mathbb{Z}+1) \\
\end{array}
\right. & (N=3) \\
(0,0), (1/2,1/2) & (N=4) \\
\left\{
\begin{array}{ll}
(0,0) & (M^{a,b} \in 2\mathbb{Z}) \\
(1/2,1/2) & (M^{a,b} \in 2\mathbb{Z}+1)
\end{array}
\right. & (N=6)
\end{array}
\right.. \label{eq:SST2ZN}
\end{align}

Here, we note that Eq.~(\ref{eq:BCPhirhoz}) is the BC around a $\mathbb{Z}_N$ fixed point $z^{{\rm fp}}_0 = 0$.
Then, let us see the behavior of $\mathbb{Z}_N$ twist around an arbitrary $\mathbb{Z}_N$ fixed point $z^{{\rm fp}}_I$.
First, let us define a new coordinate $Z_I$ such that $Z=0$ at the fixed point $z^{{\rm fp}}_I$,~i.e., 
\begin{align}
Z_I \equiv z- z^{{\rm fp}}_I. \label{eq:ZI}
\end{align}
Second, we rewrite $z$ as
\begin{align}
z=(z-z^{{\rm fp}}_I) + z^{{\rm fp}}_I = Z_I + z^{{\rm fp}}_I. \label{eq:zrewrite}
\end{align}
In particular, the second term, $z^{{\rm fp}}_I$, can be regarded as the WL phase $\zeta =  z^{{\rm fp}}_I$ ($\zeta_1= y^{{\rm fp}}_{1I},\ \zeta_2 =  y^{{\rm fp}}_{2I}$) from the point of view of the coordinate $Z$.
Then, the WL phase can be converted into SS phases by gauge transformation $U_{\zeta=z^{{\rm fp}}_I}^{-1}(Z)$,~i.e.,
\begin{align}
\Phi_{T^2/\mathbb{Z}_N}(z) = \Phi_{T^2/\mathbb{Z}_N}(Z_I + z^{{\rm fp}}_I) = U_{z^{{\rm fp}}_I}(Z) \widetilde{\Phi}(Z)  U_{z^{{\rm fp}}_I}^{-1}(Z). \label{eq:PhiZ}
\end{align}
Similarly, $\rho z$ can be rewritten as
\begin{align}
\rho z = \rho Z_I + \rho z^{{\rm fp}}_I = Z_I + z^{{\rm fp}}_I - u - v\tau, \label{eq:rhozrewrite}
\end{align}
where we use Eq.~(\ref{eq:condfp}).
Here, the third and fourth terms, $-u-v\tau$, are regarded as $T^2$ shifts while the second term, $z^{{\rm fp}}_I $, can be also regarded as the WL phase,~i.e.,
\begin{align}
&\Phi_{T^2/\mathbb{Z}_N}(\rho z) \notag \\
=& \Phi_{T^2/\mathbb{Z}_N}(\rho Z_I + z^{{\rm fp}}_I - u - v\tau ) \notag \\
=&V_{u+v\tau}^{-1}(\rho Z) \Phi_{T^2/\mathbb{Z}_N}(\rho Z_I + z^{{\rm fp}}_I) V_{u+v\tau}(\rho Z) \notag \\
=& V_{u+v\tau}^{-1}(\rho Z) U_{z^{{\rm fp}}_I}(\rho Z) \widetilde{\Phi}_{T^2/\mathbb{Z}_N}(\rho Z_I) U_{z^{{\rm fp}}_I}^{-1}(\rho Z) V_{u+v\tau}(\rho Z), \label{eq:PhirhoZ} \\
&\left( V_{u+v\tau}^{-1}(\rho Z) \equiv U_{\tau}^{-v}(\rho Z_I + z^{{\rm fp}}_I - u) U_{1}^{-u}(\rho Z_I + z^{{\rm fp}}_I ) \right). \notag
\end{align}
Combining Eqs.~(\ref{eq:PhiZ}) and (\ref{eq:PhirhoZ}), Eq.~(\ref{eq:BCPhirhoz}) can be rewritten by
\begin{align}
\widetilde{\Phi}_{T^2/\mathbb{Z}_N}(\rho Z_I ) = {\cal S}_{\Phi} U_{I,\rho} \widetilde{\Phi}_{T^2/\mathbb{Z}_N}(Z_I) U_{z^{{\rm fp}}_I}(Z) U_{I,\rho}^{-1}, \label{eq:BCPhirhoZ}
\end{align}
with
\begin{align}
U_{I,\rho}
=& U_{z^{{\rm fp}}_I}^{-1}(\rho Z) U_{1}^{u}(\rho Z_I + z^{{\rm fp}}_I ) U_{\tau}^{v}(\rho Z_I + z^{{\rm fp}}_I - u) U_{0,\rho} U_{z^{{\rm fp}}_I}(Z) \notag \\
=& U_{z^{{\rm fp}}_I}^{-1}(\rho Z) U_{\rho z^{{\rm fp}}_I}(\rho Z) U_{1}^{u}(\rho Z_I + z^{{\rm fp}}_I ) U_{\tau}^{v}(\rho Z_I + z^{{\rm fp}}_I - u) U_{0,\rho}  \notag \\
=&
\begin{pmatrix}
\rho^{\chi_{(m)I}^a} \mathbb{I}_{N_a} & \\
 & \rho^{\chi_{(m)I}^b} \mathbb{I}_{N_b}
\end{pmatrix},
\label{eq:UIrho}
\end{align}
where $\chi_{(m)I} \equiv \chi_{(m)I}^a - \chi_{(m)I}^b$ are given as
\begin{align}
\chi_{(m)I} = N\left\{u\alpha_1+v\alpha_{\tau}+\frac{M}{2} (uv+uy^{{\rm fp}}_{2I}-vy^{{\rm fp}}_{1I} ) \right\} + m\ ({\rm mod}\ N), \label{eq:winding}
\end{align}
and SS phases of $\widetilde{\Phi}_{T^2/\mathbb{Z}_N}$ are modified from $\Phi_{T^2/\mathbb{Z}_N}$ as
\begin{align}
(\beta_1,\beta_{\tau}) \equiv (\alpha_1+My^{{\rm fp}}_{2I}, \alpha_{\tau}-My^{{\rm fp}}_{1I})\ ({\rm mod}\ 1). \label{eq:SSPhitilde}
\end{align}
Here, $\chi_{(m)I}$ are called winding numbers around the fixed point $z^{{\rm fp}}_I $.
We note that the above results include $z^{{\rm fp}}_{I}=0$ case.

So far, we have seen BCs under $\mathbb{Z}_N$ twist.
On the other hand, equation of motions are not changed under the $\mathbb{Z}_N$.
Thus, $n$ th excited modes of a field $\Phi^{IJ}_{T^2/\mathbb{Z}_N^{m^{IJ}}}(z)\ (I,J=a,b)$, satisfying the following BC
\begin{align}
\Phi^{IJ}_{T^2/\mathbb{Z}_N^{m^{IJ}},n}(\rho z) = \rho^{m^{IJ}} \Phi^{IJ}_{T^2/\mathbb{Z}_N^{m^{IJ}},n}(z), \label{eq:EXBCrhoz} \\
\end{align}
in addition to BCs in Eqs.~(\ref{eq:BCPhicompz1}) and (\ref{eq:BCPhicompztau}), can be expressed by the following linear combination of the $n$ th excited modes of wave functions on the magnetized $T^2$, $\Phi^{IJ}_{T^2,n}(z)$;
\begin{align}
\Phi^{IJ}_{T^2/\mathbb{Z}_N^{m^{IJ}},n}(z) = {\cal N}_{T^2/\mathbb{Z}_N^{m^{IJ}},n} \sum_{k=0}^{N-1} (\rho^{m^{IJ}})^{-k} \Phi^{IJ}_{T^2,n}(\rho^k z), \label{eq:waveT2ZNbyT2}
\end{align}
where ${\cal N}_{T^2/\mathbb{Z}_N^{m^{IJ}},n}$ denotes the normalization factor determined similarly from Eq.~(\ref{eq:normalizationT2}).

For $IJ=aa,bb$, the following condition is useful,
\begin{align}
\psi^{aa,bb}_{T^2,+,\mathbf{n}}(\rho^k z) = \psi^{aa,bb}_{T^2,+,\rho^k \mathbf{n}}(z). \label{eq:T2ZNnaabb}
\end{align}
In particular, since the massless mode ($\mathbf{n}=0$) on $T^2$ is constant ($\psi^{aa,bb}_{T^2,+,0}(\rho^k z) = \psi^{aa,bb}_{T^2,+,0}(z)$), there is no massless mode on $T^2/\mathbb{Z}_N$ (because of $\sum_{k} (\rho^{m^{aa,bb}})^{-k}=0$) unless $m^{aa,bb}=0$.
In other words, only $\phi^{aa,bb}_{T^2/\mathbb{Z}_N}$ and $\psi^{aa,bb}_{T^2/\mathbb{Z}_N,+}$ have massless modes.
That is, the unbroken $U(N_{a,b})$ sector also becomes chiral.

For $IJ=ab\ (ba)$, the following ``modular transformation'' for wave functions $\psi^{(j+\alpha_1,\alpha_{\tau}),M}_{T^2,n}(z,\tau)$ is useful:
\begin{align} 
\psi^{(j+\alpha_1,\alpha_{\tau}),M}_{T^2,n}( e^{2\pi i/2} z,\tau)
=& \psi^{(j+\alpha_1,\alpha_{\tau}),M}_{T^2,n}(\widetilde{S}^2(z,\tau)) \notag \\
=& e^{2\pi i\alpha_{\tau}} \psi^{(M-(j+\alpha_1),\{1-\alpha_{\tau}\}),M}_{T^2,n}(z,\tau),
\label{eq:T2Z2S2} \\
\psi^{(j+\alpha,\alpha),M}_{T^2,n}(e^{2\pi i/3} z,e^{2\pi i/3})
=& \psi^{(j+\alpha,\alpha),M}_{T^2,n}(\widetilde{S}\widetilde{T}(z,e^{2\pi i/3})) \notag \\
=& \sum_{j'=0}^{M-1} \frac{e^{-\pi i/12}}{\sqrt{M}} e^{2\pi i\frac{(j+\alpha)(j'+\alpha)}{M}} e^{\pi i\frac{(j'+\alpha)^2}{M}} \psi^{(j'+\alpha,\alpha),M}_{T^2,n}(z,e^{2\pi i/3}),
\label{eq:T2Z3ST} \\
\psi^{(j+\alpha,\alpha),M}_{T^2,0}(e^{2\pi i/4} z,e^{2\pi i/4})
=& \psi^{(j+\alpha,\alpha),M}_{T^2,n}(\widetilde{S}(z,e^{2\pi i/4})) \notag \\
=& \sum_{j'=0}^{M-1} \frac{1}{\sqrt{M}} e^{2\pi i\frac{(j+\alpha)(j'+\alpha)}{M}} \psi^{(j'+\alpha,\alpha),M}_{T^2,0}(z,e^{2\pi i/4}),
\label{eq:T2Z4S} \\
\psi^{(j+\alpha,\alpha),M}_{T^2,n}(e^{2\pi i/6} z,e^{2\pi i/6})
=& \psi^{(j+\alpha,\alpha),M}_{T^2,n}(\widetilde{S}\widetilde{T}^{-1}(z,e^{2\pi i/3})) \notag \\
=& \sum_{j'=0}^{M-1} \frac{e^{\pi i/12}}{\sqrt{M}} e^{2\pi i\frac{(j+\alpha)(j'+\alpha)}{M}} e^{-\pi i\frac{(j'+\alpha)^2}{M}} \psi^{(j'+\alpha,\alpha),M}_{T^2,n}(z,e^{2\pi i/6}).
\label{eq:T2Z6ST-1}
\end{align}
They show that eigenfunctions on the magnetized $T^2/\mathbb{Z}_N$ orbifold with $\mathbb{Z}_N$ charge $m^{ab}$,~i.e. $\psi^{(j+\alpha_1,\alpha_{\tau}),M}_{T^2/\mathbb{Z}_N^{m^{ab}},n}(z,\tau)$, can be obtained from appropriate unitary transformation of wave functions on the magnetized $T^2$, $\psi^{(j+\alpha_1,\alpha_{\tau}),M}_{T^2,n}(z,\tau)$.
Thus, the degenerate numbers of $\psi^{(j+\alpha_1,\alpha_{\tau}),M}_{T^2/\mathbb{Z}_N^{m^{ab}},n}(z,\tau)$ are reduced from $M$, which is the degenerate number of $\psi^{(j+\alpha_1,\alpha_{\tau}),M}_{T^2,n}(z,\tau)$, and the sum of the degenerate numbers of $\psi^{(j+\alpha_1,\alpha_{\tau}),M}_{T^2/\mathbb{Z}_N^{m^{ab}},n}(z,\tau)$ for all $\mathbb{Z}_N$ charges must be $M$.
In particular, Ref.~\cite{Sakamoto:2020pev} shows that the chiral zero mode numbers can be written by
\begin{align}
n^{ab}_{+} - n^{ab}_{-} = n^{ab}_{+} = \frac{M}{N} - \frac{V_{(m)}}{N} + 1, \label{eq:counting}
\end{align}
where $V_{(m)}$ denotes the total winding number: $V_{(m)} \equiv \sum_{I} \chi_{(m)I}$.
In the next chapter, we show that this zero mode counting formula can be also regarded as the AS index theorem.
Note that since only $\mathbb{Z}_N$ twisted BC is further imposed while the equation of motion does not change, zero modes on the magnetized $T^2/\mathbb{Z}_N$ with $\mathbb{Z}_N$ charge $m$, in particular, can be written as
\begin{align}
\begin{array}{l}
\psi^{(j+\alpha_1,\alpha_{\tau}),M}_{T^2/\mathbb{Z}_N^{m},0}(z,\tau) = e^{-\frac{\pi M}{2{\rm Im}\tau}|z|^2} h^{(j+\alpha_1,\alpha_{\tau}),M}_{T^2/\mathbb{Z}_N^{m},0}(z), \\
h^{(j+\alpha_1,\alpha_{\tau}),M}_{T^2/\mathbb{Z}_N^{m},0}(z) = \sum_{j'} {\cal N}_{T^2/\mathbb{Z}_N^m,0,j'} \sum_{k=0}^{N-1} \rho^{-km} h^{(j'+\alpha_1,\alpha_{\tau}),M}_{T^2,0}(\rho^k z),
\end{array}
\label{eq:T2ZNzero}
\end{align}
where $h^{(j+\alpha_1,\alpha_{\tau}),M}_{T^2/\mathbb{Z}_N^{m},0}(z)$ denote the holomorphic functions of $z$ and the normalization factor ${\cal N}_{T^2/\mathbb{Z}_N^m,0,j}$ is determined by the normalization condition,
\begin{align}
\int_{T^2} dz d\bar{z} \sqrt{|{\rm det}(2h)|} \overline{\psi^{(j+\alpha_1,\alpha_{\tau}),M}_{T^2/\mathbb{Z}_N^m,0}(z,\tau)} \psi^{(k+\alpha_1,\alpha_{\tau}),M}_{T^2/\mathbb{Z}_N^{m'},0}(z,\tau) = {\cal A} (2{\rm Im}\tau)^{-1/2} \delta_{j,k} \delta_{m,m'},
\label{eq:normalizationT2ZN}
\end{align}
since we use wave functions on the magnetized $T^2$.

Similarly, we can calculate three-point couplings as well as higher-order couplings on the magnetized $T^2/\mathbb{Z}_N$ by appropriate unitary transformation for each wave function.
We have a comment that the coupling coefficient must be $\mathbb{Z}_N$ invariant.

Furthermore, in the orbifold compactifications, right-handed Majorana neutrino mass terms can also be generated by D-brane instanton effects.
Here, a D-brane instanton, which is localized on 4D space-time but wrapped on the compact space, appears as a non-perturbative effect.
Since the D-brane instanton spreads on the compact space, additional fermionic zero modes appear between the D-brane instanton and a $U(N)$ gauge D-brane.
Hereafter, we call them D-brane instanton zero modes.
First, let us consider that there are two gauge D-branes with magnetic fluxes $M^a$ and $M^b$ ($M^a > M^b$) and then the right-handed neutrino zero modes $\nu^I(z)$, which feel the magnetic flux $M = M^a - M^b$, appear between their gauge D-branes.
In addition, if D-brane instanton with magnetic flux $M_{inst}$ ($M^a > M_{inst} > M^b$) appears, D-brane instanton zero modes $\beta_j(z)$ ($\gamma_k(z)$), which feel the magnetic flux $M_{\beta} = M_a - M_{inst}$ ($M_{\gamma} = M_{inst} - M_{b}$), also appear between the D-brane instanton and the gauge D-brane.
Then, the three-point couplings $\hat{y}^{I}_{jk}$ among $\beta_j(z)$, $\gamma_k(z)$, and $\nu^I(z)$ are generated.\footnote{We assume that the contributions on the other compact spaces, $X_2$ and $X_3$, are just constants.}
Note that the D-brane instanton zero modes are localized on 4D space-time at $x$.
Then, after integrating out of the D-brane instanton zero modes, additional couplings of $\nu_I(x)$ at $x$ are generated by D-brane instanton effects.
In particular, the Majorana masses can be generated~\cite{Blumenhagen:2006xt,Ibanez:2006da,Cvetic:2007ku,Kobayashi:2015siy,Hoshiya:2021nux} by
\begin{align}
\Delta {\cal L}_{4D} 
&= M^{IJ}\nu_I(x)\nu_J(x) \notag \\
&= e^{-S_{cl}(T_\alpha,M_{inst})} \int d^2\beta d^2\gamma e^{-\hat{y}_{jk}^I \beta^j(x) \gamma^k(x) \nu_I(x)} \notag \\
&= e^{-S_{cl}(T_\alpha,M_{inst})} (\varepsilon^{ij}\varepsilon^{k\ell} \hat{y}_{ik}^{I}\hat{y}_{j\ell}^{J}) \nu_{I}(x)\nu_{J}(x) \label{eq:neutrinomass} \\
&= e^{-S_{cl}(T_\alpha,M_{inst})} m^{IJ}\nu_I(x)\nu_J(x), \notag
\end{align}
where the Majorana masses can be generated only if both the number of D-brane instanton zero modes, $\beta^{j}(x)$ and $\gamma^k(x)$ are just two $(i,j=1,2)$ since they are Grassmannian satisfying
\begin{align}
\int d\psi \psi = 1 \quad (\psi = \beta^j, \gamma^k).
\label{eq:Grassmannint}
\end{align}
On the other hand, $S_{cl}(T_\alpha,M_{inst})$ denotes the classical action of the D-brane instanton written by Dirac-Born-Infeld (DBI) action, which depends on the moduli $T_{\alpha}$, corresponding to the K\"{a}hler moduli and the dilaton as well as the axion in type-IIB string theory, and the magnetic flux $M_{inst}$ in the compact space.
Note that since D-brane instanton is independent of the gauge D-branes, we can consider an intermediate scale of Majorana neutrino masses such as $O(10^{14})$GeV.
Then, by combining their Yukawa couplings, we can also realize tiny neutrino masses.


\section{Modular symmetry in magnetized $T^2/\mathbb{Z}_N$ compactification}
\label{sec:MSmagnetizedT2ZN}

In this section, let us discuss the modular symmetry on the magnetized $T^2/\mathbb{Z}_N$ twisted orbifolds.
The following analysis is based on Refs.~\cite{Kikuchi:2020frp,Kikuchi:2021ogn}.


\subsection{Modular symmetry on magnetized $T^2/\mathbb{Z}_N$ orbifold}
\label{subsec:general}

First of all, the moduli $\tau$ of $T^2/\mathbb{Z}_N$ with $N=3,4,6$ are fixed by $\tau = \tau_N = e^{2\pi i/N}$.
Thus, they have only following modular subsymmetries,
\begin{align}
&G_N \equiv \{ \gamma \in \Gamma | \gamma \tau_N = \tau_N \}, \notag \\
\Leftrightarrow \ &
\left\{
\begin{array}{l}
G_3 = \mathbb{Z}_3 = \langle ST |(ST)^3=\mathbb{I} \rangle \\
G_4 = \mathbb{Z}_4 = \langle S | (S)^4=\mathbb{I} \rangle \\
G_6 = \mathbb{Z}_6 = \langle ST^{-1} | (ST^{-1})^6 = \mathbb{I} \rangle
\end{array},
\right. \label{eq:residualsymT2ZN}
\end{align}
and their generator $\gamma$ induces $\mathbb{Z}_N$ twist,~i.e. $\gamma : z \rightarrow \rho z$.
Since wave functions on the magnetized $T^2/\mathbb{Z}_N$ are $\mathbb{Z}_N$-eigenfunctions, they also have $G_N$ modular subsymmetries.

On the other hand, $T^2/\mathbb{Z}_2$ can be constructed for any modulus $\tau$.
Thus, there remains the full modular symmetry.
Now, let us see ``modular symmetry'' of wave functions on the magnetized $T^2/\mathbb{Z}_2$ twisted orbifold.
The difference between wave functions on magnetized $T^2$ and $T^2/\mathbb{Z}_2$ is just $\mathbb{Z}_2$ BC for wave functions on $T^2/\mathbb{Z}_2$, and it can be satisfied by the following unitary transformation for wave functions on $T^2$;
\begin{align}
\psi^{(j+\alpha_1,\alpha_{\tau}),M}_{T^2/\mathbb{Z}_N^{m},n}(z,\tau)
=& {\cal N}_{T^2/\mathbb{Z}_2^m} \left( \psi^{(j+\alpha_1,\alpha_{\tau}),M}_{T^2,n}(z,\tau) +(-1)^m \psi^{(j+\alpha_1,\alpha_{\tau}),M}_{T^2,0}(-z,\tau) \right) \notag \\
=& {\cal N}_{T^2/\mathbb{Z}_2^m} \left( \psi^{(j+\alpha_1,\alpha_{\tau}),M}_{T^2,n}(z,\tau) +(-1)^{m+2\alpha_{\tau}} \psi^{(M-(j+\alpha_1),\{1-\alpha_{\tau}\}),M}_{T^2,n}(z,\tau) \right). \label{eq:waveT2Z2}
\end{align}
Then, when $\alpha_1=\alpha_{\tau}\equiv \alpha = 0,1/2$ for $M=$even, odd, respectively, the representations of $\widetilde{S}$ and $\widetilde{T}$ are modified as
\begin{align}
\rho_{T^2/\mathbb{Z}_2^m}^{(\alpha,\alpha)}(\widetilde{S})_{j,j'} 
=&
\left\{
\begin{array}{ll}
{\cal N}_{T^2/\mathbb{Z}_2^0,n,j} {\cal N}_{T^2/\mathbb{Z}_2^0,n,j'} \frac{-4e^{\pi i/4}}{\sqrt{M}} \cos \left( 2\pi \frac{(j+\alpha)(j'+\alpha)}{M} \right) & (m=0) \\
{\cal N}_{T^2/\mathbb{Z}_2^1,n,j} {\cal N}_{T^2/\mathbb{Z}_2^1,n,j'} \frac{-4ie^{\pi i/4}}{\sqrt{M}} \sin \left( 2\pi \frac{(j+\alpha)(j'+\alpha)}{M} \right) & (m=1)
\end{array}
\right., \label{eq:rhoST2Z2} \\
\rho_{T^2/\mathbb{Z}_2^m}^{(\alpha,\alpha)}(\widetilde{T})_{j,j'} =&
\left\{
\begin{array}{ll}
e^{\pi i\frac{(j+\alpha)^2}{M}} \delta_{j,j'} & (m=0) \\
e^{\pi i\frac{(j+\alpha)^2}{M}} \delta_{j,j'} & (m=1)
\end{array}
\right., \label{eq:rhoTT2Z2}
\end{align}
and they satisfy
\begin{align}
\rho_{T^2/\mathbb{Z}_2^m}^{(\alpha,\alpha)}(\widetilde{Z})_{jj'} = \rho_{T^2/\mathbb{Z}_2^m}^{(\alpha,\alpha)}(\widetilde{S})_{jj'}^2 = (-1)^m e^{\pi i/2} \delta_{j,j'}, \label{eq:algZT2Z2}
\end{align}
in addition to Eqs.~(\ref{eq:JrhoT200S4})-(\ref{eq:rhoT200ZT}) and Eqs.~(\ref{eq:JrhoT211S4})-(\ref{eq:rhoT211ZT}), in stead of Eqs.~(\ref{eq:JrhoT200S2}) and (\ref{eq:JrhoT211S2}).
It shows that both $\mathbb{Z}_2$-even and odd mode wave functions are closed under the ``modular transformation'' each other, that is, the representation on magnetized $T^2$ can be reducible into smaller representations on the magnetized $T^2/\mathbb{Z}_2$ orbifold.
In addition, both of them with $M=$even and odd also behave as ``modular forms'' of weight $1/2$ for $\widetilde{\Gamma}(2M)$ and $\widetilde{\Gamma}(8M)$, and they transform non-trivially under $\widetilde{\Gamma}_{2M}$ and $\widetilde{\Gamma}_{8M}$, respectively.
For example, let us consider the case with $M=4$ and $(\alpha_1,\alpha_{\tau})=(0,0)$.
For wave functions on the magnetized $T^2$,
\begin{align}
\begin{pmatrix}
\psi_{T^2}^{(0,0),4}(z,\tau) \\
\psi_{T^2}^{(1,0),4}(z,\tau) \\
\psi_{T^2}^{(2,0),4}(z,\tau) \\
\psi_{T^2}^{(3,0),4}(z,\tau)
\end{pmatrix},
\label{eq:waveT2M4}
\end{align}
representations of $\widetilde{S}$ and $\widetilde{T}$ are given by
\begin{align}
\rho^{(0,0)}_{T^2}(\widetilde{S})={-e^{\pi i/4} \over 2}
\begin{pmatrix}
1 & 1 & 1 & 1 \\
1 & i & -1 & -i \\ 
1 & -1 & 1 & -1 \\
1 & -i & -1 & i \\
\end{pmatrix},
\quad
\rho^{(0,0)}_{T^2}(\widetilde{T})=
\begin{pmatrix} 
1 & 0 & 0 & 0 \\ 
0 & e^{\pi i/4} & 0 & 0 \\
 0 & 0 & -1 & 0 \\ 
0 & 0 & 0 &  e^{\pi i/4} \\
 \end{pmatrix}.
\label{eq:M4repT2}
\end{align}
On the other hand, for wave functions on magnetized $T^2/\mathbb{Z}_2$ with $m=0$,
\begin{align}
\begin{pmatrix}
\psi_{T^2/\mathbb{Z}_2^0}^{(0,0),4}(z,\tau) \\
\psi_{T^2/\mathbb{Z}_2^0}^{(1,0),4}(z,\tau) \\
\psi_{T^2/\mathbb{Z}_2^0}^{(2,0),4}(z,\tau)
\end{pmatrix}
=
\begin{pmatrix}
\psi_{T^2}^{(0,0),4}(z,\tau) \\
\frac{1}{\sqrt{2}}\left( \psi_{T^2}^{(1,0),4}(z,\tau) + \psi_{T^2}^{(3,0),4}(z,\tau) \right) \\
\psi_{T^2}^{(2,0),4}(z,\tau)
\end{pmatrix},
\label{eq:waveT2Z20M4}
\end{align}
and ones with $m=1$,
\begin{align}
\psi_{T^2/\mathbb{Z}_2^1}^{(1,0),4}(z,\tau) =
\frac{1}{\sqrt{2}} \left( \psi_{T^2}^{(1,0),4}(z,\tau) - \psi_{T^2}^{(3,0),4}(z,\tau) \right), \label{eq:waveT2Z21M4}
\end{align}
representations of $\widetilde{S}$ and $\widetilde{T}$ are given by
\begin{align}
\rho^{(0,0)}_{T^2/\mathbb{Z}_2^{0}}(\widetilde{S})={-e^{i\pi /4} \over 2}
\begin{pmatrix} 
1 & \sqrt{2} & 1 \\ 
\sqrt{2} & 0 & -\sqrt{2} \\ 
1 & -\sqrt{2} & 1 \\ 
\end{pmatrix}, \quad
&\rho^{(0,0)}_{T^2/\mathbb{Z}_2^{0}}(\widetilde{T})=
\begin{pmatrix} 
1 & 0 & 0 \\ 
0 & e^{\pi i/4} & 0 \\ 
0 & 0 & -1 \\ 
\end{pmatrix},
\label{eq:M4evenrep}
\end{align}
and
\begin{align}
\rho^{(0,0)}_{T^2/\mathbb{Z}_2^{1}}(\widetilde{S})=e^{-\pi i/4}, \quad \rho^{(0,0)}_{T^2/\mathbb{Z}_2^{1}}(\widetilde{T})=e^{\pi i/4}, \label{eq:M4oddrep}
\end{align}
respectively.
It means that the 4D-representation on magnetized $T^2$ in Eq.~(\ref{eq:M4repT2}) are decomposed into the 3D-representation on the magnetized $T^2/\mathbb{Z}_2$ with $m=0$ in Eq.~(\ref{eq:M4evenrep}) and 1D-representation on the magnetized $T^2/\mathbb{Z}_2$ with $m=1$ in Eq.~(\ref{eq:M4oddrep}).
The 1D-representation in Eq.~(\ref{eq:M4oddrep}) becomes the representation of $\mathbb{Z}_8$ group.
On the other hand, the 3D-representation in Eq.~(\ref{eq:M4evenrep}) becomes non-trivial.
In the next subsection, in particular, let us see 3D-representations on the magnetized $T^2/\mathbb{Z}_2$ twisted orbifold compatible with the modular symmetry.

Before the end of this subsection, we comment on modular symmetry for Majorana neutrino mass terms generated by D-brane instanton effects which are non-perturbative effects.
Since we can find the modular transformation
for 4D fields and three-point couplings,
we can check the modular symmetry for Majorana mass terms in Eq.~(\ref{eq:neutrinomass}).
However, we can find that they are not modular invariant by the modular transformation for measures of D-brane instanton zero modes:
\begin{align}
\widetilde{\gamma} : d^2\beta d^2 \gamma \rightarrow J_{2}(\widetilde{\gamma},\tau){\rm det}[\rho^{(\alpha_{inst},\alpha_{inst})}_{T^2/\mathbb{Z}_2^m}(\widetilde{\gamma})] d^2\beta d^2 \gamma,
\label{eq:dbetadgamma}
\end{align}
which can be obtained from Eq.~(\ref{eq:Grassmannint}) and the modular transformation for 4D D-barane instanton zero modes.
In other words, this modular symmetry anomaly\footnote{If the other moduli $T_{\alpha}$ in the classical action of the D-brane instanton $S_{cl}(T_{\alpha, M_{inst}})$ also transform appropriately under the modular transformation of $\tau$, (which may be supported by the 4D Green-Schwarz mechanism,) the modular symmetry anomaly can be canceled.} comes from integration of D-brane instanton zero modes appeared from D-brane instanton which is a non-perturbative effects.
Especially, if ${\rm det}[\rho^{(\alpha_{inst},\alpha_{inst})}_{T^2/\mathbb{Z}_2^m}(\widetilde{\gamma}_1)] \neq 1$ for the generator $\widetilde{\gamma}_1 \in \mathbb{Z}_N$, the $\mathbb{Z}_N$ subsymmetry of the modular flavor symmetry is broken by integration of D-brane instanton zero modes appeared from D-brane instanton which is a non-perturbative effects.
Thus, in order to clarify which part of the modular flavor symmetry is broken, it is important to find which elements of the modular flavor group such that the determinants of the representation of them are not equal to 1.
Note that the general analysis of the non-Abelian discrete symmetry anomaly has been discussed in Ref.~\cite{Kobayashi:2021xfs}. (See also Appendix~\ref{ap:Anomaly}.)
In particular, we can obtain the determinants of the representation of $\widetilde{T}$ and $\widetilde{S}$ transformations from Eq.~(\ref{eq:algZT2Z2}) and Eqs.~(\ref{eq:JrhoT200S4})-(\ref{eq:rhoT200ZT}) as well as Eqs.~(\ref{eq:JrhoT211S4})-(\ref{eq:rhoT211ZT});
\begin{align}
\begin{array}{l}
{\rm det}\rho^{(\alpha,\alpha)}_{T^2/\mathbb{Z}_2^m}(\widetilde{T}) = \left\{
\begin{array}{ll}
e^{\frac{\pi i}{M}\sum_{j=0}^{\frac{M}{2}}j^2} = e^{\frac{\pi i}{24}(M+1)(M+2)} & (\alpha,m) = (0,0) \\
e^{\frac{\pi i}{M}\sum_{j=1}^{\frac{M}{2}-1}j^2} = e^{\frac{\pi i}{24}(M-1)(M-2)} & (\alpha,m) = (0,1) \\
e^{\frac{\pi i}{M}\sum_{j=0}^{\frac{M-1}{2}-1}(j^2+j+\frac{1}{4})} = e^{\frac{\pi i}{24}(M-1)(M-2)} & (\alpha,m) = (\frac{1}{2},0) \\
e^{\frac{\pi i}{M}\sum_{j=0}^{\frac{M-1}{2}}(j^2+j+\frac{1}{4})} = e^{\frac{\pi i}{24}(M+1)(M+2)} & (\alpha,m) = (\frac{1}{2},1)
\end{array}
\right., \\ 
{\rm det}\rho^{(\alpha,\alpha)}_{T^2/\mathbb{Z}_2^m}(\widetilde{S}) = {\rm det}\rho^{(\alpha,\alpha)}_{T^2/\mathbb{Z}_2^m}(\widetilde{T})^{-3} {\rm det}\rho^{(\alpha,\alpha)}_{T^2/\mathbb{Z}_2^m}(\widetilde{S})^{-2} = \left\{
\begin{array}{ll}
e^{-\frac{\pi i}{8}(M+2)(M+3)} & (\alpha,m) = (0,0) \\
e^{-\frac{\pi i}{8}(M-2)(M-3)} & (\alpha,m) = (0,1) \\
e^{-\frac{\pi i}{8}M(M-1)} & (\alpha,m) = (\frac{1}{2},0) \\
e^{-\frac{\pi i}{8}M(M+1)} & (\alpha,m) = (\frac{1}{2},1)
\end{array}
\right..
\end{array}
\label{eq:detT2Z2TS}
\end{align}
Then, we can find that which combinations of $\widetilde{S}$ and $\widetilde{T}$ are anomaly-free or anomalous~\cite{Kariyazono:2019ehj}.


\subsection{Modular flavor symmetry of three-generation modes on magnetized $T^2/\mathbb{Z}_2$ twisted orbifold}
\label{subsec:T2Z2three}

In this subsection, let us see the detailed structure of modular flavor symmetry of three-generational chiral zero modes on the magnetized $T^2/\mathbb{Z}_2$ orbifold.
First of all,  as shown in Tables~\ref{tab:even} and \ref{tab:odd}, there are four three-generation modes on the magnetized $T^2/\mathbb{Z}_2$ orbifold consistent with the ``modular symmetry'': $(M;\alpha_1,\alpha_{\tau};m)=(4;0,0;0)$, $(8;0,0;1)$, $(5;1/2,1/2;1)$, and $(7;1/2,1/2;0)$,
respectively.
\begin{table}[H]
\centering
\begin{tabular}{|c|c|c|c|c|c|} \hline
 & $M$ & 2 & 4 & 6 & 8  \\ \hline
$\mathbb{Z}_2$-even: $N_0(M)$ & $\frac{M}{2}+1$ & 2 & \fbox{3} & 4 & 5 \\ \hline
$\mathbb{Z}_2$-odd: $N_1(M)$ & $\frac{M}{2}-1$ & 0 & 1 & 2 & \fbox{3} \\ \hline
order $h$ of $\widetilde{T}$ ($\widetilde{T}^h=\mathbf{1}$) & $2M$ & 4 & 8 & 12 & 16 \\ \hline
\end{tabular}
\caption{The number of the $\mathbb{Z}_2$-even ($m=0$) modes, $N_0(M)$, and the $\mathbb{Z}_2$-odd ($m=1$) modes, $N_1(M)$, on the $T^2/\mathbb{Z}_2$ twisted orbifold with $M=$even and $(\alpha_1,\alpha_2)=(0,0)$, and the order of $\widetilde{T}$.
The three generations are boxed.}
\label{tab:even}
\centering
\begin{tabular}{|c|c|c|c|c|c|} \hline
 & $M$ & 1 & 3 & 5 & 7  \\ \hline
$\mathbb{Z}_2$-even: $N_0(M)$ & $\frac{M-1}{2}$ & 0 & 1 & 2 & \fbox{3} \\ \hline
$\mathbb{Z}_2$-odd: $N_1(M)$ & $\frac{M+1}{2}$ & 1 & 2 & \fbox{3} & 4 \\ \hline
order $h$ of $\widetilde{T}$ ($\widetilde{T}^h=\mathbf{1}$) & $8M$ & 8 & 24 & 40 & 56 \\ \hline
\end{tabular}
\caption{The number of the $\mathbb{Z}_2$-even ($m=0$) modes, $N_0(M)$, and the $\mathbb{Z}_2$-odd ($m=1$) modes, $N_1(M)$, on the $T^2/\mathbb{Z}_2$ twisted orbifold with $M=$odd and $(\alpha_1,\alpha_2)=(1/2,1/2)$, and the order of $\widetilde{T}$.
The three generations are boxed.}
\label{tab:odd}
\end{table}
Here, we notice that Ref.~\cite{deAdelhartToorop:2011re} shows three-dimensional representations can be obtained from specific finite modular groups $\Gamma_N\ (N=3,4,5,7,8,16)$: $\Gamma_3 \simeq PSL(2,\mathbb{Z}_3) \simeq A_4$, $\Gamma_4 \simeq S_4$, $\Gamma_5 \simeq PSL(2,\mathbb{Z}_5) \simeq A_5$, $\Gamma_7 \simeq PSL(2,\mathbb{Z}_7)$, $\Gamma_8 \supset \Delta(96)$, and $\Gamma_{16} \supset \Delta(384)$.
In the following, we can find that the above four three-generation modes are representations of the covering or central extended groups of corresponding finite modular groups $\Gamma_N$.



\begin{itemize}
\item {\large $M=$even and $(\alpha_1,\alpha_{\tau})=(0,0)$ case}
\end{itemize}

Here, let us consider $M=$even and $(\alpha_1,\alpha_{\tau})=(0,0)$ case.
In this case, there are only two three-generation modes: $(M;\alpha_1,\alpha_{\tau};m)=(4;0,0;0), (8;0,0;1)$.
From now, we show that the three-generation modes with $M=4,8$ are representations of the quadruple covering group of $\Delta(6M^2) (\subset \Gamma_{2M})$, $\widetilde{\Delta}(6M^2) (\subset \widetilde{\Gamma}_{2M})$.

First, as in appendix~\ref{app:proof}, if the representations of $\widetilde{S}$ and $\widetilde{T}$ transformation further satisfy
\begin{align}
(\widetilde{S}^{-1}\widetilde{T}^{-1}\widetilde{S}\widetilde{T})^3=\mathbf{1}, \label{eq:X}
\end{align}
in addition to the algebraic relations of $\widetilde{\Gamma}_{2M}$ $(M \in 4\mathbb{Z})$, the following generators,
\begin{align}
a=\widetilde{S}\widetilde{T}^2\widetilde{S}^5\widetilde{T}^4, \quad a'=\widetilde{S}\widetilde{T}^2\widetilde{S}^{-1}\widetilde{T}^{-2}, \quad b=\widetilde{T}^{\frac{M}{2}+3}\widetilde{S}^{\frac{3}{2}M+3}\widetilde{T}^{M}, \quad c=\widetilde{S}\widetilde{T}^{M-2}\widetilde{S}\widetilde{T}^{\frac{3}{2}M-1}, \label{eq:DeltatildeST}
\end{align}
satisfy
\begin{align}
&a^M=a'^M=b^3=c^8=\mathbf{1}, \label{eq:algDeltatilde} \\
&aa'=a'a, \ cbc^{-1}=b^{-1}, \ bab^{-1}=a^{-1}a'^{-1}, \ ba'b^{-1}=a, \ cac^{-1}=a'^{-1}, ca'c=a^{-1}, \notag
\end{align}
which means that they become generators of $\widetilde{\Delta}(6M^2) \simeq (\mathbb{Z}_M \times \mathbb{Z}_M) \rtimes \mathbb{Z}_3 \rtimes \mathbb{Z}_8 \simeq \Delta(3M^2) \rtimes \mathbb{Z}_8$, where $a^{(')}$, $b$, $c$ denote ones of $\mathbb{Z}^{(')}_M$, $\mathbb{Z}_3$, $\mathbb{Z}_8$, respectively.
Note that if the representations of $\widetilde{S}$ and $\widetilde{T}$ transformation satisfy Eq.~(\ref{eq:X}) and the algebraic relations of $\Gamma_{2M}$ $(\Gamma'_{2M})$ $(M \in 4\mathbb{Z})$ instead of $\widetilde{\Gamma}_{2M}$, the generators in Eq.~(\ref{eq:DeltatildeST}) satisfy the relations in Eq.~(\ref{eq:algDeltatilde}) replacing $c^8=\mathbf{1}$ with $c^2=\mathbf{1}$ $(c^4=\mathbf{1})$, that is, they become generators of $\Delta(6M^2) \simeq  (\mathbb{Z}_M \times \mathbb{Z}_M) \rtimes \mathbb{Z}_3 \rtimes \mathbb{Z}_2 \simeq \Delta(3M^2) \rtimes \mathbb{Z}_2$ $(\Delta'(6M^2) \simeq  (\mathbb{Z}_M \times \mathbb{Z}_M) \rtimes \mathbb{Z}_3 \rtimes \mathbb{Z}_4 \simeq \Delta(3M^2) \rtimes \mathbb{Z}_4)$.
Thus, $\widetilde{\Delta}(6M^2)(\subset \widetilde{\Gamma}_{2M})$ ($\Delta'(6M^2) (\subset \Gamma'_{2M})$) become the quadruple (double) covering group of $\Delta(6M^2) (\subset \Gamma_{2M})$.

The representations of $\widetilde{S}$ and $\widetilde{T}$ transformation for the three-generation modes, $(M;\alpha_1,\alpha_{\tau};m)$ $=(4;0,0;0)$ and $(8;0,0;1)$ are given by
\begin{align}
\rho^{(0,0)}_{T^2/\mathbb{Z}_2^0}(\widetilde{S}) = \frac{-e^{\pi i/4}}{2}
\begin{pmatrix}
1 & \sqrt{2} & 1 \\
\sqrt{2} & 0 & -\sqrt{2} \\
1 & -\sqrt{2} & 1
\end{pmatrix},
\quad&
\rho^{(0,0)}_{T^2/\mathbb{Z}_2^0}(\widetilde{T}) = 
\begin{pmatrix}
1 & \ & \ \\
\ & e^{\pi i/4} & \ \\
\ & \ & -1
\end{pmatrix},
\label{eq:M4ST}
\end{align}
and
\begin{align}
\rho^{(0,0)}_{T^2/\mathbb{Z}_2^1}(\widetilde{S}) = \frac{-ie^{\pi i/4}}{2}
\begin{pmatrix}
1 & \sqrt{2} & 1 \\
\sqrt{2} & 0 & -\sqrt{2} \\
1 & -\sqrt{2} & 1
\end{pmatrix},
\quad&
\rho^{(0,0)}_{T^2/\mathbb{Z}_2^1}(\widetilde{T}) = e^{\pi i/8}
\begin{pmatrix}
1 & \ & \ \\
\ & e^{3\pi i/8} & \ \\
\ & \ & -1
\end{pmatrix},
\label{eq:M8ST}
\end{align}
respectively.
Both of the above $\widetilde{S}$ and $\widetilde{T}$ matrices are the same forms as
\begin{align}
\rho(\widetilde{S}) = \frac{e^{i\theta_1}}{2}
\begin{pmatrix}
1 & \sqrt{2} & 1 \\
\sqrt{2} & 0 & -\sqrt{2} \\
1 & -\sqrt{2} & 1
\end{pmatrix},
\quad&
\rho(\widetilde{T}) = e^{i\theta_2}
\begin{pmatrix}
1 & \ & \ \\
\ & e^{i\theta_3} & \ \\
\ & \ & -1
\end{pmatrix}, \quad \forall \theta_{1,2,3} \in \mathbb{R},
\label{eq:STingen}
\end{align}
and we can find that they satisfy Eq.~(\ref{eq:X}).
Thus, the representations of the three-generation modes $(M;\alpha_1,\alpha_{\tau};m)=(4;0,0;0), (8;0,0;1)$ in Eqs.~(\ref{eq:M4ST}) and (\ref{eq:M8ST}) become 
the three-dimensional representations of $\widetilde{\Delta}(6M^2)$
($\widetilde{\Delta}(96)$ and $\widetilde{\Delta}(384)$).\footnote{Eq.~(\ref{eq:M4ST}) also satisfies $\widetilde{S}^5\widetilde{T}^6\widetilde{S}\widetilde{T}^4\widetilde{S}\widetilde{T}^2\widetilde{S}\widetilde{T}^4=\mathbb{I}$~\cite{Liu:2020msy}.}
In other words, we can find that the three-generation modes with $(M;\alpha_1,\alpha_{\tau};m)=(4;0,0;0), (8;0,0;1)$ are transformed under the modular transformation as three-dimensional representations of finite modular $\widetilde{\Delta}(6M^2)$ ($\widetilde{\Delta}(96)$ and $\widetilde{\Delta}(384)$) groups with weight $1/2$.

We comment on the modular flavor symmetry anomaly.
Since we can find that the determinants of generators are
\begin{align}
{\rm det}(a) = {\rm det}(a') = {\rm det}(b) =1, \ {\rm det}(c) = {\rm det}(S)^2 {\rm det}(T)^{\frac{M}{2}-3} = e^{\pi i/4}, \ {\rm det}(c)^8=1, \label{eq:anomalygenT2Z2M48}
\end{align}
from Eq.~(\ref{eq:detT2Z2TS}),
only $\mathbb{Z}_8$ symmetry, generated by $c$, can be anomalous while the normal subgroups $\Delta(3M^2)$ ($\Delta(48)$ and $\Delta(192)$) automatically remain anomaly-free.
It is consistent with the general result of non-Abelian discrete symmetry anomaly~\cite{Kobayashi:2021xfs}.

\begin{itemize}
\item {\large $M=$odd and $(\alpha_1,\alpha_{\tau})=(1/2,1/2)$ case}
\end{itemize}

Here, let us consider $M=$odd and $(\alpha_1,\alpha_{\tau})=(1/2,1/2)$ case.
In this case, there are only two three-generation modes: $(M;\alpha_1,\alpha_{\tau};m)=(5;1/2,1/2;1), (7;1/2,1/2;0)$.
From now, we show that the three-generation modes with $M=5,7$ are representations of the $\mathbb{Z}_8$ central extension group of $\Gamma_{M} \simeq PSL(2,\mathbb{Z}_M)$, $PSL(2,\mathbb{Z}_M) \times \mathbb{Z}_8$.

First, let us see modular flavor symmetry of the three-generation modes with $(M;\alpha_1,\alpha_{\tau};m)=(5;1/2,1/2;1)$.
The representation of $\widetilde{S}$ and $\widetilde{T}$ is given by
\begin{align}
\rho^{(\frac{1}{2},\frac{1}{2})}_{T^2/\mathbb{Z}_2^1}(\widetilde{S}) = \frac{-ie^{\pi i/4}}{\sqrt{5}}
\begin{pmatrix}
2 \sin \left(\frac{\pi}{10}\right) & 2 \sin \left(\frac{3\pi}{10}\right) & \sqrt{2} \\
2 \sin \left(\frac{3\pi}{10}\right) & 2 \sin \left(\frac{\pi}{10}\right) & - \sqrt{2} \\
\sqrt{2} & - \sqrt{2} & 1
\end{pmatrix}, \ 
\rho^{(\frac{1}{2},\frac{1}{2})}_{T^2/\mathbb{Z}_2^0}(\widetilde{T}) = e^{\pi i/20}
\begin{pmatrix}
1 & \ & \ \\
\ & e^{2\pi i/5} & \ \\
\ & \ & e^{6\pi i/5}
\end{pmatrix}.
\label{eq:M5ST}
\end{align}
When we define the following generators,
\begin{align}
a=\widetilde{S}\widetilde{T}^{25}, \ b=\widetilde{S}\widetilde{T}, \ c=\widetilde{T}^5, \label{eq:M5gen}
\end{align}
they satisfy
\begin{align}
a^2=b^3=(ab)^5=c^8=\mathbf{1}, \ ac=ca, \ bc=cb, \label{eq:M5alg}
\end{align}
which mean they are the generators of $A_5 \times \mathbb{Z}_8$.
Thus, we can find that the three-generational modes with $(M;\alpha_1,\alpha_{\tau};m)=(5;1/2,1/2;1)$ are transformed as the three-dimensional representation of finite modular $A_5 \times \mathbb{Z}_8 \simeq PSL(2,\mathbb{Z}_5) \times \mathbb{Z}_8$ group with weight $1/2$.

Next, let us see modular flavor symmetry of the three-generation modes with $(M;\alpha_1,\alpha_{\tau};m)=(7;1/2,1/2;0)$.
The representation of $\widetilde{S}$ and $\widetilde{T}$ is given by
\begin{align}
\rho^{(\frac{1}{2},\frac{1}{2})}_{T^2/\mathbb{Z}_2^0}(\widetilde{S}) = \frac{-2e^{\pi i/4}}{\sqrt{7}}
\begin{pmatrix}
\cos \left(\frac{\pi}{14}\right) & \cos \left(\frac{3\pi}{14}\right) & \cos \left(\frac{5\pi}{14}\right) \\
\cos \left(\frac{3\pi}{14}\right) & \cos \left(\frac{9\pi}{14}\right) & - \cos \left(\frac{\pi}{14}\right) \\
\cos \left(\frac{5\pi}{14}\right) & - \cos \left(\frac{\pi}{14}\right) & \cos \left(\frac{3\pi}{14}\right)
\end{pmatrix}, \ 
\rho^{(\frac{1}{2},\frac{1}{2})}_{T^2/\mathbb{Z}_2^0}(\widetilde{T}) = e^{\pi i/28}
\begin{pmatrix}
1 & \ & \ \\
\ & e^{2\pi i/7} & \ \\
\ & \ & e^{6\pi i/7}
\end{pmatrix}.
\label{eq:M7ST}
\end{align}
Note that they also satisfy
\begin{align}
(\widetilde{S}^{-1}\widetilde{T}^{-1}\widetilde{S}\widetilde{T})^4=\mathbf{1}. \label{eq:add}
\end{align}
When we define the following generators,
\begin{align}
a=\widetilde{S}\widetilde{T}^{49}, \ b=\widetilde{S}\widetilde{T}, \ c=\widetilde{T}^7, \label{eq:M7gen}
\end{align}
they satisfy
\begin{align}
a^2=b^3=(ab)^7=(a^{-1}b^{-1}ab)^4=c^8=\mathbf{1}, \ ac=ca, \ bc=cb, \label{eq:M7alg}
\end{align}
which mean they are the generators of $PSL(2,\mathbb{Z}_7) \times \mathbb{Z}_8$.
Thus, we can find that the three-generational modes with $(M;\alpha_1,\alpha_{\tau};m)=(7;1/2,1/2;0)$ are transformed as the three-dimensional representation of finite modular $PSL(2,\mathbb{Z}_7) \times \mathbb{Z}_8$ group with weight $1/2$.

We comment on the modular flavor symmetry anomaly.
Since we can find that the determinants of generators are
\begin{align}
{\rm det}(a) = {\rm det}(b) =1, \ {\rm det}(c) = {\rm det}(T)^{M} = e^{3\pi i/4}, \ {\rm det}(c)^8=1, \label{eq:anomalygenT2Z2M57}
\end{align}
from Eq.~(\ref{eq:detT2Z2TS}),
only $\mathbb{Z}_8$ symmetry, generated by $c$, can be anomalous while the simple groups $PSL(2,\mathbb{Z}_M)$ ($PSL(2,\mathbb{Z}_5) \simeq A_5$ and $PSL(2,\mathbb{Z}_7)$) automatically remain anomaly-free.
It is consistent with the general result of non-Abelian discrete symmetry anomaly~\cite{Kobayashi:2021xfs}.


\subsection{Modular flavor symmetry of three-generation modes on magnetized $(T^2_1 \times T^2_2)/(\mathbb{Z}^{(t)}_2 \times \mathbb{Z}^{(p)}_2)$ orbifold}
\label{subsec:T2T2Z2tZ2pthree}

In this subsection, similarly let us see the modular flavor symmetry of three-generational chiral zero modes on the magnetized $(T^2 \times T^2)/(\mathbb{Z}^{(t)}_2 \times \mathbb{Z}^{(p)}_2)$ orbifold~\cite{Kikuchi:2020nxn}, where $\mathbb{Z}^{(t)}_2$ denotes the $\mathbb{Z}_2$ twist: $(z_1,z_2) \rightarrow (-z_1,-z_2)$, while $\mathbb{Z}^{(p)}_2$ denotes the $\mathbb{Z}_2$ permutation: $(z_1,z_2) \leftrightarrow (z_2,z_1)$.
Note that $\mathbb{Z}^{(p)}_2$ identification requires that ${\cal A}_1={\cal A}_2 \equiv {\cal A}$, $\tau_1=\tau_2 \equiv \tau$, $M_1=M_2 \equiv M$, $\alpha^{(1)}_{1,\tau} = \alpha^{(2)}_{1,\tau} \equiv \alpha_{1,\tau}$, and $m_1=m_2 \equiv m$.
In addition, since we consider the case that the wave functions have modular flavor symmetries, it requires that $\alpha_1=\alpha_{\tau} \equiv \alpha$ and $\alpha=0,1/2$ for $M=$ even, odd, respectively.
Hence, we consider the following wave functions,
\begin{align}
\Psi^{(j_1+\alpha,j_2+\alpha),M}_{(t)m(p)n,0}(z_1,z_2,\tau) 
\equiv {\cal N}^{(j_1,j_2)}_{(t,p)} \left(\Psi^{(j_1+\alpha,j_2+\alpha),M}_{(T^2_1 \times T^2_2)/\mathbb{Z}^{(t)m}_2,0}(z_1,z_2,\tau) + (-1)^n
\Psi^{(j_1+\alpha,j_2+\alpha),M}_{(T^2_1 \times T^2_2)/\mathbb{Z}^{(t)m}_2,0}(z_2,z_1,\tau) \right),
\label{eq:waveZ2tp} 
\end{align}
with
\begin{align}
\Psi^{(j_1+\alpha,j_2+\alpha),M}_{(T^2_1 \times T^2_2)/\mathbb{Z}^{(t)m}_2,0}(z_1,z_2,\tau) \equiv \psi^{(j_1+\alpha,\alpha),M}_{T^2_1/\mathbb{Z}^{(t)}_2,0}(z_1,\tau) \psi^{(j_2+\alpha,\alpha),M}_{T^2_2/\mathbb{Z}^{(t)}_2,0}(z_2,\tau), \label{eq:waveT2Z2T2Z2} 
\end{align}
where we set $j_1 \geq j_2$ and ${\cal N}^{(j_1,j_2)}_{(t,p)}=1/2$ for $j_1=j_2$, otherwise ${\cal N}^{(j_1,j_2)}_{(t,p)}=1/\sqrt{2}$.
Indeed, they satisfy
\begin{align}
\begin{array}{l}
\Psi^{(j_1+\alpha,j_2+\alpha),M}_{t(m)(p)n,0}(-z_1,-z_2,\tau) = (-1)^m \Psi^{(j_1+\alpha,j_2+\alpha),M}_{t(m)(p)n,0}(z_1,z_2,\tau), \\
\Psi^{(j_1+\alpha,j_2+\alpha),M}_{t(m)(p)n,0}(z_2,z_1,\tau) = (-1)^n \Psi^{(j_1+\alpha,j_2+\alpha),M}_{t(m)(p)n,0}(z_1,z_2,\tau).
\end{array}
\end{align}

Now, we discuss the ``modular symmetry'' for the wave functions in Eq.~(\ref{eq:waveZ2tp}).
First, they have the modular weight $1$.
On the other hand, the unitary representation,
\begin{align}
\rho_{(t)m(p)n}^{(\alpha,\alpha)}(\gamma)_{(j_1j_2)(j'_1j'_2)} \equiv
2{\cal N}^{(j_1,j_2)}_{(t,p)}{\cal N}^{(j'_1,j'_2)}_{(t,p)} \left( \rho_{(t)m}^{(\alpha,\alpha)}(\gamma)_{(j_1j_2)(j'_1j'_2)} + (-1)^n \rho_{(t)m}^{(\alpha,\alpha)}(\gamma)_{(j_1j_2)(j'_2j'_1)} \right), \label{eq:rhogammaT2T2Z2p}
\end{align}
with
\begin{align}
\rho_{(t)m}^{(\alpha,\alpha)}(\gamma)_{(j_1j_2)(j'_1j'_2)} \equiv
\rho_{T^2/\mathbb{Z}_2^m}^{(\alpha,\alpha)}(\widetilde{\gamma})_{j_1,j'_1} \rho_{T^2/\mathbb{Z}_2^m}^{(\alpha,\alpha)}(\widetilde{\gamma})_{j_2,j'_2}, \label{eq:T2T2Z2t}
\end{align}
satisfies
\begin{align}
&\rho_{(t)m(p)n}^{(\alpha,\alpha)}(Z) = \rho_{(t)m(p)n}^{(\alpha,\alpha)}(S)^2 = - \delta_{(j_1j_2),(j'_1j'_2)}, \label{eq:Z2tpZS2} \\
&\rho_{(t)m(p)n}^{(\alpha,\alpha)}(Z)^2 = \rho_{(t)m(p)n}^{(\alpha,\alpha)}(S)^4 = [\rho_{(t)m(p)n}^{(\alpha,\alpha)}(S)\rho_{(t)m(p)n}^{(\alpha,\alpha)}(T)]^3 = \delta_{(j_1j_2),(j'_1j'_2)}, \label{eq:Z2tpZ2S4ST3} \\
&[\rho_{(t)m(p)n}^{(\alpha,\alpha)}(Z)\rho_{(t)m(p)n}^{(\alpha,\alpha)}(T)] = [\rho_{(t)m(p)n}^{(\alpha,\alpha)}(T)\rho_{(t)m(p)n}^{(\alpha,\alpha)}(Z)], \label{eq:Z2tpZTTZ}
\end{align}
and
\begin{align}
\begin{array}{lll}
&\rho_{(t)m(p)n}^{(\alpha,\alpha)}(T)^{N=2M} = \delta_{(j_1j_2),(j'_1j'_2)}, & (M \in 2\mathbb{Z}) \\
&\rho_{(t)m(p)n}^{(\alpha,\alpha)}(T)^{M} = e^{\pi i/2} \delta_{(j_1j_2),(j'_1j'_2)}, & (M \in 2\mathbb{Z}+1) \\
\Rightarrow &\rho_{(t)m(p)n}^{(\alpha,\alpha)}(T)^{N=4M} = \delta_{(j_1j_2),(j'_1j'_2)}.
\end{array}
\label{eq:Z2tpTN}
\end{align}
Thus, wave functions on magnetized $(T^2_1 \times T^2_2)/(\mathbb{Z}^{(t)}_2 \times \mathbb{Z}^{(p)}_2)$ orbifold behave as ``modular forms'' of weight $1$ for $\Gamma(N)$ with $N=2M$ for $M \in 2\mathbb{Z}$, $N=4M$ for $M \in 2\mathbb{Z}+1$, and they transform non-trivially under $\Gamma'_N$.

In particular, in the following, let us see the modular flavor symmetry of three-generational chiral zero modes.
First of all, as shown in Tables~\ref{tab:Z2tZ2p} and \ref{tab:Z2tZ2pSS}, there are eight three-generation modes on the magnetized $(T^2_1 \times T^2_2)/(\mathbb{Z}^{(t)}_2 \times \mathbb{Z}^{(p)}_2)$ orbifold consistent with the ``modular symmetry'': $(M,\alpha,m,n) = (2,0,0,0)$, $(4,0,0,1)$, $(6,0,1,0)$, $(8,0,1,1)$, $(3,1/2,1,0)$, $(5,1/2,1,1)$, $(5,1/2,0,0)$, and $(7,1/2,0,1)$,
respectively.
\begin{table}[H]
  \centering
  \begin{tabular}{|c|c|c|c|c|c|} \hline
     & $M$ & 2 & 4 & 6 & 8 \\ \hline
    {\rm (even,\ even)}: $N_{(0,0)}(M)$ & $(M+2)(M+4)/8$ & \fbox{3} & 6 & 10 & 15 \\ \hline
    {\rm (even,\ odd)}: $N_{(0,1)}(M)$ & $M(M+2)/8$ & 1 & \fbox{3} & 6 & 10 \\ \hline
    {\rm (odd,\ even)}: $N_{(1,0)}(M)$ & $M(M-2)/8$ & 0 & 1 & \fbox{3} & 6 \\ \hline
    {\rm (odd,\ odd)}: $N_{(1,1)}(M)$ & $(M-2)(M-4)/8$ & 0 & 0 & 1 & \fbox{3} \\ \hline
    order $h$ of $\widetilde{T}$ ($\widetilde{T}^h=\mathbf{1}$)  & $2M$ & 4 & 8 & 12 & 16 \\ \hline
\end{tabular}
\caption{The number of ($\mathbb{Z}_2^{\rm (t)}$ twist, $\mathbb{Z}_2^{\rm (p)}$ permutation)-eigenmodes, $N_{(m,n)}(M)$, on the $(T^2_1 \times T^2_2)/(\mathbb{Z}_2^{\rm (t)}\times \mathbb{Z}_2^{\rm (p)})$ orbifold with $M=$even and $(\alpha_1,\alpha_2)=(0,0)$, and the order of $\widetilde{T}$.
The three generations are boxed.}
\label{tab:Z2tZ2p}
  \centering
  \begin{tabular}{|c|c|c|c|c|c|} \hline
     & $M$ & 1 & 3 & 5 & 7 \\ \hline
    {\rm (even,\ even)}: $N_{(0,0)}(M)$ & $(M-1)(M+1)/8$ & 0 & 1 & \fbox{3} & 6 \\ \hline
    {\rm (even,\ odd)}: $N_{(0,1)}(M)$ & $(M-1)(M-3)/8$ & 0 & 0 & 1 & \fbox{3} \\ \hline
    {\rm (odd,\ even)}: $N_{(1,0)}(M)$ & $(M+1)(M+3)/8$ & 1 & \fbox{3} & 6 & 10 \\ \hline
    {\rm (odd,\ odd)}: $N_{(1,1)}(M)$ & $(M+1)(M-1)/8$ & 0 & 1 & \fbox{3} & 6 \\ \hline
    order $h$ of $\widetilde{T}$ ($\widetilde{T}^h=\mathbf{1}$)  & $4M$ & 4 & 12 & 20 & 28 \\ \hline
\end{tabular}
\caption{The number of ($\mathbb{Z}_2^{\rm (t)}$ twist, $\mathbb{Z}_2^{\rm (p)}$ permutation)-eigenmodes, $N_{(m,n)}(M)$, on the $(T^2_1 \times T^2_2)/(\mathbb{Z}_2^{\rm (t)}\times \mathbb{Z}_2^{\rm (p)})$ orbifold with $M=$odd and $(\alpha_1,\alpha_2)=(1/2,1/2)$, and the order of $\widetilde{T}$.
The three generations are boxed.}
\label{tab:Z2tZ2pSS}
\end{table}
Similarly, in the following, we can find that those three-generation modes are representations of the covering or central extended groups of corresponding finite modular groups $\Gamma_N\ (N=3,4,5,7,8,16)$.

\begin{itemize}
\item {\large $M=$even and $\alpha=0$ case}
\end{itemize}

Here, let us consider $M=$even and $\alpha=0$ case.
In this case, there are only four three-generation modes: $(M,\alpha,m,n)=(2,0,0,0)$, $(4,0,0,1)$, $(6,0,1,0)$ $(8,0,1,1)$.
From now, we show that the three-generation modes with $M=2,4,6,8$ are representations of the double covering group of $\Delta(6M^2) (\subset \Gamma_{2M})$, $\Delta'(6M^2) (\subset \Gamma'_{2M})$.

First, the representations of $S$ and $T$ transformation for the three-generation modes, $(M,\alpha,m,n)$ $=(2,0,0,0)$, $(4,0,0,1)$, $(6,0,1,0)$ and $(8,0,1,1)$ are given by
\begin{align}
&
\rho^{(0,0)}_{(t)0(p)0}(S) = \frac{i}{2}
\begin{pmatrix}
1 & \sqrt{2} & 1 \\
\sqrt{2} & 0 & -\sqrt{2} \\
1 & -\sqrt{2} & 1
\end{pmatrix},
\quad&
\rho^{(0,0)}_{(t)0(p)0}(T) = 
\begin{pmatrix}
1 & \ & \ \\
\ & i & \ \\
\ & \ & -1
\end{pmatrix},
\label{eq:M2M2ST} \\
%
&
\rho^{(0,0)}_{(t)0(p)1}(S) = - \frac{i}{2}
\begin{pmatrix}
1 & \sqrt{2} & 1 \\
\sqrt{2} & 0 & -\sqrt{2} \\
1 & -\sqrt{2} & 1
\end{pmatrix},
\quad&
\rho^{(0,0)}_{(t)0(p)1}(T) = e^{\pi i/4}
\begin{pmatrix}
1 & \ & \ \\
\ & e^{3\pi i/4} & \ \\
\ & \ & -1
\end{pmatrix},
\label{eq:M4M4ST} \\
%
&
\rho^{(0,0)}_{(t)1(p)0}(S) = - \frac{i}{2}
\begin{pmatrix}
1 & \sqrt{2} & 1 \\
\sqrt{2} & 0 & -\sqrt{2} \\
1 & -\sqrt{2} & 1
\end{pmatrix},
\quad&
\rho^{(0,0)}_{(t)1(p)0}(T) = e^{\pi i/3}
\begin{pmatrix}
1 & \ & \ \\
\ & i & \ \\
\ & \ & -1
\end{pmatrix},
\label{eq:M6M6ST} \\
%
&
\rho^{(0,0)}_{(t)1(p)1}(S) = \frac{i}{2}
\begin{pmatrix}
1 & \sqrt{2} & 1 \\
\sqrt{2} & 0 & -\sqrt{2} \\
1 & -\sqrt{2} & 1
\end{pmatrix},
\quad&
\rho^{(0,0)}_{(t)1(p)1}(T) = e^{5\pi i/8}
\begin{pmatrix}
1 & \ & \ \\
\ & e^{5\pi i/8} & \ \\
\ & \ & -1
\end{pmatrix},
\label{eq:M8M8ST}
\end{align}
respectively.
They satisfy the algebraic relations of $\Gamma'_{2M}$ in Eqs.~(\ref{eq:Z2tpZS2})-(\ref{eq:Z2tpTN}).
In addition, since they are the same forms as ones in Eq.~(\ref{eq:STingen}), they satisfy Eq.~(\ref{eq:X}).
Then, as mentioned in previous subsection, we can prove that, in appendix~\ref{app:proof}, the generators
\begin{align}
a=ST^2ST^4, \quad a'=ST^2S^{-1}T^{-2}, \quad b=T^{\frac{M}{2}+3}S^{\frac{3}{2}M+3}T^{M}, \quad c=ST^{M-2}ST^{\frac{3}{2}M-1}, \label{eq:Delta'STeven}
\end{align}
with $M=4s\ (s \in \mathbb{Z})$, and similarly the generators
\begin{align}
a=ST^2ST^4, \quad a'=ST^2S^{-1}T^{-2}, \quad b=T^{\frac{M}{2}}S^{\frac{3}{2}M}T^{M}, \quad c=ST^{M}ST^{\frac{3}{2}M},\label{eq:Delta'STodd}
\end{align}
with $M=2(2s-1)\ (s \in \mathbb{Z})$, satisfy 
\begin{align}
&a^M=a'^M=b^3=c^4=\mathbf{1}, \label{algDelta'} \\
&aa'=a'a, \ cbc^{-1}=b^{-1}, \ bab^{-1}=a^{-1}a'^{-1}, \ ba'b^{-1}=a, \ cac^{-1}=a'^{-1}, ca'c^{-1}=a^{-1}, \notag
\end{align}
which means that they become generators of  $\Delta'(6M^2) \simeq (\mathbb{Z}_M \times \mathbb{Z}_M) \rtimes \mathbb{Z}_3 \rtimes \mathbb{Z}_4 \simeq \Delta(3M^2) \rtimes \mathbb{Z}_4$, where $a^{(')}$, $b$, $c$ denote ones of $\mathbb{Z}^{(')}_M$, $\mathbb{Z}_3$, $\mathbb{Z}_4$, respectively.
In particular, we have a comment on Eq.~(\ref{eq:M6M6ST}).
Since the $T$ matrix in Eq.~(\ref{eq:M6M6ST}) also satisfies $T^4=e^{4\pi i/3}\mathbb{I}$, this can be the $Z_3$ generator, $d=T^4$, and it commutes with all the generators in Eq.~(\ref{eq:Delta'STodd}).
In addition, in this case, the generators $a$ and $a'$ in Eq.~(\ref{eq:Delta'STodd}) satisfy $a^{2}=a'^{2}=\mathbf{1}$.
Therefore, we can find that the the three-generation modes with $(M,\alpha,m,n)=(2,0,0,0)$, $(4,0,0,1)$, $(6,0,1,0)$, and $(8,0,1,1)$ are transformed as the three-dimensional representations of finite modular $S'_4 \simeq \Delta'(24)$, $\Delta'(96)$, $S'_4 \times \mathbb{Z}_3$, and $\Delta'(384)$ groups with weight $1$, respectively.

Similarly, we comment on the modular flavor symmetry anomaly.
Since we can find that the determinants of generators are
\begin{align}
\begin{array}{l}
{\rm det}(a) = {\rm det}(a') = {\rm det}(b) =1, \\
{\rm det}(c) = \left\{
\begin{array}{ll}
{\rm det}(S)^2 {\rm det}(T)^{\frac{M}{2}-3} = e^{-\pi i/2} & (M=4,8) \\
{\rm det}(S)^2 {\rm det}(T)^{\frac{M}{2}} = e^{\pi i/2} & (M=2,6)
\end{array}
\right., \ {\rm det}(c)^4=1,
\end{array}
\label{eq:anomalygenT2Z2tpM2468}
\end{align}
and also ${\rm det}(d) = {\rm det}(T)^4 = 1$,
only $\mathbb{Z}_4$ symmetry, generated by $c$, can be anomalous while the normal subgroups, $A_4 \simeq \Delta(12)$, $\Delta(48)$, $A_4 \times \mathbb{Z}_3$, and $\Delta(192)$, automatically remain anomaly-free.
It is consistent with the general result of non-Abelian discrete symmetry anomaly~\cite{Kobayashi:2021xfs}.

\begin{itemize}
\item {\large $M=$odd and $\alpha=1/2$ case}
\end{itemize}

Next, let us consider $M=$ odd and $\alpha=1/2$ case.
In this case, there are only four three-generation modes: $(M,\alpha,m,n) = (3,1/2,1,0)$, $(5,1/2,1,1)$, $(5,1/2,0,0)$, and $(7,1/2,0,1)$.
From now, we show that the three-generation modes with $M=3,5,7$ are representations of the $\mathbb{Z}_4$ central extension group of $\Gamma_{M} \simeq PSL(2,\mathbb{Z}_M)$, $PSL(2,\mathbb{Z}_M) \times \mathbb{Z}_4$.
We note that all of the following representations of $S$ and $T$ transformation satisfy the algebraic relations of $\Gamma'_{4M}$ in Eqs.~(\ref{eq:Z2tpZS2})-(\ref{eq:Z2tpTN}).

First, from the following representation of $S$ and $T$ transformation for $(M,\alpha,m,n)=(3,1/2,1,0)$,
\begin{align}
\rho^{(\frac{1}{2},\frac{1}{2})}_{(t)1(p)0}(S) = - \frac{i}{3}
\begin{pmatrix}
1 & 2 & 2 \\
2 & 1 & - 2 \\
2  & - 2  & 1
\end{pmatrix}, \ 
\rho^{(\frac{1}{2},\frac{1}{2})}_{(t)1(p)0}(T) = 
\begin{pmatrix}
e^{\pi i/6} & \ & \ \\
\ & e^{5\pi i/6} & \ \\
\ & \ & e^{9\pi i/6}
\end{pmatrix},
\label{eq:M3M3ST}
\end{align}
we can obtain the following generators,
\begin{align}
a=ST^9, \ b=ST, \ c=T^3, \label{eq:M3M3gen}
\end{align}
satisfying
\begin{align}
a^2=b^3=(ab)^3=c^4=\mathbf{1}, \ ac=ca, \ bc=cb, \label{eq:M3M3alg}
\end{align}
which mean the generators in Eq.~(\ref{eq:M3M3gen}) are ones of $A_4 \times \mathbb{Z}_4$.
Therefore, we can find that the three-generation modes with $(M,\alpha,m,n)=(3,1/2,1,0)$ are transformed under the modular transformation as the three-dimensional representations of finite modular $A_4 \times \mathbb{Z}_4 \simeq PSL(2,\mathbb{Z}_3) \times \mathbb{Z}_4$ group with weight $1$.

Second, from the following representation of $S$ and $T$ transformation for $(M,\alpha,m,n)=(5,1/2,0,0)$,
\begin{align}
\rho^{(\frac{1}{2},\frac{1}{2})}_{(t)0(p)0}(S) &= \frac{4i}{5}
\begin{pmatrix}
A^2 & \sqrt{2} AB & B^2 \\
\sqrt{2} AB & B^2-A^2 & - \sqrt{2} AB \\
B^2 & - \sqrt{2} AB & A^2
\end{pmatrix}, \ 
\rho^{(\frac{1}{2},\frac{1}{2})}_{(t)0(p)0}(T) &= 
\begin{pmatrix}
e^{\pi i/10} & \ & \ \\
\ & e^{5\pi i/10} & \ \\
\ & \ & e^{9\pi i/10}
\end{pmatrix},
\label{eq:M5M5eST} \\
&\qquad A = \cos \left(\frac{\pi}{10}\right), \ B = \cos \left(\frac{3\pi}{10}\right), \notag
\end{align}
we can obtain the following generators,
\begin{align}
a=ST^{25}, \ b=ST, \ c=T^5, \label{eq:M5M5gen}
\end{align}
satisfying
\begin{align}
a^2=b^3=(ab)^5=c^4=\mathbf{1}, \ ac=ca, \ bc=cb, \label{eq:M5M5alg}
\end{align}
which mean the generators in Eq.~(\ref{eq:M5M5gen}) are ones of $A_5 \times \mathbb{Z}_4$.
Therefore, we can find that the three-generation modes with $(M,\alpha,m,n)=(5,1/2,0,0)$ are transformed under the modular transformation as the three-dimensional representations of finite modular $A_5 \times \mathbb{Z}_4 \simeq PSL(2,\mathbb{Z}_5) \times \mathbb{Z}_4$ group with weight $1$.

Third, similarly, from the following representation of $S$ and $T$ transformation for $(M,\alpha,m,n)$ $=(5,1/2,1,1)$,
\begin{align}
\rho^{(\frac{1}{2},\frac{1}{2})}_{(t)1(p)1}(S) &= - \frac{2i}{5}
\begin{pmatrix}
2 \left(A^2 - B^2\right) & - \sqrt{2} \left(A+B\right) & - \sqrt{2} \left(A+B\right) \\
- \sqrt{2} \left(A+B\right) & A- 1 & B + 1 \\
- \sqrt{2} \left(A+B\right) & B + 1 & A-1
\end{pmatrix},
\notag \\
&\qquad A = \sin \left(\frac{\pi}{10}\right), \ B = \sin \left(\frac{3\pi}{10}\right), \notag \\
\rho^{(\frac{1}{2},\frac{1}{2})}_{(t)1(p)1}(T) &= 
\begin{pmatrix}
e^{5\pi i/10} & \ & \ \\
\ & e^{13\pi i/10} & \ \\
\ & \ & e^{17\pi i/10}
\end{pmatrix},
\label{eq:M5M5oST}
\end{align}
we can obtain the generators in Eq.~(\ref{eq:M5M5gen}) satisfying Eq.~(\ref{eq:M5M5alg}).
Therefore, we can find that the three-generation modes with $(M,\alpha,m,n)=(5,1/2,1,1)$ are also transformed under the modular transformation as the three-dimensional representations of finite modular $A_5 \times Z_4 \simeq PSL(2,\mathbb{Z}_5) \times \mathbb{Z}_4$ group with weight $1$.

Fourth, from the following representation of $S$ and $T$ transformation for $(M,\alpha,m,n)=(7,1/2,0,1)$,
\begin{align}
&\rho^{(\frac{1}{2},\frac{1}{2})}_{(t)0(p)1}(S) = \frac{4i}{7}
\begin{pmatrix}
AD-B^2 & - \left(A^2+BC\right) & - \left(AB+CD\right) \\
- \left(A^2+BC\right) & AB-C^2 & B^2+AC \\
- \left(AB+CD\right) & B^2+AC & BD-A^2
\end{pmatrix}, \notag \\
&\qquad A = \cos \left(\frac{\pi}{14}\right), \ B = \cos \left(\frac{3\pi}{14}\right), \ C = \cos \left(\frac{5\pi}{14}\right), \ D = \cos \left(\frac{9\pi}{14}\right), \notag \\
&\rho^{(\frac{1}{2},\frac{1}{2})}_{(t)0(p)1}(T) = 
\begin{pmatrix}
e^{5\pi i/14} & \ & \ \\
\ & e^{13\pi i/14} & \ \\
\ & \ & e^{17\pi i/14}
\end{pmatrix},
\label{eq:M7M7ST}
\end{align}
which also satisfy Eq.~(\ref{eq:add}), we can obtain the following generators,
\begin{align}
a=ST^{49}, \ b=ST, \ c=T^7, \label{eq:M7M7gen}
\end{align}
satisfying
\begin{align}
a^2=b^3=(ab)^7=(a^{-1}b^{-1}ab)^4=c^4=\mathbf{1}, \ ac=ca, \ bc=cb, \label{eq:M7M7alg}
\end{align}
which mean the generators in Eq.~(\ref{eq:M7M7gen}) are ones of $PSL(2,\mathbb{Z}_7) \times \mathbb{Z}_4$.
Therefore, we can find that the three-generation modes with $(M,\alpha,m,n)=(7,1/2,0,1)$ are transformed under the modular transformation as the three-dimensional representations of finite modular $PSL(2,\mathbb{Z}_7) \times \mathbb{Z}_4$ group with weight $1$.

Similarly, we comment on the modular flavor symmetry anomaly.
Since we can find that the determinants of generators are
\begin{align}
{\rm det}(a) = {\rm det}(b) =1, \ {\rm det}(c) = {\rm det}(T)^{M} = e^{-\pi i/2}, \ {\rm det}(c)^4=1, \label{eq:anomalygenT2Z2tpM357}
\end{align}
only $\mathbb{Z}_4$ symmetry, generated by $c$, can be anomalous while the simple groups $PSL(2,\mathbb{Z}_M)$ ($PSL(2,\mathbb{Z}_3) \simeq A_4$, $PSL(2,\mathbb{Z}_5) \simeq A_5$, and $PSL(2,\mathbb{Z}_7)$) remain anomaly-free.
It is consistent with the general result of non-Abelian discrete symmetry anomaly~\cite{Kobayashi:2021xfs}.


\chapter{Magnetized blow-up manifold of $T^2/\mathbb{Z}_N$ orbifold}
\label{chap:blowup}

In the previous chapter, we have considered magnetized $T^2/\mathbb{Z}_N$ orbifold compactification.
Here, we remind that $T^2/\mathbb{Z}_N$ orbifolds have some fixed points and they become singularities with the localized curvature in Eq.~(\ref{eq:localcurvature}).
Then, let us consider a smooth manifold constructed  by blowing up the orbifold singularities as a 2D compact space.
We call the smooth manifold the blow-up manifold of $T^2/\mathbb{Z}_N$. (Hereafter, we call the blow-up manifold.)
The blow-up manifold can be constructed by replacing around singularities which become cones with smooth manifolds.
In particular, since the amount of the curvature of $T^2/\mathbb{Z}_N$ orbifolds is the same as one of $S^2$, we embed parts of $S^2$ instead cut the cones whose tops are the orbifold singularities.
In this chapter, let us consider such a blow-up manifold with magnetic fluxes (magnetized blow-up manifold).
First, in section~\ref{sec:S2}, we review the magnetized $S^2$ compactification.
Then, in section~\ref{sec:Mblowupman}, we discuss the magnetized blow-up manifold compactification.


\section{Magnetized $S^2$ compactification}
\label{sec:S2}

In this section, we review magnetized $S^2$ compactification~\cite{Conlon:2008qi,Dolan:2020sjq}.


\subsection{Geometry of $S^2$}
\label{subsec:geometryS2}

\begin{figure}[H]
\centering
\includegraphics[bb=0 0 550 500,width=6cm]{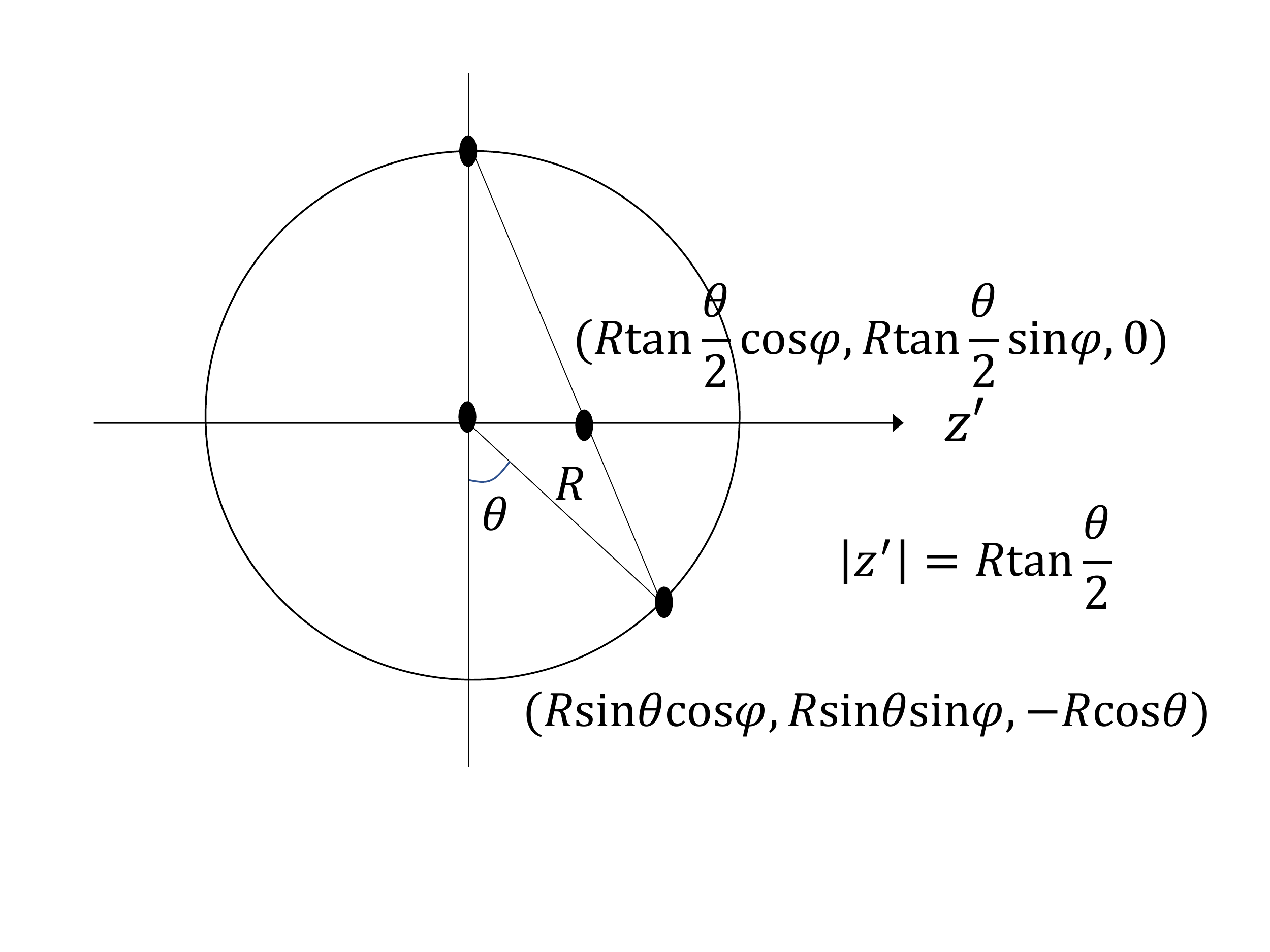}
\caption{The cross section of $S^2 \simeq \mathbb{C}\mathbb{P}^1$}
\label{fig:S2}
\end{figure}

Let us review the geometry of a 2D sphere, $S^2$.
First, as shown in Fig.~\ref{fig:S2}, we can project a point on $S^2$ whose cartesian coordinate is $(R \sin \theta \cos \varphi, R \sin \theta \sin \varphi, - R \cos \theta)$, from the north pole of $S^2$ whose cartesian coordinate is $(0,0,R)$, onto the point on the complex plane passing through the center of $S^2$ whose cartesian coordinate is $(R \tan (\theta/2) \cos \varphi, R \tan (\theta/2) \sin \varphi, 0)$,~i.e. $S^2 \simeq \mathbb{C}\mathbb{P}^1$. Here, we use the spherical parameters $(R, \theta, \varphi)$.

We define the complex coordinate on $\mathbb{C}\mathbb{P}^1$, $z'$, such that $z' = R \tan \frac{\theta}{2} e^{i\varphi}$ at the point whose cartesian coordinate is written by $(R \tan (\theta/2) \cos \varphi, R \tan (\theta/2) \sin \varphi, 0)$.

The metric on $S^2 \simeq \mathbb{C}\mathbb{P}^1$ is given by
\begin{align}
ds^2 &= g'_{ij} d{y'}^i d{y'}^j = 2 h'_{\mu\nu} dz'^{\mu} d\bar{z'}^{\nu}, \label{eq:metricS2} \\
g' &=
\begin{pmatrix}
R^2 & 0 \\
0 & R^2 \sin^2 \theta
\end{pmatrix},
\label{eq:gS2}\\
h' &=
\begin{pmatrix}
0 & 2 \cos^4 \frac{\theta}{2} \\
2 \cos^4 \frac{\theta}{2} & 0
\end{pmatrix}
=
\begin{pmatrix}
0 & \frac{2 R^4}{\left( R^2 + |z'|^2 \right)^2} \\
\frac{2 R^4}{\left( R^2 + |z'|^2 \right)^2} & 0
\end{pmatrix}, \label{eq:hS2}
\end{align}
where $d{y'}^1=d\theta$ and $d{y'}^2=d\varphi$.
Then, the are of $S^2$, ${\cal A}'$, is calculated as
\begin{align}
{\cal A}' = \int_{S^2} d{y'}^1d{y'}^2 \sqrt{|{\rm det}g'|} = \int_{S^2} dz' d\bar{z}' \sqrt{|{\rm det}(2h')|} = 4\pi R^2.
\label{eq:areaS2}
\end{align}
The gamma matrices $\gamma^{z'}$ and $\gamma^{\bar{z}'}$ satisfying $\{ \gamma^{z'}, \gamma^{\bar{z}'} \} = 2{h'}^{z'\bar{z}'}$ are similarly defined as follows.
First, the gamma matrices $\gamma^1$ and $\gamma^2$ satisfying 2D Clifford algebra $\{ \gamma^a, \gamma^b \} = 2\delta^{ab}$ ($a,b=1,2$) are given by
\begin{align}
\gamma^1 = \sigma^1 =
\begin{pmatrix}
0 & 1 \\
1 & 0
\end{pmatrix},
\quad
\gamma^2 = \sigma^2 =
\begin{pmatrix}
0 & -i \\
i & 0
\end{pmatrix}.
\label{eq:2DcliffordS2}
\end{align}
Next, we introduce a vielbein $e'$ such that $h'_{\mu\nu} = {e'}_{\mu}^a {e'}_{\nu}^b \delta_{ab}$.
It is given by
\begin{align}
e' =
\begin{pmatrix}
\frac{R^2}{R^2+|z'|^2} & \frac{R^2}{R^2+|z'|^2} \\
\frac{-iR^2}{R^2+|z'|^2} & \frac{iR^2}{R^2+|z'|^2}
\end{pmatrix}. \label{eq:eS2}
\end{align}
Then, the gamma matrices ${\gamma'}^{z'}$ and ${\gamma'}^{\bar{z}'}$ satisfying $\{ {\gamma'}^{z'}, {\gamma'}^{\bar{z}'} \} = 2{h'}^{z'\bar{z}'}$ are given as
\begin{align}
{\gamma'}^{z'} = ({e'}^{-1}) ^{z'}_a \gamma^a = \frac{1}{R^2}
\begin{pmatrix}
0 & R^2+|z'|^2 \\
0 & 0
\end{pmatrix},
\quad
{\gamma'}^{\bar{z}'} = (e^{-1}) ^{\bar{z}'}_b \gamma^b = \frac{1}{R^2}
\begin{pmatrix}
0 & 0 \\
R^2+|z'|^2 & 0
\end{pmatrix}.
\label{eq:gammazS2}
\end{align}

The non-trivial Levi-Civita connection is given by
\begin{align}
\begin{array}{l}
{\Gamma'}^{z'}_{z'z'} = {h'}^{z'\bar{z}'} \partial_{z'} {h'}_{z'\bar{z}'} = \frac{-2\bar{z}'}{R^2+|z'|^2}, \\ {\Gamma'}^{\bar{z}'}_{\bar{z}'\bar{z}'} = {h'}^{\bar{z}'z'} \partial_{\bar{z}'} {h'}_{\bar{z}'z'} = \frac{-2z'}{R^2+|z'|^2}.
\end{array}
\label{eq:LeviCivitaS2}
\end{align}
Then, we should replace $\partial_{z'}$ ($\partial_{\bar{z}'}$) with the covariant derivative $\nabla'_{z'}$ ($\nabla'_{\bar{z}'}$), which is defined for a 1-form vector field $V'_{z'}(z')$ and a 2-form tensor field $T'_{z'\bar{z}'}(z')$ as
\begin{align}
\begin{array}{ll}
\nabla'_{z'} V'_{z'}(z') \equiv (\partial_{z'} - {\Gamma'}^{z'}_{z'z'}) V'_{z'}(z'), & ( \nabla'_{\bar{z}'} V'_{\bar{z}'}(z') \equiv (\partial_{\bar{z}'} - {\Gamma'}^{\bar{z}'}_{\bar{z}'\bar{z}'}) V'_{\bar{z}'}(z') ), \\
\nabla'_{z'} T'_{z'\bar{z}'}(z') \equiv (\partial_{z'} - {\Gamma'}^{z'}_{z'z'}) T'_{z'\bar{z}'}(z'), & ( \nabla'_{\bar{z}'} T'_{z'\bar{z}'}(z') \equiv ( \partial_{\bar{z}'} - {\Gamma'}^{\bar{z}'}_{\bar{z}'\bar{z}'}) T'_{z'\bar{z}'}(z') ).
\end{array}
\label{eq:nablaVT}
\end{align}
Moreover, the spin connection is given by
\begin{align}
\begin{array}{l}
{\omega'}^a_{z' b} = ({e'}^{-1})^{z'}_b {\Gamma'}^{z'}_{z'z'} {e'}^a_{z'} - ({e'}^{-1})^{z'}_b \partial_{z'} {e'}^a_{z'} - ({e'}^{-1})^{\bar{z}'}_b \partial_{z'} {e'}^a_{\bar{z}'}\ \Rightarrow \ \omega'_{z' 12} = -\frac{i}{2}\frac{2}{R^2+|z'|^2}\bar{z}', \\
{\omega'}^a_{\bar{z}' b} = ({e'}^{-1})^{\bar{z}'}_b {\Gamma'}^{\bar{z}'}_{\bar{z}'\bar{z}'} {e'}^a_{\bar{z}'} - ({e'}^{-1})^{\bar{z}'}_b \partial_{z'} {e'}^a_{\bar{z}'} - ({e'}^{-1})^{z'}_b \partial_{\bar{z}'} {e'}^a_{z'}\ \Rightarrow \ {\omega'}_{\bar{z}' 12} = \frac{i}{2} \frac{2}{R^2+|z'|^2} z'.
\end{array}
\label{eq:spinconS2}
\end{align}
As the Lorentz generator is $\Sigma^{ab} = \frac{1}{4} [\gamma^a, \gamma^b]$, $\Sigma^{12}$ is obtained as
\begin{align}
\Sigma^{12} = \frac{i}{2} \sigma_3 = \frac{i}{2}
\begin{pmatrix}
1 & 0 \\
0 & -1
\end{pmatrix}.
\label{eq:spingen}
\end{align}
Then, the covariant derivative $\nabla'_{z'}$ ($\nabla'_{\bar{z}'}$) for a spinor $S'(z') = (S'_+(z'), S'_-(z'))^T$ is defined as
\begin{align}
\nabla'_{z'} S'(z') \equiv (\partial_{z'} + \omega'_{z'12} \Sigma^{12} ) S'(z'), \quad  ( \nabla'_{\bar{z}'} S'(z') \equiv  (\partial_{\bar{z}'} + \omega'_{\bar{z}'12} \Sigma^{12} ) S'(z')). 
\label{eq:nablaS}
\end{align}
Note that $\nabla'_{z'}=\partial_{z'}$ ($\nabla'_{\bar{z}'}=\partial_{\bar{z}'}$) for a scalar.

The curvature of $S^2$ is given by
\begin{align}
\frac{1}{2\pi i} \int_{S^2} {R'}^{z'}_{z'z'\bar{z}'} dz' \wedge d\bar{z}' = \frac{1}{2\pi} \int_{S^2} \frac{2iR^2}{(R^2+|z'|^2)^2} dz' \wedge d\bar{z}' = \chi(S^2) = 2, \label{eq:curvS2}
\end{align}
where ${R'}^{z'}_{z'z'\bar{z}'}$ is obtained by
\begin{align}
{R'}^{z'}_{z'z'\bar{z}'} = \partial_{\bar{z}'} {\Gamma'}^{z'}_{z'z'} = \frac{-2R^2}{(R^2+|z'|^2)^2}, \label{eq:RiemannS2}
\end{align}
while $\chi(S^2)$ denotes the Euler number of $S^2$.


\subsection{Wave functions on magnetized $S^2$}
\label{subsec:waveMS2}

Here, let us consider that the following Abelian homogeneous magnetic flux,
\begin{align}
\frac{1}{2\pi} \int_{S^2} \langle {F'}_{z'\bar{z}'} \rangle dz' \wedge d\bar{z}' =
\begin{pmatrix}
{M'}^a \mathbb{I}_{N_a} & \ \\
\ & {M'}^b \mathbb{I}_{N_b}
\end{pmatrix}, \ 
\left( \frac{1}{2\pi} \int_{S^2} \langle F^{aa,bb}_{z'\bar{z}'} \rangle dz' \wedge d\bar{z}' = {M'}^{a,b} \right),
\label{eq:fluxS2}
\end{align}
is inserted on $S^2$, where $N_a+N_b=N$ and ${M'}^{a,b}$ must be integers (Dirac's quantization).
The magnetic flux is given by the 2-form field strength,
\begin{align}
\frac{1}{2\pi} \langle {F'}^{aa,bb} \rangle = \frac{1}{2\pi} \langle {F'}^{aa,bb}_{z'\bar{z}'} \rangle dz' \wedge d\bar{z}' = \frac{{M'}^{a,b}}{4\pi R^2} \frac{4R^4}{\left( R^2+|z'|^2 \right)^2} \frac{i}{2} dz' \wedge d\bar{z'},
\label{eq:FS2VEV}
\end{align}
which satisfies the Yang-Mills equations
\begin{align}
\nabla'_{z'} \langle {F'}^{aa,bb}_{z'\bar{z}'} \rangle = (\partial_{z'} - \Gamma^{z'}_{z'z'} ) \langle {F'}^{aa,bb}_{z'\bar{z}'} \rangle = 0, \quad 
\nabla'_{\bar{z}'} \langle {F'}^{aa,bb}_{z'\bar{z}'} \rangle = (\partial_{\bar{z}'} - \Gamma^{\bar{z}'}_{\bar{z}'\bar{z}'})  \langle {F'}^{aa,bb}_{z'\bar{z}'} \rangle = 0.
\label{eq:YMEFVEVS2}
\end{align}
Moreover, the field strength is obtained from the 1-form background gauge field,
\begin{align}
\langle {A'}^{aa,bbb}(z') \rangle
&= \langle {A'}^{aa,bb}_{z'}(z') \rangle dz' + \langle {A'}^{aa,bb}_{\bar{z'}}(z') \rangle d\bar{z'} \notag \\
&= - \frac{i}{2} \frac{2\pi {M'}^{a,b}}{4\pi R^2}\frac{2R^2}{R^2+|z'|^2} \bar{z}' dz' + \frac{i}{2} \frac{2\pi {M'}^{a,b}}{4\pi R^2} \frac{2R^2}{R^2+|z'|^2} z' d\bar{z}', \label{eq:AS2VEV}
\end{align}
by $\langle {F'}^{aa,bb} \rangle =d' \langle {A'}^{aa,bb} \rangle$.
The covariant derivative is defined as
\begin{align}
\hat{D}' &= d' -i\langle A'(z') \rangle \notag \\
\hat{D}'_{z'}dz' + \hat{D}'_{\bar{z}'}d{\bar{z}'} &= (\nabla_{z'} -i \langle {A'}_{z'}(z') \rangle )dz' + (\nabla_{\bar{z}'} - i \langle {A'}_{\bar{z}'}(z') \rangle )d{\bar{z}'}.
\label{eq:D'}
\end{align}
%

In the following, let us see wave functions of $U(N)$ adjoint spinor, scalar, and vector fields on $S^2$ with the magnetic flux in Eq.~(\ref{eq:fluxS2}), which satisfy individual equation of motions.

%
\begin{itemize}
\item {\large Spinor fields}
\end{itemize}

First, let us see wave functions of $U(N)$ adjoint 2D MW spinor fields on the magnetized $S^2$,
\begin{align}
\begin{array}{c}
\psi_{S^2}^{(2)}(z') =
\begin{pmatrix}
\psi_{S^2,+}(z') \\ \psi_{S^2,-}(z')
\end{pmatrix}, \ 
\psi_{S^2,\pm}(z') =
\begin{pmatrix}
\psi^{aa}_{S^2,\pm}(z') & \psi^{ab}_{S^2,\pm}(z') \\
\psi^{ba}_{S^2,\pm}(z') & \psi^{bb}_{S^2,\pm}(z')
\end{pmatrix}, \\
\left( \psi_{S^2,\mp}^{JI}(z') = \sigma_1 \overline{\psi_{S^2,\pm}^{IJ(2)}(z')} \ (I,J=a,b) \right), 
\end{array}
\label{eq:spinorS2} 
\end{align}
which satisfy the Dirac equation,
\begin{align}
\left( i ({\gamma'}^{z'} \hat{D}'_{z'} + {\gamma'}^{\bar{z}'} \hat{D}'_{\bar{z}'}) - m_{n} \right) \psi_{S^2,n}^{(2)}(z') = 0.
\label{eq:DiraceqXS2}
\end{align}
Here, we define the Dirac operator,
\begin{align}
i\hat{\slashed{D}}'
&\equiv i ({\gamma'}^{z'} \hat{D}'_{z'} + {\gamma'}^{\bar{z}'} \hat{D}'_{\bar{z}'}) \notag \\
&=
\begin{pmatrix}
0 & i\frac{R^2+|z'|^2}{R^2}( \partial_{z'} - \frac{i}{2} {\omega'}_{z'12} - i \langle A'_{z'}(z') \rangle ) \\
i\frac{R^2+|z'|^2}{R^2} ( \partial_{\bar{z}'} + \frac{i}{2} {\omega'}_{\bar{z}'12} - i \langle A'_{\bar{z}'}(z) \rangle ) & 0
\end{pmatrix} \notag \\
&\equiv
\begin{pmatrix}
0 & -i{{\cal D}'}^{\dagger} \\
i{{\cal D}'} & 0
\end{pmatrix},
\label{eq:DiracOPS2}
\end{align}
and then the Dirac equation (\ref{eq:DiraceqXS2}) can be rewritten as
\begin{align}
&\begin{array}{l}
i{\cal D}' \psi_{S^2,+,n}(z') = i\frac{R^2+|z'|^2}{R^2} \left( (\partial_{\bar{z}'} + \frac{i}{2} {\omega'}_{\bar{z}'12}) \psi_{S^2,+,n}(z') -i [\langle A'_{\bar{z}'}(z') \rangle, \psi_{S^2,+,n}(z')] \right) = m_n \psi_{S^2,-,n}(z'),
\end{array}
\label{eq:Dirac+S2}  \\
&\Leftrightarrow
\left\{
\begin{array}{l}
i{\cal D}'_{aa} \psi^{aa}_{S^2,+,n}(z') = \frac{i}{R^2} \left( (R^2+|z'|^2) \partial_{\bar{z}'} - \frac{1}{2}z' \right) \psi^{aa}_{S^2,+,n}(z') = m_n \psi^{aa}_{S^2,-,n}(z') \\
i{\cal D}'_{bb} \psi^{bb}_{S^2,+,n}(z') = \frac{i}{R^2} \left( (R^2+|z'|^2) \partial_{\bar{z}'} - \frac{1}{2}z' \right) \psi^{bb}_{S^2,+,n}(z') = m_n \psi^{bb}_{S^2,-,n}(z') \\
i{\cal D}'_{ab} \psi^{ab}_{S^2,+,n}(z') = \frac{i}{R^2} \left( (R^2+|z'|^2) \partial_{\bar{z}'} - \frac{1-M'}{2}z' \right) \psi^{ab}_{S^2,+,n}(z') = m_n \psi^{ab}_{S^2,-,n}(z') \\
-i{\cal D}'_{ba} \psi^{ba}_{S^2,+,n}(z') = \frac{i}{R^2} \left( (R^2+|z'|^2) \partial_{\bar{z}'} - \frac{1+M'}{2}z' \right) \psi^{ba}_{S^2,+,n}(z') = m_n \psi^{ba}_{S^2,-,n}(z') 
\end{array}
\right., \notag \\
&\Leftrightarrow
\left\{
\begin{array}{l}
i\left( \frac{R^2}{R^2+|z'|^2} \right)^{-\frac{3}{2}} \partial_{\bar{z}'} \left[ \left( \frac{R^2}{R^2+|z'|^2} \right)^{\frac{1}{2}} \psi^{aa}_{S^2,+,n}(z') \right] = m_n \psi^{aa}_{S^2,-,n}(z') \\
i\left( \frac{R^2}{R^2+|z'|^2} \right)^{-\frac{3}{2}} \partial_{\bar{z}'} \left[ \left( \frac{R^2}{R^2+|z'|^2} \right)^{\frac{1}{2}} \psi^{bb}_{S^2,+,n}(z') \right] = m_n \psi^{bb}_{S^2,-,n}(z') \\
i\left( \frac{R^2}{R^2+|z'|^2} \right)^{-\frac{3-M'}{2}} \partial_{\bar{z}'} \left[ \left( \frac{R^2}{R^2+|z'|^2} \right)^{\frac{1-M'}{2}} \psi^{ab}_{S^2,+,n}(z') \right] = m_n \psi^{ab}_{S^2,-,n}(z') \\
i\left( \frac{R^2}{R^2+|z'|^2} \right)^{-\frac{3+M'}{2}} \partial_{\bar{z}'} \left[ \left( \frac{R^2}{R^2+|z'|^2} \right)^{\frac{1+M'}{2}} \psi^{ba}_{S^2,+,n}(z') \right] = m_n \psi^{ba}_{S^2,-,n}(z')
\end{array}
\right., \notag 
\end{align}
\begin{align}
&\begin{array}{l}
-i{{\cal D}'}^{\dagger} \psi_{S^2,-,n}(z') = i\frac{R^2+|z'|^2}{R^2} \left( (\partial_{z'} - \frac{i}{2} {\omega'}_{z'12}) \psi_{S^2,-,n}(z')  -i [\langle A'_{z'}(z') \rangle, \psi_{S^2,-,n}(z')] \right) = m_n \psi_{S^2,+,n}(z'),
\end{array}
\label{eq:Dirac-S2} \\
&\Leftrightarrow
\left\{
\begin{array}{l}
-i{{\cal D}'}^{\dagger}_{aa} \psi^{aa}_{S^2,-,n}(z') = \frac{i}{R^2} \left( (R^2+|z'|^2) \partial_{z'} - \frac{1}{2}\bar{z}' \right) \psi^{aa}_{S^2,-,n}(z') = m_n \psi^{aa}_{S^2,+,n}(z') \\
-i{{\cal D}'}^{\dagger}_{bb} \psi^{bb}_{S^2,-,n}(z') = \frac{i}{R^2} \left( (R^2+|z'|^2) \partial_{z'} - \frac{1}{2}\bar{z}' \right) \psi^{bb}_{S^2,-,n}(z') = m_n \psi^{bb}_{S^2,+,n}(z') \\
-i{{\cal D}'}^{\dagger}_{ab} \psi^{ab}_{S^2,-,n}(z') = \frac{i}{R^2} \left( (R^2+|z'|^2) \partial_{z'} - \frac{1+M'}{2}\bar{z}' \right) \psi^{ab}_{S^2,-,n}(z') = m_n \psi^{ab}_{S^2,+,n}(z') \\
i{{\cal D}'}^{\dagger}_{ba} \psi^{ba}_{S^2,-,n}(z') = \frac{i}{R^2} \left( (R^2+|z'|^2) \partial_{z'} - \frac{1-M'}{2}\bar{z}' \right) \psi^{ba}_{S^2,-,n}(z') = m_n \psi^{ba}_{S^2,+,n}(z')
\end{array}
\right., \notag \\
&\Leftrightarrow
\left\{
\begin{array}{l}
i\left( \frac{R^2}{R^2+|z'|^2} \right)^{-\frac{3}{2}} \partial_{z'} \left[ \left( \frac{R^2}{R^2+|z'|^2} \right)^{\frac{1}{2}} \psi^{aa}_{S^2,-,n}(z') \right] = m_n \psi^{aa}_{S^2,+,n}(z') \\
i\left( \frac{R^2}{R^2+|z'|^2} \right)^{-\frac{3}{2}} \partial_{z'} \left[ \left( \frac{R^2}{R^2+|z'|^2} \right)^{\frac{1}{2}} \psi^{bb}_{S^2,-,n}(z') \right] = m_n \psi^{bb}_{S^2,+,n}(z') \\
i\left( \frac{R^2}{R^2+|z'|^2} \right)^{-\frac{3+M'}{2}} \partial_{z'} \left[ \left( \frac{R^2}{R^2+|z'|^2} \right)^{\frac{1+M'}{2}} \psi^{ab}_{S^2,-,n}(z') \right] = m_n \psi^{ab}_{S^2,+,n}(z') \\
i\left( \frac{R^2}{R^2+|z'|^2} \right)^{-\frac{3-M'}{2}} \partial_{z'} \left[ \left( \frac{R^2}{R^2+|z'|^2} \right)^{\frac{1-M'}{2}} \psi^{ba}_{S^2,-,n}(z') \right] = m_n \psi^{ba}_{S^2,+,n}(z')
\end{array}
\right., \notag
\end{align}
where $M' \equiv {M'}^a - {M'}^b$ and $n$ denotes the Landau level.

In particular, the lowest modes ($n=0$) satisfying the above Dirac equation with $m_0=0$ are expressed as
\begin{align}
\begin{array}{ll}
\psi^{aa}_{S^2,+,0}(z') = \left(\frac{R^2}{R^2+|z'|^2}\right)^{-\frac{1}{2}} {h'}^{aa}_0(z'), & \psi^{aa}_{S^2,-,0}(z') = \overline{\psi^{aa}_{S^2,+,0}(z')} = \left(\frac{R^2}{R^2+|z'|^2}\right)^{-\frac{1}{2}} \bar{h}'^{aa}_0(\bar{z}'), \\
\psi^{bb}_{S^2,+,0}(z') = \left(\frac{R^2}{R^2+|z'|^2}\right)^{-\frac{1}{2}} {h'}^{bb}_0(z'), & \psi^{bb}_{S^2,-,0}(z') = \overline{\psi^{bb}_{S^2,+,0}(z')} = \left(\frac{R^2}{R^2+|z'|^2}\right)^{-\frac{1}{2}} \bar{h}'^{bb}_0(\bar{z}'), \\
\psi^{ab}_{S^2,+,0}(z') = \left(\frac{R^2}{R^2+|z'|^2}\right)^{\frac{M'-1}{2}} {h'}^{ab}_0(z'), & \psi^{ab}_{S^2,-,0}(z') = \overline{\psi^{ba}_{S^2,+,0}(z')} = \left(\frac{R^2}{R^2+|z'|^2}\right)^{\frac{M'-1}{2}} \bar{h}'^{ba}_0(\bar{z}'), \\
\psi^{ba}_{S^2,+,0}(z') = \left(\frac{R^2}{R^2+|z'|^2}\right)^{-\frac{M'+1}{2}} {h'}^{ba}_0(z'), & \psi^{ab}_{S^2,-,0}(z') = \overline{\psi^{ab}_{S^2,+,0}(z')} = \left(\frac{R^2}{R^2+|z'|^2}\right)^{\frac{M'-1}{2}} \bar{h}'^{ab}_0(\bar{z}'),
\end{array}
\label{eq:DiraczeromodeS2}
\end{align}
where ${h'}(z')$ denotes a holomorphic function.
In addition, they should be normalizable; they must be finite on $S^2 \simeq \mathbb{C}\mathbb{P}^1$.
For example, in order for $\psi^{ab}_{+}$ ($\psi^{ba}_{+}$) as well as the anti-fields $\psi^{ba}_{-} = \overline{\psi^{ab}_{+}}$ ($\psi^{ab}_{-} = \overline{\psi^{ba}_{+}}$) to have physical relevant zero mode solutions on the magnetized $S^2$, it should be required that $M' >0$ ($M'<0$) and ${h'}^{ab}_0(z')$ (${h'}^{ba}_0(z')$) as well as $\bar{h}'^{ba}_0(\bar{z}')$ ($\bar{h}'^{ab}_0(\bar{z}')$) are written by $(|M'|-1)$-polynomials, that is, there are $|M'|$ number of degenerate zero modes.
In this case, however, $\psi^{ba}_{+}$ ($\psi^{ab}_{+}$) as well as their anti-fields $\psi^{ab}_{-} = \overline{\psi^{ba}_{+}}$ ($\psi^{ba}_{-} = \overline{\psi^{ab}_{+}}$) have no physical relevant zero modes.
Indeed, this result is consistent with the AS index theorem,
\begin{align}
n^{ab}_{+} - n^{ab}_{-} &= \frac{1}{2\pi} \int_{S^2} F'_{ab} = M', \label{eq:indexFabS2} \\
n^{ba}_{+} - n^{ba}_{-} &= \frac{1}{2\pi} \int_{S^2} F'_{ba} = -M', \label{eq:indexFbaS2}
\end{align}
where $n^{ab}_{+}$, $n^{ab}_{-}$, $n^{ba}_{+}$, $n^{ba}_{-}$ denote zero mode numbers of $\psi^{ab}_{+}$, $\psi^{ab}_{-}$, $\psi^{ba}_{+}$, $\psi^{ba}_{-}$, respectively, and $F_{ab}$, $F_{ba}$ denote the magnetic fluxes which $\psi^{ab}_{\pm}$, $\psi^{ba}_{\pm}$ feel, respectively.
Note that although $S^2$ also has the curvature, it does not contribute to the AS index theorem.
Therefore, we can obtain $|M'|$ generational bi-fundamental chiral fermions from the magnetized $S^2$ compactification.
On the other hand, there are no physical relevant zero mode solutions.
Hereafter, we consider the case that $\psi^{ab}_{+}$ as well as the anti-fields $\psi^{ba}_{-} = \overline{\psi^{ab}_{+}}$ are zero modes;
\begin{align}
\begin{array}{l}
\psi^{a',M'-1}_{S^2,0}(z') = \left(\frac{R^2}{R^2+|z'|^2} \right)^{\frac{M'-1}{2}} h^{a',M'}_{S^2}(z'), \\
h^{a',M'}_{S^2}(z') = {\cal N}_{S^2,0,a'} R^{-a} z^{a'}, \quad (a \in \mathbb{Z}/M'\mathbb{Z}),
\end{array}
\label{eq:zeroS2}
\end{align}
where the normalization factor ${\cal N}_{S^2,0,a'}$ on $S^2$ is determined from the inner product,
\begin{align}
&\int_{S^2} dz'd\bar{z}' \sqrt{|{\rm det}(2h)|} \overline{\psi^{a',M'-1}_{S^2,0}(z')} \psi^{b',M'-1}_{S^2,0}(z') \notag \\
=& \delta_{a',b}' |{\cal N}_{S^2,0,a'}|^2 {\cal A}' \int_{0}^{\infty} d\left( \frac{R^2}{R^2+|z'|^2} \right) \left( 1 - \frac{R^2}{R^2+|z'|^2} \right)^{a'} \left(\frac{R^2}{R^2+|z'|^2} \right)^{M'-1-2a'} \notag \\
=& \delta_{a',b'} |{\cal N}_{S^2,0,a'}|^2 {\cal A}' \int_{0}^{1} dt t^{(M'-a')-1} (1-t)^{(a'+1)-1} \notag \\
=& \delta_{a',b'} |{\cal N}_{S^2,0,a'}|^2 {\cal A}' \beta(M'-a',a'+1), \label{eq:normalizationS2}
\end{align}
where $\beta(M'-a',a'+1)$ denotes the beta function and satisfies the following recurrence relation obtained from the partial integral;
\begin{align} 
\beta(M'-a', a'+1) = \frac{a'}{M'-a'} \beta(M'-a'+1,a'), \quad \beta(M',1) = \frac{1}{M'}, \label{eq:betafuncrecur}
\end{align}
and then it can be written by the gamma function $\Gamma(X)$ satisfying $\Gamma(X+1) = X \Gamma(X)$; 
\begin{align}
\beta(M'-a', a'+1) = \frac{\Gamma(M'-a')\Gamma(a'+1)}{\Gamma(M'+1)}. \label{eq:betagamma}
\end{align}

Here, we comment for the $n$ th excited states. (See Ref.~\cite{Dolan:2020sjq}.)
By further acting the Dirac operator on the Dirac equation, we can obtain the characteristic equations,
\begin{align}
\begin{pmatrix}
{{\cal D}'}^{\dagger} {\cal D}' & 0 \\
0 & {\cal D}' {{\cal D}'}^{\dagger}
\end{pmatrix}
\begin{pmatrix}
\psi_{S^2,+,n}(z) \\ \psi_{S^2,-,n}(z)
\end{pmatrix}
= m_n^2
\begin{pmatrix}
\psi_{S^2,+,n}(z) \\ \psi_{S^2,-,n}(z)
\end{pmatrix},
\label{eq:charactereqspinor}
\end{align}
where ${\cal D}'$ and ${{\cal D}'}^{\dagger}$ satisfy the following commutation relations:
\begin{align}
&[ {\cal D}'_{aa}, {{\cal D}'}^{\dagger}_{aa} ] = -\frac{R^2+|z'|^2}{R^4} (z'\partial_{z'} - \bar{z'} \partial_{\bar{z'}}), \label{eq:DDdaggeraaS2} \\
&[ {\cal D}'_{bb}, {{\cal D}'}^{\dagger}_{bb} ] = -\frac{R^2+|z'|^2}{R^4} (z'\partial_{z'} - \bar{z'} \partial_{\bar{z'}}),
\label{eq:DDdaggerbbS2} \\
&[ {\cal D}'_{ab}, {{\cal D}'}^{\dagger}_{ab} ] = -\frac{R^2+|z'|^2}{R^4} (z'\partial_{z'} - \bar{z'} \partial_{\bar{z'}} -M' ),
\label{eq:DDdaggerabS2} \\
&[ {\cal D}'_{ba}, {{\cal D}'}^{\dagger}_{ab} ] = -\frac{R^2+|z'|^2}{R^4} (z'\partial_{z'} - \bar{z'} \partial_{\bar{z'}} +M' ).
\label{eq:DDdaggerbaS2}
\end{align}
{
In particular, to see the solutions of:
\begin{align}
{{\cal D}'}^{\dagger}_{ab} {\cal D}'_{ab} \psi^{ab}_{S^2,+,n}(z) = m_n^2 \psi^{ab}_{S^2,+,n}(z), \label{eq:solabS2}
\end{align}
we define the following operators,
\begin{align}
J^3_{ab} 
&\equiv z'\partial_{z'} - \bar{z}'\partial_{\bar{z}'} - \frac{M'-1}{2}, \label{eq:J3} \\
J^{+}_{ab} 
&\equiv R {\cal D}'_{ab} + \frac{z'}{R} (J^3_{ab} - \frac{M'-1}{2} ) \label{eq:J+} \\
&= \frac{1}{R} \left( (R^2+|z'|^2)\partial_{\bar{z}'} - \frac{1-M'}{2}z' + z'(z'\partial_{z'} - \bar{z}'\partial_{\bar{z}'} - (M'-1)) \right) \notag \\
&= \frac{1}{R} (R^2 \partial_{\bar{z}'} + {z'}^2 \partial_{z'} - \frac{M'-1}{2}z' ), \notag \\
J^{-}_{ab} 
&\equiv R{{\cal D}'}_{ab}^{\dagger} + \frac{\bar{z}'}{R} (J^3_{ab} - \frac{M'+1}{2} ) \label{eq:J-} \\
&= -\frac{1}{R}\left((R^2+|z'|^2)\partial_{z'} - \frac{1+M'}{2}\bar{z'} - \bar{z'} (z'\partial_{z'} - \bar{z}'\partial_{\bar{z}'} -M' ) \right) \notag \\
&= -\frac{1}{R} (R^2 \partial_{z'} + \bar{z}'^2 \partial_{\bar{z}'} + \frac{M'-1}{2}\bar{z'} ), \notag \\
\mathbf{J}^2_{ab}
&= J^{-}_{ab} J^{+}_{ab} + (J^3_{ab} )^2 + J^3_{ab} \notag \\
&= R^2 {{\cal D}'}_{ab}^{\dagger} {\cal D}'_{ab} + \frac{1}{4} ({M'}^2 - 1), \label{eq:J2}
\end{align}
and they satisfy
\begin{align}
[J^{+}_{ab}, J^{-}_{ab}] = 2J^3_{ab}, \ [J^3_{ab},J^{\pm}_{ab}] = \pm J^{+}_{ab}, \ [ \mathbf{J}^2_{ab}, J^{\pm}_{ab} ] = [ \mathbf{J}^2_{ab}, J^3_{ab} ] = 0. \label{eq:J3+-2alg}
\end{align}
Then, it means that the solutions of Eq.~(\ref{eq:solabS2})
are eigenfunctions of $\mathbf{J}^2_{ab}$ and $J^3_{ab}$; they are written as
\begin{align}
\psi^{ab}_{S^2,+,n}(z') 
=& \sum_{a'=-n}^{n+M'} \psi^{a',M'-1}_{S^2,n}(z') \notag \\
=& \sum_{a'=-n}^{n+M'} \left( \frac{R^2}{R^2+|z'|^2} \right)^{\frac{M'-1}{2}} {\cal N}_{S^2,n,a'} R^{-a'} z^{a'} P_{n}^{a', M'-a'-1}\left( \frac{R^2}{R^2+|z'|^2} \right), 
\label{eq:massiveab+S2}
\end{align}
with
\begin{align}
m_n^2 = \frac{\left( n + \frac{M'}{2} \right)^2 - \left( \frac{M'}{2} \right)^2}{R^2} = \frac{\left( J+\frac{1}{2} \right)^2 - \left( \frac{M'}{2} \right)^2}{R^2}, \label{eq:massabS2}
\end{align}
where $P_{n}$ denotes the Jacobi polynomial given by
\begin{align}
P_{n}^{a',b'}(x) = \frac{(-1)^n}{2^nn!} (1-x)^{-a'}(1+x)^{-b'} \frac{d^n}{dx^n} [ (1-x)^{a'+n}(1+x)^{b'+n}], \label{eq:Yacobipol}
\end{align}
and the Landau level $n$ is related to the angular momentum $J$ as in Eq.~(\ref{eq:massabS2}), which comes from Eq.~(\ref{eq:J2}).
Moreover, by solving Eq.~(\ref{eq:Dirac+S2}) with the property:
\begin{align}
\frac{d}{dx} P_{n}^{a',M'-a'-1}(x) = \frac{n+M'}{2} P_{n-1}^{a'+1,M'-a'}(x), \label{eq:Yacobipolprop}
\end{align}
the solutions of $\psi^{ab}_{S^2,-,n}$ are expressed as
\begin{align}
\psi^{ab}_{S^2,-,n}(z') 
&= \sum_{a'=-n}^{n+M'} \psi^{a',M'+1}_{S^2,n}(z') \notag \\
&= \sum_{a'=-n}^{n+M'} \left( \frac{R^2}{R^2+|z'|^2} \right)^{\frac{M'+1}{2}} {\cal N}_{S^2,n,a'} i\sqrt{\frac{n+M'}{n}} R^{-(a'+1)} z^{a'+1} P_{n-1}^{a'+1, M'-a'}\left( \frac{R^2}{R^2+|z'|^2} \right). \label{eq:massiveab-S2}
\end{align}
Then, similarly we can obtain $\psi^{ba}_{S^2,-,n} = \overline{\psi^{ab}_{S^2,+,n}}$ and $\psi^{ba}_{S^2,+,n} = \overline{\psi^{ab}_{S^2,-,n}}$, while $\psi^{aa,bb}_{S^2,+,n}$ and $\psi^{aa,bb}_{S^2,-,n}$ are obtained from $\psi^{ab}_{S^2,+,n}$ and $\psi^{ab}_{S^2,-,n}$ with $M'=0$, respectively.

\begin{itemize}
\item {\large Scalar fields}
\end{itemize}

Second, let us see wave functions of $U(N)$ adjoint scalar fields on the magnetized $S^2$,
\begin{align}
\phi_{S^2}(z') =
\begin{pmatrix}
\phi_{S^2}^{aa}(z') & \phi_{S^2}^{ab}(z') \\
\phi_{S^2}^{ba}(z') & \phi_{S^2}^{ba}(z')
\end{pmatrix},
\label{eq:scalarS2}
\end{align}
which satisfy the Klein-Gordon equation,
\begin{align}
\left( {h'}^{z'\bar{z}'} (\hat{D}'_{z'} \hat{D}'_{\bar{z}'} + \hat{D}'_{\bar{z}'} \hat{D}'_{z'}) + m_{n}^2 \right) \phi_{S^2,n}(z') = 0.
\label{eq:KGeqS2}
\end{align}
Note that the geometrical connection does not act on the scalar fields.
Here, we define the Laplace operator,
\begin{align}
\hat{\Delta}' 
\equiv& -{h'}^{z'\bar{z}'} (\hat{D}'_{z'} \hat{D}'_{\bar{z}'} + \hat{D}'_{\bar{z}'} \hat{D}'_{z'} ) \notag \\
=& -2{h'}^{z'\bar{z}'} \hat{D}'_{z'} \hat{D}'_{\bar{z}'} + {h'}^{z'\bar{z}'} [ \hat{D}'_{z'}, \hat{D}'_{\bar{z}'} ] \notag \\
=& -2{h'}^{z'\bar{z}'} \hat{D}'_{\bar{z}'} \hat{D}'_{z'} - {h'}^{z'\bar{z}'} [ \hat{D}'_{z'}, \hat{D}'_{\bar{z}'} ],
\label{eq:LaplaceOPS2}
\end{align}
where $\hat{D}'_{z'}$ and $\hat{D}_{\bar{z}'}$ are given by
\begin{align}
\begin{array}{l}
\hat{D}'_{z'} =
\begin{pmatrix}
\hat{D}'^{aa}_{z'} & \hat{D}'^{ab}_{z'} \\
\hat{D}'^{ba}_{z'} & \hat{D}'^{bb}_{z'}
\end{pmatrix} =
\begin{pmatrix}
\partial_{z'}  & \partial_{z'} - \frac{M'}{2(R^2+|z'|^2)} \bar{z}' \\
\partial_{z'} + \frac{M'}{2(R^2+|z'|^2)} \bar{z}' & \partial_{z'} 
\end{pmatrix}, \\
\hat{D}'_{\bar{z}'} =
\begin{pmatrix}
\hat{D}'^{aa}_{\bar{z}'} & \hat{D}'^{ab}_{\bar{z}'} \\
\hat{D}'^{ba}_{\bar{z}'} & \hat{D}'^{bb}_{\bar{z}'}
\end{pmatrix} =
\begin{pmatrix}
\partial_{\bar{z}'} & \partial_{\bar{z}'} + \frac{M'}{2(R^2+|z'|^2)} z' \\
\partial_{\bar{z}'} - \frac{M'}{2(R^2+|z'|^2)} z' & \partial_{\bar{z}'} 
\end{pmatrix},
\end{array}
\label{eq:D'D'scalar}
\end{align}
and then they satisfy the commutation relations,
\begin{align}
&{h'}^{z'\bar{z}'}  [ \hat{D}'^{aa}_{z'}, \hat{D}'^{aa}_{\bar{z}'} ] = 0, \label{eq:DzDzbarcomaaS2} \\
&{h'}^{z'\bar{z}'}  [ \hat{D}'^{bb}_{z'}, \hat{D}'^{bb}_{\bar{z}'} ] = 0, \label{eq:DzDzbarcombbS2} \\
&{h'}^{z'\bar{z}'}  [ \hat{D}'^{ab}_{z'}, \hat{D}'^{ab}_{\bar{z}'} ] = \frac{M'}{2R^2}, \label{eq:DzDzbarcomabS2} \\
&{h'}^{z'\bar{z}'}  [ \hat{D}'^{ba}_{z'}, \hat{D}'^{ba}_{\bar{z}'} ] = -\frac{M'}{2R^2}. \label{eq:DzDzbarcomaaS2}
\end{align}
Thus, the solutions of the Klein-Gordon equation,
\begin{align}
\hat{\Delta}' \phi_{S^2,n}(z') = m_n^2 \phi_{S^2,n}(z'),
\label{eq:KGS2}
\end{align}
must be eigenfunctions for the operator ${h'}^{z'\bar{z}'} \hat{D}'_{z'} \hat{D}'_{\bar{z}'}$ (or $\hat{D}'_{\bar{z}'} \hat{D}'_{z'}$).
We also note that the operators in the characteristic equation for spinors can be rewritten as
\begin{align}
\begin{pmatrix}
{{\cal D}'}^{\dagger} {\cal D}' & 0 \\
0 & {\cal D}' {{\cal D}'}^{\dagger}
\end{pmatrix}
= - {h'}^{z'\bar{z}'}
\begin{pmatrix}
\widetilde{\hat{D}'_{z'}} \hat{D}'_{\bar{z}'} & 0 \\
0 & \widetilde{\hat{D}'_{\bar{z}'}} \hat{D}'_{z'}
\end{pmatrix},
\label{eq:reDDdagger}
\end{align}
where $\widetilde{\hat{D}'_{z'}}$ and $\widetilde{\hat{D}'_{\bar{z}'}}$ are defined for
\begin{align}
\begin{array}{l}
\hat{D}'_{z'} =
\begin{pmatrix}
\hat{D}'^{aa}_{z'} & \hat{D}'^{ab}_{z'} \\
\hat{D}'^{ba}_{z'} & \hat{D}'^{bb}_{z'}
\end{pmatrix} =
\begin{pmatrix}
\partial_{z'} - \frac{1}{2(R^2+|z'|^2)} \bar{z}' & \partial_{z'} - \frac{1+M'}{2(R^2+|z'|^2)} \bar{z}' \\
\partial_{z'} - \frac{1-M'}{2(R^2+|z'|^2)} \bar{z}' & \partial_{z'} - \frac{1}{2(R^2+|z'|^2)} \bar{z}'
\end{pmatrix}, \\
\hat{D}'_{\bar{z}'} =
\begin{pmatrix}
\hat{D}'^{aa}_{\bar{z}'} & \hat{D}'^{ab}_{\bar{z}'} \\
\hat{D}'^{ba}_{\bar{z}'} & \hat{D}'^{bb}_{\bar{z}'}
\end{pmatrix} =
\begin{pmatrix}
\partial_{\bar{z}'} - \frac{1}{2(R^2+|z'|^2)} z' & \partial_{\bar{z}'} - \frac{1-M'}{2(R^2+|z'|^2)} z' \\
\partial_{\bar{z}'} - \frac{1+M'}{2(R^2+|z'|^2)} z' & \partial_{\bar{z}'} - \frac{1}{2(R^2+|z'|^2)} z'
\end{pmatrix},
\end{array}
\label{eq:D'D'}
\end{align}
as
\begin{align}
\begin{array}{l}
\widetilde{\hat{D}'_{z'}} =
\begin{pmatrix}
\widetilde{\hat{D}'^{aa}_{z'}} & \widetilde{\hat{D}'^{ab}_{z'}} \\
\widetilde{\hat{D}'^{ba}_{z'}} & \widetilde{\hat{D}'^{bb}_{z'}}
\end{pmatrix} =
\begin{pmatrix}
\partial_{z'} + \frac{1}{2(R^2+|z'|^2)} \bar{z}' & \partial_{z'} + \frac{1-M'}{2(R^2+|z'|^2)} \bar{z}' \\
\partial_{z'} + \frac{1+M'}{2(R^2+|z'|^2)} \bar{z}' & \partial_{z'} + \frac{1}{2(R^2+|z'|^2)} \bar{z}'
\end{pmatrix}, \\
\widetilde{\hat{D}'_{\bar{z}'}} =
\begin{pmatrix}
\widetilde{\hat{D}'^{aa}_{\bar{z}'}} & \widetilde{\hat{D}'^{ab}_{\bar{z}'}} \\
\widetilde{\hat{D}'^{ba}_{\bar{z}'}} & \widetilde{\hat{D}'^{bb}_{\bar{z}'}}
\end{pmatrix} =
\begin{pmatrix}
\partial_{\bar{z}'} + \frac{1}{2(R^2+|z'|^2)} z' & \partial_{\bar{z}'} + \frac{1+M'}{2(R^2+|z'|^2)} z' \\
\partial_{\bar{z}'} + \frac{1-M'}{2(R^2+|z'|^2)} z' & \partial_{\bar{z}'} + \frac{1}{2(R^2+|z'|^2)} z'
\end{pmatrix},
\end{array}
\label{eq:D'D'tilde}
\end{align}
and they satisfy the following commutation relations,
\begin{align}
\begin{array}{l}
\ [\widetilde{\hat{D}'^{aa}_{z'}}, \hat{D}'^{aa}_{\bar{z}'} ] = [\widetilde{\hat{D}'^{aa}_{\bar{z}'}}, \hat{D}'^{aa}_{z'}] = -\frac{R^2}{(R^2+|z'|^2)^2}, \\
\ [\widetilde{\hat{D}'^{bb}_{z'}}, \hat{D}'^{bb}_{\bar{z}'}] = [ \widetilde{\hat{D}'^{bb}_{\bar{z}'}}, \hat{D}'^{bb}_{z'}] = -\frac{R^2}{(R^2+|z'|^2)^2}, \\
\ [\widetilde{\hat{D}'^{ab}_{z'}}, \hat{D}'^{ab}_{\bar{z}'} ] = [\widetilde{\hat{D}'^{ba}_{\bar{z}'}}, \hat{D}'^{ba}_{z'} ] = \frac{R^2}{(R^2+|z'|^2)^2}(M'-1), \\
\ [\widetilde{\hat{D}'^{ba}_{z'}}, \hat{D}'^{ba}_{\bar{z}'} ] = [\widetilde{\hat{D}'^{ab}_{\bar{z}'}}, \hat{D}'^{ab}_{z'} ] = -\frac{R^2}{(R^2+|z'|^2)^2}(M'+1).
\end{array}
\label{eq:comD'D'}
\end{align}
Thus,
when we denote the ``effective'' magnetic fluxes which spinors $\psi^{AB}$ and scalars $\phi^{AB}$ ($A,B=a,b$) feel $M'^{AB}_{\psi}$ and $M'^{AB}_{\phi}$, respectively, 
they are written by
\begin{align}
\begin{array}{ll}
M'^{aa}_{\phi} = 0, & M'^{aa}_{\psi} = -1, \\
M'^{bb}_{\phi} = 0, & M'^{bb}_{\psi} = -1, \\
M'^{ab}_{\phi} = M, & M'^{ab}_{\psi} = M'-1, \\
M'^{ba}_{\phi} = -M', & M'^{ba}_{\psi} = -M'-1,
\end{array}
\label{eq:effectiveflux}
\end{align}
and then we can find the solutions of Eq.~(\ref{eq:KGS2}):
\begin{align}
\phi^{aa}_{S^2,n}(z') = \psi^{aa,{M'}^{aa}_{\psi} \rightarrow {M'}^{aa}_{\phi}}_{S^2,+,n}(z'), &\quad m_n^2 = \frac{n(n+1)}{R^2} \ (n \geq 0), \label{eq:KGmassiveaaS2} \\
\phi^{bb}_{S^2,n}(z') = \psi^{bb,{M'}^{bb}_{\psi} \rightarrow {M'}^{bb}_{\phi}}_{S^2,+,n}(z'), &\quad m_n^2 = \frac{n(n+1)}{R^2} \ (n \geq 0), \label{eq:KGmassivebbS2} \\
\phi^{ab}_{S^2,n}(z') = \psi^{ab,{M'}^{ab}_{\psi} \rightarrow {M'}^{ab}_{\phi}}_{S^2,+,n}(z'), &\quad m_n^2 = \frac{n(n+M'+1)}{R^2} + \frac{M'}{2R^2} \ (n \geq 0), \label{eq:KGmassiveabS2} \\
\phi^{ba}_{S^2,n}(z') = \psi^{ba,{M'}^{ba}_{\psi} \rightarrow {M'}^{ba}_{\phi}}_{S^2,+,n+1}(z'), &\quad m_n^2 = \frac{n(n+M'+1)}{R^2} + \frac{M'}{2R^2} \ (n \geq 0)
. \label{eq:KGmassivebaS2}
\end{align}
In particular, $M'+1$ number of the lowest modes of $\phi^{ab}_{S^2}$ and $\phi^{ba}_{S^2}$ are massive while the lowest modes of $\phi^{aa}_{S^2}$ and $\phi^{bb}_{S^2}$ are massless and constants.
Here, we note that the curvature contribution for a spinor reduces the ``effective'' flux of the spinor than that of a scalar by $1$.

\begin{itemize}
\item {\large Vector fields}
\end{itemize}

Third, let us see wave functions of $U(N)$ adjoint vector fields on magnetized $S^2$,
\begin{align}
A_{S^2}(z') 
=& A_{S^2,z'}(z) dz' + A_{S^2,\bar{z}'} d\bar{z}' \qquad \left( A_{S^2\bar{z}'}(z') = (A_{S^2,z'}(z'))^{\dagger} \right) 
\label{eq:vectorS2} \\
=&
\begin{pmatrix}
\langle {A'}^{aa}_{z'}(z') \rangle + A^{aa}_{S^2,z'}(z') & {A'}^{ab}_{S^2,z'}(z') \\
A^{ba}_{S^2,z'}(z') & \langle {A'}^{bb}_{z'}(z') \rangle + A^{bb}_{S^2,z'}(z')
\end{pmatrix}
dz' \notag \\ &+
\begin{pmatrix}
\langle {A'}^{aa}_{\bar{z}'}(z') \rangle + A^{aa}_{S^2,\bar{z}'}(z') & A^{ab}_{S^2,\bar{z}'}(z') \\
A^{ba}_{S^2,\bar{z}'}(z') & \langle {A'}^{bb}_{\bar{z}'}(z') \rangle + A^{bb}_{S^2,\bar{z}'}(z')
\end{pmatrix}
d\bar{z}', 
\end{align}
which satisfy the Yang-Mills-Proca equation,
\begin{align}
{h'}^{z'\bar{z}'} \hat{D}'_{z} F_{S^2,\bar{z}'z'} + m_{n}^2 A_{S^2,z',n}(z') = 0, \quad \left( {h'}^{\bar{z}'z'} \hat{D}'_{\bar{z}'} F_{S^2,z'\bar{z}'} + m_{n}^2 A_{S^2,\bar{z}',n}(z') = 0 \right),
\label{eq:YMPeqS2}
\end{align}
with the gauge-fixing condition,
\begin{align}
{h'}^{z'\bar{z}'} \left(\hat{D}'_{z'}A_{S^2,\bar{z}',n}(z') + \hat{D}'_{\bar{z}'} A_{S^2,z',n}(z') \right) = 0.
\label{eq:GFCS2}
\end{align}
The field strength in Eq.~(\ref{eq:YMPeqS2}) can be rewritten as
\begin{align}
F'_{S^2,z'\bar{z}'}(z')
&= \langle F'_{z'\bar{z}'} \rangle + \hat{D}'_{z'} A_{S^2,\bar{z}'}(z) - \hat{D}'_{\bar{z}'} A_{S^2,z'}(z') \notag\\
&= \langle F'_{z'\bar{z}'} \rangle + 2 \hat{D}'_{z'} A_{S^2,\bar{z}'}(z') \label{eq:FreS2} \\
&= \langle F'_{z'\bar{z}'} \rangle - 2 \hat{D}'_{\bar{z}'} A_{S^2,z'}(z'), \notag
\end{align}
where we use the gauge fixing condition (\ref{eq:GFCS2}) in the second and third lines.
Then, the Yang-Mills-Proca equation in Eq.~(\ref{eq:YMPeqS2}) can be rewritten by
\begin{align}
&-2h'^{z'\bar{z}'} \hat{D}'_{z'} \hat{D}'_{\bar{z}'} A_{S^2,z',n}(z') + ih'^{z'\bar{z}'} [\langle F'_{z'\bar{z}'} \rangle, A_{S^2,z',n}(z')] = m_n^2 A_{S^2,z',n}(z'), \notag \\
\Leftrightarrow \ &
\left( \hat{\Delta}' - 2h'^{z'\bar{z}'} [\hat{D}'_{z'}, \hat{D}'_{\bar{z}'}] \right) A_{S^2,z',n}(z') = m_n^2 A_{S^2,z',n}(z'), \label{eq:YMPeqS2} \\
&-2h'^{z'\bar{z}'} \hat{D}'_{\bar{z}'} \hat{D}'_{z'} A_{S^2,\bar{z}',n} -i h'^{z'\bar{z}'} [\langle F'_{z'\bar{z}'} \rangle, A_{S^2,\bar{z}',n}(z)] = m_n^2 A_{S^2,\bar{z}',n}(z'), \notag \\
\Leftrightarrow \ &
\left( \hat{\Delta}' + 2h'^{z'\bar{z}'} [\hat{D}'_{z'}, \hat{D}'_{\bar{z}'}] \right) A_{S^2,\bar{z}',n}(z') = m_n^2 A_{S^2,\bar{z}',n}(z'). \label{eq:YMPeqS2}
\end{align}
Note that although only the covariant derivative in Eq.~(\ref{eq:YMPeqS2}) includes the Levi-Civita connection in Eq,~(\ref{eq:LeviCivitaS2}) due to $\Gamma'^{z'}_{\bar{z}'z}=\Gamma'^{\bar{z}'}_{\bar{z}'z} = 0$ and Eq.~(\ref{eq:nablaVT})  this contribution is absorbed in Eq.~(\ref{eq:YMEFVEVS2}).
Therefore, by using the results in Eqs.~(\ref{eq:KGmassiveaaS2})-(\ref{eq:KGmassivebaS2}), the solutions of Eq.~(\ref{eq:YMPeqS2}) are
\begin{align}
A^{aa}_{S^2,z',n}(z') = \phi^{aa}_{S^2,n}(z'), &\quad A^{aa}_{S^2,\bar{z}',n}(z') = \overline{A^{aa}_{S^2,z',n}(z')}, \quad m_n^2 = \frac{n(n+1)}{R^2}, \label{eq:YMPmassiveaaS2} \\
A^{bb}_{S^2,z',n}(z') = \phi^{bb}_{S^2,n}(z'), &\quad A^{bb}_{S^2,\bar{z}',n}(z') = \overline{A^{bb}_{S^2,z',n}(z')}, \quad m_n^2 = \frac{n(n+1)}{R^2} \ (n \geq 0),  \label{eq:YMPmassivebbS2} \\
A^{ab}_{S^2,z',n}(z') = \phi^{ab}_{S^2,n}(z'), &\quad A^{ba}_{S^2,\bar{z}',n}(z') = \overline{A^{ab}_{S^2,z',n}(z')}, \quad m_n^2 = \frac{n(n+M'+1)}{R^2} - \frac{M'}{2R^2} \ (n \geq 0), \label{eq:YMPmassiveabS2} \\
A^{ba}_{S^2,z',n}(z') = \phi^{ba}_{S^2,n}(z'), &\quad , A^{ab}_{S^2,\bar{z}',n}(z') = \overline{A^{ba}_{S^2,z',n}(z')}, \quad m_n^2 = \frac{n(n+M'+1)}{R^2} + \frac{3M'}{2R^2} \ (n \geq 0). \label{eq:YMPmassivebaS2}
\end{align}
Here, we note that in the normalization condition, we need the additional term $h^{z'\bar{z}'}$ to make the vector quanta to be scalar quanta.
Thus, the lowest massless modes of $A^{aa}_{S^2,z'}$ ($A^{aa}_{S^2,\bar{z}'}$) and $A^{bb}_{S^2,z'}$ ($A^{bb}_{S^2,\bar{z}'}$) are not physical relevant solutions.
On the other hand, $M'-1$ number of the lowest modes of $A^{ab}_{S^2,z'}$ ($A^{ba}_{S^2,\bar{z}'}$) become tachyonic while those of $A^{ba}_{S^2,z'}$ ($A^{ab}_{\bar{z}'}$) become massive.


\section{Magnetized blow-up manifold compactification}
\label{sec:Mblowupman}

In this section, let us see wave functions on magnetized blow-up manifolds of $T^2/\mathbb{Z}_N$ orbifolds.
The following analysis is based on our papers~\cite{Kobayashi:2019fma,Kobayashi:2022xsk,Kobayashi:2022tti}.


\subsection{Construction of blow-up manifold of $T^2/\mathbb{Z}_N$ orbifold}
\label{sec:blowupman}

In this subsection, we discuss construction of blow-up manifolds of $T^2/\mathbb{Z}_N$ orbifolds.
We notice that a $\mathbb{Z}_N$ fixed point becomes a singularity of the $T^2/\mathbb{Z}_N$ orbifold and around the singularity becomes cone.
The blow-up manifolds, as mentioned before, can be constructed by replacing the cones with parts of $S^2$.
First, we cut the cone around the singularity with $|z| \leq r$, that is, the slant height of the cone is $r$.
The  development of the cut-out cone is shown in the left figure of Figure~\ref{fig:ZN}.
Then, the radius of the base of the cone becomes $r/N$.
Next, we embed a part of $S^2$ smoothly instead of the cone as shown in the right figure of Figure~\ref{fig:ZN}.
The right figure shows the cross sections of the cone and the $S^2$, where the $S^2$ tangents to the surface of the cone.
Hence, the angle $\theta_0$ in the right figure of Figure~\ref{fig:ZN} satisfies $\cos \theta_0 = 1/N$ and the radius of the $S^2$ must be $R=r\cot \theta_0 = r/\sqrt{N^2-1}$.
Thus, we embed a part of $S^2$, $0 \leq \theta \leq \theta_0$, with the radius $R=r/\sqrt{N^2-1}$, that is, we embed $(N-1)/2N$-part of $S^2$, which can be checked as
\begin{align}
\frac{\int_{0}^{2\pi} R d\varphi \int_{0}^{\theta_0} d\theta R \sin \theta}{\int_{0}^{2\pi} R d\varphi \int_{0}^{\pi} d\theta R \sin \theta } = \frac{2\pi R^2 \frac{N-1}{N}}{4\pi R^2} = \frac{N-1}{2N}.
\label{eq:partS2}
\end{align}
Indeed, the curvature of the embedded part of $S^2$ becomes $2 \times (N-1)/2N = (N-1)/N$, while the curvature of the cut-out singularity has $(N-1)/N$ which comes from the deficient angle as shown in Eq.~(\ref{eq:localcurvature}).
It means that this blow-up procedure does not change the total curvature (which is a topological invariant number).
Hereafter, we call remaining region of $T^2/\mathbb{Z}_N$ and embedded regions instead of cut-out cones from $T^2/\mathbb{Z}_N$ as bulk region and blow-up regions, respectively.
 \begin{figure}[H]
\centering
  \begin{minipage}{7cm}
  \centering
     \includegraphics[bb=0 0 750 775,width=5cm]{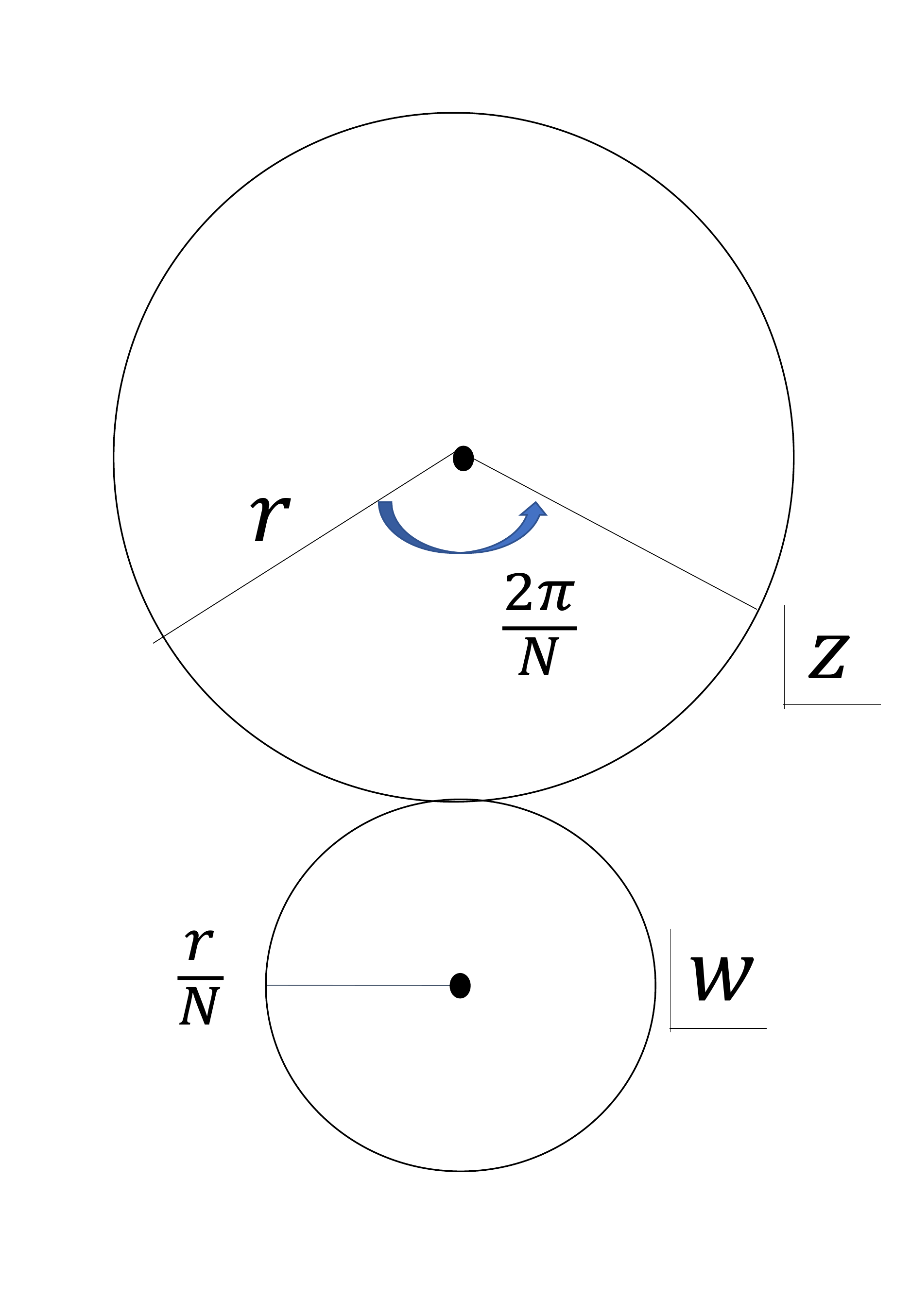}
   \end{minipage}
  \begin{minipage}{7cm}
   \centering
    \includegraphics[bb=0 0 480 425,width=5cm]{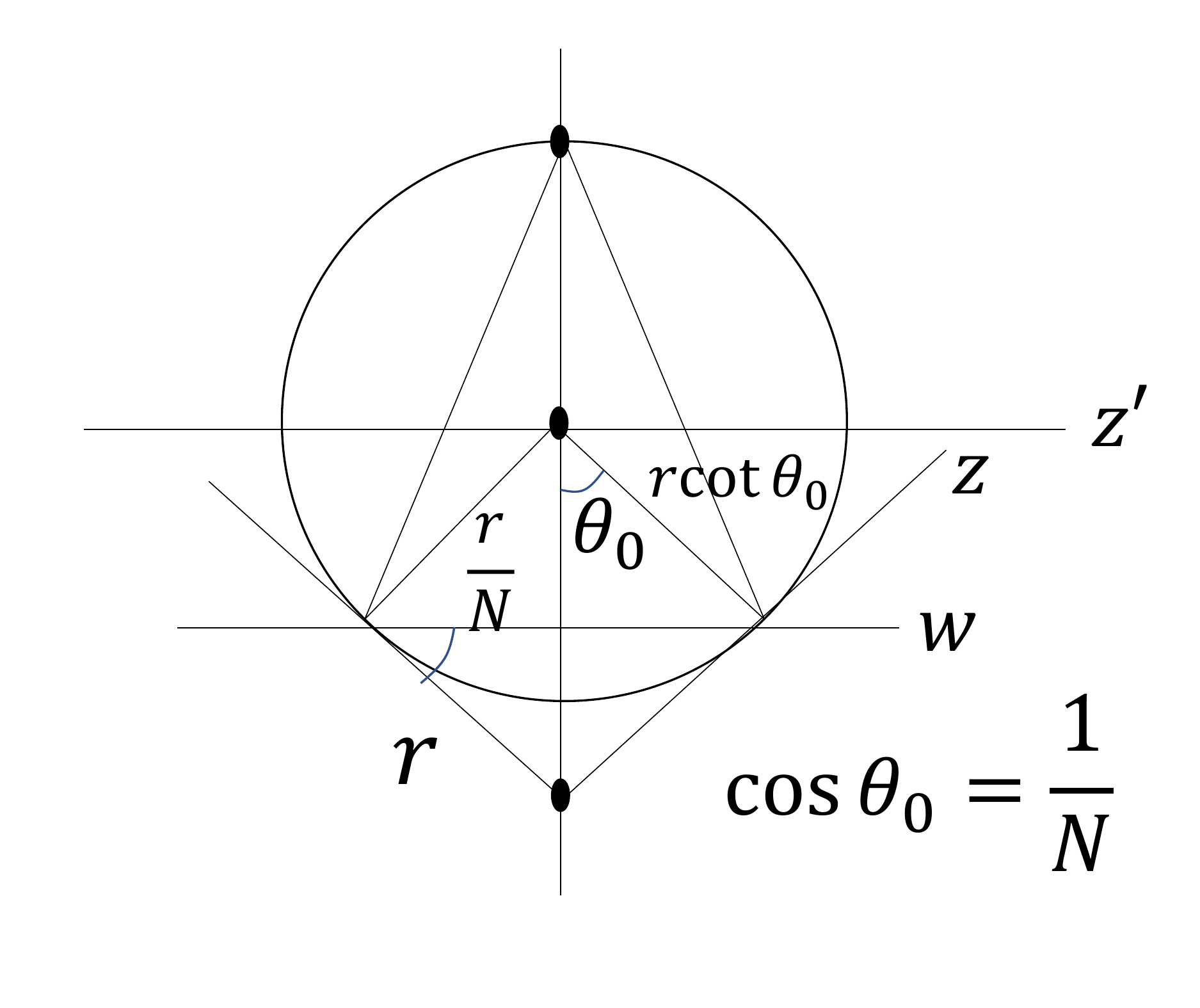}
  \end{minipage}
    \caption{The left figure shows the development of a cone around a singularity coming from a $\mathbb{Z}_N$ fixed point of $T^2/\mathbb{Z}_N$ orbifold. The right figure shows the cross sections of the cone and the $S^2$ with radius $R =r/\sqrt{N^2-1}$. The blow-up manifold of the $T^2/\mathbb{Z}_N$ orbifold can be constructed by cutting out the cone and embedding the $(N-1)/2N$-part of $S^2$ ($0 \leq \theta \leq \theta_0 = \cos^{-1}(1/N)$) instead. Here, $z$ and $z'$ denote the coordinates of $T^2/\mathbb{Z}_N$ and $S^2$, respectively, and they are related through the coordinate $w$;~i.e. at the connecting points, $z=re^{i\varphi/N}$ and $z'=\frac{r}{N+1}e^{i\varphi}$ are related through $w$ as $z \leftrightarrow w = \frac{N+1}{N} z'$.}
    \label{fig:ZN}
\end{figure}


Next, let us see coordinate on the blow-up manifold.
We use $z$ and $z'$ as the coordinates on the bulk region and blow-up regions, respectively.
From now, we show the relation of $z$ and $z'$ at the connecting points.
Hereafter, we show the case of blowing up of the singularity $z^{\rm fp}_0=0$.
The following analysis can be also applied for another singularity $z^{\rm fp}_I$ by replacing $z$ with $Z_I$.
First, at one of the connecting points, $z$ and $z'$ can be expressed as
\begin{align}
z=re^{i\varphi_{T^2}}, \quad z' = \frac{r}{\sqrt{N^2-1}} \tan \frac{\theta_0}{2} e^{i\varphi_{S^2}} = \frac{r}{N+1} e^{i\varphi_{S^2}}, \label{eq:coordinatezz'}
\end{align}
respectively.
In addition, as shown in Figure~\ref{fig:ZN}, we define a new coordinate $w$ on the complex plane on which the cut surface is and which is parallel to the complex with the coordinate $z'$ such that the coordinate of the cut edge is $w=\frac{r}{N} e^{i\varphi_{S^2}}$.
Thus, we can find that $z$ and $z'$ are related through $w$ as $z \leftrightarrow w=\frac{N+1}{N}z'$ at the connecting points in Eq.~(\ref{eq:coordinatezz'}).
Indeed, from the left figure of Figure~\ref{fig:ZN}, we obtain the relation,
\begin{align}
rd(\varphi_{T^2}) = \frac{r}{N} d \varphi(\phi_{S^2}) \ \Leftrightarrow \ \varphi_{T^2} = \frac{\varphi_{S^2}}{N}. \label{eq:angleT2ZNS2}
\end{align}
Hereafter, we denote $\varphi \equiv \varphi_{S^2}$.
On the other hand, from the right figure of Figure~\ref{fig:ZN}, we obtain the relation
\begin{align}
d|z| = Rd\theta = 2\cos^2 \frac{\theta_0}{2} d|z'| = (1+\frac{1}{N}) d|z'| = \frac{N+1}{N} d|z'|. \label{eq:amplT2ZNS2}
\end{align}
By combining these relations, we obtain the relation,
\begin{align}
e^{-i\frac{\varphi}{N}} dz \left|_{z=r e^{i\frac{\varphi}{N}}} = \frac{N+1}{N} e^{-i\varphi} dz' \right|_{z'=\frac{r}{N+1} e^{i\varphi}}, \label{eq:reldzdz'}
\end{align}
which can derived from
\begin{align}
    \begin{array}{c}
        e^{-i\frac{\varphi}{N}} dz = e^{-i\frac{\varphi}{N}} \frac{\partial z}{\partial |z|} d|z| + e^{-i\frac{\varphi}{N}} \frac{\partial z}{\partial (\frac{\varphi}{N})} d(\frac{\varphi}{N}) = d|z| + i r d(\frac{\varphi}{N}), \\
        \frac{N+1}{N} e^{-i\varphi} dz' = \frac{N+1}{N} e^{-i\varphi} \frac{\partial z'}{\partial |z'|} d|z'| + \frac{N+1}{N} e^{-i\varphi} \frac{\partial z'}{\partial \varphi} d\varphi = \frac{N+1}{N} d|z'| + i \frac{r}{N} d\varphi.
    \end{array}
\end{align}


\subsection{Singular gauge transformation}
\label{sec:blowupman}

In order to obtain wave functions on the magnetized blow-up manifold by connecting ones on magnetized $T^2/\mathbb{Z}_N$ and ones on magnetized $S^2$ smoothly, we should modify the BC under $\mathbb{Z}_N$ twist in Eq.~(\ref{eq:BCPhirhoz}).
In detail, when wave functions on $S^2$ go round the circle which is the cut edge, the values of them are the same as the original values: $\Phi_{S^2}(z'e^{2\pi i}) = \Phi_{S^2}(z')$.
Then, wave functions on $T^2/\mathbb{Z}_N$ should also satisfy $\widetilde{\Phi}_{T^2/\mathbb{Z}_N}(ze^{2\pi i/N}) = \widetilde{\Phi}_{T^2/\mathbb{Z}_N}(z)$ when they go round the circle.
Hence, we consider modification of the BCs in Eq.~(\ref{eq:BCPhirhoz}) by a (singular) gauge transformation, as in the case that SS phases and WL phases are related through a gauge transformation.

We note that since there is a freedom to insert additional magnetic fluxes independent of background magnetic fluxes at orbifold singularities (localized fluxes\footnote{For example, see Ref.~\cite{Lee:2003mc,Buchmuller:2015eya,Buchmuller:2018lkz}.}), a singular gauge transformation by which the flux is changed only at the one point can be allowed while usual gauge transformation does not change the flux.
Now, we introduce the singular gauge transformation at $z^{\rm fp}_0=0$, by the following unitary transformation\footnote{See Refs.~\cite{Lee:2003mc,Buchmuller:2015eya,Buchmuller:2018lkz,Polchinski}.},
\begin{align}
V_{\xi^F_0}(z) =
\begin{pmatrix}
V^{\xi^{F^a}_0}(z) \mathbb{I}_{N_a}  & 0 \\
0 & V^{\xi^{F^b}_0}(z) \mathbb{I}_{N_b}
\end{pmatrix},
\label{eq:eq:localgaugeunit}
\end{align}
with
\begin{align}
V(z) = \left( \frac{\psi^{(\frac{1}{2},\frac{1}{2}),1}_{T^2/\mathbb{Z}_N^1,0}(z,\tau)}{\overline{\psi^{(\frac{1}{2},\frac{1}{2}),1}_{T^2/\mathbb{Z}_N^1,0}(z,\tau)}} \right)^{1/2} = \left( \frac{\psi^{(\frac{1}{2},\frac{1}{2}),1}_{T^2,0}(z,\tau)}{\overline{\psi^{(\frac{1}{2},\frac{1}{2}),1}_{T^2,0}(z,\tau)}} \right)^{1/2} = \left( \frac{h^{(\frac{1}{2},\frac{1}{2}),1}_{T^2}(z)}{\overline{h^{(\frac{1}{2},\frac{1}{2}),1}_{T^2}(z)}} \right)^{1/2} \simeq \left( \frac{(h^{(\frac{1}{2},\frac{1}{2}),1}_{T^2})^{(1)}(0)z}{\overline{(h^{(\frac{1}{2},\frac{1}{2}),1}_{T^2})^{(1)}(0)z}} \right)^{1/2},
\label{eq:localtransV}
\end{align}
as
\begin{align}
\begin{array}{rcl}
\langle A^{aa,bb}(z) \rangle &\rightarrow& \langle \widetilde{A}^{aa,bb}(z) \rangle = \langle A^{aa,bb}(z) \rangle 
+ \delta \langle A(z) \rangle, \\
\delta \langle A^{aa,bb}(z) \rangle &=& i V^{\xi^{F^{a,b}}_0}(z) d V^{-\xi^{F^{a,b}}_0}(z) \\
&=&
 - i \frac{\xi^{F^{a,b}}_0}{2} \frac{(h^{(\frac{1}{2},\frac{1}{2}),1}_{T^2})^{(1)}(z)}{h^{(\frac{1}{2},\frac{1}{2}),1}_{T^2}(z)} +  i \frac{\xi^{F^{a,b}}_0}{2} \frac{\overline{(h^{(\frac{1}{2},\frac{1}{2}),1}_{T^2})^{(1)}(z)}}{\overline{h^{(\frac{1}{2},\frac{1}{2}),1}_{T^2}(z)}}
\simeq - i \frac{\xi^{F^{a,b}}_0}{2} \frac{1}{z} dz +  i \frac{\xi^{F^{a,b}}_0}{2} \frac{1}{\bar{z}} d\bar{z},
\end{array}
\label{eq:localgaugetrans}
\end{align}
where we showed the approximation around $z^{\rm fp}_0=0$ in the rightest sides and $(h^{(\frac{1}{2},\frac{1}{2}),1}_{T^2})^{(n)}(z) \equiv \frac{d^n h^{(\frac{1}{2},\frac{1}{2}),1}_{T^2}(z)}{dz^n}$.
Then, the singular gauge transformation modifies the field strength,
\begin{align}
\begin{array}{l}
\frac{1}{2\pi} \langle F^{aa,bb} \rangle \rightarrow \frac{1}{2\pi} \langle \widetilde{F}^{aa,bb} \rangle = \frac{1}{2\pi} \langle F^{aa,bb} \rangle + \frac{1}{2\pi} \delta \langle F^{aa,bb} \rangle, \\
\frac{1}{2\pi} \delta \langle F^{aa,bb} \rangle = i \xi^{F^{a,b}_0} \delta(z) \delta(\bar{z}) dz \wedge d\bar{z},
\end{array}
\label{eq:modifyF}
\end{align}
which induces the localized flux $\xi^{F^{a,b}}_0/N$ at $z^{\rm fp}_0=0$.
Moreover, we also consider a singular gauge transformation for the spin connection since only orbifold singularities have localized curvatures in Eq.~(\ref{eq:localcurvature}).
Here, we note that the functional form of the spin connection on $S^2$ is the same as one of the gauge connection field on $S^2$ replacing the magnetic flux $M$ by the curvature $\chi(S^2)=2$.
Then, the localized curvature at $z^{\rm fp}_0=0$, $\xi^R_0/N = (N-1)/N$, is taken in by the following singular gauge transformation for the spin connection,
\begin{align}
\begin{array}{l}
w=0 \rightarrow \widetilde{w}(z) = w
+ \delta w(z) = \delta w(z), \\
\delta w(z) = i V^{\xi^{R}_0}(z) d V^{-\xi^{R}_0}(z) =
 - i \frac{\xi^{R}_0}{2} \frac{(h^{(\frac{1}{2},\frac{1}{2}),1}_{T^2})^{(1)}(z)}{h^{(\frac{1}{2},\frac{1}{2}),1}_{T^2}(z)} +  i \frac{\xi^{R}_0}{2} \frac{\overline{(h^{(\frac{1}{2},\frac{1}{2}),1}_{T^2})^{(1)}(z)}}{\overline{h^{(\frac{1}{2},\frac{1}{2}),1}_{T^2}(z)}}
\simeq - i \frac{\xi^{R}_0}{2} \frac{1}{z} dz +  i \frac{\xi^{R}_0}{2} \frac{1}{\bar{z}} d\bar{z}.
\end{array}
\label{eq:localspintrans}
\end{align}
From those singular gauge transformation for the gauge connection field and the spin connection, wave functions on the magnetized $T^2/\mathbb{Z}_N$ are changed as
\begin{align}
\widetilde{\Phi}_{T^2/\mathbb{Z}_N}(z) = V^{-s_{\Phi}\xi^{R}_0}(z) V^{(\varphi^{IJ}_N+1/2)\xi^{F_s}_0 \delta_{\Phi,\psi}}(z) V_{\xi^{F}_0}(z) \Phi_{T^2/\mathbb{Z}_N}(z) V_{\xi^{F}_0}^{-1}(z),
\label{eq:localwave}
\end{align}
where $s_{\Phi}$ denotes the spin charge of $\Phi$: $s_{\Phi}=0$ for $\Phi =\phi$, $s_{\Phi}=1\ (-1)$ for $\Phi = A_{z}\ (A_{\bar{z}})$, and $s_{\Phi}=1/2\ (-1/2)$ for $\Phi=\psi_{+}\ (\psi_{-})$, and we also introduce the localized flux of $U(1)_s$, $\xi^{F_s}_0/N$.
For the component fields, Eq.~(\ref{eq:localwave}) can be expressed as
\begin{align}
&\left\{
\begin{array}{l}
\widetilde{\phi}_{T^2/\mathbb{Z}_N}^{aa}(z) = \phi_{T^2/\mathbb{Z}_N}^{aa}(z), \\
\widetilde{\phi}_{T^2/\mathbb{Z}_N}^{bb}(z) = \phi_{T^2/\mathbb{Z}_N}^{bb}(z), \\
\widetilde{\phi}_{T^2/\mathbb{Z}_N}^{ab}(z) = V^{\xi^F_0}(z) \phi_{T^2/\mathbb{Z}_N}^{ab}(z), \\
\widetilde{\phi}_{T^2/\mathbb{Z}_N}^{ba}(z) = V^{-\xi^F_0}(z) \phi_{T^2/\mathbb{Z}_N}^{aa}(z),
\end{array}
\right. \label{eq:phicomptilde} \\
&\left\{
\begin{array}{ll}
\widetilde{A}_{T^2/\mathbb{Z}_N,z}^{aa}(z) = V^{-\xi^R_0}(z) A_{T^2/\mathbb{Z}_N,z}^{aa}(z), & \widetilde{A}_{T^2/\mathbb{Z}_N,\bar{z}}^{aa}(z) = V^{\xi^R_0}(z) A_{T^2/\mathbb{Z}_N,\bar{z}}^{aa}(z), \\
\widetilde{A}_{T^2/\mathbb{Z}_N,z}^{bb}(z) = V^{-\xi^R_0}(z) A_{T^2/\mathbb{Z}_N,z}^{bb}(z), & \widetilde{A}_{T^2/\mathbb{Z}_N,\bar{z}}^{bb}(z) = V^{\xi^R_0}(z) A_{T^2/\mathbb{Z}_N,\bar{z}}^{bb}(z), \\
\widetilde{A}_{T^2/\mathbb{Z}_N,z}^{ab}(z) = V^{-\xi^R_0+\xi^F_0}(z) A_{T^2/\mathbb{Z}_N,z}^{ab}(z), & \widetilde{A}_{T^2/\mathbb{Z}_N,\bar{z}}^{ab}(z) = V^{\xi^R_0+\xi^F_0}(z) A_{T^2/\mathbb{Z}_N,\bar{z}}^{ab}(z), \\
\widetilde{A}_{T^2/\mathbb{Z}_N,z}^{ba}(z) = V^{-\xi^R_0-\xi^F_0}(z) A_{T^2/\mathbb{Z}_N,z}^{ba}(z), & \widetilde{A}_{T^2/\mathbb{Z}_N,\bar{z}}^{ba}(z) = V^{\xi^R_0-\xi^F_0}(z) A_{T^2/\mathbb{Z}_N,\bar{z}}^{aa}(z),
\end{array}
\right. \label{eq:Acomptilde} \\
&\left\{
\begin{array}{ll}
\widetilde{\psi}_{T^2/\mathbb{Z}_N,+}^{aa}(z) = V^{-\frac{\xi^R_0}{2}+\frac{\xi^{F_s}_0}{2}}(z) \psi_{T^2/\mathbb{Z}_N,+}^{aa}(z), & \widetilde{\psi}_{T^2/\mathbb{Z}_N,-}^{aa}(z) = V^{\frac{\xi^R_0}{2}+\frac{\xi^{F_s}_0}{2}}(z) \psi_{T^2/\mathbb{Z}_N,-}^{aa}(\rho z), \\
\widetilde{\psi}_{T^2/\mathbb{Z}_N,+}^{bb}(z) = V^{-\frac{\xi^R_0}{2}+\frac{\xi^{F_s}_0}{2}}(z) \psi_{T^2/\mathbb{Z}_N,+}^{bb}(z), & \widetilde{\psi}_{T^2/\mathbb{Z}_N,-}^{bb}(z) = V^{\frac{\xi^R_0}{2}+\frac{\xi^{F_s}_0}{2}}(z) \psi_{T^2/\mathbb{Z}_N,-}^{bb}(z), \\
\widetilde{\psi}_{T^2/\mathbb{Z}_N,+}^{ab}(z) = V^{-\frac{\xi^R_0}{2}+\frac{\xi^{F_s}_0}{2}+\xi^F_0}(z) \psi_{T^2/\mathbb{Z}_N,+}^{ab}(z), & \widetilde{\psi}_{T^2/\mathbb{Z}_N,-}^{ab}(z) = V^{\frac{\xi^R_0}{2}+\frac{\xi^{F_s}_0}{2}+\xi^F_0}(z) \psi_{T^2/\mathbb{Z}_N,-}^{ab}(z), \\
\widetilde{\psi}_{T^2/\mathbb{Z}_N,+}^{ba}(z) = V^{-\frac{\xi^R_0}{2}-\frac{\xi^{F_s}_0}{2}-\xi^F_0}(z) \psi_{T^2/\mathbb{Z}_N,+}^{ba}(z), & \widetilde{\psi}_{T^2/\mathbb{Z}_N,-}^{ba}(z) = V^{\frac{\xi^R_0}{2}-\frac{\xi^{F_s}_0}{2}-\xi^F_0}(z) \psi_{T^2/\mathbb{Z}_N,-}^{ba}(z),
\end{array}
\right. \label{eq:psicomptilde}
\end{align}
where $\xi^F_0 \equiv \xi^{F^a}_0 - \xi^{F^b}_0$.
Since $V(z)$ satisfies $V(\rho z) = \rho V(z)$, by combining Eqs.~(\ref{eq:BCphicomprhoz})-(\ref{eq:BCpsicomprhoz}), the $\mathbb{Z}_N$ twisted BCs are modified as
\begin{align}
&\left\{
\begin{array}{l}
\widetilde{\phi}_{T^2/\mathbb{Z}_N}^{aa}(\rho z) = \widetilde{\phi}_{T^2/\mathbb{Z}_N}^{aa}(z), \\
\widetilde{\phi}_{T^2/\mathbb{Z}_N}^{bb}(\rho z) = \widetilde{\phi}_{T^2/\mathbb{Z}_N}^{bb}(z), \\
\widetilde{\phi}_{T^2/\mathbb{Z}_N}^{ab}(\rho z) = \rho^{\xi^F_0+m} \widetilde{\phi}_{T^2/\mathbb{Z}_N}^{ab}(z), \\
\widetilde{\phi}_{T^2/\mathbb{Z}_N}^{ba}(\rho z) = \rho^{-\xi^F_0-m} \widetilde{\phi}_{T^2/\mathbb{Z}_N}^{aa}(z),
\end{array}
\right. \label{eq:BCphicomptilderhoz} \\
&\left\{
\begin{array}{ll}
\widetilde{A}_{T^2/\mathbb{Z}_N,z}^{aa}(\rho z) = \rho^{-\xi^R_0-1} \widetilde{A}_{T^2/\mathbb{Z}_N,z}^{aa}(z), & \widetilde{A}_{T^2/\mathbb{Z}_N,\bar{z}}^{aa}(\rho z) = \rho^{\xi^R_0+1} \widetilde{A}_{T^2/\mathbb{Z}_N,\bar{z}}^{aa}(z), \\
\widetilde{A}_{T^2/\mathbb{Z}_N,z}^{bb}(\rho z) = \rho^{-\xi^R_0-1} \widetilde{A}_{T^2/\mathbb{Z}_N,z}^{bb}(z), & \widetilde{A}_{T^2/\mathbb{Z}_N,\bar{z}}^{bb}(\rho z) = \rho^{\xi^R_0+1} \widetilde{A}_{T^2/\mathbb{Z}_N,\bar{z}}^{bb}(z), \\
\widetilde{A}_{T^2/\mathbb{Z}_N,z}^{ab}(\rho z) = \rho^{-\xi^R_0+\xi^F_0+m-1} \widetilde{A}_{T^2/\mathbb{Z}_N,z}^{ab}(z), & \widetilde{A}_{T^2/\mathbb{Z}_N,\bar{z}}^{ab}(\rho z) = \rho^{\xi^R_0+\xi^F_0+m+1} \widetilde{A}_{T^2/\mathbb{Z}_N,\bar{z}}^{ab}(z), \\
\widetilde{A}_{T^2/\mathbb{Z}_N,z}^{ba}(\rho z) = \rho^{-\xi^R_0-\xi^F_0-m-1} \widetilde{A}_{T^2/\mathbb{Z}_N,z}^{ba}(z), & \widetilde{A}_{T^2/\mathbb{Z}_N,\bar{z}}^{ba}(\rho z) = \rho^{\xi^R_0-\xi^F_0-m+1} \widetilde{A}_{T^2/\mathbb{Z}_N,\bar{z}}^{aa}(z),
\end{array}
\right. \label{eq:BCAcomptilderhoz} \\
&\left\{
\begin{array}{ll}
\widetilde{\psi}_{T^2/\mathbb{Z}_N,+}^{aa}(\rho z) = \rho^{-\frac{\xi^R_0}{2}+\frac{\xi^{F_s}_0}{2}} \widetilde{\psi}_{T^2/\mathbb{Z}_N,+}^{aa}(z), & \widetilde{\psi}_{T^2/\mathbb{Z}_N,-}^{aa}(\rho z) = \rho^{\frac{\xi^R_0}{2}+\frac{\xi^{F_s}_0}{2}+1} \widetilde{\psi}_{T^2/\mathbb{Z}_N,-}^{aa}(z), \\
\widetilde{\psi}_{T^2/\mathbb{Z}_N,+}^{bb}(\rho z) = \rho^{-\frac{\xi^R_0}{2}+\frac{\xi^{F_s}_0}{2}} \widetilde{\psi}_{T^2/\mathbb{Z}_N,+}^{bb}(z), & \widetilde{\psi}_{T^2/\mathbb{Z}_N,-}^{bb}(\rho z) = \rho^{\frac{\xi^R_0}{2}+\frac{\xi^{F_s}_0}{2}+1} \widetilde{\psi}_{T^2/\mathbb{Z}_N,-}^{bb}(z), \\
\widetilde{\psi}_{T^2/\mathbb{Z}_N,+}^{ab}(\rho z) = \rho^{-\frac{\xi^R_0}{2}+\frac{\xi^{F_s}_0}{2}+\xi^F_0+m} \widetilde{\psi}_{T^2/\mathbb{Z}_N,+}^{ab}(z), & \widetilde{\psi}_{T^2/\mathbb{Z}_N,-}^{ab}(\rho z) = \rho^{\frac{\xi^R_0}{2}+\frac{\xi^{F_s}_0}{2}+\xi^F_0+m+1} \widetilde{\psi}_{T^2/\mathbb{Z}_N,-}^{ab}(z), \\
\widetilde{\psi}_{T^2/\mathbb{Z}_N,+}^{ba}(\rho z) = \rho^{-\frac{\xi^R_0}{2}-\frac{\xi^{F_s}_0}{2}-\xi^F_0-m-1} \widetilde{\psi}_{T^2/\mathbb{Z}_N,+}^{ba}(z), & \widetilde{\psi}_{T^2/\mathbb{Z}_N,-}^{ba}(\rho z) = \rho^{\frac{\xi^R_0}{2}-\frac{\xi^{F_s}_0}{2}-\xi^F_0-m} \widetilde{\psi}_{T^2/\mathbb{Z}_N,-}^{ba}(z).
\end{array}
\right. \label{eq:BCpsicomptilderhoz}
\end{align}
Therefore, in order for $\widetilde{\Phi}_{T^2/\mathbb{Z}_N}(z)$ to satisfy $\widetilde{\Phi}_{T^2/\mathbb{Z}_N}(\rho z) = \widetilde{\Phi}_{T^2/\mathbb{Z}_N}(z)$, the following relations,
\begin{align}
\xi^R_0 = N-1, \quad \xi^F_0 \equiv -m\ ({\rm mod}\ N), \quad \xi^{F_s}_0 \equiv \xi^R_0\ ({\rm mod}\ N),
\label{eq:localcurvflux}
\end{align}
are needed.
In particular, the total localized flux which $\Phi^{ab}$ feels at $z^{\rm fp}_0=0$ becomes
\begin{align}
\frac{\xi^{F_{\rm total}}_0}{N} = \frac{1}{2}\frac{\xi^{F_s}_0}{N}\delta_{\Phi,\psi} + \frac{\xi^{F}_0}{N} = \frac{N-1}{2N} - \frac{m}{N} + \ell_0, \quad \ell_0 \in \mathbb{Z},
\label{eq:totalflux0}
\end{align}
where $\ell_0$ comes from mod $N$ property.
Similarly, around a $\mathbb{Z}_N$ singularity $z^{\rm fp}_I$, in order for $\widetilde{\Phi}_{T^2/\mathbb{Z}_N}(Z)$ to satisfy $\widetilde{\Phi}_{T^2/\mathbb{Z}_N}(\rho Z) = \widetilde{\Phi}_{T^2/\mathbb{Z}_N}(Z)$,
the following relations
\begin{align}
\xi^R_I = N-1, \quad \xi^F_I \equiv -\chi_{(m)I}\ ({\rm mod}\ N), \quad \xi^{F_s}_I \equiv \xi^R_I\ ({\rm mod}\ N),
\label{eq:localcurvflux}
\end{align}
are needed and the total localized flux which $\Phi^{ab}$ feels at $z^{\rm fp}_I$ becomes
\begin{align}
\frac{\xi^{F_{\rm total}}_I}{N} = \frac{1}{2} \frac{\xi^{F_s}_I}{N}\delta_{\Phi,\psi} + \frac{\xi^{F}_I}{N} = \frac{N-1}{2N} - \frac{\chi_{(m)I}}{N} + \ell_I, \quad \ell_I \in \mathbb{Z}.
\label{eq:totalfluxI}
\end{align}
We note that the shift BCs are modified by
\begin{align}
\begin{array}{l}
V(z+1) = e^{2\pi i(1/2)} e^{\pi i\frac{{\rm Im}z}{{\rm Im}\tau}} V(z), \\
V(z+\tau) = e^{2\pi i(1/2)} e^{\pi i\frac{{\rm Im}\bar{\tau}z}{{\rm Im}\tau}} V(z).
\end{array}
\label{eq:BCVz1tau}
\end{align}

Hereafter, we consider the following lowest modes on the magnetized $T^2/\mathbb{Z}_N$ with $\mathbb{Z}_N$ charge $m$,
\begin{align}
\widetilde{\psi}^{(j+\alpha_1,\alpha_{\tau}),M}_{T^2/\mathbb{Z}_N^{m},0}(z,\tau) 
=& V^{-m+\ell_0N}(z) \psi^{(j+\alpha_1,\alpha_{\tau}),M}_{T^2/\mathbb{Z}_N^{m},0}(z,\tau) \notag \\
=& e^{-\frac{\pi M}{2{\rm Im}\tau}|z|^2} \left| h^{(\frac{1}{2},\frac{1}{2}),1}_{T^2}(z) \right|^{m-\ell_0N} \left( h^{(\frac{1}{2},\frac{1}{2}),1}_{T^2}(z) \right)^{-m+\ell_0N} h^{(j+\alpha_1,\alpha_{\tau}),M}_{T^2/\mathbb{Z}_N^{m},0}(z) \notag \\
\simeq& e^{-\frac{\pi M}{2{\rm Im}\tau}|z|^2} |z|^{m-\ell_0N} C^{j}_0 N z^{\ell_0N},
\label{eq:T2ZNzerotilde} \\
C^{j}_0 =& \left( \frac{(h^{(\frac{1}{2},\frac{1}{2}),1}_{T^2})^{(1)}(0)}{\left| (h^{(\frac{1}{2},\frac{1}{2}),1}_{T^2})^{(1)}(0) \right|} \right)^{-m+\ell_0N} \sum_{j'} {\cal N}'_{T^2/\mathbb{Z}_N^m,0,j'} \frac{(h^{(j'+\alpha_1,\alpha_{\tau}),M}_{T^2,0})^{(m)}(0)}{m!},
\label{eq:coefficientT2ZNtilde}
\end{align}
where we take the normalization condition,
\begin{align}
\int_{T^2} dz d\bar{z} \sqrt{|{\rm det}(2h)|} \overline{\widetilde{\psi}^{(j+\alpha_1,\alpha_{\tau}),M}_{T^2/\mathbb{Z}_N^m,0}(z,\tau)} \widetilde{\psi}^{(k+\alpha_1,\alpha_{\tau}),M}_{T^2/\mathbb{Z}_N^{m},0}(z,\tau) = \delta_{j,k}.
\label{eq:normalizationT2ZNtilde}
\end{align}


\subsection{Wave functions on magnetized blow-up manifold}
\label{subsec:junction}

Now, let us see wave functions of the lowest modes on the magnetized blow-up manifold by connecting smoothly ones on the magnetized $T^2/\mathbb{Z}_N$ in Eq.~(\ref{eq:normalizationT2ZNtilde}), which become ones on the bulk region, and ones on magnetized $S^2$ in Eq.~(\ref{eq:zeroS2}), which become ones on blow-up regions.
In the following, we consider blow-up of the singularity $z^{\rm fp}_0=0$.

Due to the coordinate relation in Eq.~(\ref{eq:reldzdz'}), the junction condition is given by
\begin{align}
    \begin{array}{c}
        \tilde{\psi}_{T^2/Z^m_N,0}^{(j+\alpha_1,\alpha_{\tau}),M}(z) \biggl|_{z=r e^{i\varphi/N}} = \psi_{S^2,0}^{M'(-1)}(z') \biggl|_{z'=\frac{r}{N+1}e^{i\varphi}}, \\
        \frac{1}{e^{-i\frac{\varphi}{N}}} \frac{d \tilde{\psi}_{T^2/Z^m_N,0}^{(j+\alpha_1,\alpha_{\tau}),M}(z)}{d z} \biggl|_{z=r e^{i\varphi/N}} = \frac{1}{\frac{N+1}{N}e^{-i\varphi}} \frac{d \psi_{S^2,0}^{M'(-1)}(z')}{d z'}\biggl|_{z'=\frac{r}{N+1}e^{i\varphi}}.
    \end{array}
    \label{eq:concon}
\end{align}
From the non-holomorphic parts of wave functions, we obtain the following relations:
\begin{align}
    \frac{\pi r^2}{N {\rm Im}\tau} M + \frac{N-1}{2N} - \frac{m}{N} + \ell_0 = \frac{N-1}{2N}M',
    \label{eq:fluxcondspinor}
\end{align}
for spinor fields and
\begin{align}
    \frac{\pi r^2}{N {\rm Im}\tau} M - \frac{m}{N} + \ell_0 = \frac{N-1}{2N}M',
    \label{eq:fluxcondscalarvector}
\end{align}
for scalar and vector fields.
(Note that in the above result as well as the following analysis, we can similarly obtain the results for scalar and vector cases by replacing the magnetic flux which the spinor feels, $M'$, with $M'+1$, and then hereafter, we mainly discuss the spinor cases.)
On the other hand, from holomorphic parts, the holomorphic function on the blow-up region can be determined as
\begin{align}
    h_{S^2}^{M'}(z') = {C'}^{j}_{0} R^{-\ell_0} z'^{\ell_0}, \quad {C'}^{j}_{0} = C^{j}_{0} N r^{m} e^{-\frac{\pi M}{2{\rm Im}\tau}r^2} \left( \frac{N+1}{N-1} \right)^{\frac{\ell_0}{2}} \left( \frac{N+1}{2N} \right)^{-\frac{M'-1}{2}}. \label{eq:holomorphicrel}
\end{align}
Therefore, wave functions on the magnetized blow-up manifold can be written as
\begin{align}
    \psi^{j}_{{\rm blow-up},0} = \left\{
    \begin{array}{ll}
	\left( \frac{R^2}{R^2+|z'|^2} \right)^{\frac{M'-1}{2}} {C'}^{j}_{0} R^{-\ell_0} z'^{\ell_0} & (|z'| \leq \frac{r}{N+1}) \\
        e^{-\frac{\pi M}{2{\rm Im}\tau}|z|^2} \left| h^{(\frac{1}{2},\frac{1}{2}),1}_{T^2}(z) \right|^{m-\ell_0N} \left( h^{(\frac{1}{2},\frac{1}{2}),1}_{T^2}(z) \right)^{-m+\ell_0N} h^{(j+\alpha_1,\alpha_{\tau}),M}_{T^2/\mathbb{Z}_N^{m},0}(z) & (r \leq |z|) \\
        \ \simeq  e^{-\frac{\pi M}{2{\rm Im}\tau}|z|^2} |z|^{m-\ell_0N} C^{j}_{0} N z^{\ell_0N}
    \end{array}
    \right.. \label{eq:wavbulk}
\end{align}
To determine the normalization, we first calculate the following inner product,
\begin{align}
    G_{ij}
    =& \int_{{\rm blow-up\ manifold}} dz d\bar{z} \sqrt{|{\rm det}(2h)|} \overline{\psi^{i}_{{\rm blow-up},0}} \psi^{j}_{{\rm blow-up},0} \notag \\
    =&\, \delta_{i,j} - \int_{0}^{r} d|z| |z| \int_{0}^{\frac{2\pi}{N}} d\varphi (C^{i}_{0})^{\ast} C^{j}_{0} N^2 |z|^{2m} e^{-\frac{\pi M}{{\rm Im}\tau}|z|^2} \notag \\
    &+ \int_{0}^{\frac{r}{N+1}} d|z'| |z'| \int_{0}^{2\pi} d\varphi  \left( \frac{2R^2}{R^2+|z'|^2} \right)^2 \left( \frac{R^2}{R^2+|z'|^2} \right)^{M'-1} ({C'}^{i}_{0})^{\ast} {C'}^{j}_{0} R^{-2\ell_0} |z'|^{2\ell_0} \notag \\
    \simeq&\, \delta_{i,j} +  (C^{i}_{0})^{\ast} C^{j}_{0} \pi (r^2)^{m+1} B_{0}\,, 
    \label{eq:normalbulk}
\end{align}
with 
\begin{align}
    B_{0} \simeq& \left( \frac{N-1}{2N}(M'-\ell_0) \right)^{-1} \frac{1-\sum_{p=0}^{\ell_0} \frac{\Gamma(M'+1)}{\Gamma(M'-p+1) \Gamma(p+1)}  \left( \frac{N+1}{2N} \right)^{M'-p} \left( \frac{N-1}{2N} \right)^{p}}{\frac{\Gamma(M'+1)}{\Gamma(M'-\ell_0+1) \Gamma(\ell_0+1)}  \left( \frac{N+1}{2N} \right)^{M'-\ell_0} \left( \frac{N-1}{2N} \right)^{\ell_0}} + \left( - \frac{m+1}{N} \right)^{-1}. \notag
\end{align}
We show the detailed calculation in Appendix~\ref{ap:normalbulk}.
Then, we next perform the unitary transformation for flavor index $j$,
\begin{align}
    \begin{array}{l}
        \psi^{j'}_{{\rm blow-up},0} = U_{j'j} \psi^{j}_{{\rm blow-up},0}, \\
        U = \prod_{J} \left( U^{J(J+1)} \right) {\rm diag}(e^{-i{\rm arg}(C^{j}_{0})}), \\
        U^{J(J+1)} =
        \begin{pmatrix}
        1 & \ & \ & \ & \ \\
        \ & \ddots & \ & \ & \ \\
        \ & \ &\begin{array}{cc}
        \cos \theta_{J(J+1)} & - \sin \theta_{J(J+1)} \\
        \sin \theta_{J(J+1)} & \cos \theta_{J(J+1)}
        \end{array}& \ & \ \\
        \ & \ & \ & \ddots & \ \\
        \ & \ & \ & \ & \ & 1
        \end{pmatrix}, \quad \tan^2 \theta_{J(J+1)} =
        \frac{\sum_{I=1}^{J} |C^{I}_{0}|^2}{|C^{J+1}_{0}|^2}.
    \end{array} \label{eq:unitary}
\end{align}
Then, the inner product $(G)_{i'j'}$ can be rewritten as
\begin{align}
    \begin{array}{l}
        G\simeq 
        \begin{pmatrix}
        1 & \ & \ \\
        \ & \ddots & \ \\
        \ & \ & 1 + \sum_{j} |C^{j}_{0}|^2 \pi (r^2)^{m+1} B_0
        \end{pmatrix}
        . \label{eq:reG}
    \end{array}
\end{align}
Thus, by redefining the normalization factor for the last mode $j'=j'_{\text{max}}$ as ${\cal N'}_{T^2/\mathbb{Z}_N}^{j'_{\text{max}}} = {\cal N}_{T^2/\mathbb{Z}_N}^{j'_{\text{max}}} (1 + O((r^2)^{m+1}))^{-1/2}$, all of the above modes can be orthonormal basis.

So far, we have obtained zero mode wave functions on the magnetized blow-up manifold by blowing up the singularity $z^{\rm fp}_0=0$ in Eq.~(\ref{eq:wavbulk}).
(Hereafter, we call them bulk zero mode wave functions.)
We can also apply for blowing up another singularity $z^{\rm fp}_I$ by replacing $z$, $m$, and $(\alpha_1,\alpha_{\tau})$ with $Z$, $\chi_{(m)I}$, and $(\beta_1,\beta_{\tau})$, respectively.
In particular, Eqs.~(\ref{eq:fluxcondspinor}) and (\ref{eq:fluxcondscalarvector}) can be rewritten as
\begin{align}
    \frac{\pi r_I^2}{N {\rm Im}\tau} M + \frac{N-1}{2N} - \frac{\chi_{(m)I}}{N} + \ell_I = \frac{N-1}{2N}M',
    \label{eq:fluxcondspinorI}
\end{align}
for spinor fields and
\begin{align}
    \frac{\pi r_I^2}{N {\rm Im}\tau} M - \frac{\chi_{(m)I}}{N} + \ell_0 = \frac{N-1}{2N}M',
    \label{eq:fluxcondscalarvectorI}
\end{align}
for scalar and vector fields.
We discuss their physical meaning in the next subsection.


\subsection{Physical meaning of the result from junction condition}
\label{subsec:phymeanjunccond}

In this subsection, we discuss the physical meaning of the results from the junction condition in Eq.~(\ref{eq:fluxcondspinorI}) as well as Eq.~(\ref{eq:fluxcondscalarvectorI}).
By using the result in Eq.~(\ref{eq:totalflux0}), they can be rewritten as
\begin{align}
\frac{\pi r_I^2}{N{\rm Im}\tau} M + \frac{\xi^{F_{\rm total}}_I}{N} = \frac{N-1}{2N} M'. \label{eq:BUfluxcond}
\end{align}
Now, we can find the physical meaning: the left-hand side shows the magnetic flux on the cut-out cones from $T^2/\mathbb{Z}_N$ including localized flux\footnote{Since wave functions on the blow-up region are different for spinor case and scalar and vector cases because of the curvature contribution, the results in Eqs.~(\ref{eq:fluxcondspinorI}) and (\ref{eq:fluxcondscalarvectorI}) seem to be inconsistent. However, due to the additional $U(1)_s$ localized flux which only spinors feel, the contribution of it and the curvature contribution for spinors are cancelled each other, and then it becomes consistent with scalar and vector cases.}, while the right-hand shows the magnetic flux on the embedded part of $S^2$.
In other words, not only the curvature but also the magnetic flux (which are topological invariant numbers) are not changed under the blow-up procedure.
In particular, in the orbifold limit $r_I \rightarrow 0$, the magnetic flux on the blow-up region corresponds to the localized flux at $z^{\rm fp}_I$ indeed:
\begin{align}
\frac{\xi^{F_{\rm total}}_I}{N} = \frac{N-1}{2N} M' \biggl|_{r_I \rightarrow 0}. \label{eq:BUlocalflux}
\end{align}

Moreover, since the blow-up manifold is a smooth manifold, we can apply the AS index theorem for the magnetized blow-up models.
We notice that the AS index theorem on a general 2D compact smooth manifold ${\cal M}_2$~\cite{Witten:1984dg,Green:1987mn} has only flux contribution:
\begin{align}
n_{+} - n_{-} = \frac{1}{2\pi} \int_{{\cal M}_2} F.
\end{align}
Then, the AS index theorem on the blow-up manifold can be expressed as
\begin{align}
n_{+}^{ab}-n_{-}^{ab}&= \frac{1}{2\pi}\int_{{\rm{blow-up\, manifold}}} F_{ab} \label{eq:Ind01} \\
&= \frac{1}{2\pi}\int_{{{T^2/{\mathbb{Z}}_N}\,\rm{bulk}}} F_{ab}+
\sum_{I} \frac{1}{2\pi}\int_{\frac{N-1}{2N}\times S^2} F'_{ab} \label{eq:Ind02} \\
&=\left( \frac{M}{N} -\sum_{I} \frac{\pi r_I^2}{N{\rm{Im}}\tau}M \right) +\sum_{I}\frac{N-1}{2N} M^{\prime}_I(r_I) \label{eq:Ind03} \\
&=\left( \frac{M}{N} -\sum_{I} \frac{\pi r_I^2}{N{\rm{Im}}\tau}M \right) + \sum_{I} \left( \frac{\pi r^2_I}{N{\rm{Im}}\tau}M + \frac{\xi^{F_{\rm total}}_I}{N} \right) \label{eq:Ind04} \\
&= \frac{M}{N} + \sum_{I} \frac{\xi^{F_{\rm total}}_I}{N}. \label{eq:Ind05}
\end{align}
We emphasize that the result does not depend on the blow-up radius $r_I$.
In other words, the result still holds in the orbifold limit $r_I \rightarrow 0$:
\begin{align}
n_{+}^{ab} - n_{-}^{ab} 
&= \frac{1}{2\pi}\int_{T^2/\mathbb{Z}_N} \widetilde{F}_{ab} \label{eq:Ind01limit} \\
&= \frac{1}{2\pi}\int_{T^2/\mathbb{Z}_N} F_{ab} + 
\sum_{I} \frac{1}{2\pi}\int_{T^2/\mathbb{Z}_N} \delta F_{Iab} \label{eq:Ind02limit} \\
&= \frac{M}{N} + \sum_{I} \frac{\xi^{F_{\rm total}}_I}{N}, \label{eq:Ind05limit}
\end{align}
where Eqs.~(\ref{eq:Ind01limit}), (\ref{eq:Ind02limit}), and (\ref{eq:Ind05limit}) correspond to the limit of Eqs.~(\ref{eq:Ind01}), (\ref{eq:Ind02}), and (\ref{eq:Ind04}), nothing but (\ref{eq:Ind05}), respectively.\footnote{We note that in the orbifold limit $r_I \rightarrow 0$ ($R \rightarrow 0$), the second term of Eq.~(\ref{eq:Ind02limit}) with the field strength in Eq.(\ref{eq:FS2VEV}) can be written by
\begin{align}
\int_{\frac{N-1}{2N}\times S^2} i M^{\prime}_I \delta(z^{\prime})\delta(\bar{z}^{\prime}) dz^{\prime} \wedge d{\bar{z}}^{\prime}. \notag
\end{align}}
Therefore, this can be regarded as the AS theorem on the $T^2/\mathbb{Z}_N$ orbifold and it is important that the index can be given by only the amount of magnetic fluxes including localized fluxes which $\psi^{ab}_{\pm}$ feels.
Here, the localized fluxes are determined by Eq.~(\ref{eq:totalfluxI}), which are related to the localized curvature and the winding numbers.
Then, this AS theorem can be rewritten as
\begin{align}
n_{+}^{ab} - n_{-}^{ab} = \frac{M}{N} + \sum_{I} \left( - \frac{\chi_{(m)I}}{N} + \frac{1}{2}\frac{\xi^R_I}{N} + \ell_I \right) = \frac{M}{N} - \frac{V_{(m)}}{N} + 1 + \sum_{I} \ell_I.
\label{eq:recounting}
\end{align}
Thus, when we take $\ell_I=0$, it corresponds to Eq.~(\ref{eq:counting}), which shows that the above result of AS theorem is correct.
However, the degree of freedom of localized fluxes $\ell_I$, which comes from mod $N$ property in Eq.~(\ref{eq:localcurvflux}), shows that there also exists $\sum_{I} \ell_I$ number of chiral zero modes on the blow-up manifold as well as the $T^2/\mathbb{Z}_N$ orbifold.
We will discuss the detailed physical meaning in the next subsection.


\subsection{Localized zero-mode wave functions}
\label{subsec:localmode}

Now, let us see the new chiral zero mode wave functions due to the degree of freedom of localized flux $\ell_I$.
In particular, we discuss $\ell_0 \neq 0$ case.

First, we recall that
the bulk zero mode wave functions on the bulk region near the fixed point $z^{{\rm fp}}_0=0$ and the blow-up region are proportional to $z^{\ell_0N}$ and $z'^{\ell_0}$, respectively. 
Then, it indicates that the new zero mode wave functions on the bulk region near $z^{{\rm fp}}_0=0$ and the blow-up region will be proportional to $z^{a_0 N}$ and $z'^{a_0}$ for $a_0=0,...,\ell_0-1$, respectively.
We also remind that the factor $z^{\ell_0N}$ comes from the fact that the holomorphic function of the following wave function,
\begin{align}
    \psi^N_{T^2/\mathbb{Z}_N^1}(z) \equiv \left( \psi^{(\frac{1}{2}, \frac{1}{2}),1}_{T^2/\mathbb{Z}_N^1}(z) \right)^{N} = \left( \psi^{(\frac{1}{2}, \frac{1}{2}),1}_{T^2}(z) \right)^{N},
    \label{eq:psiN1}
\end{align}
is proportional to $z^N$ near $z^{{\rm fp}}_0=0$ though it is $\mathbb{Z}_N$ invariant.
Here, BCs of Eq.~(\ref{eq:psiN1}) are the same as ones of wave functions with $M=N$, $(\alpha_1, \alpha_{\tau}) \equiv (\frac{N}{2}-[\frac{N}{2}], \frac{N}{2}-[\frac{N}{2}])$, and $m=0$,~i.e., $\psi^{(j+\frac{N}{2}-[\frac{N}{2}], \frac{N}{2}-[\frac{N}{2}]),N}_{T^2/\mathbb{Z}_N^0}(z)$, where $[x]$ denotes the floor function.
It means that wave functions in Eq.~(\ref{eq:psiN1}) can be expanded by  $\psi^{(j+\frac{N}{2}-[\frac{N}{2}], \frac{N}{2}-[\frac{N}{2}]),N}_{T^2/\mathbb{Z}_N^0}(z)$.
Thus, if we construct other wave function $\psi^N_{T^2/\mathbb{Z}_N^0}(z)$, which has the same BCs of $\psi^{(j+\frac{N}{2}-[\frac{N}{2}], \frac{N}{2}-[\frac{N}{2}]),N}_{T^2/\mathbb{Z}_N^0}(z)$, from a $\mathbb{Z}_N$ invariant ($m=0$) mode,
we can obtain new zero mode wave function whose holomorphic function is proportional to $z^{a_0 N}$ near $z^{{\rm fp}}_0=0$ by replacing $(\psi^N_{T^2/\mathbb{Z}_N^1}(z))^{\ell_0-a_0}$ with $(\psi^N_{T^2/\mathbb{Z}_N^0}(z))^{\ell_0-a_0}$.
Actually, we can know the zero mode number of $\psi^{(j+\frac{N}{2}-[\frac{N}{2}], \frac{N}{2}-[\frac{N}{2}]),N}_{T^2/\mathbb{Z}_N^0}(z)$ is just two, which means that there exists the other zero mode different from Eq.~(\ref{eq:psiN1}).
The result of $\psi^N_{T^2/\mathbb{Z}_N^0}(z)$ is the following:
\begin{align}
    \psi^N_{T^2/\mathbb{Z}_N^0}(z)
    &\equiv e^{-\frac{\pi N}{2{\rm Im}\tau}|z|^2} h_0^N(z) \notag \\
    &\equiv \left\{
    \begin{array}{ll}
        \left( \psi^{(0,0),1}_{T^2/\mathbb{Z}_N^0}(z) \right)^{N} = \left( \psi^{(0,0),1}_{T^2}(z) \right)^{N} & (N=2,4) \\
        \left( \psi^{(\frac{1}{6},\frac{1}{6}),1}_{T^2/\mathbb{Z}_N^0}(z) \right)^{N} = \left( \psi^{(\frac{1}{6},\frac{1}{6}),1}_{T^2}(z) \right)^{N} & (N=3) \\
        \left( \psi^{(0,0),2}_{T^2/\mathbb{Z}_N^0}(0) \psi^{(0,0),2}_{T^2/\mathbb{Z}_N^0}(z) \right)^{N/2} & (N=6) 
    \end{array}
    \right.,
\end{align}
with
\begin{align}
    \psi^{(0,0),2}_{T^2/\mathbb{Z}_N^0}(z) = \sqrt{\frac{\sqrt{3}+1}{2\sqrt{3}}} e^{-\pi i/8} \psi^{(0,0),2}_{T^2}(z) + \sqrt{\frac{\sqrt{3}-1}{2\sqrt{3}}} e^{\pi i/8} \psi^{(1,0),2}_{T^2}(z). \notag
\end{align}
Therefore, the $\ell_0$ number of new chiral zero mode wave functions can be expressed as
\begin{align}
    \begin{array}{rl}
        \tilde{\psi}^{a_0}_{T^2/\mathbb{Z}_N,+,0} &\equiv {\cal N}^{a_0}_{T^2/\mathbb{Z}_N} \left( \frac{\psi^N_{T^2/\mathbb{Z}_N^0}(z)}{\psi^N_{T^2/\mathbb{Z}_N^1}(z)} \right)^{\ell_0-a_0} \tilde{\psi}_{T^2/\mathbb{Z}_N^m,+,0}^{(\alpha_1,\alpha_{\tau})}(z)  \\
         &\simeq C^{a_0}_{0} N |z|^{m-\ell_0N} e^{-\frac{\pi M}{2{\rm Im}\tau}|z|^2} z^{a_0 N}, 
    \end{array}
    \label{eq:wavloc}
\end{align}
where the coefficient $C^{a_0}$ is given by
\begin{align}
    C^{a_0}_{0} \equiv {\cal N}^{a_0}_{T^2/\mathbb{Z}_N} \left( \frac{h_{0}^N(0)}{((h^{(\frac{1}{2},\frac{1}{2})}_{T^2})^{(1)}(0))^N} \right)^{\ell_0-a_0} \sum_{j} C^{j}_{0}, \quad a_0 \in \mathbb{Z}/\ell_{0} \mathbb{Z}.
\end{align}
Note that the non-holomorphic part of Eq.~(\ref{eq:wavloc}) does not change from that of Eq.~(\ref{eq:T2ZNzerotilde}), which means that the new chiral zero mode wave functions (\ref{eq:wavloc}) satisfy not only the same BCs but also the same equation of motion with the bulk zero mode wave functions.
These new zero-modes diverge at the singular point $z^{{\rm fp}}_0=0$, while they are suppressed as they go away from the singular point, as shown in Figure~\ref{fig:localizedmode}.
\begin{figure}[H]
    \centering
    \includegraphics[bb=0 0 400 320,width=6.5cm]{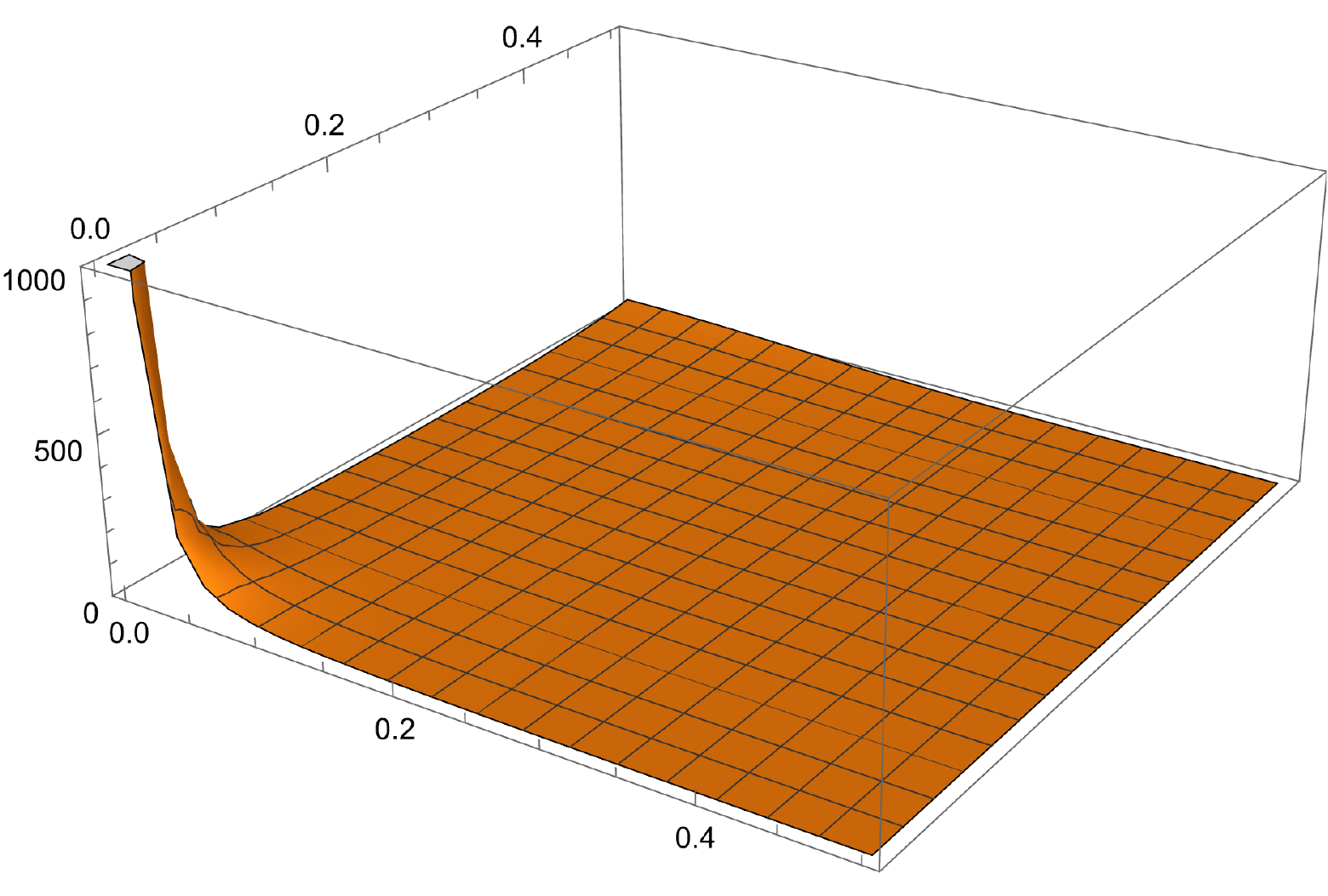}
    \caption{Probability density of unnormalized zero mode wave function $|\tilde{\psi}^{a_0=\ell_0-1}_{T^2/\mathbb{Z}_4,+,0}|^2$.}
    \label{fig:localizedmode}
\end{figure}
That is, the $\ell_0$ number of new zero modes correspond to localized modes around the singular point $z^{{\rm fp}}_0=0$.
Similarly, there are $\ell_I$ number of zero modes around the singular point $z^{{\rm fp}}_I$ as the localized modes.
Therefore, there also exists $\sum_{I} \ell_I$ number of localized zero modes.

Although these localized modes in Eq.~(\ref{eq:wavloc}) diverge at $z^{{\rm fp}}_0=0$, they can be regularized by replacing the cone around $z^{{\rm fp}}_0=0$ with the part of $S^2$.
Then, to calculate their normalization, we consider their wave functions on the magnetized blow-up manifold.
Similarly, by applying the junction condition in Eq.~(\ref{eq:concon}),
the wave functions, corresponding to localized modes, on the magnetized blow-up manifold can be obtained as
\begin{align}
    \psi^{a_0}_{{\rm blow-up},0} = \left\{
    \begin{array}{ll}
      \left( \frac{R^2}{R^2+|z'|^2} \right)^{\frac{M'-1}{2}} {C'}^{a_0}_{0} R^{-a_0} z'^{a_0} & (|z'| \leq \frac{r}{N+1}) \\
         e^{-\frac{\pi M}{2{\rm Im}\tau}|z|^2} \left| h^{(\frac{1}{2},\frac{1}{2}),1}_{T^2}(z) \right|^{m-\ell_0N}  {\cal N}^{a_0}_{T^2/\mathbb{Z}_N} \left( \frac{h_{0}^N(z)}{(h^{(\frac{1}{2},\frac{1}{2})}_{T^2}(z))^N} \right)^{\ell_0-a_0} \sum_{j}\tilde{h}^j(z) & (r \leq |z|) \\
        \ \simeq e^{-\frac{\pi M}{2{\rm Im}\tau}|z|^2}  |z|^{m-\ell_0N} C^{a_0}_{0} N z^{a_0 N} 
    \end{array}
    \right., \label{eq:wavlocalblowup}
\end{align}
where the coefficient ${C'}^{a_0}$ is given by
\begin{align}
    {C'}^{a_0}_{0} = C^{a_0}_{0} N r^{m-(\ell_0-a_0)N} e^{-\frac{\pi M}{2{\rm Im}\tau}r^2} \left( \frac{N+1}{N-1} \right)^{\frac{a_0}{2}} \left( \frac{N+1}{2N} \right)^{-\frac{M'-1}{2}}.
\end{align}
Furthermore, since these wave functions are suppressed as they go away from the orbifold singular point and then the contributions of wave functions on the bulk region are tiny, we assume that the approximation of wave functions on the bulk region near the singular point $z^{{\rm fp}}_0=0$ is valid on bulk region anywhere and also we expand the integral region to $|z| \rightarrow \infty$.
Under the assumptions, it turns out that the $\ell_0$ number of new zero modes are orthogonal to each other and also orthogonal to all of the bulk zero modes by using the following results:
\begin{align}
    \int_{0}^{\frac{2\pi}{N}} d{\rm arg}(z)\ z^{k N} = 0, \quad \int_{0}^{2\pi} d{\rm arg}(z')\ {z'}^{k} = 0, \quad (k \neq 0).
    \label{eq:orthogonality}
\end{align}
Thus, the normalization of localized modes can be determined in the following way:
\begin{align}
    1
    =& \int_{{\rm blow-up\ manifold}} dz d\bar{z} \sqrt{|{\rm det}(g)|} |\psi^{a_0}_{{\rm blow-up},0}|^2 \notag \\
    \simeq& \int_{r}^{\infty} d|z| |z| \int_{0}^{\frac{2\pi}{N}} d\varphi |C^{a_0}_{0}|^2 N^2 |z|^{2(m-(\ell_0-a_0)N)} e^{-\frac{\pi M}{{\rm Im}\tau}|z|^2} \notag \\
    &+ \int_{0}^{\frac{r}{N+1}} d|z'| |z'| \int_{0}^{2\pi} d\varphi \left( \frac{2R^2}{R^2+|z'|^2} \right)^2
 \left( \frac{R^2}{R^2+|z'|^2} \right)^{M'-1} |{C'}^{a_0}_{0}|^2 R^{-2a_0} |z'|^{2a_0} \notag \\
    \simeq& \left| C^{a_0}_{0} \right|^2 \pi \left(\frac{1}{r^2} \right)^{(\ell_0-a_0)N-(m+1)} \left[ N \frac{\left( -\frac{\pi M}{{\rm Im}\tau} r^2 \right)^{(\ell_0-a_0)N-(m+1)}}{[(\ell_0-a_0)N-(m+1)]!} E_1\left( \frac{\pi M}{{\rm Im}\tau}r^2 \right) +  L_0 \right],
    \label{eq:normallocal}
\end{align}
with 
\begin{align}
        L_0 \simeq& \left( \frac{N-1}{2N}(M'-a_0) \right)^{-1} \frac{1 - \sum_{p=0}^{a_0} \frac{M'!}{(M'-p)! p!} \left( \frac{N+1}{2N} \right)^{M'-p} \left( \frac{N-1}{2N} \right)^{p}}{\frac{M'!}{(M'-a_0)! a_0!} \left( \frac{N+1}{2N} \right)^{M'-a_0} \left( \frac{N-1}{2N} \right)^{a_0}} + \left( (\ell_0 - a_0) - \frac{m+1}{N} \right)^{-1}, \notag
\end{align}
where $E_1$ denotes the exponential integral.
The detailed calculation of Eq.~(\ref{eq:normallocal}) is shown in Appendix~\ref{ap:normallocal}.
Therefore, we obtained normalizable zero mode wave functions in Eq.~(\ref{eq:wavlocalblowup}) and they correspond to localized modes under the orbifold limit $r \rightarrow 0$.
Similarly, we can apply the above analysis for localized modes around other orbifold singular points by replacing $z$, $m$, and $(\alpha_1,\alpha_{\tau})$ with $Z$, $\chi_{(m)I}$, and $(\beta_1,\beta_{\tau})$, respectively.

Finally, we comment for three-point couplings of bulk zero modes as well as localized modes on the magnetized blow-up manifold.
When we denote bulk zero-modes and localized zero-modes shortly as B and L, respectively,
only three patterns of three-point couplings: (i) B$_1$-B$_2$-B$_3$ coupling, (ii) L$_1$-L$_2$-L$_3$ coupling, and (iii) B$_1$-L$_2$-L$_3$ coupling, can become non-vanishing three-point couplings\footnote{This selection rule may come from conservation of angular momentum.}.
These three-point couplings can be similarly calculated by applying the calculation of normalization.


\chapter{Summary and discussions}
\label{chap:sumdis}

We summarize this paper.
We have considered ten-dimensional ${\cal N}=1$ $U(N)$ supersymmetric Yang-Mills theory on ${\cal M}_4 \times X_2 \times X_2 \times X_3$ with background magnetic fluxes on each two-dimensional compact space $X_i\ (i=1,2,3)$ as the effective field theory of magnetized D-brane models in type-IIB superstring theory.
As the two-dimensional compact space, we have considered two-dimensional torus $T^2$ in chapter~\ref{chap:Mtorusmodel}, its $\mathbb{Z}_N$ twisted orbifold $T^2/\mathbb{Z}_N$ in chapter~\ref{chap:Mtorusorbifold}, and furthermore blow-up manifolds of the $T^2/\mathbb{Z}_N$ orbifold which are constructed by replacing the orbifold singularities with parts of two-dimensional sphere $S^2$ in chapter~\ref{chap:blowup}.

In chapters~\ref{chap:Mtorusmodel} and \ref{chap:Mtorusorbifold}, we have discussed the modular symmetry in magnetized torus and torus orbifold models. (The detailed calculations related to the modular symmetry are in Appendix~\ref{app:MTT2T2Z2}.)
In particular, we have found the followings.
\begin{itemize}
\item When the magnetic flux $M$ on the magnetized $T^2$ is even (odd) and the Scherk-Schwarz phases $(\alpha_1,\alpha_{\tau}) = (0,0)$ ($(\alpha_1,\alpha_{\tau})=(1/2,1/2)$), $M$ number of bi-fundamental chiral zero modes behave as ``modular forms'' of weight $1/2$ for $\widetilde{\Gamma}(2M)$ ($\widetilde{\Gamma}(8M)$) and then they transform under $\widetilde{\Gamma}_{2M} \equiv \widetilde{\Gamma}/\widetilde{\Gamma}(2M)$ ($\widetilde{\Gamma}_{8M} \equiv \widetilde{\Gamma}/\widetilde{\Gamma}(8M)$) non-trivially.
\item In addition, $T^2/\mathbb{Z}_2$ orbifold also has the same modular symmetry as $T^2$ and both $\mathbb{Z}_2$-even ($m=0$) and odd modes ($m=1$) are closed independently under the ``modular transformation''.
In other words, the $M$ number of zero modes on the magnetized $T^2$ can be decomposed into smaller representations on the magnetized $T^2/\mathbb{Z}_2$ orbifold.
\item In particular, three-generational zero modes with $(M;\alpha_1,\alpha_{\tau};m) = (4;0,0;0), (8;0,0;1)$ transform as three-dimensional representations of modular $\widetilde{\Delta}(6M^2)$ group (which is the quadruple covering group of $\Delta(6M^2)$), while three-generational zero modes with $(M;\alpha_1,\alpha_{\tau})$ $= (5;1/2,1/2;1), (7;1/2,1/2;0)$ transform as three-dimensional representations of modular $PSL(2,\mathbb{Z}_M) \times \mathbb{Z}_8$ group.
\item Along with the ``modular transformation'' for wave functions on the magnetized $T^2$ as well as $T^2/\mathbb{Z}_2$ orbifold, the four-dimensional fields also transform non-trivially under the metapletic finite modular group $\widetilde{\Gamma}_{2M}$ ($\widetilde{\Gamma}_{8M}$) with modular weight $-k=-1/2$. Then, we can find that the kinetic terms are modular invariant.
\item Furthermore, since coupling coefficients in the four-dimensional effective field theory such as Yukawa couplings can be obtained from overlap integration of wave functions on the magnetized $T^2$ as well as $T^2/\mathbb{Z}_2$ orbifold, we can also find the modular transformation for them. Note that the obtained couplings are couplings in a global supersymmetric theory and they are related to the (holomorphic) couplings in supergravity theory through K\"{a}hler potential. In particular, holomorphic three-point couplings become modular forms of weight $1/2$ for $\widetilde{\Gamma}(2{\rm lcm}(M^{ab},M^{bc},M^{ca}))$ and then they transform under $\widetilde{\Gamma}_{2{\rm lcm}(M^{ab},M^{bc},M^{ca})}$ non-trivially, where $M^{AB}\ (A,B=a,b,c)$ denote the magnetic fluxes which coupling fields feel. By combining the modular transformation for the four-dimensional fields, we have found that the holomorphic superpotential of the three-point couplings has modular weight $-1$, which is consistent within the supergravity theory. Through the K\"{a}hler potential of the modulus on $T^2$ as well as $T^2/\mathbb{Z}_2$, we can obtain the couplings in the global symmetric four-dimensional effective theory and then we can find that their coupling terms are also modular invariant.
\item However, when we consider non-perturbative effects such as Majorana neutrino mass terms derived from D-brane instanton effects, some parts of the modular symmetry which a classical action has can be broken. For example, the Majorana neutrino mass terms are not modular invariant by the modular transformation for measures of D-brane instanton zero modes which are integrated out. (We discuss, in Appendix~\ref{ap:Anomaly}, which parts of discrete symmetries can be anomalous.)
\end{itemize}

In addition, we have discussed the magnetized blow-up manifold compactification in chapter~\ref{chap:blowup}.
In particular, we have obtained wave functions on the blow-up manifold by replacing around orbifold singularities like cones with parts of $S^2$ not to modify the topological invariant numbers: curvature and magnetic fluxes.
In addition, since the blow-up manifold is a smooth two-dimensional manifold, we can apply the Atiyah-Singer index theorem on the blow-up manifold, and then we have found that the chiral zero mode numbers on not only the blow-up manifold but also the orbifold, by taking orbifold limits, are determined only by the magnetic fluxes including localized fluxed at the orbifold singularities.
Moreover, the Atiyah-Singer index theorem on the $T^2/\mathbb{Z}_N$ orbifold also shows that the additional degree of freedom of localized fluxes gives additional new chiral zero modes.
We have found that the additional new zero modes correspond to localized modes around orbifold singularities.
(The detailed calculations of normalizations of their zero mode wave functions on the magnetized blow-up manifold are in Appendix~\ref{ap:NWMBU}.)

Since coupling coefficients such as Yukawa couplings are obtained from overlap integration of wave functions, it is important to clear the wave functions on the compact space and their properties such as symmetries to understand the structure of couplings such as flavor structure.
Then, I would like to study followings for the future.
\begin{itemize}
\item We have found the modular groups and the representations as well as modular weight of four-dimensional fields obtained from the magnetized $T^2$ as well as $T^2/\mathbb{Z}_2$ orbifold compactifications. They are determined by the magnetic flux and the fact that the modular weight is $1/2$. However, it has not been clear that the reason why the modular weight is $1/2$. I would like to reveal it elsewhere.
\item I would also like to explore the detailed properties of wave functions on magnetized blow-up manifolds such as the modular symmetry.
\item It is important to extend those analyses to more higher-dimensional toroidal orbifolds such as $T^4/\mathbb{Z}_N$ and $T^6/\mathbb{Z}_N$, and also their blow-up manifolds. Then, we may analytically find phenomena on the Calabi-Yau manifold indirectly for the future.
\item Moreover, to determine the flavor structure, we should fix the value of the modulus, called the modulus stabilization. Actually, the modulus stabilization was studied in supergravity theory (with some assumptions) in Refs.~\cite{Ferrara:1990ei,Cvetic:1991qm}. Then, it is important to consider not only ten-dimensional supersymmetric Yang-Mills theory but also ten-dimensional supergravity theory as the effective field theory of the superstring theory. I would also like to reveal properties such as the modular symmetry of the four-dimensional effective field theory of not only ten-dimensional supersymmetric Yang-Mills theory but also the supergravity theory in concrete compactifications such as toroidal orbifolds as well as their blow-up manifolds with background magnetic fluxes and three-form fluxes. Furthermore, by applying those results, I would like to determine the structure of coupling coefficients such as Yukawa couplings without any assumptions for the future.
\end{itemize}


\appendix


\chapter{Modular symmetry on magnetized $T^2$ and $T^2/\mathbb{Z}_2$ orbifold}
\label{app:MTT2T2Z2}


\section{Modular transformation for wave functions on magnetized $T^2$}
\label{app:MTW}

Here, we show the detailed calculations of ``modular transformation'' for wave functions on the magnetized $T^2$.

First, let us show the explicit calculation of Eq.~(\ref{eq:waveT2}) in the following:
\begin{align}
&\psi^{(j+\alpha_{1},\{\alpha_{\tau}+\alpha_{1}+M/2\}),M}_{T^2,0}(z,\tau+1) \notag \\
=& e^{-\frac{\pi M}{2{\rm Im}(\tau+1)}|z|^2} (M)^{1/4} e^{2\pi i\frac{j+\alpha_1}{M} \{\alpha_{\tau}+\alpha_1+M/2\}} e^{\frac{\pi M}{2{\rm Im}(\tau+1)}z^2} \vartheta
\begin{bmatrix}
\frac{j+\alpha_1}{M} \\ -\{\alpha_{\tau}+\alpha_1+M/2\}
\end{bmatrix}
(Mz, M(\tau+1)) \notag \\
=& e^{-\frac{\pi M}{2{\rm Im}\tau}|z|^2} (M)^{1/4} e^{2\pi i\frac{j+\alpha_1}{M} \{\alpha_{\tau}+\alpha_1+M/2\}} e^{\frac{\pi M}{2{\rm Im}\tau}z^2} \sum_{l \in \mathbb{Z}} e^{\pi iM(\tau+1) \left( \frac{j+\alpha_1}{M} + l \right)^2} e^{2\pi i(Mz - \{\alpha_{\tau}+\alpha_1+M/2\}) \left( \frac{j+\alpha_1}{M} + l \right)} \notag \\
=& e^{\pi iM \left( \frac{j+\alpha_1}{M} \right)^2} e^{-\frac{\pi M}{2{\rm Im}\tau}|z|^2} (M)^{1/4} e^{2\pi i\frac{j+\alpha_1}{M} \alpha_{\tau}} e^{\frac{\pi M}{2{\rm Im}\tau}z^2} \notag \\
&\times \sum_{l \in \mathbb{Z}} e^{\pi iM\tau \left( \frac{j+\alpha_1}{M} + l \right)^2} e^{2\pi i(Mz - \alpha_{\tau}) \left( \frac{j+\alpha_1}{M} + l \right)} e^{2\pi i \left((j+\alpha_1)+M/2+\alpha_{\tau}-(\alpha_{\tau}+\alpha_1+M/2) + [\alpha_{\tau}+\alpha_1+M/2] \right) l} \notag \\
=& e^{\pi iM \left( \frac{j+\alpha_1}{M} \right)^2} e^{-\frac{\pi M}{2{\rm Im}\tau}|z|^2} (M)^{1/4} e^{2\pi i\frac{j+\alpha_1}{M} \alpha_{\tau}} e^{\frac{\pi M}{2{\rm Im}\tau}z^2} \sum_{l \in \mathbb{Z}} e^{\pi iM\tau \left( \frac{j+\alpha_1}{M} + l \right)^2} e^{2\pi i(Mz - \alpha_{\tau}) \left( \frac{j+\alpha_1}{M} + l \right)} \notag \\
=& e^{\pi iM \left( \frac{j+\alpha_1}{M} \right)^2} \left( e^{-\frac{\pi M}{2{\rm Im}\tau}|z|^2} (M)^{1/4} e^{2\pi i\frac{j+\alpha_1}{M} \alpha_{\tau}} e^{\frac{\pi M}{2{\rm Im}\tau}z^2} \vartheta
\begin{bmatrix}
\frac{j+\alpha_1}{M} \\ -\alpha_{\tau}
\end{bmatrix}
(Mz,M\tau) \right) \notag \\
=& \sum_{j'=0}^{M-1} e^{\pi i\frac{(j+\alpha_1)^2}{M}} \delta_{j,j'} \psi^{(j'+\alpha_1,\alpha_{\tau}),M}_{T^2,0}(z,\tau).\label{eq:MTwaveT2T}
\end{align}

On the other hand, the explicit calculation of Eq.~(\ref{eq:waveS2}) is the following:
\begin{align}
&\psi^{(j+\{1-\alpha_{\tau}\},\alpha_{1}),M}_{T^2,0}\left( -\frac{z}{\tau},-\frac{1}{\tau} \right) \notag \\
=& e^{-\frac{\pi M}{2{\rm Im}\left(-\frac{1}{\tau} \right)} \left|-\frac{z}{\tau}\right|^2} (M)^{1/4} e^{2\pi i\frac{j+\{1-\alpha_{\tau}\}}{M} \alpha_1} e^{\frac{\pi M}{2{\rm Im}\left(-\frac{1}{\tau} \right)} \left(-\frac{z}{\tau} \right)^2} \vartheta
\begin{bmatrix}
\frac{j+\{1-\alpha_{\tau}\}}{M} \\ -\alpha_1
\end{bmatrix}
\left(-M\frac{z}{\tau},-M\frac{1}{\tau} \right) \notag \\
=& e^{-\frac{\pi M}{2{\rm Im}\tau}|z|^2} (M)^{1/4} e^{2\pi i\frac{j+\{1-\alpha_{\tau}\}}{M} \alpha_1} e^{\frac{\pi M}{2{\rm Im}\tau}z^2} e^{-\pi iM\frac{z^2}{\tau}}
\sum_{\l \in \mathbb{Z}} e^{-\pi iM \frac{1}{\tau} \left(\frac{j+\{1-\alpha_{\tau}\}}{M}+l \right)^2} e^{2\pi i\left(-M\frac{z}{\tau} -\alpha_1\right) \left(\frac{j+\{1-\alpha_{\tau}\}}{M}+l \right) } \notag \\
=& e^{-\frac{\pi M}{2{\rm Im}\tau}|z|^2} (M)^{1/4} e^{\frac{\pi M}{2{\rm Im}\tau}z^2} \sum_{l \in \mathbb{Z}} e^{-\pi i\frac{M}{\tau} \left( z+\frac{j+\{1-\alpha_{\tau}\}}{M}+l \right)^2} e^{-2\pi i\alpha_1 l} \notag \\
=& -(-\tau)^{1/2} \sum_{j'=0}^{M-1} \frac{-e^{\pi i/4}}{\sqrt{M}} e^{-\frac{\pi M}{2{\rm Im}\tau}|z|^2} (M)^{1/4} e^{\frac{\pi M}{2{\rm Im}\tau}z^2} \sum_{l' \in \mathbb{Z}} e^{\pi iM\tau \left( \frac{j'+\alpha_1}{M}+l' \right)^2} e^{2\pi iM\left(z + \frac{j+\{1-\alpha_{\tau}\}}{M} \right) \left( \frac{j'+\alpha_1}{M}+l' \right)} \label{eq:Smiddle} \\
=& -(-\tau)^{1/2} \sum_{j'=0}^{M-1} \frac{-e^{\pi i/4}}{\sqrt{M}} e^{2\pi i\frac{(j+\{1-\alpha_{\tau}\})(j'+\alpha_1)}{M}} e^{-\frac{\pi M}{2{\rm Im}\tau}|z|^2} (M)^{1/4} e^{2\pi i\frac{j'+\alpha_1}{M}\alpha_{\tau}} e^{\frac{\pi M}{2{\rm Im}\tau}z^2} \notag \\
&\times \sum_{l' \in \mathbb{Z}} e^{\pi iM\tau \left( \frac{j'+\alpha_1}{M}+l' \right)^2} e^{2\pi i(Mz -\alpha_{\tau}) \left( \frac{j'+\alpha_1}{M}+l' \right)} e^{2\pi i(j+1-\alpha_{\tau} - [1-\alpha_{\tau}] + \alpha_{\tau})l' } \notag \\
=& -(-\tau)^{1/2} \sum_{j'=0}^{M-1} \frac{-e^{\pi i/4}}{\sqrt{M}} e^{2\pi i\frac{(j+\{1-\alpha_{\tau}\})(j'+\alpha_1)}{M}} \notag \\
&\times e^{-\frac{\pi M}{2{\rm Im}\tau}|z|^2} (M)^{1/4} e^{2\pi i\frac{j'+\alpha_1}{M}\alpha_{\tau}} e^{\frac{\pi M}{2{\rm Im}\tau}z^2} \sum_{l' \in \mathbb{Z}} e^{\pi iM\tau \left( \frac{j'+\alpha_1}{M}+l' \right)^2} e^{2\pi i(Mz -\alpha_{\tau}) \left( \frac{j'+\alpha_1}{M}+l' \right)} \notag \\
=& - (-\tau)^{1/2} \sum_{j'=0}^{M-1} \frac{-e^{\pi i/4}}{\sqrt{M}} e^{2\pi i\frac{(j+\{1-\alpha_{\tau}\})(j'+\alpha_1)}{M}} \left( e^{-\frac{\pi M}{2{\rm Im}\tau}|z|^2} (M)^{1/4} e^{2\pi i\frac{j'+\alpha_1}{M}\alpha_{\tau}} e^{\frac{\pi M}{2{\rm Im}\tau}z^2} \vartheta
\begin{bmatrix}
\frac{j'+\alpha_1}{M} \\ -\alpha_{\tau}
\end{bmatrix}
(Mz,M\tau) \right) \notag \\
=& - (-\tau)^{1/2} \sum_{j'=0}^{M-1} \frac{-e^{\pi i/4}}{\sqrt{M}} e^{2\pi i\frac{(j+\{1-\alpha_{\tau}\})(j'+\alpha_1)}{M}} \psi^{(j'+\alpha_1,\alpha_{\tau}),M}_{T^2,0}(z,\tau).
\label{eq:MTwaveT2S}
\end{align}
Here, in Eq.~(\ref{eq:Smiddle}), we used the Poisson resummation:
\begin{align}
\sum_{l \in \mathbf{Z}} f(l) = \sum_{n \in \mathbf{Z}} \hat{f}(n), \label{eq:PS}
\end{align}
where $\hat{f}$ denotes the Fourier mode of a function $f$,
\begin{align}
\hat{f}(k) = \int_{-\infty}^{\infty} dx f(x) e^{-2\pi ikx}. \notag
\end{align}
Specifically, when we set
\begin{align}
f(x) = e^{-\pi i\frac{M}{\tau} \left( z+\frac{j+\{1-\alpha_{\tau}\}}{M}+x \right)^2} e^{-2\pi i\alpha_1 x}, \notag
\end{align}
we obtain
\begin{align}
\hat{f}(k) 
=& \int_{-\infty}^{\infty} dx  e^{-\pi i\frac{M}{\tau} \left( z+\frac{j+\{1-\alpha_{\tau}\}}{M}+x \right)^2} e^{-2\pi i(k+\alpha_1) x} \notag \\
=& e^{\pi i\frac{\tau}{M}(k+\alpha_1)^2} e^{2\pi i \left( z+\frac{j+\{1-\alpha_{\tau}\}}{M} \right) (k+\alpha_1) } \int_{-\infty}^{\infty} dx  e^{-\pi i\frac{M}{\tau}\left( z+\frac{j+\{1-\alpha_{\tau}\}}{M}+x +k+\alpha_1 \right)^2 } \notag \\
=& e^{\pi i\tau \frac{(k+\alpha_1)^2}{M}} e^{2\pi i \left( z+\frac{j+\{1-\alpha_{\tau}\}}{M} \right) (k+\alpha_1) } \int_{-\infty}^{\infty} dy  e^{-\pi i\frac{M}{\tau}y^2 } \notag \\
=& -(-\tau)^{1/2} \frac{-e^{\pi i/4}}{\sqrt{M}} e^{\pi iM\tau \left(\frac{k+\alpha_1}{M} \right)^2} e^{2\pi i \left( z+\frac{j+\{1-\alpha_{\tau}\}}{M} \right) (k+\alpha_1) }, \notag
\end{align}
and then we can obtain
\begin{align}
&\sum_{l \in \mathbf{Z}} e^{-\pi i\frac{M}{\tau} \left( z+\frac{j+\{1-\alpha_{\tau}\}}{M}+l \right)^2} e^{-2\pi i\alpha_1 l} \notag \\
=& -(-\tau)^{1/2} 
\sum_{j'=0}^{M-1} \sum_{l' \in \mathbf{Z}} \frac{-e^{\pi i/4}}{\sqrt{M}} e^{\pi iM\tau \left(\frac{j'+\alpha_1}{M}+l' \right)^2} e^{2\pi i M\left( z+\frac{j+\{1-\alpha_{\tau}\}}{M} \right) \left(\frac{j'+\alpha_1}{M}+l' \right) }, \label{eq:Smiddle2}
\end{align}
where we redefined $n \equiv Ml'+j'\ (j'=0,...,M-1)$.

The detailed calculations of Eqs.~(\ref{eq:JrhoT200S2})-(\ref{eq:rhoT200ZT}) and (\ref{eq:JrhoT211S2})-(\ref{eq:rhoT211ZT}) are as follows:
\begin{align}
\rho_{T^2}^{(\alpha,\alpha)}(\widetilde{S})_{jj''}^2 =& \frac{e^{\pi i/2}}{M} \sum_{j'=0}^{M-1} e^{2\pi i\frac{(j'+\alpha)\left( (j+\alpha) + (j''+\alpha) \right)}{M}} = e^{2\pi i\alpha} e^{\pi i/2} \delta_{M-(j+\alpha),(j''+\alpha)}, \label{eq:rhoT2aaS2} \\
\rho_{T^2}^{(\alpha,\alpha)}(\widetilde{S})_{jj'}^4 =& - \sum_{j''=0}^{M-1} \delta_{M-(j+\alpha),(j''+\alpha)} \delta_{(j''+\alpha),M-(j'+\alpha)} = - \delta_{j,j'}, \label{eq:rhoT2aaS4} \\
\rho_{T^2}^{(\alpha,\alpha)}(\widetilde{S})_{jj'}^8 =& \delta_{j,j'}, \label{eq:rhoT2aaS8} 
\end{align}
\begin{align}
&[\rho_{T^2}^{(\alpha,\alpha)}(\widetilde{S}) \rho_{T^2}^{(\alpha,\alpha)}(\widetilde{T})]^3_{jj'''} \notag \\
=& \frac{-e^{3\pi i/4}}{M\sqrt{M}} \sum_{j',j''}
e^{2\pi i\frac{(j+\alpha)(j'+\alpha)}{M}} e^{\pi i\frac{(j'+\alpha)^2}{M}} e^{2\pi i\frac{(j'+\alpha)(j''+\alpha)}{M}} e^{\pi i\frac{(j''+\alpha)^2}{M}} e^{2\pi i\frac{(j''+\alpha)(j'''+\alpha)}{M}} e^{\pi i\frac{(j'''+\alpha)^2}{M}} \notag \\
=& \frac{-e^{3\pi i/4}}{M\sqrt{M}} \sum_{j',j''=0}^{M-1} e^{\pi i \frac{\left( (j'+\alpha)+(j''+\alpha)+(j'''+\alpha) \right)^2}{M}} e^{2\pi i\frac{(j'+\alpha)(j-j''')}{M}} \notag \\
=& \frac{1}{M} \sum_{j'=0}^{M-1} e^{2\pi i\frac{(j'+\alpha)(j-j''')}{M}} \label{eq:ST3middle} \\
=& \delta_{j,j'''}, \label{eq:rhoT2aaST3} 
\end{align}
\begin{align}
\rho_{T^2}^{(0,0)}(\widetilde{T})_{jj'}^{2M} =& e^{2\pi iM\frac{j^2}{M}} \delta_{j,j'} = \delta_{j,j'}, \quad (\alpha=0), \label{eq:rhoT200T2M} \\
\rho_{T^2}^{(\frac{1}{2},\frac{1}{2})}(\widetilde{T})_{jj'}^{M} =& e^{\pi iM\frac{(j+\frac{1}{2})^2}{M}} \delta_{j,j'} = e^{\pi i/4} \delta_{j,j'}, \quad \rho_{T^2}^{(\frac{1}{2},\frac{1}{2})}(\widetilde{T})_{jj'}^{8M} = \delta_{j,j'}, \quad (\alpha=\frac{1}{2}), \label{eq:rhoT211TM8M} 
\end{align}
\begin{align}
[\rho_{T^2}^{(\alpha,\alpha)}(\widetilde{S})^2 \rho_{T^2}^{(\alpha,\alpha)}(\widetilde{T})]_{jj'} 
=& \sum_{j''=0}^{M-1} e^{\pi i/2} \delta_{M-(j+\alpha),(j''+\alpha)} e^{\pi i\frac{(j''+\alpha)^2}{M}} \delta_{j'',j'} \notag \\
=& \sum_{j''=0}^{M-1} e^{\pi i\frac{(j''+\alpha)^2}{M}} \delta_{j',j''} e^{\pi i/2} \delta_{(j''+\alpha),M-(j+\alpha)} \notag \\
=& [ \rho_{T^2}^{(\alpha,\alpha)}(\widetilde{T})\rho_{T^2}^{(\alpha,\alpha)}(\widetilde{S})^2 ]_{j'j}, \label{eq:rhoT2aaZT}
\end{align}
where $\alpha = \alpha_1 = \alpha_{\tau} = 0, 1/2$ for $M=$even, odd, respectively.
Here, in Eq.~(\ref{eq:ST3middle}), we used the Landsberg-Schaar relation:
\begin{align}
\frac{1}{\sqrt{p}} \sum_{n=0}^{p-1} e^{\frac{\pi in^2q}{p}} = \frac{e^{i\pi /4}}{\sqrt{q}} \sum_{n=0}^{q-1} e^{-\frac{\pi in^2p}{q}}. \label{eq:LSrel}
\end{align}


\section{$\widetilde{\Delta}(6M^2)$ as subgroup of $\widetilde{\Gamma}_{2M}$}
\label{app:proof}

Here, we prove that when the generators of $\widetilde{\Gamma}_{2M}$:
\begin{align}
\widetilde{S}^2\widetilde{T} = \widetilde{T}\widetilde{S}^2, \quad \widetilde{S}^4 = e^{2\pi i (k/2)} \mathbf{1}, \quad \widetilde{S}^8 = (\widetilde{S}\widetilde{T})^3 = \widetilde{T}^{2M} = \mathbf{1}, \label{eq:reGammatilde2Malg}
\end{align}
further satisfy Eq.~(\ref{eq:X}):
\begin{align}
(\widetilde{S}^{-1}\widetilde{T}^{-1}\widetilde{S}\widetilde{T})^3=\mathbf{1}, \label{eq:reX}
\end{align}
the generators in Eq.~(\ref{eq:DeltatildeST}):
\begin{align}
a=\widetilde{S}\widetilde{T}^2\widetilde{S}^5\widetilde{T}^4, \quad a'=\widetilde{S}\widetilde{T}^2\widetilde{S}^{-1}\widetilde{T}^{-2}, \quad b=\widetilde{T}^{\frac{M}{2}+3}\widetilde{S}^{\frac{3}{2}M+3}\widetilde{T}^{M}, \quad c=\widetilde{S}\widetilde{T}^{M-2}\widetilde{S}\widetilde{T}^{\frac{3}{2}M-1}, \label{eq:reDeltatildeST}
\end{align}
in particular for $M \in 4\mathbb{Z}$, become the generators of the subgroup of $\widetilde{\Gamma}_{2M}$, $\widetilde{\Delta}(6M^2) \simeq (\mathbb{Z}_M \times \mathbb{Z}_M) \rtimes \mathbb{Z}_3 \rtimes \mathbb{Z}_8$:
\begin{align}
&a^M=a'^M=b^3=c^8=\mathbf{1}, \label{eq:realgDeltatilde} \\
&aa'=a'a, \ cbc^{-1}=b^{-1}, \ bab^{-1}=a^{-1}a'^{-1}, \ ba'b^{-1}=a, \ cac^{-1}=a'^{-1}, ca'c=a^{-1}, \notag
\end{align}
where $a^{(')}$, $b$, $c$ denote ones of $\mathbb{Z}^{(')}_M$, $\mathbb{Z}_3$, $\mathbb{Z}_8$, respectively.
Note that when $k/2=$ integer [even], that is, $\widetilde{S}$ and $\widetilde{T}$ are generators of $\Gamma'_{2M}$ [$\Gamma_{2M}$], we can similarly prove that the generators in Eq.~(\ref{eq:reDeltatildeST}) satisfy Eq.~(\ref{eq:realgDeltatilde}) replacing $c^8=\mathbf{1}$ with $c^4=\mathbf{1}$ [$c^2=\mathbf{1}$], which means that they become the generators of the subgroup of $\Gamma'_{2M}$ [$\Gamma_{2M}$], $\Delta'(6M^2) \simeq  (\mathbb{Z}_M \times \mathbb{Z}_M) \rtimes \mathbb{Z}_3 \rtimes \mathbb{Z}_4$ [$\Delta(6M^2) \simeq  (\mathbb{Z}_M \times \mathbb{Z}_M) \rtimes \mathbb{Z}_3 \rtimes \mathbb{Z}_2$].

First, by using Eqs.~(\ref{eq:reGammatilde2Malg}), Eq.~(\ref{eq:reX}) can be rewritten\footnote{When $k/2 \in \mathbb{Z}$ and $M=1,2$, we can check that Eq.~(\ref{eq:reX}) is already satisfied.} as
\begin{align}
\mathbf{1}
&= (\widetilde{S}^{-1}\widetilde{T}^{-1}\widetilde{S}\widetilde{T})^3 \notag \\
&=(\widetilde{T}\widetilde{S}\widetilde{T}\widetilde{S}\widetilde{S}\widetilde{T})^3 \notag \\
&=(\widetilde{T}\widetilde{S}^3\widetilde{T}^2)^3 \notag \\
&= (S^3T^3)^3.
\label{eq:S3T3}
\end{align}
By using Eqs.~(\ref{eq:reGammatilde2Malg}) and (\ref{eq:S3T3}), the generator $a'$ can be rewritten as
\begin{align}
a'
&= \widetilde{S}\widetilde{T}^2\widetilde{S}^{-1}\widetilde{T}^{-2} \notag \\
&= \widetilde{S}\widetilde{T}^3\widetilde{T}^{-1} \widetilde{S}^{-1}\widetilde{T}^{-1}\widetilde{T}^{-1} \notag \\
&= \widetilde{S}\widetilde{T}^3\widetilde{S}\widetilde{T}\widetilde{S}\widetilde{T}^{-1} \notag \\
&= \widetilde{S}^{-2} \widetilde{S}^3\widetilde{T}^3\widetilde{S}^3\widetilde{T}^3 \widetilde{T}^{-1} \widetilde{T}^{-1}\widetilde{S}^{-1}\widetilde{T}^{-1} \notag \\
&= \widetilde{T}^{-3}\widetilde{S}^{-3} \widetilde{T}^{-1} \widetilde{S}^{-1} \widetilde{T}\widetilde{S} \notag \\
&= \widetilde{T}^{-3} \widetilde{T}\widetilde{S}\widetilde{T} \widetilde{T}\widetilde{S}^{-1} \notag \\
&= \widetilde{T}^{-2} \widetilde{S}\widetilde{T}^2\widetilde{S}^{-1}.
\label{eq:a're}
\end{align}
Then, we can obtain the relation,
\begin{align}
\widetilde{S}\widetilde{T}^{2p}\widetilde{S}^{-1}\widetilde{T}^{2q} =(\widetilde{S}\widetilde{T}^2\widetilde{S}^{-1})^{p}\widetilde{T}^{2q} = \widetilde{T}^{2q}\widetilde{S}\widetilde{T}^{2p}S^{-1}, \quad p,q \in \mathbb{Z},  \label{eq:T2p2q}
\end{align}
in general.
Similarly, by using this relation, the generator $a$ can be rewritten as
\begin{align}
a
&= \widetilde{S}\widetilde{T}^2\widetilde{S}^5\widetilde{T}^4 \notag \\
&= \widetilde{T}^4\widetilde{S}\widetilde{T}^2\widetilde{S}^5.
\label{eq:are}
\end{align}
Thus, we can prove the following relations,
\begin{align}
&a^M = \widetilde{S}^{-2M}\widetilde{T}^{4M}\widetilde{S}\widetilde{T}^{2M}\widetilde{S}^{-1} = \mathbf{1}, \quad a'^M = \widetilde{T}^{-2M}\widetilde{S}\widetilde{T}^{2M}\widetilde{S}^{-1} = \mathbf{1}, \label{eq:aa'M} \\
&aa' = \widetilde{S}\widetilde{T}^4\widetilde{S}^5\widetilde{T}^2 = a'a, \label{eq:aa'a'a}
\end{align}
where we also use Eq.~(\ref{eq:reGammatilde2Malg}) with $M \in 4\mathbb{Z}$.~\footnote{It is because $\widetilde{S}^{-2M}=\mathbf{1}$, which is satisfied only if $M \in 4\mathbb{Z}$. However, when $k/2 \in \mathbb{Z}$, it is satisfied even if $M=2(2s-1)\ (s \in \mathbb{Z})$.}
Furthermore, from Eqs.~(\ref{eq:reGammatilde2Malg}) and (\ref{eq:S3T3}), we also obtain
\begin{align}
\mathbf{1}
&= (\widetilde{T}^3\widetilde{S}^3\widetilde{T}^3\widetilde{S}^3\widetilde{T}^3\widetilde{S}^3) (\widetilde{S}\widetilde{T}\widetilde{S}\widetilde{T}\widetilde{S}\widetilde{T}) \notag \\
&= \widetilde{T}^3 \widetilde{S}^3 \widetilde{T}^3 \widetilde{S}^3 \widetilde{T}^4 \widetilde{S}\widetilde{T}\widetilde{S}\widetilde{T} \widetilde{S}^4 \notag \\
&= \widetilde{T}^{-2} \widetilde{T}^5 \widetilde{S}^5 \widetilde{T}^5 \widetilde{S}^5 \widetilde{T}^5 \widetilde{T}^{-1} \widetilde{S}^{-1}\widetilde{T}^{-1}\widetilde{S}^{-1} \widetilde{T} \widetilde{S}^4 \notag \\
&= \widetilde{T}^{-2} \widetilde{T}^5 \widetilde{S}^5\widetilde{T}^5 \widetilde{S}^5 \widetilde{T}^5 \widetilde{S}^5 \widetilde{T}^2 \notag \\
&= (\widetilde{S}^5\widetilde{T}^5)^3.
\label{eq:S5T5}
\end{align}
Then, we can prove
\begin{align}
(\widetilde{S}^{2n-1}\widetilde{T}^{2n-1})^3=\mathbf{1}, \ n \in \mathbb{N}, \label{eq:S2n1T2n1}
\end{align}
on the mathematical induction.
Thus, we can prove the other relations in Eq.~(\ref{eq:realgDeltatilde}),
\begin{align}
b^3
&= (\widetilde{T}^{\frac{M}{2}+3} \widetilde{S}^{\frac{3}{2}M+3}\widetilde{T}^M)^3 \notag \\
&= \widetilde{T}^{\frac{M}{2}+3} (\widetilde{S}^{\frac{3}{2}M+3}\widetilde{T}^{\frac{3}{2}M+3})^3 \widetilde{T}^{-\frac{M}{2}-3} \notag \\
&= \mathbf{1}, \label{eq:b3} \\
c^2
&= \widetilde{S}\widetilde{T}^{M-2}\widetilde{S}\widetilde{T}^{\frac{3}{2}M-1} \widetilde{S}\widetilde{T}^{M-2}\widetilde{S}\widetilde{T}^{\frac{3}{2}M-1} \notag \\
&= \widetilde{S}\widetilde{T}^{M-2}\widetilde{S}^{-1}\widetilde{T}^{-1}\widetilde{S}^{-1}\widetilde{T}^{-1}\widetilde{T}^{M-1}\widetilde{S}^{M-1}\widetilde{T}^{M-1} \widetilde{S}^{-M+6} \notag \\
&= \widetilde{S}\widetilde{T}^{M-2} \widetilde{T}\widetilde{S} \widetilde{S}^{-M+1}\widetilde{T}^{-M+1}\widetilde{S}^{-M+1} \widetilde{S}^{-M+6} \notag \\
&= \widetilde{S}^{-M+2}, \label{eq:c2} \\
c^4 &= \widetilde{S}^4, \label{eq:c4} \\
c^8 &= \mathbf{1}, \label{eq:c8} \\
cbc^{-1}
&= (\widetilde{S}\widetilde{T}^{M-2}\widetilde{S}\widetilde{T}^{\frac{3}{2}M-1}) (\widetilde{T}^{\frac{M}{2}+3}\widetilde{S}^{\frac{3}{2}M+3}\widetilde{T}^{M}) (\widetilde{T}^{-\frac{3}{2}M+1}\widetilde{S}^{-1}\widetilde{T}^{-M+2}\widetilde{S}^{-1}) \notag \\
&= \widetilde{S}\widetilde{T}^{M-2}\widetilde{S}\widetilde{T}^{2} \widetilde{S}^{\frac{3}{2}M+3} \widetilde{T}^{-\frac{M}{2}+1} \widetilde{S}^{-1}\widetilde{T}^{-M+2}\widetilde{S}^{-1} \notag \\
&= \widetilde{S}^{\frac{3}{2}M+4} \widetilde{T}^2 \widetilde{S} \widetilde{T}^{\frac{M}{2}-1} \widetilde{S}^{-1}\widetilde{T}^{-M+2}\widetilde{S}^{-1} \notag \\
&= \widetilde{S}^{\frac{3}{2}M+6} \widetilde{T}^2 \widetilde{S}^{-1} \widetilde{T}^{-1} \widetilde{S}^{-1}\widetilde{T}^{-M+2}\widetilde{S}^{-1} \widetilde{T}^{\frac{M}{2}} \notag \\
&= \widetilde{S}^{\frac{3}{2}M+6} \widetilde{T}^3 \widetilde{S} \widetilde{T}^{-M+3} \widetilde{S}^{-1} \widetilde{T}^{\frac{M}{2}} \notag \\
&= \widetilde{S}^{\frac{3}{2}M} \widetilde{T}^{-M} \widetilde{T}^{-M+3} \widetilde{S}^{-M+3} \widetilde{T}^{-M+3} \widetilde{S}^{-M+3} \widetilde{T}^{-M+3} \widetilde{T}^{-\frac{M}{2}-3} \notag \\
&= \widetilde{T}^{-M} \widetilde{S}^{-\frac{3}{2}M-3} \widetilde{T}^{-\frac{M}{2}-3} \notag \\
&= b^{-1}, \label{eq:cbc-1b-1}
\end{align}
\begin{align}
bab^{-1}
&= (\widetilde{T}^{\frac{M}{2}+3} \widetilde{S}^{\frac{3}{2}M+3} \widetilde{T}^{M}) (\widetilde{S}\widetilde{T}^2\widetilde{S}^5\widetilde{T}^{4}) (\widetilde{T}^{-M} \widetilde{S}^{-\frac{3}{2}M-3} \widetilde{T}^{-\frac{M}{2}-3}) \notag \\
&= \widetilde{S}^{\frac{3}{2}M+4} \widetilde{T}^{\frac{M}{2}+5} \widetilde{S}^5 \widetilde{T}^{4} \widetilde{S}^{-\frac{3}{2}M-3} \widetilde{T}^{-\frac{M}{2}-3} \notag \\
&= \widetilde{T}^{5} \widetilde{S}^5 \widetilde{T}^{5} \widetilde{T}^{-1} \widetilde{S}^{-1} \widetilde{T}^{-1} \widetilde{T}^{-2} \widetilde{S}^2 \notag \\
&= \widetilde{S}^{-5}\widetilde{T}^{-5}\widetilde{S}^{-5} \widetilde{S}\widetilde{T}\widetilde{S} \widetilde{T}^{-2} \widetilde{S}^2 \notag \\
&= \widetilde{S}^{-5} \widetilde{T}^{-4} \widetilde{S}^{-1} \widetilde{T}^{-2} \notag \\
&= \widetilde{T}^{-2} \widetilde{S}^{-5} \widetilde{T}^{-4} \widetilde{S}^{-1} \notag \\
&= a^{-1}a'^{-1}, \label{eq:bab-1a-1a'-1} \\
ba'b^{-1}
&= (\widetilde{T}^{\frac{M}{2}+3}\widetilde{S}^{\frac{3}{2}M+3}\widetilde{T}^{M}) (\widetilde{S}\widetilde{T}^2\widetilde{S}^{-1}\widetilde{T}^{-2}) (\widetilde{T}^{-M}\widetilde{S}^{-\frac{3}{2}M-3}\widetilde{T}^{-\frac{M}{2}-3}) \notag \\
&= \widetilde{T}^{\frac{M}{2}+3} \widetilde{S}^{-1} \widetilde{T}^{-2} \widetilde{S} \widetilde{T}^{-\frac{M}{2}-1} \notag \\
&= \widetilde{T}^4 \widetilde{T}^{-1} \widetilde{S}^{-1} \widetilde{T}^{-1} \widetilde{T}^{-1} \widetilde{S}^{-1} \widetilde{T}^{-1} \widetilde{S}^2 \notag \\
&= \widetilde{T}^4 \widetilde{S}\widetilde{T}\widetilde{S} \widetilde{S}\widetilde{T}\widetilde{S} \widetilde{S}^2 \notag \\
&= \widetilde{T}^4 \widetilde{S}\widetilde{T}^2\widetilde{S}^5 \notag \\
&= \widetilde{S}\widetilde{T}^2\widetilde{S}^5\widetilde{T}^4 \notag \\
&= a, \label{eq:ba'-1ba} \\
cac^{-1}
&= (\widetilde{S}\widetilde{T}^{M-2}\widetilde{S}\widetilde{T}^{\frac{3}{2}M-1}) (\widetilde{S}\widetilde{T}^2\widetilde{S}^5\widetilde{T}^4) (\widetilde{T}^{1-\frac{3}{2}M}\widetilde{S}^{-1}\widetilde{T}^{2-M}\widetilde{S}^{-1}) \notag \\
&= \widetilde{S}\widetilde{T}^{M-2}\widetilde{S}^{-1}\widetilde{T}^{-1}\widetilde{S}^{-1}\widetilde{T}^2\widetilde{S}^5\widetilde{T}^5\widetilde{S}\widetilde{T}^{2-M}\widetilde{S} \notag \\
&= \widetilde{S}\widetilde{T}^{M-1}\widetilde{S}\widetilde{T}^3\widetilde{S}^5\widetilde{T}^5\widetilde{S}\widetilde{T}^{2-M}\widetilde{S} \notag \\
&= \widetilde{S}\widetilde{T}^{M-1}\widetilde{S}\widetilde{T}^{-2} \widetilde{S}^{-5}\widetilde{T}^{-3-M}\widetilde{S}^{-3} \notag \\
&= \widetilde{S}^{-1}\widetilde{T}^{-1}\widetilde{S}\widetilde{T} \widetilde{T}^{-3}\widetilde{S}^{-3}\widetilde{T}^{-3}\widetilde{S}^{-3} \notag \\
&= \widetilde{T}\widetilde{S}\widetilde{T}^2 \widetilde{S}^5 \widetilde{T}^3 \notag \\
&= \widetilde{T}^2 \widetilde{T}\widetilde{S}\widetilde{T}^2\widetilde{S}\widetilde{T} \widetilde{S}^4 \notag \\
&= \widetilde{T}^2 \widetilde{S} \widetilde{T}^{-2}\widetilde{S}^{-1} \notag \\
&= a'^{-1}, \label{eq:cac-1a'-1} \\
ca'c^{-1}
&= (\widetilde{S}\widetilde{T}^{M-2}\widetilde{S}\widetilde{T}^{\frac{3}{2}M-1}) (\widetilde{S}\widetilde{T}^2\widetilde{S}^{-1}\widetilde{T}^{-2}) (\widetilde{T}^{1-\frac{3}{2}M}\widetilde{S}^{-1}\widetilde{T}^{2-M}\widetilde{S}^{-1}) \notag \\
&= \widetilde{S}\widetilde{T}^{M-2}\widetilde{S}^{-1}\widetilde{T}^{-1}\widetilde{S}^{-1}\widetilde{T}^2\widetilde{S}^{-1}\widetilde{T}^{-1}\widetilde{S}^{-1}\widetilde{T}^{2-M}\widetilde{S}^3 \notag \\
&= \widetilde{S}\widetilde{T}^{M-1}\widetilde{S}\widetilde{T}^4\widetilde{S}\widetilde{T}^{3-M}\widetilde{S}^3 \notag \\
&= \widetilde{S}^3\widetilde{T}^3\widetilde{S}^3\widetilde{T}^3 \widetilde{T}\widetilde{S}\widetilde{T}^{-1}\widetilde{S}^{-1} \notag \\
&= \widetilde{T}^{-3}\widetilde{S}^{-3}\widetilde{T}\widetilde{S}\widetilde{T}^{-1}\widetilde{S}^{-1} \notag \\
&= \widetilde{T}^{-4}\widetilde{S}^{-5}\widetilde{T}^{-2}\widetilde{S}^{-1} \notag \\
&= a^{-1} \label{eq:ca'c-1a-1}.
\end{align}
Therefore, when Eq.~(\ref{eq:reX}) is also satisfied in addition to Eq.~(\ref{eq:reGammatilde2Malg}), particularly for $M \in 4\mathbb{Z}$, the generators in Eq.~(\ref{eq:reDeltatildeST}) become generators of $\widetilde{\Delta}(6M^2) \simeq (\mathbb{Z}_M \times \mathbb{Z}_M) \rtimes \mathbb{Z}_3 \rtimes \mathbb{Z}_8$, which is the subgroup of $\widetilde{\Gamma}_{2M}$.


\chapter{Normalization of wave functions on magnetized blow-up manifold}
\label{ap:NWMBU}


\section{Normalization of bulk zero modes}
\label{ap:normalbulk}

Here, we show the detailed calculation of Eq.~(\ref{eq:normalbulk}).
It consists of three terms.
The first term comes from
all regions of the original $T^2/\mathbb{Z}_N$ orbifold.
The second term comes from
the region of the cone around $z^{\rm fp}_0=0$ which is cut out from the $T^2/\mathbb{Z}_N$ orbifold.
The third term comes from
the region of the part of $S^2$ which is embedded instead of the cone.
In the following, we show the detailed calculation of the second and third terms.

The second term can be calculated as
\begin{align}
    G_{ij}^{(2)}
    &\equiv \int_{0}^{r} d|z| |z| \int_{0}^{\frac{2\pi}{N}} d\varphi (C^{i}_{0})^{\ast} C^{j}_{0} N^2 |z|^{2m} e^{-\frac{\pi M}{{\rm Im}\tau}|z|^2} \notag \\
    &=  (C^{i}_{0})^{\ast} C^{j}_{0} \pi N \left(\frac{\pi M}{{\rm Im}\tau}\right)^{-(m+1)} \int_{0}^{\frac{\pi M}{{\rm Im}\tau}r^2} d\left( \frac{\pi M}{{\rm Im}\tau} |z|^2 \right) \left( \frac{\pi M}{{\rm Im}\tau} |z|^2 \right)^{m} e^{-\left(\frac{\pi M}{{\rm Im}\tau} |z|^2\right)} \notag \\
    &=  (C^{i}_{0})^{\ast} C^{j}_{0} \pi N \left(\frac{\pi M}{{\rm Im}\tau}\right)^{-(m+1)} \int_{0}^{\frac{\pi M}{{\rm Im}\tau}r^2} dt\ t^{(m+1)-1} e^{-t} \notag \\
    &=  (C^{i}_{0})^{\ast} C^{j}_{0} \pi N \left(\frac{\pi M}{{\rm Im}\tau}\right)^{-(m+1)} \gamma \left(m+1, \frac{\pi M}{{\rm Im}\tau}r^2 \right), \notag
\end{align}
where $\gamma(m+1, \frac{\pi M}{{\rm Im}\tau}r^2)$ denotes the lower incomplete gamma function.
It satisfies the following recurrence relation:
\begin{align}
    \begin{array}{l}
        \gamma \left(m+1, \frac{\pi M}{{\rm Im}\tau}r^2 \right) = m \gamma \left(m, \frac{\pi M}{{\rm Im}\tau}r^2 \right) -\left( \frac{\pi M}{{\rm Im}\tau}r^2 \right)^m e^{-\left( \frac{\pi M}{{\rm Im}\tau}r^2 \right)}  \\
        \gamma \left(1,\frac{\pi M}{{\rm Im}\tau}r^2 \right) = 1 - e^{-\left( \frac{\pi M}{{\rm Im}\tau}r^2 \right)} 
    \end{array}. \notag
\end{align}
By solving this recurrence relation, $\gamma(m+1, \frac{\pi M}{{\rm Im}\tau}r^2)$ can be expressed as
\begin{align}
    \gamma \left(m+1, \frac{\pi M}{{\rm Im}\tau}r^2 \right)
    &= m! e^{-\frac{\pi M}{{\rm Im}\tau}r^2} \left[ e^{\frac{\pi M}{{\rm Im}\tau}r^2}  - \sum_{p=0}^{m} \frac{1}{p!} \left( \frac{\pi M}{{\rm Im}\tau}r^2 \right)^{p} \right] \notag \\
    &= e^{-\frac{\pi M}{{\rm Im}\tau}r^2} \frac{1}{m+1} \left( \frac{\pi M}{{\rm Im}\tau}r^2 \right)^{m+1} \sum_{p=0}^{\infty} \frac{(m+1)!}{(m+1+p) !} \left( \frac{\pi M}{{\rm Im}\tau}r^2 \right)^{p}. \notag
\end{align}
Thus, the second term $G_{ij}^{(2)}$ can be written by
\begin{align}
    G_{ij}^{(2)} = (C^{i}_{0})^{\ast} C^{j}_{0} \pi (r^2)^{m+1} e^{-\frac{\pi M}{{\rm Im}\tau}r^2} \left( \frac{m+1}{N} \right)^{-1} \sum_{p=0}^{\infty} \frac{(m+1)!}{(m+1+p) !} \left( \frac{\pi M}{{\rm Im}\tau}r^2 \right)^{p}.
\end{align}
On the other hand, the third term can be calculated as
\begin{align}
    G_{ij}^{(3)}
    \equiv& \int_{0}^{\frac{r}{N+1}} d|z'| |z'| \int_{0}^{2\pi} d\varphi  \left( \frac{2R^2}{R^2+|z'|^2} \right)^2 \left( \frac{R^2}{R^2+|z'|^2} \right) ({C'}^{i}_{0})^{\ast} {C'}^{j}_{0} R^{-2\ell_0} |z'|^{2\ell_0} \notag \\
    =&  ({C'}^{i}_{0})^{\ast} {C'}^{j}_{0} 4\pi R^2 \notag \\
    &\times \int_{\frac{N+1}{2N}}^{1} d\left( \frac{R^2}{R^2+|z'|^2} \right) \left( 1 - \frac{R^2}{R^2+|z'|^2} \right)^{\ell_0} \left( \frac{R^2}{R^2+|z'|^2} \right)^{M'-\ell_0-1} \notag \\
    =& (C^{i}_{0})^{\ast} C^{j}_{0} N^2 (r^2)^{m} e^{-\frac{\pi M}{{\rm Im}\tau}r^2} \left( \frac{N+1}{N-1} \right)^{\ell_0} \left( \frac{2N}{N+1} \right)^{M'-1} 4\pi \left( \frac{r^2}{(N-1)(N+1)} \right) \notag \\
    &\times \left( \int_{0}^{1} dt\ t^{(M'-\ell_0)-1} (1-t)^{(\ell_0+1)-1} - \int_{0}^{\frac{N+1}{2N}} dt\ t^{(M'-\ell_0)-1} (1-t)^{(\ell_0+1)-1} \right)  \notag \\
    =&  (C^{i}_{0})^{\ast} C^{j}_{0} \pi (r^2)^{m+1} e^{-\frac{\pi M}{{\rm Im}\tau}r^2} \left( \frac{2N}{N+1} \right)^{M'-\ell_0} \left( \frac{2N}{N-1} \right)^{\ell_0+1} \notag \\
    &\times \left( \beta(M'-\ell_0, \ell_0+1) - \beta_{\frac{N+1}{2N}}(M'-\ell_0,\ell_0+1) \right), \notag
\end{align}
where $\beta(M'-\ell_0, \ell_0+1)$ and $\beta_{\frac{N+1}{2N}}(M'-\ell_0,\ell_0+1)$ denote the beta function and the incomplete beta function, respectively.
They satisfy the following recurrence relations:
\begin{align}
    &\begin{array}{l}
        \beta(M'-\ell_0,\ell_0+1) = \frac{\ell_0}{M'-\ell_0} \beta_{\frac{N+1}{2N}}(M'-\ell_0+1,\ell_0) \\
        \beta(M',1) = \frac{1}{M'}
    \end{array}, \notag \\
    &\begin{array}{l}
        \beta_{\frac{N+1}{2N}}(M'-\ell_0,\ell_0+1) = \frac{1}{M'-\ell_0} \left( \ell_0 \beta_{\frac{N+1}{2N}}(M'-\ell_0+1,\ell_0) + \left( \frac{N+1}{2N} \right)^{M'-\ell_0} \left( \frac{N-1}{2N} \right)^{\ell_0} \right) \\
        \beta_{\frac{N+1}{2N}}(M',1) = \frac{1}{M'} \left( \frac{N+1}{2N} \right)^{M'}
    \end{array}. \notag
\end{align}
By solving these recurrence relations, they can be expressed as
\begin{align}
    \beta(M'-\ell_0,\ell_0-1) =& \frac{\Gamma(M'-\ell_0) \Gamma(\ell_0+1)}{\Gamma(M'+1)}, \notag \\
    \beta_{\frac{N+1}{2N}}(M'-\ell_0,\ell_0-1) =& \frac{\Gamma(M'-\ell_0) \Gamma(\ell_0+1)}{\Gamma(M'+1)} \notag \\
	&\times \sum_{p=0}^{\ell_0} \frac{\Gamma(M'+1)}{\Gamma(M'-p+1) \Gamma(p+1)} \left( \frac{N+1}{2N} \right)^{M'-p} \left( \frac{N-1}{2N} \right)^{p}, \notag
\end{align}
respectively.
Here, $\Gamma(X)$ denotes the gamma function, which satisfies the recurrence relation
\begin{align}
    \Gamma(X+1) = X\Gamma(X). \notag
\end{align}
Thus, the third term $G_{ij}^{(3)}$ can be written by
\begin{align}
    G_{ij}^{(3)}
    =& (C^{i}_{0})^{\ast} C^{j}_{0} \pi (r^2)^{m+1} e^{-\frac{\pi M}{{\rm Im}\tau}r^2} \left( \frac{N-1}{2N}(M'-\ell_0) \right)^{-1} \notag \\
&\times \frac{1-\sum_{p=0}^{\ell_0} \frac{\Gamma(M'+1)}{\Gamma(M'-p+1) \Gamma(p+1)}  \left( \frac{N+1}{2N} \right)^{M'-p} \left( \frac{N-1}{2N} \right)^{p}}{\frac{\Gamma(M'+1)}{\Gamma(M'-\ell_0+1) \Gamma(\ell_0+1)}  \left( \frac{N+1}{2N} \right)^{M'-\ell_0} \left( \frac{N-1}{2N} \right)^{\ell_0}}.
\end{align}
By combining those results, we obtain Eq.~(\ref{eq:normalbulk}).


\section{Normalization of localized zero modes}
\label{ap:normallocal}

Here, we show the detailed calculation in Eq.~(\ref{eq:normallocal}).
The first term comes from
the bulk region, while the second term comes from
the blow-up region.
The first term can be calculated as
\begin{align}
    & \int_{r}^{\infty} d|z| |z| \int_{0}^{\frac{2\pi}{N}} d\varphi |C^{a_0}_{0}|^2 N^2 |z|^{2(m-(\ell_0-a_0)N)} e^{-\frac{\pi M}{{\rm Im}\tau}|z|^2} \notag \\
    =& \left| C^{a_0}_{0} \right|^2 \pi N \left( \frac{\pi M}{{\rm Im}\tau} \right)^{(\ell_0-a_0)N-(m+1)} \int_{\frac{\pi M}{{\rm Im}\tau}r^2}^{\infty} d\left( \frac{\pi M}{{\rm Im}\tau}|z|^2 \right) \left( \frac{\pi M}{{\rm Im}\tau}|z|^2 \right)^{m-(\ell_0-a_0)N} e^{-\left( \frac{\pi M}{{\rm Im}\tau}|z|^2 \right)} \notag \\
    =& \left| C^{a_0}_{0} \right|^2 \pi N \left( \frac{\pi M}{{\rm Im}\tau} \right)^{(\ell_0-a_0)N-(m+1)} \int_{\frac{\pi M}{{\rm Im}\tau}r^2}^{\infty} dt\ t^{m-(\ell_0-a_0)N} e^{-t} \notag \\
    =& \left| C^{a_0}_{0} \right|^2 \pi N \left( \frac{\pi M}{{\rm Im}\tau} \right)^{(\ell_0-a_0)N-(m+1)} \Gamma \left( 1+m-(\ell_0-a_0)N, \frac{\pi M}{{\rm Im}\tau}r^2 \right), \notag
\end{align}
where $\Gamma \left( 1+m-(\ell_0-a_0)N, \frac{\pi M}{{\rm Im}\tau}r^2 \right)$ denotes the upper incomplete gamma function.
Note that $1+m-(\ell_0-a_0)N<0$.
It satisfies the following recurrence relation:
\begin{align}
    \begin{array}{l}
        \Gamma \left( 1+m-(\ell_0-a_0)N, \frac{\pi M}{{\rm Im}\tau}r^2 \right) = \\
        \frac{1}{1+m-(\ell_0-a_0)N} \left( \Gamma \left( 2+m-(\ell_0-a_0)N, \frac{\pi M}{{\rm Im}\tau}r^2 \right) - \left( \frac{\pi M}{{\rm Im}\tau}r^2 \right)^{1+m-(\ell_0-a_0)N} e^{-\left( \frac{\pi M}{{\rm Im}\tau}r^2 \right)} \right) \\
        \Gamma \left( 0, \frac{\pi M}{{\rm Im}\tau}r^2 \right) = E_1\left( \frac{\pi M}{{\rm Im}\tau}r^2 \right) 
    \end{array}, \notag
\end{align}
where $E_1\left( \frac{\pi M}{{\rm Im}\tau}r^2 \right)$ denotes the exponential integral.
We note that if $\frac{\pi M}{{\rm Im}\tau}r^2$ is sufficiently large, 
$E_1\left( \frac{\pi M}{{\rm Im}\tau}r^2 \right)$ can be expanded as
\begin{align}
    E_1\left( \frac{\pi M}{{\rm Im}\tau}r^2 \right) \simeq e^{-\left( \frac{\pi M}{{\rm Im}\tau}r^2 \right)} \sum_{p=0} (-1)^{p} p! \left( \frac{\pi M}{{\rm Im}\tau}r^2 \right)^{-(p+1)}. \notag
\end{align}
By solving this recurrence relation, $\Gamma \left( 1+m-(\ell_0-a_0)N, \frac{\pi M}{{\rm Im}\tau}r^2 \right)$ can be expressed as
\begin{align}
    &\Gamma \left( 1+m-(\ell_0-a_0)N, \frac{\pi M}{{\rm Im}\tau}r^2 \right) \notag \\
    =& \frac{(-1)^{(\ell_0-a_0)N-(m+1)}}{[(\ell_0-a_0)N-(m+1)]!} \left[ E_1\left( \frac{\pi M}{{\rm Im}\tau}r^2 \right) - e^{-\left( \frac{\pi M}{{\rm Im}\tau}r^2 \right)} \sum_{p=0}^{(\ell_0-a_0)N-(m+2)} (-1)^{p} p! \left( \frac{\pi M}{{\rm Im}\tau}r^2 \right)^{-(p+1)}  \right]. \notag 
\end{align}
On the other hand, the second term is the same as $G_{ij}^{(3)}$ in the previous appendix by replacing $\ell_0$ with $a_0$.
Thus, by combining those results, we obtain Eq.~(\ref{eq:normallocal}).


\chapter{Anomaly structure of discrete symmetries}
\label{ap:Anomaly}


\section{Anomalies of discrete symmetry}
\label{sec:Anomalyrev}

Here, we review anomalies of discrete symmetry~\cite{Araki:2008ek,Chen:2015aba}.
First of all, let us assume that a classical action $S$ is invariant under unitary transformation by a discrete group $G$ for chiral fermions $\psi_L = P_L \Psi$, $\psi_L \rightarrow \rho(g) \psi_L$, where $\rho(g)$ denotes a unitary representation of $\forall g \in G$.
In this case, we say that the theory has $G$ symmetry at least at classical level.
However, the classical chiral symmetry can be broken by quantum effects.
In the following, we see quantum anomalies of the chiral discrete symmetry by the Fujikawa's method~\cite{Fujikawa:1979ay,Fujikawa:1980eg}.

First, let us see the case with the global $G=\mathbb{Z}_N$ symmetry under background non-Abelian gauge fields as well as gravity, where the chiral fermions have ${\bf R}$ representation under the non-Abelian gauge group $G_{\rm gauge}$.
Note that the generator $g \in \mathbb{Z}_N$ satisfies $g^N=e$ and then the unitary representation can be expressed as $\rho(g)_{jk}=e^{i\alpha q_j}\delta_{jk}$ with the phase parameter, $\alpha=2\pi/N$, and the $\mathbb{Z}_N$ charge of $j$ th component of $\psi_L$, $q_j \in \mathbb{Z}/N\mathbb{Z}$.
In the Fujikawa's method, the measure in the path integral, $\int D\Psi D\bar{\Psi} e^{iS}$, transforms as
\begin{align}
D\Psi D\bar{\Psi} \rightarrow J(\alpha) D\Psi D\bar{\Psi}.
\end{align}
The Jacobian $J(\alpha)$ can be written as~\cite{Alvarez-Gaume:1983ihn,Alvarez-Gaume:1984zlq},
\begin{align}
J(\alpha) = {\rm exp}\left[ i\int d^4x ~(A(x;\alpha)_{\rm gauge} + A(x;\alpha)_{\rm grav})\right],
\end{align}
with $\alpha=2\pi/N$.
The anomaly functions are written by
\begin{align}
A(x;\alpha)_{\rm gauge} =\frac{1}{32\pi^2} \epsilon^{\mu\nu\rho\sigma} {\rm Tr}(\alpha q_j [F_{\mu\nu} F_{\rho\sigma}]), \  A(x;\alpha)_{\rm grav}=-\frac{1}{2} \frac{1}{384\pi^2} \frac{1}{2} \epsilon^{\mu\nu\rho\sigma} R_{\mu\nu}^{\lambda\gamma} R_{\rho\sigma\lambda\gamma} {\rm tr} (\alpha q_j{\bf R}),
\end{align}
where ${\rm Tr}$ denotes the summation over all internal indices.
The index theorems~\cite{Alvarez-Gaume:1983ihn,Alvarez-Gaume:1984zlq} give
\begin{align}
\int d^4x \frac{1}{32\pi^2} \epsilon^{\mu\nu\rho\sigma} F_{\mu\nu}^a F_{\rho\sigma}^b
{\rm tr}[t^a t^b] \in \mathbb{Z}, \quad
\frac{1}{2} \int d^4x \frac{1}{384\pi^2} \frac{1}{2} \epsilon^{\mu\nu\rho\sigma} R_{\mu\nu}^{\lambda\gamma} R_{\rho\sigma\lambda\gamma} \in \mathbb{Z},
\label{eq:index}
\end{align}
where $t^{a,b}$ denote generators of $G_{\rm gauge}$ in the ${\bf R}$ representation.
We use the normalization of Dynkin index $T_2({\bf R})$, 
\begin{align}
T_2({\bf R}) \delta_{ab}= {\rm tr}[t^a t^b] ,
\end{align}
such that $T_2({\bf R})=1/2$ for $N$ fundamental representation of $SU(N)$ 
and $T_2({\bf R})=1$ for $2N$ vector representation of $SO(2N)$.
In the case of ${\bf R} ={\bf 27}$ representation of $E_6$, for example, we obtain $T_2({\bf R})=3$.
Thus, the anomaly-free condition for the mixed anomaly $\mathbb{Z}_N-G_{\rm gauge}-G_{\rm gauge}$ 
is given by
\begin{align}
J(\alpha) = e^{2\pi i \sum q_j 2T_2({\bf R})n/N} = 1, \ \forall n \in \mathbb{Z} \quad \Leftrightarrow \quad
\sum q_j 2T_2({\bf R}) \equiv 0\ ({\rm mod}\ N).
\end{align}
Otherwise, the $\mathbb{Z}_N$ symmetry can be anomalous.

Next, let us see the case with the global non-Abelian discrete $G$ symmetry.
Since each element $g \in G$ satisfies $g^{N(g)}=e$, where $N(g)$ is the order of $g$, we can study its anomalies similar to the case with $\mathbb{Z}_N$ symmetry.
We note that chiral fermions construct a multiplet under the non-Abelian $G$ symmetry in general, and then the unitary representation of $g \in G$ for such a multiplet, $\rho(g)$, forms as unitary matrix.
However, we can always make $\rho(g)$ diagonalized as $\rho(g)_{jk}=e^{i\alpha(g)q_j(g)}\delta_{jk}$
with the phase parameter of the $g$ transformation, $\alpha(g)=2\pi/N(g)$, and the charge of $j$ th component of the multiplet for $g$ transformation, $q_j \in \mathbb{Z}/N(g)\mathbb{Z}$, by taking the appropriate base of the fermions.
Here, we comment that, in such a base, the unitary representations of some of the other elements $g' \in G$, $\rho(g')$, form as non-diagonalized matrices.
Then, we can apply the analysis of the $\mathbb{Z}_N$ symmetry anomalies to the non-Abelian discrete $G$ symmetry anomalies.
The anomaly-free condition for the mixed anomalies 
$G-G_{\rm gauge}-G_{\rm gauge}$ is given by 
\begin{align}
J(\alpha(g)) = e^{2\pi i \sum q_j 2T_2({\bf R})n/N(g)} = (e^{2\pi i Q(g)/N(g)})^{\sum_{\bf R}2T_2({\bf R})n} = ({\rm det}\rho(g))^{\sum_{\bf R}2T_2({\bf R})n} = 1, \ \forall n \in \mathbb{Z},
\end{align}
where $Q(g) \equiv \sum_{j} q_j(g)$ and it is preserved even if the representation $\rho(g)$ is non-diagonalized matrix.
Hence, the symmetries given by elements $g$ with
\begin{align}
{\rm det}\rho(g)=1, \quad (\Leftrightarrow  Q(g) \equiv 0\ ({\rm mod}\ N(g))),
\end{align}
are always anomaly free.
Other parts of $G$ can be anomalous.
The anomalies of symmetries given by elements $g$ with
$\rho(g) \neq 1$ depend on matter contents.
That is, for $\sum_{\bf R}2T_2({\bf R})=P$, 
the subgroup constructed by elements $g$ with 
$({\rm det}\rho (g))^P =1$ is anomaly free, although 
the subgroup constructed by elements $g$ with 
${\rm det}\rho (g) =1$ is always anomaly free.
Thus, the determinant ${\rm det}\rho (g)$ is the key point 
in the analysis of following sections.


\section{
Determinant of representations}
\label{sec:rep}

Here, let us see the anomaly structure of non-Abelian discrete $G$ symmetry by using concrete representations $\rho(g)$ for the elements $g \in G$.
In this section, in particular, we concentrate on the theory with $\sum_{\bf R}2T_2({\bf R})=1$, and then the anomaly-free condition is ${\rm det}\rho (g) =1$.
We comment on the theory with $\sum_{\bf R}2T_2({\bf R})=P > 1$ and anomaly-free condition $({\rm det}\rho (g))^P =1$ in section~\ref{sec:comment}.

When we know representations of $\forall g \in G$, $\rho(g)$, we can calculate ${\rm det}\rho(g)$ explicitly.
Let us suppose that $g^{N(g)}=e$ and then ${\rm det}\rho(g)^{N(g)}=1$ are satisfied for a fixed element $g$, and also $({\rm det}\rho(g))^{N}=1$ is satisfied for any element $g$ in $G$.
Then, the determinant ${\rm det}\rho(g)$ for $\forall g \in G$ can be written as ${\rm det}\rho(g)=e^{2\pi i Q'(g)/N}$, where $Q'(g)$ is written by $Q'(g)=Q(g)N/N(g)$.
As shown in the previous section, if ${\rm det}\rho(g)=1\ (Q'(g) \equiv 0\ ({\rm mod}\ N))$ is satisfied, 
the $g$ transformation can be regarded as anomaly-free transformation.
Here, we define
\begin{align}
G_0 \equiv \{ g_0 \in G | {\rm det}\rho(g_0)=1 \} ,
\end{align}
as the subset of $G$.
In the following proof, we can find that $G_0$ becomes a normal subgroup of $G$, $G_0 \triangleleft G$.
Thus, if all of anomalous transformations are broken by quantum effects, the $G$ symmetry is broken into the normal subgroup $G_0$ at quantum level.
\begin{pf}
We can prove that $G_0$ is a subgroup of $G$, $G_0 \subset G$, from (I), and also a normal subgroup of $G$, $G_0 \triangleleft G$, from (II).
\begin{enumerate}
\renewcommand{\labelenumi}{(\Roman{enumi})}
\item Let us take $\forall g_0 \in G_0$ and $\forall g'_0 \in G_0$ (${\rm det}\rho(g_0)={\rm det}\rho(g'_0)=1$).
Then, the element $g_0g'_0$ is also included in $G_0$, $g_0g'_0 \in G_0$ (${\rm det}\rho(g_0g'_0)=1$). 
In particular, the identity element $e$ is included in $G_0$ (${\rm det}\rho(e)=1$), and then the inverse element $g_0^{-1}$ for $\forall g_0 \in G_0$ (${\rm det}\rho(g_0)=1$) is also included in $G_0$ (${\rm det}\rho(g_0^{-1})=1$).
\item When we take $\forall g_0 \in G_0$ (${\rm det}\rho(g_0)=1$) and $\forall g \in G$, 
the conjugate element $gg_0g^{-1}$ is also included in $G_0$, $gg_0g^{-1} \in G_0$ (${\rm det}\rho(gg_0g^{-1})={\rm det}\rho(g_0)=1$).
\end{enumerate}
\end{pf}
Now, we can rewrite $\forall g \in G$ with ${\rm det}\rho(g) = e^{2\pi i k/N}\ (Q'(g) \equiv k\ ({\rm mod}\ N))$,
by $g_1$ with ${\rm det}\rho(g_1) = e^{2\pi i/N}\ (Q'(g_1) \equiv 1\ ({\rm mod}\ N))$ and $\exists g_0 \in G_0$,
as $g=g_0g_1^k$.
That is, the coset, whose element $g$ satisfies ${\rm det}\rho(g) = e^{2\pi i k/N}$, can be expressed as $G_0g_1^k$. (See also Fig.~\ref{fig:anomalousZN}.)
In addition, in the following proof, we can find that such cosets generate the residue class group $G/G_0 \simeq \mathbb{Z}_N$.
\begin{pf}
We can prove that the residue class group $G/G_0$ is Abelian from (I) and also isomorphic to $\mathbb{Z}_N$ from (II).
\begin{enumerate}
\renewcommand{\labelenumi}{(\Roman{enumi})}
\item
$G_0g_1^{k_1}$ and $G_0g_1^{k_2}$ satisfy the relation $(G_0g_1^{k_1})(G_0g_1^{k_2}) = (G_0g_1^{k_2})(G_0g_1^{k_1}) = G_0g_1^{k_1+k_2}$.
\item
$G_0g_1^{N-k}$ becomes inverse coset of $G_0g_1^{k}$ since $g_1^N \in G_0$ (${\rm det}\rho(g_1^N)=1$).
\end{enumerate}
\end{pf}
We comment that the element $g_1 \in G$ generally satisfies $g_1^N = g_0\ (\exists g_0 \in G)$, while it satisfies $g_1^{N(g_1)}=e$.
Hence, we find $G_0 \cap \mathbb{Z}_{N(g_1)} = \mathbb{Z}_{N(g_1)/N}$ in general, where $\mathbb{Z}_{N(g_1)}$ ($\mathbb{Z}_{N(g_1)/N}$), whose generator is $g_1$ ($g_1^N$), is the subgroup of $G$ ($G_0$).
By applying the isomorphism theorem $2$ in subsection~\ref{app:isomorphism}, indeed, we obtain
\begin{align}
G/G_0 \simeq G_0\mathbb{Z}_{N(g_1)}/G_0 \simeq \mathbb{Z}_{N(g_1)}/\mathbb{Z}_{N(g_1)/N} \simeq \mathbb{Z}_N.
\end{align}

In particular, if $g_1$ satisfies $g_1^N=e \in G_0\ (N(g_1)=N)$, $g_1$ generates $\mathbb{Z}_N$ subgroup of $G$ and also the $\mathbb{Z}_N$ subgroup satisfies $G=G_0\mathbb{Z}_N$ and $G_0 \cap \mathbb{Z}_N = \{e\}$.
Hence, in this case, $G$ can be decomposed\footnote{See also subsection~\ref{app:semidirect}.} as
\begin{align}
G \simeq G_0 \rtimes \mathbb{Z}_N. \label{eq:G0ZN}
\end{align}
It means that the anomaly-free and anomalous parts of $G$ can be separated.
In more general, if there exists $\exists g \in G$ with $N(g)=N$ and ${\rm gcd}(Q(g),N(g))=1$, $G$ can be written as Eq.~(\ref{eq:G0ZN}) since the element $g$ generates $\mathbb{Z}_N$ subgroup of $G$ and it satisfies $G=G_0\mathbb{Z}_N$ and $G_0 \cap \mathbb{Z}_N = \{e\}$.
For example, when $N$ is a prime number, they are automatically satisfied.
\begin{figure}[H]
\centering
\includegraphics[bb=0 0 650 510,width=9cm]{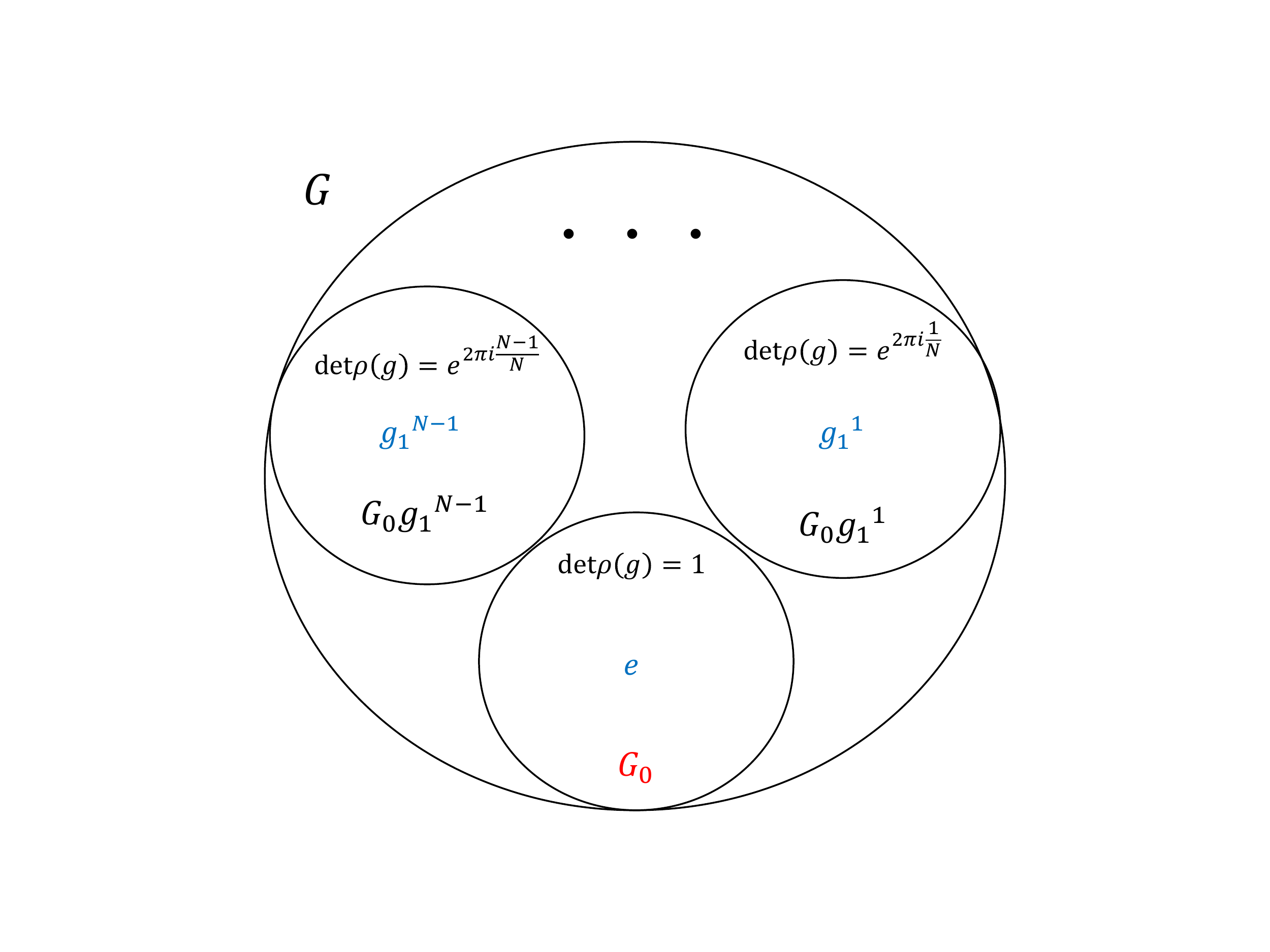}
\caption{Image of cosets $G_0g_1^k$ 
to which
elements $g$ 
with
${\rm det}\rho(g)=e^{2\pi ik/N}$ belong. 
Here, $g_1^k$ (written by blue) denotes the representative element of the coset $G_0g_1^k$}
\label{fig:anomalousZN}
\end{figure}

We summarize the important points in this section.
\begin{itemize}
\item Generally, the anomaly-free subset of $G$, $G_0$, becomes a normal subgroup of $G$, $G_0 \triangleleft G$, and then the anomalous part becomes 
$G/G_0 \simeq \mathbb{Z}_N$, where $N$ can be found from ${\rm det}\rho(g)=e^{2\pi i Q'(g)/N}\ (\forall g \in G)$.
\item In particular, if there exists $\exists g \in G$ with $N(g)=N$ and ${\rm gcd}(Q(g),N(g))=1$, $G$ can be decomposed as $G \simeq G_0 \rtimes \mathbb{Z}_N$.
It means that the anomaly-free and anomalous parts can be separated.
\end{itemize}


\section{
Group structure}
\label{sec:group}

In the previous section, we have obtained $G/G_0 \simeq \mathbb{Z}_N$ generally when we know the representation of $\forall g \in G$, $\rho(g)$.
However, even if we do not use explicit representations, the result is still useful although we do not know the explicit number $N$ from the beginning.
In this section, we study the detailed structure of $G_0$ and $G/G_0 \simeq \mathbb{Z}_N$ from the group structure of $G$.
We emphasize that the following analysis can be applied for any representations.

First, we introduce the derived subgroup of $G$, 
\begin{align}
D(G) \equiv <xyx^{-1}y^{-1} \in G | x, y \in G >. \label{eq:D(G)}
\end{align}
(It is also called the commutator subgroup.)
The derived subgroup $D(G)$ is a normal subgroup $G$, $D(G) \triangleleft G$. It can be checked by
\begin{align}
g(xyx^{-1}y^{-1})g^{-1}=(gxg^{-1})(gyg^{-1})(gxg^{-1})^{-1}(gyg^{-1})^{-1} \in D(G), \notag
\end{align}
for $\forall xyx^{-1}y^{-1} \in D(G)$ and $\forall g \in G$.
The quotient group $G/D(G)$ becomes Abelian. It can be checked by
\begin{align}
(D(G)X)(D(G)Y)=(D(G)Y)(D(G)X), \notag
\end{align}
for any cosets  $D(G)X$ and $D(G)Y$ with $X,Y \notin D(G)$, because of $XYX^{-1}Y^{-1} \in D(G)$.
We note that, among normal subgroups of $G$, $K_G$, such that $G/K_G$ becomes Abelian, 
the derived subgroup $D(G)$ is the smallest normal subgroup.\footnote{If a discrete group $G$ itself is an Abelian group, we obtain $D(G)=\{e\}$.}
Thus, since $G/G_0 \simeq \mathbb{Z}_N$, we can find that
\begin{align}
G_0 \supseteq D(G) \label{eq:G0D(G)}.
\end{align}
Indeed, we can check it by
\begin{align}
{\rm det} \rho(xyx^{-1}y^{-1}) = {\rm det}[\rho(x)\rho(y)\rho(x)^{-1}\rho(y)^{-1}] = 1. \notag
\end{align}
Therefore, the derived subgroup of $G$, $D(G)$, is always anomaly free.
The whole anomaly-free subgroup $G_0$ is either the same as $D(G)$ or larger than $D(G)$. 
In the following, we study how large $G_0$ is compared with $D(G)$.

When we factorize the order of $G/D(G)$ into prime numbers $p_i$,~i.e. $|G/D(G)|=\prod_{i=1}^{r} p_i^{A_i}=p_1^{A_1} \cdots p_r^{A_r}$, due to the fundamental structure theorem of finite Abelian group\footnote{See also subsection \ref{app:Abelian}.}, $G/D(G)$ can be generally written as
\begin{align}
G/D(G) \simeq
(\mathbb{Z}_{p_1^{a_{1,1}}}
\times \cdots \times \mathbb{Z}_{p_1^{a_{1,n_1}}}) \times
\cdots \times
(\mathbb{Z}_{p_r^{a_{r,1}}}
\times \cdots \times \mathbb{Z}_{p_r^{a_{r,n_r}}}),
\label{eq:generalGD(G)}
\end{align}
where $a_{i,j}$ satisfy
\begin{align}
A_i = \sum_{j=1}^{n_i} a_{i,j}, \quad a_{i,j} \geq a_{i,j+1}. \notag
\end{align}
On the other hand, $G/G_0$ is isomorphic to a single cyclic group $\mathbb{Z}_N$.
Thus, the structure of the Abelian group $G/D(G)$ in Eq.~(\ref{eq:generalGD(G)}) is clue to 
study how large $G_0$ is compared with $D(G)$.

Now, let us classify our particle theories by the determinant of the representation (including reducible representation), ${\rm det}\rho(X_{i,j})$, for the element $X_{i,j} \in \mathbb{Z}_{p_i^{a_{i,j}}} \subset G/D(G)$.\footnote{Although $\mathbb{Z}_{p_i^{a_{i,j}}}$ is a subgroup of $G/D(G)$, it is not a subgroup of $G$ in general. In other words, the element $X_{i,j} \in \mathbb{Z}_{p_i^{a_{i,j}}}$ corresponds to an element of the coset $D(G)X_{i,j}$ in $G$. However, since we obtain ${\rm det}\rho(g_D)=1\ (\forall g_D \in D(G))$, the determinant ${\rm det}\rho(X_{i,j})$ can also be reflected in $G$ as the same way.}
%
\begin{itemize}
\item[(i)]
In the theory in which
the element $\forall X_{i,j} \in \mathbb{Z}_{p_i^{a_{i,j}}}$ for $\forall i,j$ satisfies
\begin{align}
{\rm det}\rho(X_{i,j})=1,
\end{align}
we can easily find that
\begin{align}
G_0=G.
\end{align}
\item[(ii)]
In the theory in which
the element $\exists X_{i,j} \in \mathbb{Z}_{p_i^{a_{i,j}}}-\{e\}$ for $\exists i,j$ satisfies
\begin{align}
{\rm det}\rho(X_{i,j})=1,
\end{align}
while any other element $\forall X'_{i,j} \in \mathbb{Z}_{p_i^{a_{i,j}}}-\{e,X_{i,j}\}$ for $\forall i,j$ satisfies
\begin{align}
{\rm det}\rho(X'_{i,j}) \neq 1,
\end{align}
we can find that
\begin{align}
G \supset G_0 \supset D(G).
\end{align}
Since there are several patterns belong to this class in some discrete groups $G$, we do not discuss this class in detail.
However, we can also apply the following analysis to this class.
\item[(iii)]
In
the theory in which
the element $\forall X_{i,j} \in \mathbb{Z}_{p_i^{a_{i,j}}}-\{e\}$ for $\forall i,j$ satisfies
\begin{align}
{\rm det}\rho(X_{i,j}) \neq 1, \label{eq:not1}
\end{align}
we can find that
\begin{align}
G \supset G_0 \supseteq D(G),
\end{align}
in the following.
First, if there exists $a_{i,2} \neq 0$ for $\exists i$, $p_i^{a_{i,2}}-1$ numbers of combination elements $X_{i,1}^{m_1}X_{i,2}^{m_2}$ satisfy 
\begin{align}
{\rm det}\rho(X_{i,1}^{m_1}X_{i,2}^{m_2})=1,
\end{align}
where $m_1$ and $m_2$ satisfy the following relation;
\begin{align}
\begin{array}{ll}
&Q(X_{i,1}^{m_1}X_{i,2}^{m_2}) = m_1 Q(X_{i,1}) + m_2 Q(X_{i,2}) p_i^{a_{i,1}-a_{i,2}} \equiv 0 \quad ({\rm mod}\ p_i^{a_{i,1}}), \\
\Leftrightarrow
&m_1 Q(X_{i,1}) = p_i^{a_{i,1}-a_{i,2}} n, \quad  m_2 Q(X_{i,2}) = p_i^{a_{i,2}} - n, \quad \forall n \in \mathbb{Z}/p_i^{a_{i,2}}\mathbb{Z}-\{0\}.
\end{array}
\end{align}
Thus, those $X_{i,1}^{m_1}X_{i,2}^{m_2}$ as well as $e$ are also included in $G_0$, which means that
\begin{align}
G \supset G_0 \supset D(G),
\end{align}
and also they construct $\mathbb{Z}_{p_i^{a_{i,2}}}$ subgroup of $G_0/D(G)$.
Generally,
if there are $\mathbb{Z}_{N_1}$ and $\mathbb{Z}_{N_2}$ symmetries in $G/D(G)$, where $N_1$ and $N_2$ are not coprime to each other, ${\rm gcd}(N_1,N_2)-1$ numbers of elements $X_{1}^{m_1}X_{2}^{m_2}\ (X_1 \in \mathbb{Z}_{N_1}, X_2 \in \mathbb{Z}_{N_2})$ as well as $e$ satisfy ${\rm det}\rho(X_1X_2)=1$ and then they construct $\mathbb{Z}_{{\rm gcd}(N_1,N_2)}$ subgroup of $G_0/D(G)$.
In a similar way, we can find that
\begin{align}
G_0/D(G) \simeq
(\mathbb{Z}_{p_1^{a_{1,2}}}
\times \cdots \times \mathbb{Z}_{p_1^{a_{1,n_1}}}) \times
\cdots \times
(\mathbb{Z}_{p_r^{a_{r,2}}}
\times \cdots \times \mathbb{Z}_{p_r^{a_{r,n_r}}}).
\label{eq:generalG0D(G)}
\end{align}
Then, by applying the isomorphism theorem $3$ in subsection~\ref{app:isomorphism} with Eqs.~(\ref{eq:generalGD(G)}) and (\ref{eq:generalG0D(G)}), we can obtain
\begin{align}
\mathbb{Z}_N \simeq G/G_0 \simeq (G/D(G))/(G_0/D(G)) \simeq \mathbb{Z}_{(p_1^{a_{1,1}} \cdots p_r^{a_{r,1}})}. \label{eq:GG0iii}
\end{align}
Indeed, when Eq.~(\ref{eq:not1}) is satisfied, the determinant of the representation for $\forall g \in G$, ${\rm det}\rho(g)$, can be written as ${\rm det}\rho(g)=e^{2\pi iQ'(g)/(p_1^{a_{1,1}} \cdots p_r^{a_{r,1}})}$. It also shows Eq.~(\ref{eq:GG0iii}).
Notice that $N= p_1^{a_{1,1}} \cdots p_r^{a_{r,1}} = \prod_{i=1}^{r} p_i^{a_{i,1}}$ is the least common multiple of orders of each $Z_{p_i^{a_{i,j}}}$ in $G/D(G)$, which becomes the maximum order of the anomalous $G/G_0 \simeq \mathbb{Z}_N$.
In other words, the maximum order of the anomalous $G/G_0 \simeq \mathbb{Z}_N$ can be found from $G/D(G)$, which is determined by $G$.

In particular, if and only if 
$A_i=a_{i,1}\ (a_{i,2}=0)$ for all $i$, any element $X$ in the Abelian group $G/D(G)$ leads to
\begin{align}
{\rm det}\rho(X) \neq 1.
\end{align}
It means that
\begin{align}
G_0=D(G),
\end{align}
and then, by applying Eq.~(\ref{eq:mn}) in subsection~\ref{app:Abelian}, we can obtain
\begin{align}
\mathbb{Z}_N \simeq G/G_0 = G/D(G) \simeq \mathbb{Z}_{(p_1^{A_1} \cdots p_r^{A_r})}.
\end{align}
Therefore, $D(G)$ and $G/D(G)$ are important clues to understand the structure of $G_0$ and $G/G_0 \simeq \mathbb{Z}_N$.
In the following analysis, we mainly discuss this class (iii) in which Eq.~(\ref{eq:not1}) is satisfied.
\end{itemize}


\subsection{Various examples of discrete groups}
\label{subsec:example}

Here, let us see the detail structure of $G_0$ and $G/G_0 \simeq \mathbb{Z}_N$ from the structure of $D(G)$ and $G/D(G)$ in specific examples of discrete group $G$.

First, let $G$ be a perfect group, which is defined as a group satisfying $D(G)=G$.
In this case, we obviously find that $G = G_0 = D(G)$,
that is, the whole $G$ symmetry is always anomaly free.
One of the example is non-Abelian simple group such as $A_n\ (n \geq 5)$ and $PSL(2,\mathbb{Z}_p)\ (p \neq 2,3,\ p \in \mathbb{P})$.\footnote{Since a simple group is defined as a group $G$ whose normal subgroups are just $\{e\}$ and $G$ itself, in order for $G/D(G)$ to be Abelian, $D(G)$ must be $G$ itself for a non-Abelian simple group $G$.}

On the other hand, an Abelian simple group, which is just isomorphic to $\mathbb{Z}_p\ (p \in \mathbb{P})$, is not a perfect group since 
$D(G)=\{e\}$.
In this case, the flavor model can be classified as
either the class (i), $G_0=G \simeq \mathbb{Z}_p$, or the class (iii), $G_0=D(G)=\{e\}$, $G/G_0 \simeq \mathbb{Z}_N=\mathbb{Z}_p$.
One of the example is $G = A_3 \simeq \mathbb{Z}_3$ group.

Next, let us see the case that the group $G$ can be written by a semidirect product,~i.e. $G \simeq K_G \rtimes G^{(1)}$, in the following four steps.

\begin{step}[$G \simeq \mathbb{Z}_A \rtimes \mathbb{Z}_B$]
Let us start from the simplest group, $G \simeq \mathbb{Z}_A \rtimes \mathbb{Z}_B\ (A \geq 3)$.
Here, we find that $D(G) \subseteq \mathbb{Z}_A$ from $G/\mathbb{Z}_A \simeq \mathbb{Z}_B$.
The algebraic relations for generators $\alpha \in \mathbb{Z}_A$ and $\beta \in \mathbb{Z}_B$ are given by
\begin{align}
\alpha^A = \beta^B = e,
\end{align}
and also 
\begin{align}
&\beta \alpha \beta^{-1} = \alpha^m \in \mathbb{Z}_A, \quad (\beta \alpha \beta^{-1} \alpha^{-1} = \alpha^{m-1} \in D(G)), \quad m \in \mathbb{Z}/A\mathbb{Z} - \{0,1\}, \label{eq:semidircond1ZAZB} \\
\Rightarrow \ 
&\beta^b \alpha^a \beta^{-b} = a^{am^b}, \quad (\beta^b \alpha^a \beta^{-b} \alpha^{-a} = a^{a(m^b-1)} \in D(G)),
\quad a \in \mathbb{Z}/A\mathbb{Z}, \ b \in \mathbb{Z}/B\mathbb{Z}, \label{eq:semidircond2ZAZB}
\end{align}
where $m$ satisfies\footnote{If $m=1$, in particular, the group $G$ can be written as $G \simeq \mathbb{Z}_A \times \mathbb{Z}_B$. This is the specific case of the above general analysis.} the following conditions,
\begin{align}
\begin{array}{l}
(m^b - 1) = (m-1)(\sum_{r=0}^{b-1}m^{r}) \not\equiv 0 \quad ({\rm mod}\ A) \quad {\rm for}\ \forall b \\
(m^B - 1) = (m-1)(\sum_{r=0}^{B-1}m^{r}) \equiv 0 \quad ({\rm mod}\ A)
\end{array}. \label{eq:constm}
\end{align}
From Eq,~(\ref{eq:semidircond2ZAZB}), we find that
\begin{align}
D(G) = \{ \alpha^{a'{\rm gcd}(m-1,A)} | a' \in \mathbb{Z}/(A/{\rm gcd}(m-1,A))\mathbb{Z} \} = \mathbb{Z}_{A/{\rm gcd}(m-1,A)} \subseteq G_0. \label{eq:D(G)ZAZB}
\end{align}
Note that, in the case of $G \simeq \mathbb{Z}_A$,  $D(G)=\{e\}$.
Then, we obtain that
\begin{align}
G/D(G) \simeq \mathbb{Z}_{{\rm gcd}(m-1,A)} \times \mathbb{Z}_B, \label{eq:GD(G)ZAZB}
\end{align}
since cosets $D(G)\alpha$ and $D(G)\beta$ satisfy 
\begin{align}
\begin{array}{l}
(D(G)\alpha)^{{\rm gcd}(m-1,A)}=(D(G)\beta)^B=D(G), \\
(D(G)\beta)(D(G)\alpha)(D(G)\beta)^{-1}=D(G)(\beta\alpha\beta^{-1}\alpha^{-1})\alpha=D(G)\alpha.
\end{array}
\notag
\end{align}

The above result can also be understood from the viewpoint of $\mathbb{Z}_A$ charge constraint.
Suppose that the chiral fermions are $\mathbb{Z}_A$ eigenstates, $\rho(\alpha)_{jk}=e^{2\pi iq_j/A}\delta_{jk}$, and $q_j$ is the $\mathbb{Z}_A$ charge of $j$ th component.
Eq.~(\ref{eq:semidircond1ZAZB}) means that there exists a state with $\mathbb{Z}_A$ charge $mq_j$ and 
$\beta$ transforms the $j$ th component with $q_j$ to the state with $mq_j$.
For example,
when we consider a fundamental irreducible representation, it forms $B$-dimensional representation with $\mathbb{Z}_A$ charge $ ^t(q_1,q_2,q_3,...,q_B)$ $=$ $ ^t(q, mq, m^2q,...,m^{B-1}q)$.
It means that $\mathbb{Z}_A$ charges in a multiplet are constrained by the semidirect product by $\mathbb{Z}_B$, while there is no constraint for $\mathbb{Z}_B$ charges.
Indeed, we obtain
\begin{align}
{\rm det}\rho(\alpha) = e^{2\pi i\sum_{r=0}^{B-1}m^rq/A} = e^{2\pi iqn/{\rm gcd}(m-1,A)},
\end{align}
and then we can check Eqs.~(\ref{eq:D(G)ZAZB}) and (\ref{eq:GD(G)ZAZB}) from it.

Now, as discussed in the above general analysis, let us study the class (iii).
First of all, we denote $A'\equiv {\rm gcd}(m-1,A)$.
When we consider
the elements, $\alpha^x\beta^y$, which satisfy
\begin{align}
&Q(\alpha^x)B/{\rm gcd}(A',B) + Q(\beta^y)A'/{\rm gcd}(A',B) \equiv 0 \quad ({\rm mod}\ {\rm lcm}(A',B)), \label{eq:QxQy} \\
\Leftrightarrow \quad
&xQ(\alpha) = ({\rm gcd}(A',B) - s) A'/{\rm gcd}(A',B), \ yQ(\beta) = s B/{\rm gcd}(A',B), \ s \in \mathbb{Z}/{\rm gcd}(A',B)\mathbb{Z}, \notag
\end{align}
they satisfy ${\rm det}\rho(\alpha^x\beta^y)=1$.
It means that the elements of ${\rm gcd}(A',B)$ numbers of cosets $D(G)\alpha^x\beta^y$ (including $D(G)\alpha^{A'}=D(G)$) become the elements of $G_0$.
In addition, since those cosets satisfy 
$(D(G)\alpha^x\beta^y)^{{\rm gcd}(A',B)}=D(G)$, 
they give
\begin{align}
G_0/D(G) \simeq \mathbb{Z}_{{\rm gcd}(A',B)},
\end{align}
which is the (normal) subgroup of $G/D(G)$.
We note that the generators of $G_0$ are $\alpha^{A'}$ and $\alpha^x\beta^y$ with Eq.~(\ref{eq:QxQy}),
and they satisfy
\begin{align}
(\alpha^{A'})^{A/A'}=e, \ 
(\alpha^x\beta^y)^{{\rm gcd}(A',B)} = (\alpha^{A'})^{k=\frac{x(m^{y{\rm gcd}(A',B)}-1)}{A'(m^y-1)}},\ 
(\alpha^x\beta^y)(\alpha^{A'})(\alpha^x\beta^y)^{-1}=(\alpha^{A'})^{m^y}. 
\label{eq:alphabetacondition}
\end{align}
In particular, if $k \equiv 0\ ({\rm mod}\ A/A')$, that is, if $\alpha^x\beta^y$ generates $\mathbb{Z}_{{\rm gcd}(A',B)}$ subgroup of $G_0$, $G_0$ can be decomposed as
\begin{align}
G_0 \simeq D(G) \rtimes \mathbb{Z}_{{\rm gcd}(A',B)} = \mathbb{Z}_{A/A'} \rtimes \mathbb{Z}_{{\rm gcd}(A',B)}.
\end{align}
Then, by applying the isomorphism theorem $3$ in subsection~\ref{app:isomorphism}, we can obtain
\begin{align}
\mathbb{Z}_N \simeq G/G_0 \simeq (G/D(G))/(G_0/D(G)) \simeq (\mathbb{Z}_{A'} \times \mathbb{Z}_B)/\mathbb{Z}_{{\rm gcd}(A',B)} \simeq Z_{{\rm lcm}(A',B)},
\end{align}
where we also apply Eq.~(\ref{eq:gcdlcm}) in subsection~\ref{app:Abelian} for Eq.~(\ref{eq:GD(G)ZAZB}),
\begin{align}
G/D(G) \simeq \mathbb{Z}_{A'} \times \mathbb{Z}_{B} \simeq \mathbb{Z}_{{\rm gcd}(A',B)} \times \mathbb{Z}_{{\rm lcm}(A,B)}. \label{eq:GD(G)gcdlcm}
\end{align}
Similar to Eq.~(\ref{eq:QxQy}), there exists $\alpha^{x'}\beta^{y'}$ with $\exists (x',y')$, which satisfies
\begin{align}
Q(\alpha^{z'})B/{\rm gcd}(A',B) + Q(\beta^{y'})A'/{\rm gcd}(A',B) \equiv 1 \quad ({\rm mod}\ {\rm lcm}(A',B)), \label{eq:Qx'Qy'}
\end{align}
since $A'/{\rm gcd}(A',B)$ and $B/{\rm gcd}(A',B)$ are coprime to each other.
It means that the element $\alpha^{x'}\beta^{y'}$ corresponds to $g_1$ in the previous section.
Indeed, the coset $D(G)\alpha^{x'}\beta^{y'} \subset G_0\alpha^{x'}\beta^{y'}$ satisfies $(D(G)\alpha^{x'}\beta^{y'})^{{\rm lcm}(A',B)}=D(G) \subset G_0$.
We note that, similar to Eq.~(\ref{eq:alphabetacondition}), 
the element $\alpha^{x'}\beta^{y'}$ with Eq.~(\ref{eq:Qx'Qy'}) generally satisfies 
\begin{align}
\begin{array}{l}
(\alpha^{x'}\beta^{y'})^{{\rm lcm}(A',B)} = (\alpha^{A'})^{k'=\frac{x'(m^{y'{\rm lcm}(A',B)}-1)}{A'(m^{y'}-1)}}, \\
(\alpha^{x'}\beta^{y'})(\alpha^{A'})(\alpha^{x'}\beta^{y'})^{-1}=(\alpha^{A'})^{m^{y'}}, \\
(\alpha^{x'}\beta^{y'})(\alpha^x\beta^y)(\alpha^{x'}\beta^{y'})^{-1}=(\alpha^{A'})^{\ell'=[x(m^{y'}-1)-x'(m^y-1)]/A'}(\alpha^x\beta^y).
\end{array}
\end{align}
If $k' \equiv 0\ ({\rm mod}\ A/A')$, that is, if $\alpha^{x'}\beta^{y'}$ generates $Z_{{\rm lcm}(A',B)}$ subgroup of $G$, $G$ can be written as $G \simeq G_0 \rtimes \mathbb{Z}_{{\rm lcm}(A',B)}$.
In addition, if $\ell' \equiv 0\ ({\rm mod}\ A/A')$, $\alpha^{x'}\beta^{y'}$ commutes with $\alpha^x\beta^y$.
Thus, if $k, k', \ell'$ are multiples of $A/A'$, that is, if $\mathbb{Z}_{{\rm gcd}(A',B)} \times \mathbb{Z}_{{\rm lcm}(A,B)}$ in Eq.~(\ref{eq:GD(G)gcdlcm}) is actually a subgroup of $G$, $G$ can be written as
\begin{align}
\begin{array}{rl}
G
&\simeq D(G) \rtimes ( \mathbb{Z}_{{\rm gcd}(A',B)} \times \mathbb{Z}_{{\rm lcm}(A,B)} ) \simeq \mathbb{Z}_{A/A'} \rtimes (\mathbb{Z}_{A'} \times \mathbb{Z}_B) \\
&\simeq (D(G) \rtimes \mathbb{Z}_{{\rm gcd}(A',B)}) \rtimes \mathbb{Z}_{{\rm lcm}(A,B)} \\
&\simeq G_0 \rtimes \mathbb{Z}_{{\rm lcm}(A,B)}, \label{eq:GD(G)G0}
\end{array}
\end{align}
where the second line can be found from Eqs.~(\ref{eq:decomposiG1}) and (\ref{eq:decomposiK}) in subsection~\ref{app:semidirect}.

Now, let us see examples.
First, if $A=p \in \mathbb{P}$, 
we obtain $A'=1$, and then we find that
$D(G)=\mathbb{Z}_A \subseteq G_0$.
In the class (iii), 
we obtain $G_0=D(G)=\mathbb{Z}_A$ and $G/G_0 \simeq \mathbb{Z}_N = \mathbb{Z}_B$.

Second, let us see $G = D_A \simeq \mathbb{Z}_A \rtimes \mathbb{Z}_2$ case.
In this case, we obtain $m=A-1$ from Eq.~(\ref{eq:constm}), and then we can find that
\begin{align}
&D(G)
= \left\{
\begin{array}{ll}
< \tilde{\alpha} | \tilde{\alpha}^{A/2} = e > = \mathbb{Z}_{A/2} & (A \in 2\mathbb{Z}) \\
< \alpha | \alpha^A = e > = \mathbb{Z}_A & (A \in 2\mathbb{Z}+1)
\end{array}
\right., \\
&G/D(G)
\simeq \left\{
\begin{array}{ll}
\mathbb{Z}_{2} \times \mathbb{Z}_2 = < (d_{\alpha},d_{\beta}) | d_{\alpha}^2=d_{\beta}^2=d_{e}, d_{\alpha}d_{\beta}=d_{\beta}d_{\alpha}=d_{\alpha\beta} > & (A \in 2\mathbb{Z}) \\
\mathbb{Z}_2 = < d_{\beta} | d_{\beta}^2=d_{e} > & (A \in 2\mathbb{Z}+1)
\end{array}
\right.,
\end{align}
where $\tilde{\alpha} \equiv \alpha^2$, $d_{X} \equiv D(G)X$.
In the class (iii), 
we can obtain
\begin{align}
&G_0 = \left\{
\begin{array}{ll}
< \tilde{\alpha}, \tilde{\beta} | \tilde{\alpha}^{A/2}=\tilde{\beta}^2=e, \tilde{\beta}\tilde{\alpha}\tilde{\beta}^{-1}=\tilde{\alpha}^{-1} > = \mathbb{Z}_{A/2} \rtimes Z_2 \simeq D_{A/2} & (A \in 2\mathbb{Z}) \\
< \alpha | \alpha^A=e > = \mathbb{Z}_A & (A \in 2\mathbb{Z}+1)
\end{array}
\right., \\
&G/G_0 \simeq \mathbb{Z}_N = \mathbb{Z}_2 = < g_{\beta} | g_{\beta}^2=g_{e} >, \quad \forall A,
\end{align}
where $\tilde{\beta} \equiv \alpha\beta \ (x=y=1)$, $g_{\beta} \equiv G_0\beta \ (x'=0, y'=1)$.
In particular, since $\beta^2=\tilde{\beta}^2=e$,
$G=D_A$ can be written as
\begin{align}
\begin{array}{rl}
D_A=G
&\simeq G_0 \rtimes \mathbb{Z}_2 \\
&\simeq \left\{
\begin{array}{ll}
(D(G) \rtimes \mathbb{Z}_2) \rtimes \mathbb{Z}_2 &  (A \in 2\mathbb{Z}) \\
D(G) \rtimes \mathbb{Z}_2 &  (A \in 2\mathbb{Z}+1)
\end{array}
\right. \\
&\simeq \left\{
\begin{array}{ll}
D_{A/2} \rtimes \mathbb{Z}_2 &  (A \in 2\mathbb{Z}) \\
\mathbb{Z}_A \rtimes \mathbb{Z}_2 &  (A \in 2\mathbb{Z}+1)
\end{array}
\right..
\end{array}
\end{align}
Here, we comment on the case with $A=2(2m-1)\ (m \in \mathbb{Z})$ especially.
In this case, we have the coset relation, $g_{\beta}=g_{\alpha^{A/2}}\ (x'=A/2,y'=0)$, and then $k, k', \ell'$ are multiples of $A/A'=A/2$.
Thus, Eq.~(\ref{eq:GD(G)G0}) is satisfied;
\begin{align}
D_A=G
&\simeq D(G) \rtimes (\mathbb{Z}_2 \times \mathbb{Z}_2) \simeq  \mathbb{Z}_{A/2} \rtimes (\mathbb{Z}_2 \times \mathbb{Z}_2) \notag \\
&\simeq G_0 \times \mathbb{Z}_2 \simeq D_{A/2} \times \mathbb{Z}_2. \notag
\end{align}
One of the example is $G= S_3 \simeq D_3 \simeq A_3 \rtimes \mathbb{Z}_2$.
In this case,
the $S_3$ flavor model can be classified as either the class (iii), $G_0=D(G)=\mathbb{Z}_3 \simeq A_3$, $G/G_0 \simeq \mathbb{Z}_N=\mathbb{Z}_2$, or the class (i), $G_0=G=S_3$.
It depends on representations (including reducible representations) in the model.
Indeed, $S_3$ group has three irreducible representations, ${\bf 1}\ ({\rm det}\rho_{{\bf 1}}(\beta)=1)$, ${\bf 1'}\ ({\rm det}\rho_{{\bf 1'}}(\beta)=-1)$, and ${\bf 2}\ ({\rm det}\rho_{{\bf 2}}(\beta)=-1)$~\cite{Ishimori:2010au,Ishimori:2012zz}.
Then,
the whole $S_3$ symmetry is anomaly free in flavor models in which even numbers of ${\bf 1'}$ and ${\bf 2}$ are included.
Otherwise, the $\mathbb{Z}_2$ subsymmetry can be anomalous.

We comment on $G=Q_A$ case.
The algebraic relations of $Q_A$ are same as those of $D_A$ with $A \in 2\mathbb{Z}$, except that $\beta^2=\alpha^{A/2}$ instead of $\beta^2=e$.
Then, we find that $D(Q_A)=D(D_A)$, while the representation of $\beta$ depends on that of $\alpha$.
In the case of $A \in 4\mathbb{Z}$, since both $\alpha^{A/2}$ and $e$ are elements of $D(Q_A)$, the analysis of $Q_A$ is the same as that of $D_A$.
On the other hand,
in the case of $A \in 2(2\mathbb{Z}+1)$, since $\alpha^{A/2} \notin D(Q_A)$, $\beta$ becomes anomalous $\mathbb{Z}_4$ generator in general.
This is the different point from $D_A$.

Third, let us see $G=T_{p^k} \simeq \mathbb{Z}_{p^k} \rtimes \mathbb{Z}_3$ with $p \neq 3$ and $p \in \mathbb{P}$.
If ${\rm gcd}(m-1,p^k) \neq 1$, $m$ can be expressed as $m=p^{\ell}+1$ with $\ell \in \mathbb{Z}/k\mathbb{Z}-\{0\}$.
It needs to satisfy that
\begin{align}
m^3-1 = p^{\ell} (p^{2\ell}+3p^{\ell}+3) \equiv 0 \ ({\rm mod}\ p^k), \notag \\
\Rightarrow \ 
p^{2\ell}+3p^{\ell}+3 = p^{k-\ell}x, \quad (x \in \mathbb{Z}), \notag \\
3 = p^{k-\ell}x - p^{2\ell} - 3p^{\ell} = p^{\ell'}y .
\end{align}
However, it cannot be satisfied if $p\neq3$.
Hence, we find that ${\rm gcd}(m-1,p^k) = 1$ and then $D(G)=\mathbb{Z}_{p^k} \subseteq G_0$.
Therefore, the $T_{p^k}$ flavor model corresponds to either the class (iii), $G_0=D(G)=\mathbb{Z}_{p^k}$, $G/G_0 \simeq \mathbb{Z}_N=\mathbb{Z}_3$, or the class (i), $G_0=G=T_{p^k}$.
In more general, when we consider $G \simeq \mathbb{Z}_{p^k} \rtimes \mathbb{Z}_B$ with ${\rm gcd}(p,B)=1$, 
we find that ${\rm gcd}(m-1,p^k)=1$ and then $D(G)=\mathbb{Z}_{p^k} \subseteq G_0$.
\end{step}

\begin{step}[$G \simeq (\mathbb{Z}_A \times \mathbb{Z}'_A) \rtimes \mathbb{Z}_B$]
Next, we study more complicated case, $G \simeq (\mathbb{Z}_A \times \mathbb{Z}'_{A}) \rtimes \mathbb{Z}_B$.
%
First, let us see $G=\Sigma(2n^2) \simeq (\mathbb{Z}_n \times \mathbb{Z}'_n) \rtimes \mathbb{Z}_2$.
The algebraic relations for generators $\alpha \in \mathbb{Z}_n$, $\alpha' \in \mathbb{Z}'_n$, and $\beta \in \mathbb{Z}_2$ are given by
\begin{align}
\alpha^n = \alpha^{\prime n} = \beta^2 = e,
\end{align}
and also
\begin{align}
&\alpha' \alpha \alpha^{\prime -1} = \alpha \in \mathbb{Z}_n, \quad (\alpha \alpha' \alpha^{-1} = \alpha' \in \mathbb{Z}'_n) \\
&\beta \alpha \beta^{-1} = \alpha^{m_1} \alpha^{\prime m_2} \in \mathbb{Z}_n \times \mathbb{Z}'_n, \quad \beta \alpha^{-1} \beta^{-1} = \alpha^{m_3} \alpha^{\prime m_4} \in \mathbb{Z}_n \times \mathbb{Z}'_n. \label{eq:semidircond1ZnZnZ2}
\end{align}
Here, $m_i\ (i=1,2,3,4)$ can be determined by the constraints, $\beta^2 \alpha \beta^{-2}=\alpha$, $\beta^2 \alpha' \beta^{-2}=\alpha'$,
and then we obtain $m_1=m_4=0,\ m_2=m_3=1$.
Namely, Eq.~(\ref{eq:semidircond1ZnZnZ2}) can be rewritten as
\begin{align}
\beta \alpha \beta^{-1} = \alpha', \quad \beta \alpha' \beta^{-1} = \alpha. \label{eq:semidircond2ZnZnZ2}
\end{align}
Hence, we can find that
\begin{align}
&D(G) = < \tilde{\alpha} | \tilde{\alpha}^n=e > = \mathbb{Z}_n, \label{eq:D(G)ZnZnZ2} \\
&G/D(G) \simeq \mathbb{Z}_n \times \mathbb{Z}_2 = < (d_{\tilde{\alpha}'}, d_{\beta}) | d_{\tilde{\alpha}'}^n=d_{\beta}^2=d_e, d_{\tilde{\alpha}'}d_{\beta}=d_{\beta}d_{\tilde{\alpha}'}=d_{\alpha\beta} >, \label{eq:GD(G)ZnZnZ2}
\end{align}
where $\tilde{\alpha} \equiv \alpha \alpha^{\prime -1}$, $d_X \equiv D(G)X$.

The above result can also be understood from the viewpoint of $\mathbb{Z}_n$ and $\mathbb{Z}'_n$ charge constraints.
Suppose that the chiral fermions are $\mathbb{Z}_n$ and $\mathbb{Z}'_n$ eigenstates, where we denote $[\mathbb{Z}_n, \mathbb{Z}'_n]$ charges of the $j$ th component field as $[q_j, q'_j]$.
Eq.~(\ref{eq:semidircond2ZnZnZ2}) means that there exists a state with charge $[q'_j, q_j]$ and
$\beta$ transforms the $j$ th component with charge $[q_j, q'_j]$ to the state with charge $[q'_j, q_j]$.
For example, when we consider a fundamental irreducible representation, it forms a doublet with charge $ ^t([q_1,q'_1], [q_2, q'_2])$ $=$ $ ^t([q, q'], [q', q])$. Then, we obtain 
\begin{align}
{\rm det}\rho(\alpha) = {\rm det}\rho(\alpha') = e^{2\pi i (q+q')/n},
\end{align}
and we can actually check
Eqs.~(\ref{eq:D(G)ZnZnZ2}) and (\ref{eq:GD(G)ZnZnZ2}).
We note that there is no constraint for $\mathbb{Z}_2$ charges.

In the class (iii), we can obtain
\begin{align}
&G_0 = \left\{
\begin{array}{ll}
< \tilde{\alpha}, \tilde{\beta} | \tilde{\alpha}^n=e,  \tilde{\beta}^2=\tilde{\alpha}^{n/2}, \tilde{\beta}\tilde{\alpha}\tilde{\beta}^{-1}=\tilde{\alpha}^{-1} > = Q_n & (n \in 2\mathbb{Z}) \\
< \tilde{\alpha} | \tilde{\alpha}^n=e > = \mathbb{Z}_n & (n \in 2\mathbb{Z}+1)
\end{array}
\right., \\
&G/G_0 \simeq \mathbb{Z}_N = \left\{
\begin{array}{ll}
\mathbb{Z}_n = < g_{\alpha} | g_{\alpha}^n=g_e > & (n \in 2\mathbb{Z}) \\
\mathbb{Z}_n \times \mathbb{Z}_2 \simeq \mathbb{Z}_{2n} = <g_{\tilde{\gamma}} | g_{\tilde{\gamma}}^{2n}=g_e > & (n \in 2\mathbb{Z}+1)
\end{array}
\right.,
\end{align}
where $\tilde{\beta} \equiv \alpha^{-n/2}\beta$, $\tilde{\gamma} \equiv \alpha^{-(n-1)/2}\beta$, $g_X \equiv G_0X$.
We also introduce $\tilde{\alpha}' \equiv \tilde{\gamma}^2 = \tilde{\alpha}^{-(n-1)/2}\alpha'$ and $\tilde{\beta}' \equiv \tilde{\gamma}^n = \tilde{\alpha}^{-(n^2-1)/4}\beta$, and then
since $\alpha^n=\tilde{\gamma}^{2n}(=\tilde{\alpha}^{\prime n}=\tilde{\beta}^{\prime 2})=e$,
$G=\Sigma(2n^2)$ can be written as
\begin{align}
\begin{array}{rl}
\Sigma(2n^2)=G
&\simeq G_0 \rtimes \mathbb{Z}_N \\
&\simeq \left\{
\begin{array}{ll}
Q_n \rtimes \mathbb{Z}_n & (n \in 2\mathbb{Z}) \\
\mathbb{Z}_n \rtimes \mathbb{Z}_{2n} & (n \in 2\mathbb{Z}+1)
\end{array}
\right. \\
&\simeq \left\{
\begin{array}{ll}
Q_n \rtimes \mathbb{Z}_n & (n \in 2\mathbb{Z}) \\
(\mathbb{Z}_n \rtimes \mathbb{Z}_{2}) \times \mathbb{Z}_n \simeq D_n \times \mathbb{Z}_n & (n \in 2\mathbb{Z}+1)
\end{array}
\right.,
\end{array}
\end{align}
where the semidirect products for $n \in 2\mathbb{Z}$ case and $n \in 2\mathbb{Z}+1$ case come from the relations, $\alpha \tilde{\beta} \alpha^{-1} = \tilde{\alpha}\tilde{\beta}$, and $\tilde{\beta}' \tilde{\alpha} \tilde{\beta}^{\prime -1} = \tilde{\alpha}^{-1}$, $\tilde{\alpha}' \tilde{\alpha} \tilde{\alpha}^{\prime -1} = \tilde{\alpha}$, $\tilde{\alpha}' \tilde{\beta}' \tilde{\alpha}^{\prime -1} = \tilde{\beta}'$, respectively.

This analysis can be easily applied to $G=\Sigma(3n^3) \simeq (\mathbb{Z}_n \times \mathbb{Z}'_n \times \mathbb{Z}''_n) \rtimes \mathbb{Z}_3$.
The algebraic relations for generators $\alpha \in \mathbb{Z}_n$, $\alpha' \in \mathbb{Z}'_n$, $\alpha'' \in \mathbb{Z}''_n$, and $\beta \in \mathbb{Z}_3$ are given by
\begin{align}
\begin{array}{l}
\alpha^n = \alpha^{\prime n} = \alpha^{\prime\prime n} = \beta^3 = e, \\
\alpha\alpha' = \alpha'\alpha, \ \alpha'\alpha'' = \alpha''\alpha', \ \alpha''\alpha = \alpha\alpha'', \\
\beta \alpha \beta^{-1} = \alpha', \ \beta \alpha' \beta^{-1} = \alpha'', \ \beta \alpha'' \beta^{-1} = \alpha.
\end{array}
\end{align}
Then, we can find that
\begin{align}
&D(G) = < \tilde{\alpha}, \tilde{\alpha}' | \tilde{\alpha}^n=\tilde{\alpha}^{\prime n}=e, \tilde{\alpha}\tilde{\alpha}'=\tilde{\alpha}'\tilde{\alpha} > = \mathbb{Z}_n \times \mathbb{Z}_n, \label{eq:D(G)ZnZnZnZ3} \\
&G/D(G) \simeq \mathbb{Z}_n \times \mathbb{Z}_3 = < (d_{\alpha}, d_{\beta}) | d_{\alpha}^n=d_{\beta}^3=d_e, d_{\alpha}d_{\beta}=d_{\beta}d_{\alpha}=d_{\alpha\beta} >, \label{eq:GD(G)ZnZnZnZ3} 
\end{align}
where $\tilde{\alpha} \equiv \alpha'\alpha^{\prime\prime -1}$, $\tilde{\alpha}' \equiv \alpha\alpha^{\prime -1}$, $d_X \equiv D(G)X$.
In the class (iii), we can obtain
\begin{align}
&G_0 = \left\{
\begin{array}{ll}
< \tilde{\alpha}, \tilde{\alpha}', \tilde{\beta} | \tilde{\beta}^3=(\tilde{\alpha}\tilde{\alpha}^{\prime -1})^{n/3}, \tilde{\beta}\tilde{\alpha}\tilde{\beta}^{-1}=\tilde{\alpha}^{-1}\tilde{\alpha}^{\prime -1}, \tilde{\beta}\tilde{\alpha}'\tilde{\beta}^{-1}=\tilde{\alpha} > \equiv
R_n & (n \in 3\mathbb{Z}) \\
< \tilde{\alpha}, \tilde{\alpha}' > =
\mathbb{Z}_n \times \mathbb{Z}_n & ({\rm otherwise})
\end{array}
\right., \\
&G/G_0 \simeq \mathbb{Z}_N = \left\{
\begin{array}{ll}
\mathbb{Z}_n = < g_{\alpha} | g_{\alpha}^n=g_e > & (n \in 3\mathbb{Z}) \\
\mathbb{Z}_n \times \mathbb{Z}_3 \simeq \mathbb{Z}_{3n} = <g_{\tilde{\gamma}} | g_{\tilde{\gamma}}^{2n}=g_e > & ({\rm otherwise})
\end{array}
\right.,
\end{align}
where $\tilde{\beta} \equiv \alpha^{-n/3}\beta$, $\tilde{\gamma} \equiv \alpha^{-(n \mp 1)/3}\beta$, $g_X=G_0X$.
Here, since the relations between $\tilde{\alpha}$ and $\tilde{\alpha}'$ are same as Eq.~(\ref{eq:D(G)ZnZnZnZ3}), we omit them.
We also introduce $\tilde{\alpha}'' \equiv \tilde{\gamma}^3$ and $\tilde{\beta}' \equiv \tilde{\gamma}^n$,
and then
since $\alpha^n=\tilde{\gamma}^{3n}(=\tilde{\alpha}^{\prime\prime n}=\tilde{\beta}^{\prime 3})=e$,
$G=\Sigma(3n^3)$ can be written as
\begin{align}
\begin{array}{rl}
\Sigma(3n^3)=G
&\simeq G_0 \rtimes \mathbb{Z}_N \\
&\simeq \left\{
\begin{array}{ll}
R_n \rtimes \mathbb{Z}_n & (n \in 3\mathbb{Z}) \\
(\mathbb{Z}_n \times \mathbb{Z}_n) \rtimes Z_{3n} & ({\rm otherwise})
\end{array}
\right. \\
&\simeq \left\{
\begin{array}{ll}
R_n \rtimes \mathbb{Z}_n & (n \in 3\mathbb{Z}) \\
((\mathbb{Z}_n \times \mathbb{Z}_n) \rtimes \mathbb{Z}_{3}) \times \mathbb{Z}_n \simeq \Delta(3n^2) \times \mathbb{Z}_n & ({\rm otherwise})
\end{array}
\right..
\end{array}
\end{align}
where 
$R_n$ is
related to the following $\Delta(3n^2)$ as with the case that $Q_n$ is related to $D_n$.
The semidirect products for $n \in 3\mathbb{Z}$ case and $n \in 3\mathbb{Z} \pm 1$ cases come from the relations,
$\alpha \tilde{\beta} \alpha^{-1} = \tilde{\alpha}'\tilde{\beta}$, and $\tilde{\beta}'\tilde{\alpha}\tilde{\beta}^{\prime -1}=\tilde{\alpha}^{-1}\tilde{\alpha}^{\prime -1}$, $\tilde{\beta}'\tilde{\alpha}'\tilde{\beta}^{\prime -1}=\tilde{\alpha}$, $\tilde{\alpha}'' \delta \tilde{\alpha}^{\prime\prime -1} = \delta \ (\delta = \tilde{\alpha}, \tilde{\alpha}', \tilde{\beta}')$, respectively.

Next, let us see another example, $G=\Delta(3n^2) \simeq (\mathbb{Z}_n \times \mathbb{Z}'_n) \rtimes \mathbb{Z}_3$.
The algebraic relations for generators $\alpha \in \mathbb{Z}_n$, $\alpha' \in \mathbb{Z}'_n$, and $\beta \in \mathbb{Z}_3$ are given by
\begin{align}
\alpha^n = \alpha^{\prime n} = \beta^3 = e,
\end{align}
and also
\begin{align}
&\alpha' \alpha \alpha^{\prime -1} = \alpha \in \mathbb{Z}_n, \quad (\alpha \alpha' \alpha^{-1} = \alpha' \in Z'_n) \\
&\beta \alpha \beta^{-1} = \alpha^{m_1} \alpha^{\prime m_2} \in \mathbb{Z}_n \times \mathbb{Z}'_n, \quad \beta \alpha^{-1} \beta^{-1} = \alpha^{m_3} \alpha^{\prime m_4} \in \mathbb{Z}_n \times \mathbb{Z}'_n. \label{eq:semidircond1ZnZnZ3}
\end{align}
Here, $m_i\ (i=1,2,3,4)$ can be determined  by the constraints, $\beta^3 \alpha \beta^{-3}=\alpha$, $\beta^3 \alpha' \beta^{-3}=\alpha'$,
and then  we obtain $m_1=m_2=-1,\ m_3=1,\ m_4=0$.
Namely, Eq.~(\ref{eq:semidircond1ZnZnZ3}) can be rewritten as
\begin{align}
\beta \alpha \beta^{-1} = \alpha^{-1} \alpha^{\prime -1}, \quad \beta \alpha' \beta^{-1} = \alpha. \label{eq:semidircond2ZnZnZ3}
\end{align}
Hence, we can obtain
\begin{align}
&D(G) = \left\{
\begin{array}{ll}
< \tilde{\alpha}, \tilde{\alpha}' | \tilde{\alpha}^n=\tilde{\alpha}^{\prime n/3}=e, \tilde{\alpha}\tilde{\alpha}'=\tilde{\alpha}'\tilde{\alpha} > = \mathbb{Z}_n \times \mathbb{Z}_{n/3} & (n \in 3\mathbb{Z}) \\
< \tilde{\alpha}, \tilde{\alpha}' | \tilde{\alpha}^n=\tilde{\alpha}^{\prime n}=e, \tilde{\alpha}\tilde{\alpha}'=\tilde{\alpha}'\tilde{\alpha} > = \mathbb{Z}_n \times \mathbb{Z}_n & ({\rm otherwise})
\end{array}
\right., \label{eq:D(G)ZnZnZ3} \\
&G/D(G) \simeq \left\{
\begin{array}{ll}
\mathbb{Z}_3 \times \mathbb{Z}_3 = < (d_{\alpha}, d_{\beta}) | d_{\alpha}^3=d_{\beta}^3=d_e, d_{\alpha}d_{\beta}=d_{\beta}d_{\alpha}=d_{\alpha\beta} > & (n \in 3\mathbb{Z}) \\
\mathbb{Z}_3 = < d_{\beta} | d_{\beta}^3=d_e > & ({\rm otherwise})
\end{array}
\right., \label{eq:GD(G)ZnZnZ3}
\end{align}
where $\tilde{\alpha} \equiv \alpha \alpha^{\prime -1}$, $\tilde{\alpha}' \equiv \alpha^{-3}$, $d_X \equiv D(G)X$.

The above result can also be understood from the viewpoint of $\mathbb{Z}_n$ and $\mathbb{Z}'_n$ charge constraints.
Suppose that the chiral fermions are $\mathbb{Z}_n$ and $\mathbb{Z}'_n$ eigenstates, where we denote the $[\mathbb{Z}_n, \mathbb{Z}'_n]$ charges of the $j$ th component field as $[q_j, q'_j]$.
Eq.~(\ref{eq:semidircond2ZnZnZ3}) means that there exists a state with charge $[-(q_j+q'_j), q_j]$ and 
$\beta$ transforms the $j$ th component with charge $[q_j, q'_j]$ to the sate with charge $[-(q_j+q'_j), q_j]$.
For example, when we consider a fundamental irreducible representation, it forms a triplet with charge $ ^t([q_1, q'_1], [q_2, q'_2], [q_3, q'_3])$ $=$ $ ^t([q, q'], [-(q+q'), q], [q', -(q+q')])$. Then, we obtain
\begin{align}
{\rm det}\rho(\alpha) = {\rm det}\rho(\alpha') = 1, \label{eq:aa'}
\end{align}
and we can actually check
Eqs.~(\ref{eq:D(G)ZnZnZ3}) and (\ref{eq:GD(G)ZnZnZ3}).
We note that for the triplet, Eq.~(\ref{eq:aa'}) and also ${\rm det}\rho(\beta)=1$ are satisfied, even if $n \in 3\mathbb{Z}$.

In the class (iii), we can obtain
\begin{align}
&G_0 = \left\{
\begin{array}{ll}
< \tilde{\alpha}, \tilde{\alpha}', \tilde{\beta} | \tilde{\beta}^3=e, \tilde{\beta}\tilde{\alpha}\tilde{\beta}^{-1}=\tilde{\alpha}\tilde{\alpha}', \tilde{\beta}\tilde{\alpha}'\tilde{\beta}^{-1}=\tilde{\alpha}^{-3}\tilde{\alpha}^{\prime -2} > = (\mathbb{Z}_n \times \mathbb{Z}_{n/3}) \rtimes \mathbb{Z}_3 & (n \in 3\mathbb{Z}) \\
< \tilde{\alpha}, \tilde{\alpha}' > = \mathbb{Z}_n \times \mathbb{Z}_n & ({\rm otherwise})
\end{array}
\right., \\
&G/G_0 \simeq \mathbb{Z}_N = \mathbb{Z}_3 = < g_{\beta} | g_{\beta}^3=g_e >, \quad \forall n,
\end{align}
where $\tilde{\beta} \equiv \alpha^{-1}\beta$, $g_{\beta} \equiv G_0\beta$.
Here, since the relations between $\tilde{\alpha}$ and $\tilde{\alpha}'$ are same as Eq.~(\ref{eq:D(G)ZnZnZ3}), we omit them.
Then, since $\beta^3=\tilde{\beta}^3=e$,
$G=\Delta(3n^2)$ can be written as
\begin{align}
\begin{array}{rl}
\Delta(3n^2)=G
&\simeq G_0 \rtimes \mathbb{Z}_3 \\
&\simeq \left\{
\begin{array}{ll}
(D(\Delta(3n^2)) \rtimes \mathbb{Z}_3) \rtimes \mathbb{Z}_3 &  (n \in 3\mathbb{Z}) \\
D(\Delta(3n^2)) \rtimes \mathbb{Z}_3 & ({\rm otherwise})
\end{array}
\right. \\
&\simeq \left\{
\begin{array}{ll}
((\mathbb{Z}_n \times \mathbb{Z}_{n/3}) \rtimes \mathbb{Z}_3) \rtimes \mathbb{Z}_3 &  (n \in 3\mathbb{Z}) \\
(\mathbb{Z}_n \times \mathbb{Z}_n) \rtimes \mathbb{Z}_3 & ({\rm otherwise})
\end{array}
\right.,
\end{array}
\end{align}
where the last $\mathbb{Z}_3$ semidirect product for $n \in 3\mathbb{Z}$ comes from the relations,
$\beta\tilde{\alpha}\beta^{-1}=\tilde{\alpha}\tilde{\alpha}'$, $\beta\tilde{\alpha}'\beta^{-1}=\tilde{\alpha}^{-3}\tilde{\alpha}^{\prime -2}$, $\beta\tilde{\beta} \beta^{-1} = \tilde{\alpha}^{-1}\tilde{\alpha}^{\prime -1}\tilde{\beta}$.
One of the examples is $G=A_4 \simeq \Delta(12)$.
In this case,
the $A_4$ flavor model can be classified as either the class (iii), $G_0=D(G)=\mathbb{Z}_n \times \mathbb{Z}'_n$, $G/G_0 \simeq \mathbb{Z}_N=\mathbb{Z}_3$, or the class (i), $G_0=G=A_4$.
Indeed, the $A_4$ symmetry has four irreducible representations, ${\bf 1}\ ({\rm det}\rho_{\bf 1}(\beta)=1)$, ${\bf 1'}\ ({\rm det}\rho_{\bf 1'}(\beta)=e^{2\pi i/3})$, ${\bf 1''}\ ({\rm det}\rho_{\bf 1''}(\beta)=e^{4\pi i/3})$, and ${\bf 3}\ ({\rm det}\rho_{\bf 3}(\beta)=1)$~\cite{Ishimori:2010au,Ishimori:2012zz}.
Then,
the whole $A_4$ symmetry is anomaly free in flavor models in which proper numbers of ${\bf 1'}$ and ${\bf 1''}$ are included. Otherwise, the $\mathbb{Z}_3$ subsymmetry can be anomalous.
Note that, in the double covering group $T'$, there is no modification from $A_4$ case except the double covering.

\end{step}

\begin{step}[$G \simeq K_G \rtimes \mathbb{Z}_B$]
All of discrete groups discussed in the steps $1$ and $2$ are specific cases of $G \simeq K_G \rtimes \mathbb{Z}_B$ type.
Now, we study a more generic case, $G \simeq K_G \rtimes \mathbb{Z}_B$.
Since $G/K_G \simeq \mathbb{Z}_B$, we find that $D(G) \subseteq K_G$.
Then, we can find that
\begin{align}
G/D(G) \simeq (K_G/D(G)) \times \mathbb{Z}_B, \label{eq:GD(G)KGZB}
\end{align}
from the following proof.
\begin{pf}
We can prove Eq.~(\ref{eq:GD(G)KGZB}).
\begin{enumerate}
\renewcommand{\labelenumi}{(\Roman{enumi})}
\item From the isomorphism theorem $3$ in subsection~\ref{app:isomorphism},
we obtain
\begin{align}
\mathbb{Z}_B \simeq G/K_G \simeq (G/D(G))/(K_G/D(G)).
\end{align}
\item Since $G \simeq K_G \rtimes \mathbb{Z}_B$, $\mathbb{Z}_B$ is a subgroup of $G$ and 
the relation $K_G \cap \mathbb{Z}_B = \{e\}$ is satisfied.
Then, the relation $(K_G/D(G)) \cap \mathbb{Z}_B = \{e\}$ is also satisfied.
Hence, we can write $G/D(G) \simeq (K_G/D(G)) \rtimes \mathbb{Z}_B$.
\item In particular, since $G/D(G)$ is Abelian, $G/D(G) \simeq (K_G/D(G)) \times \mathbb{Z}_B$.
\end{enumerate}
\end{pf}
In the class (iii), the order $N$ of $\mathbb{Z}_N \simeq G/G_0$ can be calculated as the least common multiple of all orders of cyclic groups in $G/D(G)$. 

One of the example which is not mentioned in the steps $1$ and $2$ is $G=S_n \simeq A_n \rtimes \mathbb{Z}_2\ (n \geq 5)$.
Then, it should be satisfied that $D(G) \subseteq A_n$.
On the other hand,
since $A_n\ (n \geq 5)$ is a perfect group, $A_n$ should be included in $D(G)$, $A_n \subseteq D(G)$.
Hence, we find that $D(G)=A_n$.
In this case,
the $S_n$ flavor model can be classified as either the class (iii), $G_0=D(G)=A_n$, $G/G_0 \simeq \mathbb{Z}_N=\mathbb{Z}_2$, or the class (i), $G_0=G=S_n$.
%
\end{step}

\begin{step}[General $G \simeq K_G \rtimes G^{(1)}$]
Finally, we study general case $G \simeq K_G \rtimes G^{(1)}$.
We note that the case that $G^{(1)}$ is Abelian comes down to the step $3$ by applying the fundamental structure theorem of finite Abelian group in subsection~\ref{app:Abelian} and Eqs.~(\ref{eq:decomposiG1})-(\ref{eq:decomposiK}) in subsection~\ref{app:semidirect}.
In this case, we can generally obtain
\begin{align}
G/D(G) \simeq (K_G/D(G)) \times G^{(1)}({\rm Abelian}).
\end{align}

Now, let us consider the case that $G^{(1)}$ is non-Abelian.
Since $G/K_G \simeq G^{(1)}$ and $G^{(1)} \triangleright D(G^{(1)}) \neq \{e\}$, by applying the correspondence theorem in subsection~\ref{app:isomorphism}, we find that $K_G \subset D(G) \subseteq G_0$.
Here, similar to $G_0$, we define $G_0^{(1)} \equiv \{ g_0^{(1)} \in G^{(1)} | {\rm det}\rho(g_0^{(1)})=1 \}$, and then 
we find $D(G^{(1)}) \subseteq G_0^{(1)}$.
By applying the isomorphism theorem $3$ in subsection~\ref{app:isomorphism}, we can find that
\begin{align}
\begin{array}{rl}
\mathbb{Z}_N
&\simeq G/G_0 \\
&\simeq (G/D(G))/(G_0/D(G)) \\
&\simeq [(G/K_G)/(D(G)/K_G)]/[(G_0/K_G)/(D(G)/K_G)] \\
&\simeq (G^{(1)}/D(G^{(1)}))/(G_0^{(1)}/D(G^{(1)})) \\
&\simeq G^{(1)}/G_0^{(1)}.
\end{array}
\end{align}
Therefore, in this case, the structures of $D(G)$, $G/D(G)$, $G_0$, and $G/G_0$ depend on $D(G^{(1)})$, $G^{(1)}/D(G^{(1)})$, $G_0^{(1)}$, and $G^{(1)}/G_0^{(1)}$, respectively.

For example, let us see $G=\Delta(6n^2)$,
\begin{align}
\begin{array}{rl}
\Delta(6n^2)
&\simeq (\mathbb{Z}_n \times \mathbb{Z}'_n) \rtimes S_3
\simeq (\mathbb{Z}_n \times \mathbb{Z}'_n) \rtimes (\mathbb{Z}_3 \rtimes \mathbb{Z}_2) \\
&\simeq ((\mathbb{Z}_n \times \mathbb{Z}'_n) \rtimes \mathbb{Z}_3) \rtimes \mathbb{Z}_2
\simeq \Delta(3n^2) \rtimes \mathbb{Z}_2,
\end{array}
\end{align}
where we use Eqs.~(\ref{eq:decomposiG1}) and (\ref{eq:decomposiK}).
Then, we can find that $\mathbb{Z}_n \times \mathbb{Z}'_n \subset G_0$ and $\mathbb{Z}_N \simeq S_3/G_0^{(1)}$.
In addition, since $S_3 \simeq A_3 \rtimes \mathbb{Z}_2 \simeq \mathbb{Z}_3 \rtimes \mathbb{Z}_2$, we obtain $G_0^{(1)} \supseteq A_3 \simeq \mathbb{Z}_3$.
Thus, in this case,
$\Delta(6n^2)$ flavor model can be classified as either the class (iii), $G_0=D(G)=\Delta(3n^2)$, $G/G_0 \simeq \mathbb{Z}_N=\mathbb{Z}_2$, or the class (i), $G_0=G=\Delta(6n^2)$.
One of the examples is $G = S_4 \simeq \Delta(24) \simeq \Delta(12) \rtimes \mathbb{Z}_2 \simeq A_4 \rtimes \mathbb{Z}_2$.
Indeed, the $S_4$ group has five irreducible representations, ${\bf 1}\ ({\rm det}\rho_{\bf 1}(\beta)=1)$, ${\bf 1'}\ ({\rm det}\rho_{\bf 1'}(\beta)=-1)$, ${\bf 2}\ ({\rm det}\rho_{\bf 2}(\beta)=-1)$, ${\bf 3}\ ({\rm det}\rho_{\bf 3}(\beta)=-1)$, and ${\bf 3'}\ ({\rm det}\rho_{\bf 3'}(\beta)=1)$~\cite{Ishimori:2010au,Ishimori:2012zz}.
Then,
the whole $S_4$ symmetry is anomaly free in flavor models in which even numbers of ${\bf 1'}$, ${\bf 2}$, and ${\bf 3}$ are included.
Otherwise, the $\mathbb{Z}_2$ subsymmetry can be anomalous.
\end{step}

So far, we have seen the detailed structure of the anomaly-free subgroup $G_0$ and the anomalous part $G/G_0 \simeq \mathbb{Z}_N$ for various typical discrete groups $G$ from the structure of the derived subgroup $D(G)$ and its residue class group $G/D(G)$.
Here, we list $D(G)$ and $G/D(G)$ for those typical discrete groups $G$ in Table~\ref{tab:D(G)GD(G)}.
We note again that no matter what representations of $G$ our model has, the derived subgroup $D(G)$ automatically becomes the subgroup of the anomaly-free group $G_0$.
Then,
in the case of $G=S_n \simeq A_n \rtimes \mathbb{Z}_2$ and $G=\Delta(6n^2) \simeq \Delta(3n^2) \rtimes \mathbb{Z}_2$, in particular, we can find that $G_0 \supseteq A_n$ and $G_0 \supseteq \Delta(3n^2)$ at least, respectively.
\begin{table}[H]
\centering
\begin{tabular}{|c|c|c|} \hline
$G$ & $D(G) (\subseteq G_0)$ & $G/D(G)$ \\ \hline \hline
$\mathbb{Z}_p$ & $\{e\}$ & $\mathbb{Z}_p$ \\
$(A_3 \simeq \mathbb{Z}_3)$ & $(\{e\}$) & $(\mathbb{Z}_3 \simeq A_3)$ \\
\hline
$D_n \simeq \mathbb{Z}_n \rtimes \mathbb{Z}_2$ &
$\left\{
\begin{array}{cc}
\mathbb{Z}_{n/2} & (n \in 2\mathbb{Z}) \\
\mathbb{Z}_n & (n \in 2\mathbb{Z}+1)
\end{array}
\right.$ &
$\left\{
\begin{array}{cc}
\mathbb{Z}_2 \times \mathbb{Z}_2 & (n \in 2\mathbb{Z}) \\
\mathbb{Z}_2 & (n \in 2\mathbb{Z}+1)
\end{array}
\right.$ \\
$(S_3 \simeq D_3 \simeq A_3 \rtimes \mathbb{Z}_2)$ & $(\mathbb{Z}_3 \simeq A_3)$ & $(\mathbb{Z}_2)$ \\
\hline
$T_{p^k} \simeq \mathbb{Z}_{p^k} \rtimes \mathbb{Z}_3\ (p \neq 3)$ & $\mathbb{Z}_{p^k}$ & $\mathbb{Z}_3$ \\
\hline
$\Sigma(2n^2) \simeq (\mathbb{Z}_n \times \mathbb{Z}_n) \rtimes \mathbb{Z}_2$ & $\mathbb{Z}_n$ & $\mathbb{Z}_n \times \mathbb{Z}_2$ \\
\hline
$\Sigma(3n^3) \simeq (\mathbb{Z}_n \times \mathbb{Z}_n \times \mathbb{Z}_n) \rtimes \mathbb{Z}_3$ & $\mathbb{Z}_n \times \mathbb{Z}_n$ & $\mathbb{Z}_n \times \mathbb{Z}_3$ \\
\hline
$\Delta(3n^2) \simeq (\mathbb{Z}_n \times \mathbb{Z}_n) \rtimes \mathbb{Z}_3$ & 
$\left\{
\begin{array}{cc}
\mathbb{Z}_n \times \mathbb{Z}_{n/3} & (n \in 3\mathbb{Z}) \\
\mathbb{Z}_n \times \mathbb{Z}_n & ({\rm otherwise})
\end{array}
\right.$ &
$\left\{
\begin{array}{cc}
\mathbb{Z}_3 \times \mathbb{Z}_3 & (n \in 3\mathbb{Z}) \\
\mathbb{Z}_3 & ({\rm otherwise})
\end{array}
\right.$ \\
$(A_4 \simeq \Delta(12))$ & $(\mathbb{Z}_2 \times \mathbb{Z}_2)$ & $(\mathbb{Z}_3)$ \\
\hline
$\begin{array}{rl}
\Delta(6n^2)
&\simeq (\mathbb{Z}_n \times \mathbb{Z}_n) \rtimes S_3 \\
&\simeq (\mathbb{Z}_n \times \mathbb{Z}_n) \rtimes (\mathbb{Z}_3 \rtimes \mathbb{Z}_2) \\
&\simeq ((\mathbb{Z}_n \times \mathbb{Z}_n) \rtimes \mathbb{Z}_3) \rtimes \mathbb{Z}_2 \\
&\simeq \Delta(3n^2) \rtimes \mathbb{Z}_2
\end{array}$ & $\Delta(3n^2)$ & $\mathbb{Z}_2$ \\
$\begin{array}{rl}
(S_4
&\simeq \Delta(24) \\
&\simeq \Delta(12) \rtimes \mathbb{Z}_2 \\
&\simeq A_4 \rtimes \mathbb{Z}_2)
\end{array}$
& $(\Delta(12) \simeq A_4)$ & $(\mathbb{Z}_2)$ \\
\hline
$PSL(2,\mathbb{Z}_p)\ (p \neq 2,3)$ & $PSL(2,\mathbb{Z}_p)\ (p \neq 2,3)$ & - \\
\hline
$A_n\ (n \geq 5)$ & $A_n\ (n \geq 5)$ & - \\
\hline
$S_n \simeq A_n \rtimes \mathbb{Z}_2\ (n \geq 5)$ & $A_n\ (n \geq 5)$ & $\mathbb{Z}_2$ \\
\hline
\end{tabular}
\caption{The derived subgroup $D(G)$ and its residue class group $G/D(G)$ for various typical discrete groups $G$.}
\label{tab:D(G)GD(G)}
\end{table}

Finally, we summarize the important points in this section again.
\begin{itemize}
\item
The derived subgroup of $G$, $D(G)$, defined in Eq.~(\ref{eq:D(G)}), is always included in $G_0$,~i.e., $D(G) \subseteq G_0\ (D(G) \triangleleft G_0)$, even if we have any representations of $G$.
In addition,
$G/D(G)$ becomes Abelian as in Eq.~(\ref{eq:generalGD(G)}) and each cyclic groups can be anomalous, while $G/G_0 \simeq \mathbb{Z}_N$.
The order $N$ is given by the least common multiple of orders of the anomalous cyclic subgroups (which is a divisor of the least common multiple of orders of all cyclic subgroups of $G/D(G)$).
Therefore, $D(G)$ 
is important to explore the structure of  the anomaly-free subgroup $G_0$ and the anomalous part $G/G_0 \simeq \mathbb{Z}_N$.
\item
The detailed structure of $D(G)$ depends on the structure of $G$.
In particular,
when $G$ can be written by semidirect product,~i.e. $G \simeq K_G \rtimes G^{(1)}$, we can obtain some information as discussed in the above step $4$;
we can obtain
$D(G) \subseteq K_G$ and $G/D(G) \simeq (K_G/D(G)) \times G^{(1)}$ in the case that $G^{(1)}$ is Abelian,
while we can obtain
$K_G \subset D(G) \subseteq G_0$ and $G/D(G) \simeq G^{(1)}/D(G^{(1)})$ in the case that $G^{(1)}$ is non-Abelian.
For example, in the case of
$G=S_n \simeq A_n \rtimes \mathbb{Z}_2$ and $G=\Delta(6n^2) \simeq \Delta(3n^2) \rtimes \mathbb{Z}_2$, in particular,
we find that $G_0 \supseteq A_n$ and $G_0 \supseteq \Delta(3n^2)$ at least.
\end{itemize}


\section{Comment on generic theories with $P>1$}
\label{sec:comment}

Here, we extend our analysis to the theory with $\sum_{\bf R}2T_2({\bf R})=P > 1$ and 
anomaly-free condition $({\rm det}\rho (g))^P =1$.

Suppose that the representation of any element $\forall g \in G$ satisfies $({\rm det}\rho(g))^N=1$.
In this case, the representation of the anomaly-free element $g_n$, which satisfies $({\rm det}\rho (g_n))^P =1$, also satisfies $({\rm det}\rho(g_n))^n=1$ with $n={\rm gcd}(N,P)$.
Namely, the determinant can be written as ${\rm det}\rho(g_n)=e^{2\pi iQ''(g_n)/n}$.
Here, we define the subset of $G$,
\begin{align}
G_n \equiv \{ g_n \in G | {\rm det}\rho(g_n) = e^{2\pi i Q'(g_n)/N} = e^{2\pi i Q''(g_n)/n} \}, 
\end{align}
where $Q'(g_n)(=Q(g_n)N/N(g_n))=Q''(g_n)N/n$.
Similar to $G_0$, we find that 
$G_n$ is also a normal subgroup of $G$, $G_n \triangleleft G$.
We note that $G_0$ is included as the normal subgroup of $G_n$, $G_0 \triangleleft G_n$.
Then, we can similarly find that $G/G_n \simeq \mathbb{Z}_{N/n}$ and $G_n/G_0 \simeq \mathbb{Z}_n$. Indeed, by applying the isomorphism theorem $3$ in subsection~\ref{app:isomorphism}, we obtain
\begin{align}
G/G_n \simeq (G/G_0)/(G_n/G_0) \simeq \mathbb{Z}_N/\mathbb{Z}_n \simeq \mathbb{Z}_{N/n}.
\end{align}
Therefore, in the theory, the subgroup $G_n$ is anomaly free while $G/G_n \simeq \mathbb{Z}_{N/n}$ can be anomalous.
We notice that the anomalous symmetry becomes the single cyclic group.
Furthermore, if there exists $\exists g \in G$ with $N(g)=N/n$ and ${\rm gcd}(Q(g),N(g))=1$, $G$ can be decomposed as
\begin{align}
G \simeq G_n \rtimes \mathbb{Z}_{N/n}.
\end{align}
It means that the anomaly-free and anomalous parts of $G$ can be separated.

There is an interesting example.
When $P$ is a multiple of $N$, the whole $G$ symmetry is anomaly free.
As the example, let us suppose that our particle theory has $E_6$ gauge symmetry\footnote{$E_6$ gauge symmetry is automatically anomaly free.} and all chiral fermions in the theory transform as ${\bf 27}^i$ representation under $E_6$ transformation, where $i$ denotes the flavor index.
(For example, the $i$ th generational standard model quarks and leptons are embedded in ${\bf 27}^i$ representation.)
Furthermore, we also suppose discrete $G$ flavor symmetry for those chiral fermions at least at classical level.
In this case, we obtain $P=2T_2({\bf 27})=6$.
Then, from Eq.~(\ref{eq:GG0iii}), when $G$ corresponds to either of the groups listed in Table~\ref{tab:D(G)GD(G)} except $\Sigma(2n^2)$ and $\Sigma(3n^2)$, at least, we can find that whole $G$ flavor symmetry can be automatically anomaly free no matter what representations of $G$ the fermions have.


\section{Property of group theory}
\label{ap:propGT}


\subsection{Isomorphism Theorems}
\label{app:isomorphism}

We give the fundamental homomorphism and then the isomorphism theorems as well as the correspondence theorem.
\begin{thmhom}
Let $K$ be a normal subgroup of $G$, $K \triangleleft G$. Then there is a natural homomorphism $\pi: G \rightarrow G/K$.
In addition,
let $f: G \rightarrow G'$ be a group homomorphism. Then, ${\rm Ker}(f) \triangleleft G$ and we can consider $G'={\rm Im}(f)$ without loss of generality.
If $K$ is a subset of ${\rm Ker}(f)$, $K \subseteq {\rm Ker}(f)$, there exists a unique homomorphism $F: G/K \rightarrow G'$ such that $F \circ \pi = f$.
\end{thmhom}
\begin{thmiso}
In particular, if $K={\rm Ker}(f)$, $F: G/K \rightarrow G'(={\rm Im}(f))$ becomes a isomorphism;
\begin{align}
G/{\rm Ker}(f) \simeq {\rm Im}(f).
\end{align}
This is always satisfied whenever we consider $f: G \rightarrow G'$.
\end{thmiso}
\begin{thmcor}
It can be applied even if $K \subset {\rm Ker}(f)$.
In this case, we obtain $G/K \triangleright {\rm Ker}(F) \neq \{e\}$ and then
\begin{align}
(G/K)/{\rm Ker}(F) \simeq G' \simeq G/{\rm Ker}(f).
\end{align}
Here, $K \subset {\rm Ker}(f) = {\rm Ker}(F \circ \pi) = \pi^{-1}({\rm Ker}(F)) \triangleleft G$.
Furthermore, there exists a group homomorphism $\phi: G \rightarrow G''$ such that ${\rm Ker}(\phi)=K$, and then $G'' \simeq G/K$.
Accordingly, there exists $\tilde{K}'' \triangleleft G''$ such that $\tilde{K}'' \simeq {\rm Ker}(F)$.
Thus, for $\phi: G \rightarrow G''$, we obtain
\begin{align}
G/\tilde{K} \simeq G''/\tilde{K}'',
\end{align}
in general, where $\tilde{K}'' \triangleleft G''$ and $\tilde{K} \equiv \phi^{-1}(\tilde{K}'') (\supset {\rm Ker}(\phi)) \triangleleft G$.
This is often called the corresponding theorem.
\end{thmcor}
\begin{thmiso}
Let $K$ be a normal subgroup of $G$, $K \triangleleft G$. Let $H$ be a subgroup of $G$, $H \subset G$.
In this case, $K$ is also a normal subgroup of $KH$. 
In particular, 
$H \cap K$ is also a normal subgroup of $H$.
When we consider $\phi: KH \rightarrow H$, we obtain $\phi(K)=H \cap K$ and then
\begin{align}
KH/K \simeq H/(H \cap K).
\end{align}
\end{thmiso}
\begin{thmiso}
Let both $K_1$ and $K_2$ be normal subgroups of $G$ with $K_1 \subset K_2$.
In this case, $K_1 \triangleleft K_2$ and $K_2/K_1 \triangleleft G/K_1$.
By applying the corresponding theorem, we obtain
\begin{align}
G/K_2 \simeq (G/K_1)/(K_2/K_1).
\end{align}
\end{thmiso}


\subsection{Semidirect Product}
\label{app:semidirect}

We comment on semidirect product.
If a normal subgroup of $G$, $K_G$, and a subgroup of $G$, $G^{(1)}$ satisfy the following conditions,
\begin{align}
G=K_G G^{(1)}, \quad K_G \cap G^{(1)} = \{e\}, \label{eq:condsemidirect}
\end{align}
$G$ can be written by $K_G$ and $G^{(1)}$ as
\begin{align}
G \simeq K_G \rtimes G^{(1)}.\label{eq:semidirect}
\end{align}
In particular, if any elements in $G^{(1)}$ commute all elements in $K_G$, $G$ can be written as $G \simeq K_G \times G^{(1)}$.
Applying the second isomorphism theorem in subsection~\ref{app:isomorphism}, we find that $G/K_G \simeq G^{(1)}$.
Note that $G$ cannot be always decomposed as $G \simeq K_G \rtimes G^{(1)}$ just because $G/K_G \simeq G^{(1)}$.
If this $G^{(1)}$ is actually a subgroup of $G$, Eq.~(\ref{eq:condsemidirect}) can be satisfied and then $G$ can be written as Eq.~(\ref{eq:semidirect}).
In terms of group elements in Eq.~(\ref{eq:semidirect}), since $K_G$ is the normal subgroup of $G$, the following relation,
\begin{align}
g_1 k g_1^{-1} = k^{(g_1)} \in K_G,
\end{align}
should be satisfied, where $k, k^{(g_1)} \in K_G$ and $g_1 \in G^{(1)}$.

Now, let us consider the case that $G^{(1)}$ can be further decomposed as $G^{(1)} \simeq K_{G^{(1)}} \rtimes G^{(2)}$.
In this case, $G$ can be written as
\begin{align}
&G \simeq K_G \rtimes G^{(1)} \simeq K_G \rtimes (K_{G^{(1)}} \rtimes G^{(2)}), \label{eq:decomposiG1} \\
\Rightarrow \quad
&G \simeq (K_G \rtimes K_{G^{(1)}}) \rtimes G^{(2)} \simeq K'_{G} \rtimes G^{(2)}. \label{eq:decomposiK}
\end{align}
This can be found, in the following, by considering relations among their elements:
$k$, $k^{(k_1)}$, $k^{(g_2)}$, $k^{(g_2 \rightarrow k_1)} \in K_G$, $k_1$, $k_1^{(g_2)}$, $k_1^{(g_2 \rightarrow k)} \in K_{G^{(1)}}$, and $g_2 \in G^{(2)}$.
In the case of Eq.~(\ref{eq:decomposiG1}),
\begin{align}
k_1 k k_1^{-1} = k^{(k_1)}, \quad g_2 k g_2^{-1} = k^{(g_2)}, \quad g_2 k_1 g_2^{-1} = k_1^{(g_2)}, \label{eq:relG1}
\end{align}
are satisfied, while in the case of Eq.~(\ref{eq:decomposiK}),
\begin{align}
k_1 k k_1^{-1} = k^{(k_1)}, \quad g_2 k g_2^{-1} = k^{(g_2)} k_1^{(g_2 \rightarrow k)}, \quad g_2 k_1 g_2^{-1} = k^{(g_2 \rightarrow k_1)} k_1^{(g_2)}, \label{eq:relK}
\end{align}
are satisfied.
Thus, Eq.~(\ref{eq:relG1}) is a sufficient condition for Eq.~(\ref{eq:relK}), but Eq.~(\ref{eq:relK}) does not always satisfy Eq.~(\ref{eq:relG1}).


\subsection{Finite Abelian Groups}
\label{app:Abelian}

We give several theorems of finite Abelian groups.
\begin{thmAbel}
Every finite Abelian group $G$ with the order $|G|=p_1^{A_1} \cdots p_r^{A_r}=\prod_{i=1}^{r} p_i^{A_i}$ can be written as
\begin{align}
G \simeq
(\mathbb{Z}_{p_1^{a_{1,1}}}
\times \cdots \times \mathbb{Z}_{p_1^{a_{1,n_1}}}) \times
\cdots \times
(\mathbb{Z}_{p_r^{a_{r,1}}}
\times \cdots \times \mathbb{Z}_{p_r^{a_{r,n_r}}}),
\end{align}
where each $p_i$ is a distinct prime number and $a_{i,j}$ satisfy
\begin{align}
A_i = \sum_{j=1}^{n_i} a_{i,j}, \quad a_{i,j} \geq a_{i,j+1}.
\end{align}
Note that $a_{i,j}$ is uniquely determined by $G$.
\end{thmAbel}
Here, in the above theorem, the following theorem is applied.
\begin{thmcirc}
When $m$ and $n$ are coprime to each other, it is satisfied that
\begin{align}
\mathbb{Z}_{mn} \simeq \mathbb{Z}_{m} \times \mathbb{Z}_{n}. \label{eq:mn}
\end{align}
\end{thmcirc}
Note that, by using this theorem, the following relation,
\begin{align}
\mathbb{Z}_m \times \mathbb{Z}_n \simeq \mathbb{Z}_{{\rm gcd}(m,n)} \times \mathbb{Z}_{{\rm lcm}(m,n)}, \label{eq:gcdlcm}
\end{align}
is generally obtained.


\end{document}